\documentclass[a4paper,11pt,reqno]{amsart}
\usepackage{fullpage} 
\usepackage[utf8]{inputenc}
\usepackage{mathrsfs}
\usepackage{dsfont}
\usepackage{hyperref}
\usepackage{amsmath}
\usepackage{amssymb}
\usepackage{amsthm}
\usepackage{amsfonts}
\usepackage{amstext}
\usepackage{amsxtra}
\usepackage{mathrsfs}
\usepackage{dsfont}
\usepackage{esint}
\usepackage{enumitem}
\usepackage{braket}

\allowdisplaybreaks

\newtheorem{lemma}{Lemma}
\newtheorem{theorem}[lemma]{Theorem}

\newtheorem{remark}[lemma]{Remark}

\newcommand{\R}{\mathbb{R}}

\newcommand\1{{\ensuremath {\mathds 1} }}
\renewcommand\phi{\varphi}
\newcommand{\gH}{\mathfrak{H}}

\newcommand{\cV}{\mathcal{V}}

\newcommand{\cK}{\mathcal{K}}
\newcommand{\cE}{\mathcal{E}}
\newcommand{\cF}{\mathcal{F}}
\newcommand{\cN}{\mathcal{N}}

\newcommand{\cH}{\mathcal{H}}
\newcommand{\cL}{\mathcal{L}}

\newcommand\pscal[1]{{\ensuremath{\left\langle #1 \right\rangle}}}
\newcommand{\norm}[1]{ \left\| #1 \right\|}
\newcommand{\tr}{{\rm Tr}\,}
\newcommand{\supp}{{\rm Supp}}
\renewcommand{\geq}{\geqslant}
\renewcommand{\leq}{\leqslant}
\renewcommand{\hat}{\widehat}

\newcommand{\eps}{\varepsilon}
\usepackage{color}

\newcommand{\nn}{\nonumber}

\newcommand{\dd}{\mathrm{d}}
\newcommand{\Vext}{V_{\rm ext}}
\newcommand{\ch}{\mathrm{ch}}
\newcommand{\sh}{\mathrm{sh}}

\newcommand{\ad}{\mathrm{ad}}
\newcommand{\FM}{\mathcal{F}^{\leq M}_+}
\newcommand{\F}{\mathcal{F}}

\newcommand{\ao}{\mathfrak{a}_0}
\newcommand{\hc}{{\rm h.c.}}

\begin{document}

\title{Bogoliubov excitation spectrum of trapped Bose gases in the Gross--Pitaevskii regime}

\author[P.T. Nam]{Phan Th\`anh Nam}
\address{Department of Mathematics, LMU Munich, Theresienstrasse 39, 80333 Munich, and Munich Center for Quantum Science and Technology, Schellingstr. 4, 80799 Munich, Germany} 
\email{nam@math.lmu.de}

\author[A. Triay]{Arnaud Triay}
\address{Department of Mathematics, LMU Munich, Theresienstrasse 39, 80333 Munich, and Munich Center for Quantum Science and Technology, Schellingstr. 4, 80799 Munich, Germany}  
\email{triay@math.lmu.de}

\date{\today}

\begin{abstract} We consider an inhomogeneous system of $N$ bosons in $\R^3$ confined by an external potential and interacting via a repulsive potential of the form $N^2 V(N(x-y))$. We prove that the low-energy excitation spectrum of the system is determined by the eigenvalues of an effective one-particle operator, which agrees with Bogoliubov's approximation. 

%\noindent 
%{\bf Keywords:} Trapped Bose gases, Gross--Pitaevskii  equation, Bogoliubov's theory.  
%
%\noindent 
%{\bf 2020 Mathematics Subject Classification:} 81V70, 81V73. 
\end{abstract}

\date{\today}
 
 \maketitle

%\tableofcontents

\section{Introduction}

The rigorous understanding of the macroscopic behaviors of Bose gases has been a central problem in mathematical physics, especially since the first experimental observations of the Bose--Einstein condensation in 1995 \cite{Wieman-Cornell-95, Ketterle-95}. While the ideal Bose gas was already well understood in 1924 due to the pioneer works of Bose \cite{Bose-24} and Einstein  \cite{Einstein-24}, the theory of interacting Bose gases was only started in 1947 with Bogoliubov's seminal  paper \cite{Bogoliubov-47} where he proposed an approximation of the excitation spectrum and used that to explain Landau's criterion for superfluidity \cite{Landau-41}.   

In the last decade, there has been remarkable progress on the justification of Bogoliubov's excitation spectrum from first principles, namely from many-body Schr\"odinger equations. In \cite{Seiringer-11,GreSei-13}, Seiringer and Grech--Seiringer  have for the first time justified Bogoliubov's theory for a class of trapped Bose gases in the mean-field regime. Their results have been extended in various directions, including singular interaction potentials \cite{LewNamSerSol-15}, large volume settings  \cite{DerNap-14}, multiple-condensations \cite{NamSei-15,RouSpe-18}, and higher order expansions \cite{Pizzo-15,BosPetSei-20,NamNap-21}. On the other hand, in the Gross--Pitaevskii regime, which is most relevant to the physical setup in \cite{Wieman-Cornell-95, Ketterle-95}, the analysis is significantly more challenging since Bogoliubov's theory admits a subtle correction due to the strong correlation between particles at short distances. Recently, this prediction has been established for translation-invariant systems in $\mathbb{T}^3$  by Boccato--Brennecke--Cenatiempo--Schlein \cite{BocBreCenSch-19} (scaling limits interpolating between the mean-field and the Gross--Pitaevskii regimes were previously considered in \cite{BocBreCenSch-17}). In the present paper, we will verify this conjecture for general trapped systems in $\R^3$.

\subsection{Model}

We consider a system of $N$ bosons in $\R^3$ described by the Hamiltonian 
\begin{equation} \label{eq:HN-GP}
H_N = \sum_{i=1}^N (-\Delta_{x_i}+\Vext(x_i)) + \sum_{1\leq i<j\leq N} N^2 V(N(x_i-x_j)) 
\end{equation}
which acts on the symmetric space $\gH^N=\bigotimes_{\text{sym}}^N L^2(\R^3)$. Here $x_i\in \R^3$ stands for the coordinate of the $i$-th particle (we ignore the spin for simplicity). We assume that the external potential $\Vext$ and the interaction potential $V$  satisfy 
\begin{align} \label{eq:Vext-ass}
0 &\leq \Vext(x)\leq  Ce^{C|x|} \text{ for some constant }C>0,\quad \lim_{|x|\to \infty} \Vext(x)=\infty,\\
0 &\leq V\in L^1(\R^3), \quad V \text{ is radially symmetric and compactly supported.} \label{eq:V-ass}
\end{align}
Under these conditions, $H_N$ is a positive operator with  the core domain $\bigotimes_{\text{sym}}^N C_c^\infty(\R^3)$. Hence, it can be extended to be a self-adjoint operator on $\gH^N$ by Friedrichs' method. The extension, still denoted by $H_N$, is positive and has compact resolvent. We are interested in the asymptotic behavior of the eigenvalues of  $H_N$ when $N\to \infty$.

The explicit scaling  of the interaction potential $N^2 V(N\cdot)$ places us in the Gross--Pitaevskii regime, where the system is very dilute since the range of the interaction is much smaller than the average distance of particle ($N^{-1} \ll N^{-1/3}$), but the interaction potential is very strong within its range. The strong correlation at short distances leads to nonlinear corrections to the mean-field approximation, making the rigorous understanding of the system highly challenging.

In \cite{LieSeiYng-00}, Lieb--Seiringer--Yngvason proved that the ground state energy $\lambda_1(H_N)$ of $H_N$ satisfies 
\begin{equation} \label{eq:BEC-EN}
\lim_{N\to \infty}\frac{\lambda_1(H_N)}{N} = \inf_{\|u\|_{L^2(\R^3)}=1} \cE_{\rm GP}(u)
\end{equation}
where $\cE_{\rm GP}$ is the {Gross--Pitaevskii functional} %\cite{Gross-61,Pitaevskii-61} 
\begin{equation} \label{eq:EGP}
\cE_{\rm GP}(u)=  \int_{\R^3} \Big(|\nabla u(x)|^2 + \Vext(x)|u(x)|^2 + 4\pi \ao |u(x)|^4 \Big) \dd x.
\end{equation}
Here the scattering length $\ao$  of $V$ is defined by 
\begin{align} \label{eq:def-a}
8\pi \ao = \inf\left\{ \int_{\R ^3} (2 |\nabla f(x)| ^2 + V(x) |f(x)| ^2) \dd x, \quad \lim_{|x|\to \infty}f(x)=1 \right\}. 
\end{align}
In the Gross--Pitaevskii scaling regime, the scattering length of $N^2 V(N\cdot)$ is exactly $\ao/N$.  
Note that in the usual mean-field approximation, one would simply restrict to the uncorrelated ansatz $u^{\otimes N}$, but this leads to a {\em wrong} functional with $\ao$ replaced by the larger value  $(8\pi)^{-1}\widehat V(0)$ (which is the first Born approximation of $\ao$). Thus correlations already play a crucial role for the leading order behavior of the system.

It was also proved in \cite{LieSeiYng-00} that the Gross--Pitaevskii functional has a minimizer $\varphi>0$, which is unique up to a complex phase and under the constraint $\|\varphi\|_{L^2}=1$.  Moreover, $\varphi$ is uniformly bounded and solves the Gross--Pitaevskii equation 
 \begin{equation} \label{eq:GPe} 
D \varphi =0, \quad D:=-\Delta+\Vext+ 8\pi \ao \varphi^2-\mu, \quad \mu:= \int_{\R^3} \left(  |\nabla \varphi|^2 + V_{\textrm{ext}} |\varphi|^2 + 8\pi \ao \varphi^4 \right).   
\end{equation}

In \cite{LieSei-06}, Lieb--Seiringer proved that the  ground state of $H_N$ exhibits the complete Bose--Einstein condensation (BEC) in the Gross--Pitaevskii minimizer $\varphi$. More precisely, they proved that if $\Psi_N$ is a ground state of $H_N$, or more generally an approximate ground state in the sense that $\langle \Psi_N, H_N \Psi_N \rangle = \lambda_1(H_N) + o(N)$, then 
\begin{equation} \label{eq:BEC}
\lim_{N\to \infty} \frac{\langle \Psi_N, a^*(\varphi) a(\varphi)\Psi_N\rangle}{N}  = 1.
\end{equation}
Here $a^*,a$ are the creation and annihilation operators on Fock space (see (\ref{eq:cF-intro}) and below for a definition). The convergence \eqref{eq:BEC} means that most of particles occupy the one-body state $\varphi$. See also \cite{LieSei-02} for an earlier result on the homogeneous system on $\mathbb{T}^3$, and see \cite{NamRouSei-16} for an alternative proof of  \eqref{eq:BEC-EN} and \eqref{eq:BEC}.

Recently, the {optimal convergence rate} of \eqref{eq:BEC-EN} and \eqref{eq:BEC} has been obtained in \cite{NamNapRicTri-21} for interactions with sufficiently small scattering length, and extended later in \cite{BreSchSch-21} to large interactions, in both cases with slightly stronger conditions on $V$ than what we assume in the present paper (see also \cite{BocBreCenSch-17b,BocBreCenSch-20,Hainzl-20} for optimal estimates in the translation-invariant case). From these works, we know that the ground state energy $\lambda_1(H_N)$ is equal to $N\cE_{\rm GP}(\varphi)+ \mathcal{O}(1)$. In the present paper, we are interested in the excitation spectrum of $H_N$, namely the low-lying eigenvalues of $H_N-\lambda_1(H_N)$.

\subsection{Main result} We prove that in the limit $N\to \infty$, the {\em elementary excitations} of $H_N$ are determined by the positive eigenvalues of the one-body  operator on $L^2(\R^3)$,
\begin{align} \label{eq:E-inf}
E_\infty=(D^{1/2} (D+16 \pi \ao \varphi^2) D^{1/2})^{1/2}
\end{align}
where $D$ is given in \eqref{eq:GPe}. Note that $D$ has compact resolvent  and a unique ground state $\varphi$ with energy $0$, and the same holds for $E_\infty$.  Consequently, $E_\infty$ has eigenvalues $0=e_0<e_1\leq e_2\leq ...$ with $e_n\to \infty$ as $n\to \infty$. Our main result is  the following theorem.  

\begin{theorem}[Excitation spectrum in the Gross--Pitaevskii regime] \label{thm:main} Let $\Vext$ and $V$ satisfy \eqref{eq:Vext-ass} and \eqref{eq:V-ass}, respectively. Let $\lambda_1(H_N)$ be the ground state energy of the Hamiltonian $H_N$ in \eqref{eq:HN-GP}. Then the spectrum of $H_N -\lambda_1(H_N)$ below an energy $\Lambda \in [1, N^{1/16}]$ is equal to finite sums of the form  
$$\sum_{i \geq 1} n_i e_i + \mathcal{O}(\Lambda N^{-1/16})$$
where $n_i\in \{0,1,2,...\}$ and $\{e_i\}_{i=1}^\infty$ are the positive eigenvalues of the operator $E_\infty$ in \eqref{eq:E-inf}. 
\end{theorem}

We can improve the error $\mathcal{O}(\Lambda N^{-1/16})$ under some additional assumption on the growth of $\Vext$, but we do not pursue this here for simplicity. 

We have some immediate remarks on related results in the litterature.

\begin{itemize}

\item Our result in Theorem \ref{thm:main} verifies a well-known conjecture which was formulated explicitly by Grech--Seiringer in \cite[Conjecture 1]{GreSei-13}. In the case of translation-invariant systems in $\mathbb{T}^3$, this result has been proved by Boccato--Brennecke--Cenatiempo--Schlein \cite{BocBreCenSch-19}, see also \cite{HaiSchTri-22} for an alternate proof. In this case, the condensate is simply $\varphi=1$ and the elementary excitations are given explicitly by $e_p=(|p|^4+16\pi \ao |p|^2)^{1/2}$ with $p\in 2\pi\mathbb{Z}^3$. Note that $e_p$ behaves linearly at small momentum $|p|$, which is consistent with Landau's criterion for superfluidity as predicted by Bogoliubov in 1947 \cite{Bogoliubov-47} and experimentally confirmed in 2002 \cite{Ketterle-02,Davidson-02}. 

\medskip

\item After our paper had been finished, a result similar to Theorem 1 was obtained independently by Brennecke, Schlein and Schraven in \cite{BreSchSch-21b}. In this work, an explicit expansion to the second order of the ground state energy was also provided, see (1.8) and (1.9) in \cite{BreSchSch-21b}. They required however slightly stronger assumptions on both the interaction and the external potentials, namely $V \in L^3(\mathbb{R}^{3})$, $V_{\rm ext} \in C^{1}$ satisfies $V_{\rm ext}(x+y) \leq C(V_{\rm ext}(x)+C)(V_{\rm ext}(y)+C)$ (which in particular requires $V_{\rm ext}$ to grow at most exponentially), and that $|\nabla V_{\rm ext}|$ grows at most exponentially (while we only assume that $V\in L^1(\mathbb{R}^{3})$ and that $V_{\rm ext}$ grows at most exponentially). 
\end{itemize}

%Recently, in \cite{BreSchSch-21}, Brennecke, Schlein, and Schraven  developed the techniques from \cite{BocBreCenSch-19} to prove the optimal rate of BEC for the inhomogeneous systems in $\R^3$. They also announced that they are able to push further the method to achieve the excitation spectrum and that the proof will be provided in a separate paper. 

%{\bf Note added in proof.}  

%While writing up the last details of our
%manuscript, a result similar to Theorem 1.1 was posted in [23]. While the strategy used
%in [23] is similar to ours (and to the one previously developed in [6]), the details of their
%implementation are different.

The aim of the present paper is not only to  prove Theorem \ref{thm:main}, but also to provide with a general, transparent approach to the theory of dilute Bose gases. While the line of our proof is inspired by \cite{BocBreCenSch-19}, we introduce several conceptual simplifications and generalizations (discussed after (\ref{eq:T_1'''-intro}) below) that could be useful in other contexts. In fact, recently a simplified proof of the result in \cite{BocBreCenSch-19} has been given in \cite{HaiSchTri-22} using some ideas introduced in the present paper. Here are some potential extensions of our result.

%Their work is restricted to the case of a torus, where the additional translation invariance simplifies things quite a bit. They have very recently also extended the results to the trapped case (arXiv 2108.11129) but their work appeared several weeks after the one by Nam-Triay, and is valid only under stronger regularity assumptions on the interaction potential.
% 
% The value of the present work by Nam-Triay is not just to extend the work by Schlein et al. to the non-translation-invariant setting, though. While the main strategy is certainly based on the key contributions by Schlein et al., they introduce several conceptual simplifications that, I expect, will prove to be of importance in future research in this rapidly developing field. Hence, in summary, while Schlein et al. certainly did the heavy lifting on this problem, the contributions by Nam-Triay are highly non-trivial and constitute an important new ingredient in the puzzle. Hence I recommend considering the article seriously.

\begin{itemize}

\item Our proof works for a large class of interaction potentials.  The assumptions in \eqref{eq:V-ass} are already rather general, but they could be relaxed further in several ways. In particular, the conditions that $V$ is radial and compactly supported were used in \cite{LieSei-06,NamRouSei-16} to get  \eqref{eq:BEC}, but these conditions can  be relaxed if we can justify \eqref{eq:BEC} without them (it is indeed possible and we will come to this issue in a separate work). The condition $V\in L^1(\R^3)$ is also technical; we expect that the same result  holds as soon as $V\geq 0$ has a finite scattering length (including hard sphere potentials), but additional arguments will be needed.  

\medskip

\item Our strategy may be useful to handle systems `beyond Gross--Pitaevskii limit', namely when the interaction potential $N^2V(Nx)$ is replaced by $N^{2(1-\kappa)} V(N^{1-\kappa}x)$ for some $\kappa>0$. In the translation-invariant case, when $\kappa>0$ is small, the BEC and the excitation spectrum have been recently established in \cite{AdhBreSch-21,Fournais-20} and in \cite{BreCapSch-21}, respectively. Our approach should be applicable to inhomogeneous systems in $\R^3$, or to systems in a bounded domain with a general boundary condition, in particular the problem with Neumann boundary condition serves as an intermediate step to understand the thermodynamic limit, thus extending the Lee-Huang-Yang formula studied in \cite{YauYin-09,FouSol-20,FouSol-21,BasCenSch-21} to the free energy at positive temperatures.
\end{itemize}

In the present paper, we will not pursue these generalizations in order to keep the ideas transparent. In the following, we will recall heuristic arguments from Bogoliubov's theory and explain main ingredients of our proof. 

\subsection{Bogoliubov's approximation} 
Heuristically, the excitation spectrum of $H_N$ can be predicted quickly using the Gross--Pitaevskii functional. Let $Q=1-|\varphi\rangle \langle \varphi|$. For $v\in \gH_+ =Q \gH$, we have the Taylor expansion  
$$
\cE_{\rm GP}\Bigg(\frac{\varphi+v}{\sqrt{1+\norm{v}_{L^2}^2}}\Bigg)=\cE_{\rm GP}(\varphi)+\frac{1}{2}\pscal{\begin{pmatrix}v\\  {v}\end{pmatrix}, \cE_{\rm GP}''(\varphi) \begin{pmatrix}v\\  {v}\end{pmatrix}}  +o\big(\norm{v}^2_{H^1(\R^3)}\big)
$$ 
with the Hessian matrix $\cE_{\rm GP}''(\varphi)$ on $\gH_+ \oplus \gH_+$ given by 
$$    \;\cE_{\rm GP}''(\varphi) = \begin{pmatrix}
D+ Q (8\pi \ao \varphi^2) Q & Q (8\pi \ao \varphi^2) Q \\ Q (8\pi \ao \varphi^2) Q & {D}+  Q (8\pi \ao \varphi^2) Q\end{pmatrix}.  
%=\int_{\R^d} \overline{v(x)} (h+K) v(x) \d x + \frac{1}{2} \iint_{\R^d\times \R^d} K(x,y) \Big( {v(x)}  v(y) + \overline{v(x)} \overline{v(y)} \Big) \d x \d y
$$
Here we identity $D=QDQ$ since $D\varphi=0$. The Hessian matrix can be diagonalized as 
$$
\mathcal{V}^* \cE_{\rm GP}''(\varphi) \mathcal{V} = \begin{pmatrix}
E_\infty & 0 \\ 0 &  E_\infty\end{pmatrix}, \quad E_\infty= \Big(D^{1/2} (D+ 16\pi \ao \varphi^2)D^{1/2}\Big)^{1/2} 
$$
where
\begin{align} \label{eq:cV-def-intro}
\mathcal{V}=  \begin{pmatrix}
\sqrt{1+s^2} & s \\ s &  \sqrt{1+s^2} \end{pmatrix}
\end{align}
with  a Hilbert-Schmidt, self-adjoint operator $s$ on $\gH_+$ (see \cite{GreSei-13,NamNapSol-16}). Note that $\cV$ is a symplectic matrix on $\gH_+\oplus \gH_+$, namely 
$$
 \cV^*  \begin{pmatrix}1 & 0 \\ 0 &  -1  \end{pmatrix} \cV = \begin{pmatrix}1 & 0 \\ 0 &  -1  \end{pmatrix},
$$
which is associated with a unitary transformation on Fock space $\cF_+=\cF(\gH_+)$ that we will explain below. As explained in \cite{LewNamSerSol-15}, Bogoliubov's theory suggests that the particles outside of the condensate can be described by the second quantization of the Hessian $\cE_{\rm GP}''(\varphi)$ on the Fock space $\cF(\gH_+)$. In particular, the spectrum of $H_N-\lambda_1(H_N)$ can be described by the spectrum of $\dd\Gamma(E_\infty)$, the quantization of $E_\infty$ as in \eqref{eq:dGA-intro}, namely
\begin{align} \label{eq:ext-spe-heu-intro}
\sigma( H_N-\lambda_1(H_N) ) \approx \sigma( \dd\Gamma(E_\infty) ) = \left\{ \sum_{i\geq 1} n_i e_i: n_i \in \{0,1,...\}, e_i \in \sigma(E_\infty) \right\}. 
\end{align}

Next, let us put the above heuristic explanation in a more concrete manner. We will recall the Fock space formalism and explain the original arguments of Bogoliubov in \cite{Bogoliubov-47}.

\medskip
\noindent{\bf Fock space formalism.}  For $\gH=L^2(\R^3)$, we define the bosonic Fock space  
\begin{align} \label{eq:cF-intro}
\cF=\cF(\gH)= \bigoplus_{n=0}^\infty \gH^{n}, \quad \gH^n=\bigotimes_{\text{sym}}^n \gH.  
\end{align}
For $g\in \gH$, we can define the creation and annihilation operators $a^*(g)$, $a(g)$ on $\cF$ by 
\begin{align*}
	(a^* (g) \Psi )(x_1,\dots,x_{n+1})&= \frac{1}{\sqrt{n+1}} \sum_{j=1}^{n+1} g(x_j)\Psi(x_1,\dots,x_{j-1},x_{j+1},\dots, x_{n+1}), \\
	(a(g) \Psi )(x_1,\dots,x_{n-1}) &= \sqrt{n} \int_{\R^3} \overline{g(x_n)}\Psi(x_1,\dots,x_n) \dd x_n, \quad \forall \Psi \in \gH^n,\, \forall n. 
\end{align*}
We may also define 
the operator-valued distributions $a_x^*$ and $a_x$, with $x\in \R^3$, such that 
%by 
%$$
%a_x^*= \sum_{n=1}^\infty \overline{f_n(x)} a^*(f_n), \quad a_x= \sum_{n=1}^\infty f_n(x) a(f_n)
%$$
%where $\{f_n\}_{n=1}^\infty$ is an orthonormal basis of $\gH$ (the definition is independent of the choice of the basis). Equivalently, we have
\[
	a^*(g)=\int_{\R^3}   g(x) a_x^* \dd x, \quad a(g)=\int_{\R^3}  \overline{g(x)} a_x \dd x, \quad \forall g\in \gH.
\]
These operators satisfy the canonical commutation relations (CCR)
\begin{align*}
	[a(g_1),a(g_2)]&=[a^*(g_1),a^*(g_2)]=0,\quad [a(g_1), a^* (g_2)]= \langle g_1, g_2 \rangle, \quad \forall g_1,g_2 \in \gH, \\ 
 [a^*_x,a^*_y]&=[a_x,a_y]=0, \quad [a_x,a^*_y]=\delta(x-y), \quad \forall x,y\in \R^3.
 \end{align*}
The creation and annihilation operators can be used to express several important observables.  For example, for any one-body self-adjoint operator $A$ we can write its second quantization as 
\begin{align} \label{eq:dGA-intro}
	\dd \Gamma(A):= \bigoplus_{n=0}^\infty \left(  \sum_{i=1}^n A_{x_i} \right) =  \int_{\R^3} a^*_x A_x a_x \dd x.
\end{align}
In particular, $\cN= \dd \Gamma(1)$ is the number operator. The Hamiltonian in \eqref{eq:HN-GP} can be written as 
\begin{align} \label{eq:2nd-Q}
	H_N = \int_{\R^3} a^*_x (-\Delta_x + \Vext(x)) a_ x \dd x   + \frac{1}{2}\int_{\R^3}\int_{\R^3} N^2V_N(x-y) a_x^* a_y^* a_x a_y \dd x \dd y
\end{align}
with $V_N(x)=V(Nx)$. The right-hand side of \eqref{eq:2nd-Q} coincides with  \eqref{eq:HN-GP} in the $N$-body sector. To shorten notations, we will often omit the integration variables, like $\dd x$, when there is no ambiguity.

%In the following, we will also use the excited Fock space $\cF_+=\cF(\gH_+)$ with $\gH_+=\{\varphi\}^\bot$. We can think of $\cF_+$ as a subspace of $\cF$ and always use the notation $a_x^*$ and $a_x$ for creation and annihilation operators on $\cF$. 

\bigskip
\noindent
{\bf Ingredients of Bogoliubov's approximation.} The original arguments of Bogoliubov in \cite{Bogoliubov-47} were written in the translation-invariant setting, but they can be translated to our setting as follows. To avoid confusing the notations, let us consider the Hamiltonian
\begin{align} \label{eq:2nd-Q-H}
	H = \int_{\R^3} a^*_x (-\Delta_x + \Vext(x)) a_ x \dd x   + \frac{1}{2}\int_{\R^3}\int_{\R^3} N^{-1}W(x-y) a_x^* a_y^* a_x a_y \dd x \dd y
\end{align}
with a general interaction potential  $N^{-1}W: \R^3\to \R$ instead of $N^2V(N\cdot)$. Roughly speaking, Bogoliubov's approximation contains four key steps. 

\bigskip
\noindent{\bf Step 1 (c-number substitution).} The contribution of the condensate can be replaced by a scalar field, namely the annihilation operator $a_x$ on $\cF$ can be formally  decomposed as 
$$a_x \approx  N_0^{1/2}  g(x) + c_x$$
where $g$ is a normalized function in  $L^2(\R^3)$ that stands for the condensate, $N_0$ is the number of particles in the condensate, and $c_x$ is the annihilation operator on the excited Fock space $\cF_+= \cF(\gH_+)$ with $\gH_+=Q_g \gH$ and $Q_g=1-|g\rangle \langle g|$.  Thus from \eqref{eq:2nd-Q-H} we have 
\begin{align} \label{eq:Bog-intro-1}
	&H \approx \int_{\R^3} (N_0^{1/2} \overline{g(x)}  + c^*_x) (-\Delta_x + \Vext(x)) (N_0^{1/2}g(x)+c_ x)  \dd x  \\
	& + \frac{1}{2}\int_{\R^3}\int_{\R^3} N^{-1}W(x-y) (N_0^{1/2}\overline{g(x)} + c_x^*)(N_0^{1/2} \overline{g(y)} + c_y^*) (N_0^{1/2}g(x) + c_x) (N_0^{1/2} g(y)+ c_y) \dd x \dd y. \nn
\end{align}

\bigskip
\noindent{\bf Step 2 (Quadratic reduction).} Thanks to the complete BEC, 
%$$\int_{\R^3} c_x^* c_x= N - N_0 = o(N),$$
we may use $c_x \ll N_0^{1/2} \approx N^{1/2}$ and expand \eqref{eq:Bog-intro-1} to second order. More precisely, we ignore all terms containing more than 2 operators $c_x$ or $c_x^{*}$ and obtain  
\begin{align} \label{eq:Bog-intro-2}
H &\approx  N\mathcal{E}_{\rm H}(g) - \frac{1}{2} \int (W*|g|^2)|g|^2 + \sqrt{N} \Big[ c^* ( (D_H +\mu_H ) g )+ \hc \Big] +  \int_{\R^3} c^*_x (D_H)_x  c_x \dd x \\
&+  \int_{\R^3}\int_{\R^3} W(x-y) g(x)  \overline{g(y)} c_x^* c_y \dd x \dd y + \frac{1}{2}\int_{\R^3}\int_{\R^3} W(x-y) \Big [g(x) g(y) c_x^* c^*_y +\hc \Big] \dd x \dd y \nn
\end{align}
where
\begin{align*}
\mathcal{E}_{\rm H}(g) &=  \int_{\R^3} \left( |\nabla g|^2 + \Vext |g|^2 +\frac{1}{2} | g| ^2 (W* |g|^2)\right),\\
D_H &= -\Delta +\Vext + W* |g|^2 -\mu_H,\\
 \mu_H &=   \int_{\R^3} \left( |\nabla g|^2 + \Vext |g|^2 + | g| ^2 (W* |g|^2)\right).
\end{align*}
To describe the low-lying spectrum of $H$, it is natural to take $g$ as the minimizer of the Hartree functional  $\mathcal{E}_{\rm H}(g)$ under the constraint $\|g\|_{L^2}=1$. Note that the Hartree minimizer exists uniquely up to a phase and can be chosen positive. Since $D_H g=0$ and $c(g)=c^*(g)=0$), the linear terms in \eqref{eq:Bog-intro-2} vanish.  

\bigskip
\noindent{\bf Step 3 (Diagonalization).} The quadratic operator  
$$
 \int_{\R^3} c^*_x (D_H)_x  c_x \dd x +  \frac{1}{2}\int_{\R^3}\int_{\R^3} W(x-y) g(x)  g(y) ( 2 c_x^* c_y + c_x^* c^*_y + c_x c_y)  \dd x \dd y
 $$
 is known to be diagonalized by a unitary operator $T$ on $\cF_+$, which is characterized by the actions 
 $$
 T^* c(v) T = c(\sqrt{1+s^2} v) + c^*(sv), \quad   T^* c^*(v) T = c^*(\sqrt{1+s^2} v) + c(sv), \quad \forall v\in \gH_+
 $$
 for some well-chosen operator $s$ on $\gH_+$ (c.f. the form of $\cV$ in \eqref{eq:cV-def-intro}). Thus from \eqref{eq:Bog-intro-2} we deduce that, up to a unitary transformation, 
%Combining with \eqref{eq:Bog-intro-2} are quadratic in $c_x$ and $c_x^*$, which can be diagonalized by a unitary operator $T$, leading to 
%by a quadratic transformation of the form 
%$$
%T = \exp \left( \int_{\R^3}\int_{\R^3} ( \mathfrak{K}(x,y) c_x^* c_y ^* - \hc ) \dd x \dd y \right)
%$$ 
%with an appropriate kernel $\mathfrak{K}(x,y)$. This gives
\begin{align} \label{eq:Bog-MF}
H  \approx N\mathcal{E}_{\rm H}(g) - \frac{1}{2} \int (W*g^2)g^2 + \frac{1}{2} \tr (E_H- D_H-K_H) + \dd \Gamma(E_{H})
\end{align}
where $E_H= ( D_H^{1/2} (D_H+2K_{H})D_{H}^{1/2})^{1/2}$ and $K_H=Q_g\widetilde K_HQ_g$ with $\widetilde{K}_H(x,y)= g(x) W(x-y) g(y)$. 

\medskip

It turns out that in the mean-field regime, when $W$ is independent of $N$, the excitation spectrum of $H$ is described correctly by the one-body operator $E_H$, see \cite{GreSei-13,LewNamSerSol-15} for rigorous results\footnote{In \cite{GreSei-13,LewNamSerSol-15},  $N^{-1}W$ is replaced by $(N-1)^{-1}W$ , and hence there is no $-\frac{1}{2}\int (W*|g|^2)g^2$ in \eqref{eq:Bog-MF}}. However, in the Gross--Pitaevskii regime, when $W=N^3 V(N\cdot)$, the approximation \eqref{eq:Bog-MF} is wrong. It can be seen already in the leading order constant, since formally replacing $N^3 V(Nx) \approx \hat V(0) \delta_x$ in the Hartree functional gives 
\begin{align*}
\mathcal{E}_{\rm H}(g) \approx \int_{\R^3} \left( |\nabla g|^2 + \Vext |g|^2 +\frac{\hat V(0)}{2} | g| ^4 \right)
\end{align*}
while  in \eqref{eq:BEC-EN} we obtain the Gross--Pitaevskii functional instead.  
%
%gives the effective interaction potential $N^2 V(Nx) \mapsto N^{-1}\hat V(0) \delta_x$, which is different from $8\pi \ao N^{-1}\delta_x$ from the Gross--Pitaevskii functional. 
This suggests a subtle correction in the Gross--Pitaevskii regime. 

\bigskip
\noindent{\bf Step 4 (Landau's correction\footnote{This step goes back to an  important remark of Landau, see \cite[p. 31]{Bogoliubov-47}.}).} $\hat V(0)$ should be replaced by $8\pi \ao$ everywhere, with $\ao$ the scattering length of $V$. Consequently, \eqref{eq:Bog-MF} becomes
\begin{align}  
 H  \approx N\mathcal{E}_{\rm GP}(\varphi) - 4\pi \ao \int_{\R^3} \varphi^4 + \frac{1}{2} \tr (E_\infty- D- Q (16 \pi \ao \varphi^2)Q) + \dd \Gamma(E_\infty),
\end{align}
which is consistent with our Theorem \ref{thm:main}. 

%on the right side of \eqref{eq:Bog-MF}
%
%$\cE_{H}(g)$, $D_H$, $K_H$,  $E_H$ should be replaced by $\cE_{\rm GP}(\varphi)$, $D$, $8 \pi \ao \varphi^2$, $(D^{1/2} (D+16\pi \ao \varphi^2)D^{1/2})^{1/2}$, respectively, namely  This adjustment suggests that the Hamiltonian 
%%\begin{align}  
%%T^* H T \approx N\mathcal{E}_{\rm GP(\varphi) - 4\pi \ao \int \varphi^4 + \frac{1}{2} \tr (E_\infty- D- Q (16 \pi \ao \varphi^2)Q) + % \dd \Gamma(E_\infty)
%%\end{align}

%leads to an approximation which is consistent with our Theorem \ref{thm:main}. 
%
%
%eq:ext-spe-heu-intro

Obviously the last step is very formal.  Implementing it rigorously is the main challenge to understand the Gross--Pitaevskii regime. 

\subsection{Proof strategy} Roughly speaking, our proof is based on the rigorous approximation
\begin{align} \label{eq:all-uni-tran-intro}
T_2^* T_c^* T_1^* U H_N U^* T_1 T_c T_2 \approx \lambda_1(H_N) + \dd\Gamma(E_\infty) + o(1)_{N\to \infty}  
\end{align}
on the excited Fock space $\cF_+=\cF(\gH_+)$ with $\gH_+=\{\varphi\}^\bot = Q L^2(\R^3)$ with $Q=1-|\varphi\rangle\langle \varphi|$. Here $U,T_1, T_c,T_2$ are suitable unitary transformations that we will explained below. Our overall approach is similar to that of \cite{BocBreCenSch-19}, but our choice of the transformations is different.

\bigskip
\noindent
{\bf Excitation operator $U$.} Our starting point is the operator $U=U(\varphi)$ introduced in  \cite{LewNamSerSol-15}. As explained in \cite{LewNamSerSol-15}, any function $\Psi_N\in \gH^N$ admits a unique decomposition
\begin{equation*}
	\Psi_N = \varphi^{\otimes N} \xi_0 + \varphi^{\otimes N-1} \otimes_s \xi_1 + \varphi^{\otimes N-2} \otimes_s \xi_2 + ... + \xi_N
\end{equation*}
with $\xi_k \in \gH_+^{k}$. The operator  
\begin{align} \label{eq:U-intro}
U : \Psi_N \mapsto (\xi_0, \xi_1, ..., \xi_N)
\end{align}
is a unitary transformation from $\gH^N$ to $\cF_+^{\leq N}=\1(\cN\leq N)\cF_+$.  This operator maps $a(\varphi)$ and $a^*(\varphi)$ to $\sqrt{N-\cN}$, which is close to $\sqrt{N}$ due to the complete BEC. Thus the operator $U$ provides with a rigorous tool to implement the first step (c-number substitution) in Bogoliubov's approximation. Up to an error $o(1)$,  we can write (see Lemma \ref{lem:UHU*})
\begin{align} \label{eq:UHU-intro}
U H_N U^* = \1_+^{\leq N} \mathcal H \1_+^{\leq N} \approx \1_+^{\leq N} \left(  \sum_{i=0}^4 \cL_i  \right) \1_+^{\leq N}  
\end{align}
where $\1_+^{\leq N}$ is the projection on $\cF_+^{\leq N}$ and  $\cL_i$'s are the operators on the full Fock space $\cF(\gH)$:
\begin{align*}
	\cL_0&=  (N-\cN) \int \left(|\nabla\varphi|^2 + \Vext |\varphi|^2 + \frac{1}{2}(N^3 V_N \ast \varphi^2)  \varphi^2 \right) - \frac{1}{2} (\cN +1) \int \left(N^3 V_N \ast \varphi^2\right)  \varphi^2 \\
	&\qquad+ \dd\Gamma \Big(-\Delta+V_{\rm ext} + N^3 V_N *\varphi^2 + \varphi(x) N^3 V_N(x-y) \varphi(y)\Big),\\
	\cL_1&=\sqrt{N}   a  \left(  \left(-\Delta + V_{\rm ext} + N^3 V_N*\varphi^2\right)\varphi\right) + \hc , \\
	\cL_2&=\frac{1}{2} \left( 1 - \frac{\cN}{N} - \frac{1}{2N} \right)  \iint N^3 V_N (x-y) \varphi(x) \varphi(y) a_x a_y  \dd x\, \dd y + \hc  ,\\
	\cL_3&=  \sqrt{(1-\cN/N)_+} \iint N^{5/2}V_N(x-y) \varphi(x) a^*_y a_{x} a_{y}  \dd x \dd y + \hc  , \\
	\cL_4&=  \iint N^2 V_N(x-y)  a_x^* a^*_y a_{x} a_{y}   \dd x \dd y.
\end{align*}

In the expression of $\mathcal{L}_0$, we identify $\varphi(x) N^3 V_N(x-y) \varphi(y)$ with the operator whose kernel is given by this formula. In the mean-field regime, $\mathcal{L}_3$ and $\mathcal{L}_4$ are of order $o(1)$ and the second step of Bogoliubov's approximation is carried out rigorously in \cite{GreSei-13,LewNamSerSol-15}. However, in the Gross--Pitaevskii regime, $\mathcal{L}_4$ is of order $N$ and $\mathcal{L}_3$ is of order $1$. As proposed in \cite{BocBreCenSch-19}, we will renormalize $\mathcal{L}_4$ and  $\mathcal{L}_3$ by two unitary transformations $T_1$ and $T_c$, which together implement  the second and fourth steps in Bogoliubov's approximation simultaneously. Then the resulting Hamiltonian can be diagonalized by a further unitary transformation $T_2$ as in the third step in Bogoliubov's argument. % and thus essentially places us in the mean-field regime. 

Note that while $\cL_i$'s are defined as the operators on the full Fock space $\cF=\cF(\gH)$, all the unitary transforms $T_1,T_c,T_2$ preserve the excited Fock space  $\cF_+=\cF(\gH_+)$. This allows us to apply relevant estimates for states on the excited Fock space $\cF_+$ at the end. 

\bigskip
\noindent
{\bf First quadratic transformation $T_1$.} In order to extract the correct energy contribution of order $N$ from $\cH$, we have to extract the correlation structure of particles which is encoded in the scattering problem. Under our assumption on $V$, the variational problem \eqref{eq:def-a} has a unique minimizer $0\leq f\leq 1$ and it solves the scattering equation 
\begin{align}\label{eq:scattering_equation-intro}
-2\Delta f + V f = 0 \text{ in }\R^3, \quad \lim_{|x|\to \infty} f(x)=1.
\end{align}
For $N^{-1} \ll \ell \leq 1$, we denote $\omega=1-f$ and the truncated functions 
$$\omega_{\ell,N} (x) = \chi ( x /\ell) \omega(Nx),\quad \varepsilon_{\ell,N}= 2\Delta ( \omega_{\ell,N} (x) -\omega(Nx))$$
where $0\leq \chi \leq 1$ is a smooth function satisfying $\chi(x) =1$ if $|x|\leq 1/2$ and $\chi(x) = 0$ if $|x| \geq 1$. Note that $\varepsilon_{\ell,N}$ is supported in $\{\ell/2 \leq |x| \leq \ell \}$ and
\begin{align} \label{eq:eps-lN-intro}
N^3 \int_{\R^3} \varepsilon_{\ell,N}= 8\pi \ao,
\end{align}
see Section \ref{sec:scattering}. We choose the first quadratic transformation via the actions
\begin{align} \label{eq:T1-intro}
T_1^* a^*(g) T_1 = a^*(\sqrt{1 + s_1^2} g) + a(s_1g), \quad \forall g\in \gH
\end{align}
where
$$
s_1 = Q^{\otimes 2} \widetilde s_1 \in \gH_+^2, \quad \widetilde s_1 (x,y) = - N \omega_{\ell,N}(x-y) \varphi(x) \varphi(y).
$$
In Lemma \ref{lemma:main_lemma_quadratic_transform}, we prove that if $N^{-1} \ll \ell\ll 1$,  then up to an error $o(1)$, we have
\begin{align}\label{eq:T1H-intro}
T^*_1 \mathcal  H T_1 
	&\approx  N \cE_{\rm GP}(\varphi) -\frac{1}{2}  \int N^3 ( V_Nf_N\ast \varphi^2) \varphi^2  -  \frac{N^4}{2} \int (\omega_{\ell,N} \varepsilon_{\ell,N} \ast \varphi^2) \varphi^2 \nn \\
	&\quad + \dd\Gamma\Big(-\Delta + \Vext + N^3 V_N*\varphi^2 + \varphi(x)N^{3}V_N(x-y)\varphi(y)  - \mu\Big)  \nn\\
	&\quad+ \frac{1}{2} \iint N^{3}\varepsilon_{\ell,N}(x-y) \varphi(x) \varphi(y) ( a^*_x a^*_y + a_x a_y ) \dd x \dd y+   \cL_3  +  \cL_4.  
\end{align}
%with 
%\begin{align*}
%h_N &= -\Delta + V_{\textrm{ext}} + N^3 V_N \ast \varphi^2 + \varphi(x)\widehat V(p/N) \varphi(x) \\
%\mu_N &= \int |\nabla \varphi|^2 + V_{\textrm{ext}} |\varphi|^2 + \int N^3 (V_N f_N) \ast \varphi^2 \varphi^2. 
%\end{align*}

Thus the transformation $T_1$ extracts the exact contribution of order $N$ from $\cH$ and at the same time replaces the short range potential $V_N(x-y)$ in $\mathcal{L}_2$ by the longer range potential $\varepsilon_{\ell,N}(x-y)$. %, except that on the right side of \eqref{eq:T1H-intro} the quadratic terms are still of order $\ell^{-1}$ instead of order $1$. 

\bigskip
{\em Comparison to the literature.} The transformation $T_1$ is different from existing tools in the literature and gives us some major advantages. Let us explain the improvements here.  

The idea of renormalizing the many-body Hamiltonian in the Gross--Pitaevskii limit by a Bogoliubov transformation was initiated by Benedikter--de Oliveira--Schlein \cite{BenOliSch-15b} to derive the Gross--Pitaevskii dynamics with an optimal convergence rate.  They focused on initial quasi-free states on Fock space and introduced a Bogoliubov transformation of the form 
\begin{align} \label{eq:T_1'-intro}
{T}_1' = \exp \left(\frac{1}{2} \iint \mathfrak{K}_1(x,y) a^*_x a^*_y \dd x \dd y- \hc  \right)
\end{align}
which would do the same job as $T_1$ if $\mathfrak{K}_1=\sh^{-1} s_1$ (see Section \ref{sec:quadratic-transforms}). In \cite{BreSch-19},  Brennecke--Schlein adapted the approach in  \cite{BenOliSch-15b} to study the quantum dynamics on $\gH^N$. In this canonical setting, after conjugated by $U$ in \eqref{eq:U-intro}, the Schr\"odinger generator acts on the truncated Fock space $\cF_+^{\leq N}$. On the other hand, in general a  Bogoliubov transformation of the form \eqref{eq:T_1'-intro} does not leave $\cF_+^{\leq N}$ invariant, leading to an incompatibility problem. In \cite{BreSch-19} the authors resolved this by introducing a generalized Bogoliubov transformation of the form  
\begin{align}\label{eq:T_1''-intro}
{T}_1'' = \exp \left(\frac{1}{2} \iint \mathfrak{K}_1(x,y) b^*_x b^*_y \dd x \dd y- \hc  \right)\quad\text{ with }b_x= \sqrt{1-\cN/N} a_x.
\end{align}
The operator $T_1''$ is a proper unitary transformation on $\cF_+^{\leq N}$, but it is not a standard Bogoliubov transformation since $b_x$ is not an exact annihilation operator. Consequently,  $T_1''$ does not satisfy an exact formula similar to \eqref{eq:T1-intro} and its actions have to be analyzed very carefully, see \cite[Section 3]{BreSch-19} for details. The operator $T_1''$ was also an essential tool to study the spectral problem in a series of papers  \cite{BocBreCenSch-17,BocBreCenSch-17b,BocBreCenSch-20,BocBreCenSch-19,BreSchSch-21,BreSchSch-21b}. 

The idea of using the operators $b_x$'s  goes back to the work of Lieb--Solovej \cite{LieSol-01}, where they introduced $\widetilde b_x=N^{-1/2}a^*(\varphi) a_{x}$, which is the same to $b_x$ up to the unitary transformation $U$. In \cite[Theorem 6.3]{LieSol-01}, they proved that the commutator relation $[\widetilde b_x, \widetilde b_x^*]\leq \delta_{0}(x)$ is sufficient to derive a lower bound for the ground state energy via Bogoliubov's diagonalization argument. %This argument was also used in \cite{NamNapRicTri-21} to derive the optimal rate of BEC. 
In the pioneer works on the excitation spectrum in the mean-field regime, Seiringer \cite{Seiringer-11} and Grech--Seiringer \cite{GreSei-13} also diagonalized a quadratic Hamiltonian in $\widetilde{b}_x$ by a generalized Bogoliubov transformation of the form
 \begin{align}\label{eq:T_1'''-intro}
{T}_1''' = \exp \left(\frac{1}{2} \iint \widetilde {\mathfrak{K}}_1 (x,y) \widetilde{b}^*_x \widetilde{b}^*_y \dd x \dd y- \hc  \right). 
\end{align}
Note that ${T}_1'''$ is similar to $T_1''$ up to the transformation $U$. In fact, ${T}_1'''$ is a proper unitary operator on $\gH^N$, which helps to avoid the incompatibility problem problem caused by the fact that standard Bogoliubov transformations do not preserve the number of particles.  

In the present paper, we solve the incompatibility  issue at the level of the Hamiltonian rather than the unitary:  first we extend $U H_N U^*$ to an operator on the full Fock space as in \eqref{eq:UHU-intro}, and then we choose $T_1$ as a standard Bogoliubov transformation. In this way, the exact formula \eqref{eq:T1-intro} simplifies several computations. At the very end (after implementing further transformations $T_c,T_2$), we will project the resulting operator back to the truncated Fock space $\cF_+^{\leq N}$ by a localization method on Fock space which was proposed by Lieb--Solovej  in \cite{LieSol-01} and developed further in \cite{LewNamSerSol-15}. This argument is inspired by the general approach in \cite{LewNamSerSol-15} which was initially introduced to study the excitation spectrum in the mean-field regime. 

%All this leads to a substantial  simplification compared to the use of $T_1''$. 

Another simplification of $T_1$ comes from the choice of the kernel $s_1(x,y)$, which is based on the simple truncated function $\omega_{\ell,N}$ of the global scattering solution. In  \cite{BenOliSch-15b,BreSch-19,BocBreCenSch-17,BocBreCenSch-17b,BocBreCenSch-20,BocBreCenSch-19,BreSchSch-21,BreSchSch-21b}, the quadratic kernels of $T_1',T_1''$ are constructed using a localized scattering solution on $|x|\leq \ell$ with Neumann boundary condition. This is conceptually the same with our choice, but technically more complicated since controlling the Neumann solution requires some additional regularity on $V$ (for example $V\in L^3$). Here our simple truncated scattering solution can be used for a larger class of potentials.  Moreover, our truncated solution exactly agrees with the real one on the support of the interaction, namely $N^2 V_N\omega_{\ell,N} = N^2 V_N \omega_N$, so that the real scattering energy appears immediately, thus making the computation simpler. % (possibly extendable to hard sphere potentials). 

The last, and conceptually the most important advantage of $T_1$ is based on the choice $\ell \ll 1$. In \cite{BenOliSch-15b,BreSch-19,BocBreCenSch-17,BocBreCenSch-17b,BocBreCenSch-20,BocBreCenSch-19,BreSchSch-21,BreSchSch-21b}, the cut-off of the scattering solution is taken as $\ell \sim 1$, for which the Bogoliubov transformation $T_1'$, $T_1''$ renormalize the pairing term $\mathcal{L}_2$ to a new term of order $1$, but also change the cubic term $\mathcal{L}_3$ by a quantity of order $1$. This is sufficient to get the optimal rate of BEC, but complicates the analysis of the excitation spectrum. Our choice of $\ell\ll 1$ ensures that the transformation $T_1$ renormalizes $\mathcal{L}_2$ in a more economical way (the new quadratic  term will be of order $\ell^{-1}$ which is larger than 1 but sufficient for us) and at the same time leaves the cubic terms invariant. This allows us to  simplify greatly not only the analysis of $T_1$, but also the choices of $T_c$ and $T_2$ in the next steps. In fact a similar choice was exploited in the papers \cite{CarCenSch-21,CarCenSch-22} dealing with the two dimensional Gross-Pitaevskii regime, and it also allows to analyse the three dimensional Bose gas beyond the Gross-Pitaevskii limit, as it has been done in \cite{BreCapSch-21,AdhBreSch-21}.

\bigskip
\noindent
{\bf Cubic transformation $T_c$.}  To remove the cubic term $\mathcal{L}_3$ from \eqref{eq:T1H-intro}, 
%$$\mathcal{L}_3\approx \int N^{5/2} V_N (x-y) \varphi(x) a^*_x a^*_y a_y,$$ 
we apply a cubic transformation of the form  
\begin{align*}
T_c = e^{S}, \quad S= \theta_M \iint k_{c}(x,y,y') a^*_x a^*_y a_{y'} \dd x \dd y \dd y' - \hc   
\end{align*}
where $\theta_M \approx \1(\cN\leq M)$ and 
\begin{align*}
k_{c} = Q^{\otimes 2} \widetilde k_{c} Q, \quad \widetilde k_c (x,y,y') = - N^{1/2}\varphi(x) \omega_{\ell,N}(x-y)\delta_{y,y'}. 
\end{align*}
The cut-off parameter $1\ll M \ll N$ is introduced to ensure that the action of $T_c$ can be extracted  by simple commutator estimates of the form
$$
T_c^* A T_c \approx A - [S, A] + \frac{1}{2}[S,[S,A]].
$$
In particular, the kernel $S$ of  $T_c$ has been chosen to gain a cancelation from 
$$\mathcal{L}_3 - [ S, \dd\Gamma(-\Delta) + \mathcal{L}_4] \approx 0.$$
Moreover, $T_c$ also replaces the potential $V_N$ in the second line of \eqref{eq:T1H-intro} by $\varepsilon_{\ell,N}$, namely
\begin{align*}
&\dd\Gamma\Big(N^3 V_N*\varphi^2+\varphi(x)  N^{3}V_N(x-y) \varphi(y)\Big)   + [S, \mathcal{L}_3] + \frac{1}{2}[S,[S, \dd \Gamma(-\Delta) + \mathcal L_4]] \\
&\qquad  \approx \dd\Gamma \Big( N^{3}\varepsilon_{\ell,N}*\varphi^2 + \varphi(x) N^{3}\varepsilon_{\ell,N} (x-y)  \varphi(y)\Big). 
\end{align*}
In Lemma \ref{prop:recap_cubic_transform}, we prove that up to an error $o(1)$, 
 \begin{align} \label{TcT1H-intro}
T_{c}^*T^*_1\mathcal H T_1 T_{c} &\approx N \cE_{\rm GP}(\varphi) -4\pi \ao \int \varphi^4  -  \frac{N^4}{2} \int (\omega_{\ell,N} \varepsilon_{\ell,N} \ast \varphi^2) \varphi^2 \\
&\quad + \dd\Gamma(D) + \frac{1}{2}  \int N^{3}\varepsilon_{\ell,N} (x-y) \varphi(x) \varphi(y)  (2a^*_x a_y + a^*_x a^*_y + a_x a_y )  \dd x \dd y. \nn
\end{align}
This essentially places us in the mean-field regime.

{\em Comparison to the literature.} The cubic transformation is a decisive tool introduced by Boccato--Brennecke--Cenatiempo--Schlein \cite{BocBreCenSch-19} to study the excitation spectrum for dilute Bose gases. It is well-known that quadratic transformations alone are insufficient to access the second order expansion of the ground state energy of dilute gases, even for an upper bound,  see \cite{ErdSchYau-08,NapReuSol-18}. The idea of using a cubic generator was initiated by Yau--Yin \cite{YauYin-09} when they constructed a trial state for the sharp second order upper bound to the Lee--Huang--Yang formula in the thermodynamic limit (see \cite{BasCenSch-21} for a simplified proof of the upper bound, and \cite{FouSol-20} for the matching lower bound). The nice property of the cubic transformation in \cite{BocBreCenSch-19} is that it is a unitary transformation, thus bringing up the upper and lower bounds simultaneously. Note that the cubic transformation used in \cite{BocBreCenSch-19} is slightly different from ours, since in \cite{BocBreCenSch-19} the cubic term $\mathcal{L}_3$ has already been modified by their first quadratic transformation $T_1''$. Here our choice of $T_c=e^S$ is simpler since we do not need to add terms with three creation operators in the kernel $S$. Moreover, we also introduce the cut-off $\theta_M \approx \1(\cN\leq M)$ in $S$ which helps a lot to control several estimates. On the other hand, choosing $\ell \ll 1$ in the kernel $\varphi(x) \omega_{\ell,N}(x-y)$ in $S$ is conceptually the same with separating high and low momenta as in \cite{BocBreCenSch-19}.\footnote{In principle, the cut-off parameter $\ell$ in $T_c$ is not necessarily the same with that of $T_1$, but we take the same choice to simplify the notation.}

\bigskip
\noindent
{\bf Second quadratic transformation $T_2$.} Finally we diagonalize the quadratic Hamiltonian
$$
\dd\Gamma(D)   + \frac{1}{2}  \int N^{3}\varepsilon_{\ell,N} (x-y) \varphi(x) \varphi(y)  (2a^*_x a_y + a^*_x a^*_y + a_x a_y ) \dd x \dd y.
$$
Since we are essentially in the mean-field regime, we will take a standard Bogoliubov transformation $T_2$ similarly as in  \cite{GreSei-13}. However, in our case the analysis is more complicated since the convergence of the pairing operator $\varphi(x) N^3 \varepsilon_{N,\ell}(x-y) \varphi(y) \to 8\pi \ao \varphi^2$ holds in operator norm but not in the Hilbert-Schmidt norm (the limit is not a Hilbert-Schmidt operator). Hence, the comparison of the quadratic Hamiltonians has to be carried out  carefully. In Lemma \ref{prop:second_quadratic_transform}, we prove that  up to an error $o(1)$, 
\begin{align}\label{eq:T2TcT1H-intro}
T_2^* T_{c}^* T_1^* \mathcal H T_1 T_{c} T_2 
	&\approx N \cE_{\rm GP}(\varphi) -4\pi \ao \int \varphi^2  -  \frac{N^4}{2} \int (\omega_{\ell,N} \varepsilon_{\ell,N} \ast \varphi^2) \varphi^2 \nn \\
	&\quad +\frac{1}{2} \tr (E-D-K) + \dd\Gamma(E)   
	 \end{align}
where 
$$E= (D^{1/2}(D+2K)D^{1/2})^{1/2}, \quad K=Q\widetilde K Q, \quad \widetilde K(x,y)= \varphi(x) N^3  \varepsilon_{\ell,N} (x-y)  \varphi(y).$$
%Since the potential $\varepsilon_{\ell,N}$ is long-range ($\ell^1 N^{-1} \gg N^{-1}$),  the computation in this step is similar to that in the mean-field regime \cite{GreSei-13}. Note that the quadratic expression in \eqref{TcT1H-intro} is of order $O\ell^{-1}$, which is larger than that in the mean-field regime due to our choice $\ell\ll N$. However, this is controllable thanks  to the explicit diagonalization by $T_2$. 
This implies that the ground state energy  of $H_N$ satisfies 
\begin{align}
\lambda_1(H_N) &=  N \cE_{\rm GP}(\varphi) -\frac{1}{2}  \int N^3 ( V_Nf_N\ast \varphi^2) \varphi^2  -  \frac{N^4}{2} \int (\omega_{\ell,N} \varepsilon_{\ell,N} \ast \varphi^2) \varphi^2 \nn \\
	&\quad+\frac{1}{2} \tr (E-D-K)  + o(1)
	\end{align}
and that the eigenvalues  of $H_N$ satisfy 
$$
\lambda_L(H_N) - \lambda_1(H_N) = \lambda_L (\dd \Gamma(E))+ o(1), \quad L=1,2,...
$$
Note that the operator $E=(D^{1/2}(D+2K)D^{1/2})^{1/2}$ still depends on $N$ via $K$. However, thanks to the convergence $\varphi(x) N^3 \varepsilon_{N,\ell}(x-y) \varphi(y) \to 8\pi \ao \varphi^2$, we find that $E\to E_\infty$ in an appropriate sense (which is enough to deduce the convergence of the eigenvalues). Again the choice $\ell \ll 1$ is crucial in this step. This completes the overview of our proof of Theorem \ref{thm:main}.

\bigskip
\noindent
{\bf Optimal BEC.} Note that our approach implies easily the operator inequality 
\begin{align} \label{eq:opt-BEC-intro}
U H_N U^* \geq N\cE_{\rm GP}(\varphi) + C^{-1} \cN - C
\end{align}
for some constant $C>0$, which gives the optimal rate of BEC. To obtain this bound, we can use the above estimates with $\ell$ being of order $1$, and $M$ of order $N$ ($\ell$ and $M$ are cut-off parameter in $T_1,T_c$). As mentioned before, the bound \eqref{eq:opt-BEC-intro} has been proved in \cite{NamNapRicTri-21,BreSchSch-21} under slightly stronger conditions on $V$.  For the completeness, we will quickly derive this bound in Section \ref{sec:BEC}, which is useful to conclude the excitation spectrum in Section \ref{sec:conclusion}. Hence our proof only needs the standard BEC in \eqref{eq:BEC} as an input.

\bigskip

In summary, in view of \eqref{eq:all-uni-tran-intro}, the  excitation transformation $U$ implements Bogoliubov's c-number substitution, while the quadratic transformation $T_1$ and the cubic transformation $T_c$ renormalize the short-range potential $N^2 V(Nx)$ by a longer-range one. Thus we are essentially placed in the mean-field regime where the Hamiltonian can be effectively diagonalized by the second quadratic transformation $T_2$, leading to a rigorous implementation of Bogoliubov's approximation. 

\bigskip

As a final remark, let us mention that the idea of renormalizing short-range potentials by longer-range ones is  very natural and has been employed for a long time. In 1957, Dyson \cite{Dyson-57} proved that if $W$ is supported inside a ball $B(0,R)$, $U$ is radial, supported outside $B(0,R)$, and $\int U=8\pi \ao(W)$, then 
\begin{equation} 
\label{eq:Dyson} -2\Delta + W(x) \geq U(x)  \quad \text{ on }L^2(\R^3).
 \end{equation}
 Dyson's lemma and its generalizations are crucial tools to derive the leading order of the ground state energy in the thermodynamic limit in  \cite{Dyson-57,LieYng-98}, as well as to prove the complete BEC \eqref{eq:BEC} in \cite{LieSei-02,LieSei-06,NamRouSei-16}. Heuristically, the approximation \eqref{eq:all-uni-tran-intro} does the same job of Dyson's lemma, but it also gives access to the second order contribution of the Hamiltonian. We hope that the approach presented in the present paper will be applicable to a wide range of dilute Bose gases.

\bigskip
\noindent
{\bf Organization of the paper.} The proof of Theorem \ref{thm:main} occupies the rest of the paper. In Section \ref{sec:pre}, we will quickly recall some basic tools, including the property of the Gross--Pitaevskii theory, the scattering solutions, and quadratic transformations. Then we analyze the actions of the transformations $U$, $T_1$, $T_c$, $T_2$ in Sections \ref{sec:U}, \ref{sec:T1}, \ref{sec:Tc}, \ref{sec:T2}, respectively. In Section \ref{sec:BEC}, we quickly explain the implication of the optimal rate of BEC. Finally, we conclude the asymptotic behavior of the excitation spectrum in Section \ref{sec:conclusion}.

\bigskip
\noindent
{\bf Acknowledgements.} PTN thanks S\o ren Fournais and Christian Hainzl for helpful discussion. We thank the referees for many useful comments and suggestions. We received funding from the Deutsche Forschungsgemeinschaft (DFG, German Research Foundation) under Germany's Excellence Strategy (EXC-2111-390814868).

\section{Preliminaries}\label{sec:pre}

In this section we collect some well-known results which are useful for the rest of the paper. 

\subsection{Gross--Pitaevskii theory} \label{sec:GP} The following results are essentially taken from \cite{LieSeiYng-00}.

\begin{lemma}[Gross--Pitaevskii minimizer] \label{lemma:GPmin} Let $\Vext$ satisfy \eqref{eq:Vext-ass}. Then the followings hold true. 

\begin{itemize}

\item [(i)] The Gross--Pitaevskii functional in \eqref{eq:EGP} has a minimizer $\varphi \in H^1(\R^3)$, $\|\varphi\|_{L^2(\R^3)}=1$. The minimizer is unique up to a phase and can be chosen such that $\varphi>0$.  

\item [(ii)] The Gross--Pitaevskii equation \eqref{eq:GPe} holds in the distributional sense. 

\item[(iii)]
We have the uniform boundedness and exponential decay 
$$\max \left\{ \varphi(x), \Vext(x) \varphi(x), |\Delta \varphi(x)| \right\}  \leq M_t e^{-t|x|}, \quad \forall t>0.$$ 
\end{itemize}
\end{lemma}

\begin{proof} In \cite[Theorem 2.1 and Lemma A.5]{LieSeiYng-00} it was proved that if  $0\leq \Vext \in L^\infty_{\rm loc}(\R^3)$ and $\Vext(x)\to \infty$ as $|x|\to \infty$, then we have (i), (ii) and $\varphi(x) \leq M_t e^{-t|x|}$ for all $t>0$. Under the additional condition that $\Vext$ grows at most exponentially, we find that $\Vext (x)\varphi(x) \leq M_t e^{-t|x|}$ for all $t>0$. From the Gross--Pitaevskii equation, we get $|\Delta \varphi(x)| \leq M_t e^{-t|x|}$ for all $t>0$
\end{proof}

In the rest of the paper, we will repeatedly use that $\varphi,\Vext\varphi,  \Delta \varphi \in L^2 (\R^3) \cap L^\infty(\R^3)$. Moreover, since $\varphi>0$ is a ground state of $D=-\Delta+\Vext+ 8\pi \ao \varphi^2-\mu$ and $D$ has compact resolvent, $\varphi$ is a unique non-degenerate ground state for $D$. Hence, $D$ leaves $\gH_+=\{\varphi\}^\bot$ invariant and $D>0$ on  $\gH_+$.

\subsection{Truncated scattering solution} \label{sec:scattering}

It is well-known (see e.g. \cite{LieSeiSolYng-05})  that under our assumption on $V$, the variational problem \eqref{eq:def-a} has a unique minimizer $f$. Then the function $\omega=1-f$ solves the zero-scattering  equation 
\begin{align}
-\Delta \omega = \frac{1}{2} V (1-\omega) \text{ in }\R^3, \quad \lim_{|x|\to \infty} \omega (x)=0 \label{eq:scattering_equation}.
\end{align}
Consequently, from the definition  \eqref{eq:def-a} and the equation \eqref{eq:scattering_equation}, the scattering length can be formulated as
\begin{align} \label{eq:scattering_length_int}
8\pi \ao &= \int_{\R^3} \Big( 2|\nabla \omega|^2 +  V |1-\omega|^2 \Big) = \int_{\R^3} \Big( 2\omega (-\Delta \omega) +  V |1-\omega|^2 \Big)\nonumber\\
&=\int_{\R^3} \Big( \omega V (1-\omega) +  V |1-\omega|^2 \Big) = \int_{\R^3} V (1-\omega). 
\end{align}
Moreover, we can rewrite \eqref{eq:scattering_equation} as
\begin{align} \label{eq:omega-full-a}
\omega = \frac{1}{2}(-\Delta)^{-1} ( V(1-\omega)) = \frac{1}{8\pi |x|} * ( V(1-\omega)). 
\end{align}
Since $V(1-\omega)$ is radial, by Newton's theorem and \eqref{eq:scattering_length_int} we find that 
\begin{align} \label{eq:omega-full-a}
\omega(x) = \frac{\ao}{|x|} \quad \text{ for all $x$ outside the support of $V$}. 
\end{align}

We will need a modified version of $\omega$ with a cut-off. For $1\geq\ell \gg N^{-1}$, we define 
$$\omega_{\ell,N} (x) = \chi ( x/\ell) \omega(Nx)$$
where $0\leq \chi \leq 1$ is a  smooth radial function satisfying $\chi(x) =1$ if $|x|\leq 1/2$ and $\chi(x) = 0$ if $|x| \geq 1$. Using the notation $V_N=V(N\cdot)$, $\omega_N=\omega(N\cdot)$ and $\chi_\ell = \chi(\cdot /\ell)$ we can write 
\begin{align}
	\label{eq:scattering_equation_truncated}
-2\Delta \omega_{\ell,N} =  N^2 V_N (1-\omega_{\ell,N}) -  N^2 \varepsilon_{\ell,N}
\end{align}
with
\begin{align}
	\label{eq:def_omega_ell}
N^2 \varepsilon_{\ell,N} = 2\Delta (\omega_{\ell,N} -   \omega_N) = 4 \nabla \omega_N \cdot \nabla \chi_\ell + 2 \omega_N \Delta \chi_\ell. 
\end{align}
%
%
%,\quad \varepsilon_{\ell,N}= 2\Delta ( \omega_{\ell,N} (x) -\omega(Nx))
%
%$$\omega_{\ell} = \omega \chi_\ell, \quad \chi_\ell=\chi(|\cdot| \ell^{-1})$$
%with $\chi \in C^2(\mathbb{R}_+,[0,1])$ satisfying $\chi(t) =1$ if $t \leq 1/2$ and $\chi(t) = 0$ if $t \geq 1$. Thus
%\begin{align}
%	\label{eq:scattering_equation_truncated}
%-2\Delta \omega_\ell =  V (1-\omega) -  \varepsilon_\ell
%\end{align}
%with
%\begin{align}
%	\label{eq:def_omega_ell}
%\varepsilon_\ell = 2\Delta (\omega_\ell -   \omega) = 4 \nabla \omega \cdot \nabla \chi_\ell + 2 \omega \Delta \chi_\ell.
%\end{align}
%By scaling, one can check that $f_N:= f(N\cdot)$ satisfies
%\begin{align}
%-\Delta f_N + \frac{1}{2} N^2 V_N f_N = 0 \label{eq:scattering_equation_N}.
%\end{align}
%We denote $$\omega_{\ell,N} = \omega_{N\ell}(N\cdot), \quad \varepsilon_{\ell,N} = \varepsilon_{N\ell}(N \cdot).$$
%$$\omega_{\ell} = \omega \chi_\ell, \quad \chi_\ell=\chi(|\cdot| \ell^{-1})$$
%with $\chi \in C^2(\mathbb{R}_+,[0,1])$ satisfying $\chi(t) =1$ if $t \leq 1/2$ and $\chi(t) = 0$ if $t \geq 1$. Thus
%\begin{align}
%	\label{eq:scattering_equation_truncated}
%-2\Delta \omega_\ell =  V (1-\omega) -  \varepsilon_\ell
%\end{align}
%with
%\begin{align}
%	\label{eq:def_omega_ell}
%\varepsilon_\ell = 2\Delta (\omega_\ell -   \omega) = 4 \nabla \omega \cdot \nabla \chi_\ell + 2 \omega \Delta \chi_\ell.
%\end{align}
%
%
%
Note that both $\omega_{\ell,N}$ and $\varepsilon_{\ell,N}$ are radial. Moreover, they satisfy the following properties.

\begin{lemma}[Properties of $\varepsilon_{\ell,N}$ and $\omega_{\ell,N}$] 
	\label{prop:prop_epsilon}
	Assume that $V$ is supported in $B(0,R_0)$. Then for all $\ell > 2 R_0 N^{-1}$, we have for all $x\in \R^3$, 
\begin{align}
	\label{eq:w-pointwise}
|\omega_{\ell,N} (x)| \leq  \frac{C \1_{ \{|x| \leq \ell \} }}{|Nx|+1},\quad |\nabla \omega_{\ell,N} (x) | \leq \frac{C N \1_{\{|x|\leq \ell \}}}{|Nx|^2+1}, \quad N^3 |\varepsilon_{\ell,N} (x)| \leq C \ell^{-3} \mathds{1}_{ \{|x| \leq \ell \}}.
\end{align}
Moreover, 
\begin{align}
\|\omega_{\ell,N} \|_{L^1} &\leq C N^{-1} \ell^2 , \qquad \|\omega_{\ell,N} \|_{L^2} \leq C N^{-1} \ell^{1/2},  \label{item:prop:prop_epsilon_0} \\
\|\nabla \omega_{\ell,N} \|_{L^1} &\leq C N^{-1} \ell, \qquad \|\nabla \omega_{\ell,N} \|_{L^2} \leq C N^{-1/2} \label{item:prop:prop_epsilon_1},
\end{align}
and
\begin{align}\label{item:prop:prop_epsilon_1-Vwe} \int_{\R^3} N^3 \varepsilon_{\ell,N}  =  8 \pi \ao\end{align}
\end{lemma}

\begin{proof}[Proof of Lemma \ref{prop:prop_epsilon}] From \eqref{eq:omega-full-a} we get 
$$
0\leq \omega (x) \leq \frac{C}{|x|+1}, \quad
|\nabla \omega (x) | \leq \frac{C}{|x|^2+1}.
$$
Hence, using $\supp \chi_\ell \subset B(0,\ell)$ and $|\nabla \chi_\ell| \leq C\ell^{-1} \1_{\{\ell/2 \leq |x| \leq \ell \}}$ we get
%
%
%$\omega_{\ell} = \omega \chi(|\cdot| \ell^{-1})$ satisfies
%$$
%0\leq \omega_\ell (x) \leq \frac{C \1_{|x| \leq  \ell}}{|x|+1}, \quad
%|\nabla \omega_\ell (x) | \leq \frac{C \1_{|x|\leq \ell}}{|x|^2+1}.
%$$
%Here we used that $\supp \chi \subset B(0,\ell)$ and $|\nabla \chi_\ell| \leq C\ell^{-1} \1_{\ell/2 \leq |x| \leq \ell}$. By scaling $\omega_{\ell,N} = \omega_{N\ell}(N\cdot)$ satisfies
\begin{align} \label{item:prop:prop_epsilon_0-pw} 
0 \leq \omega_{\ell,N} (x) \leq \frac{C \1_{\{|x| \leq  \ell \}}}{|Nx|+1}, \quad
|\nabla \omega_{\ell,N} (x) | \leq \frac{C N \1_{\{|x|\leq \ell \}}}{|Nx|^2+1}.
\end{align} 
Similarly, combining this with $|\Delta \chi_\ell| \leq C\ell^{-2} \1_{\{\ell/2 \leq |x| \leq \ell\}}$ we obtain the last bound in \eqref{eq:w-pointwise} immediately from \eqref{eq:def_omega_ell}. The bounds \eqref{item:prop:prop_epsilon_0} and  \eqref{item:prop:prop_epsilon_1} follow from \eqref{item:prop:prop_epsilon_0-pw} and some simple computations. For example, 
\begin{align*}
\|\omega_{\ell,N} \|_{L^1} &\leq CN^{-1} \int \frac{\mathds{1}_{\{|x|\leq \ell \}}}{|x|} \dd x \leq C N^{-1} \ell^2.
\end{align*}
The other inequalities in \eqref{item:prop:prop_epsilon_0} and  \eqref{item:prop:prop_epsilon_1} are similar and we skip the details. 
%From the definition \eqref{eq:def_omega_ell}, we can combine \eqref{item:prop:prop_epsilon_0-pw} and 
%$$|\nabla \chi_\ell| \leq C \ell^{-1} \1_{\ell/2\leq |x|\leq \ell},\quad |\Delta \chi_\ell| \leq C\ell^{-2} \1_{\ell/2 \leq |x| \leq \ell}.$$
% All this gives, for $x\in \mathbb{R}^3$,
%\begin{align*}
%|\varepsilon_\ell (x)| \leq C\mathds{1}_{\ell/2 \leq |x| \leq  \ell} \left( \frac{\ell^{-1}}{|x|^2+1} +  \frac{\ell^{-2}}{|x|+1}\right) \leq C \ell^{-3}\mathds{1}_{\ell/2 \leq |x| \leq \ell}.
%\end{align*}
%By scaling, we obtain the pointwise bound for $\varepsilon_{\ell,N} = \varepsilon_{N\ell} (N\cdot)$ as in \eqref{eq:w-pointwise}. 
Finally, from the choice 
 $\ell>2R_0 N^{-1}$ we have the scattering equation
 $$
 (-2 \Delta \omega_N) \chi_\ell  = N^2 V_N (1-\omega_N)  \chi_\ell = N^2 V_N (1-\omega_N).
 $$
 Combining with  \eqref{eq:def_omega_ell}, \eqref{eq:scattering_length_int} and  the integration by parts, we get
\begin{align*}
N^3 \int_{\R^3}  \varepsilon_{\ell,N} &= N  \int_{\R^3} (4 \nabla \omega_N \cdot \nabla \chi_\ell + 2 \omega_N \Delta \chi_\ell)  = N  \int_{\R^3} (-2 \Delta \omega_N) \chi_\ell   \\
&=  N^3 \int_{\R^3} V_N (1-\omega_N)   =   \int_{\R^3} V (1-\omega)  = 8\pi \ao.
\end{align*}
%Here we have used the scattering equation $-2\Delta\omega_N = V_N (1-\omega_N) = V_N (1-\omega_N)  \chi_\ell $, where the latter identity follows from the choice 
% $\ell>2R_0 N^{-1}$. 
% 
%  we have $-\Delta \omega_N (1-\chi_{\ell}) = \frac{1}{2} Vf  (1-\chi_{N\ell})  = 0$ and also $-\Delta \omega \chi_{N\ell} = \frac{1}{2} V f$. Hence using integration by parts, we have
%\begin{align*}
%\int_{\R^3} N^3 \varepsilon_{\ell,N}  &= \int_{\R^3} 2 \Delta( \omega (\chi_{N\ell}(x)-1)) =  \int_{\R^3}  \Delta \omega (\chi_{N\ell}-1) + 2 \nabla\omega \nabla \chi_{N\ell} + \omega \Delta \chi_{N\ell}  \\
%&=  \int_{\R^3} \nabla\omega \nabla \chi_{N\ell} = \int_{\R^3} (-\Delta \omega) \chi_{N\ell} = \int_{\R^3} V f = 8\pi \ao.
%\end{align*}
%
\end{proof}

\subsection{General quadratic transforms} \label{sec:quadratic-transforms}

It is well-known (see e.g. \cite{Solovej-ESI-2014,BenPorSch-15}) that for every real, symmetric, Hilbert--Schmidt operator $s$ on $\gH$, there exists a unitary transformation $T$ on Fock space $\cF(\gH)$ such that 
	\begin{align} \label{item:prop:T_1_1}
	T^* a^*(f) T = a^*(c(f)) + a(s(f)), \quad \forall f\in \gH
	\end{align}
where $c=\sqrt{1+s^2}$. In fact, we can take the explicit formula 
\begin{align} \label{eq:def-T-exp}
T = \exp \left(\frac{1}{2} \int k(x,y) a^*_x a^*_y - \hc  \right),
\end{align}
where $k = \sh^{-1} s$. To prove (\ref{item:prop:T_1_1}) we use the Duhamel formula, noting $K = K^* - K^{\circ}$ with 
\begin{align*}
K^* = \frac{1}{2} \int k(x,y) a^*_x a^*_y, \qquad K^{\circ} = \frac{1}{2} \int k(x,y) a_x a_y,
\end{align*} 
we have for $f \in L^2(\mathbb{R}^3)$
\begin{align*}
T^* a^*(f) T = \sum_{n\geq 0} \frac{(-1)^n}{n!} \ad_{K}^{(n)}(a^*(f)),
\end{align*}
it remains to compute
\begin{align*}
\ad_{K}(a^*(f)) 
	&= [K,a^*(f)] = [K^* - K^\circ,a^*(f)] \\
	&= -\frac{1}{2}\int k(x,y) [a_x a_y, a^*(f)] = - a(k(f)).
\end{align*}
Repeating the argument and separating even and odd ranks, we obtain the Taylor expansion of $\sh(k)$ and $\ch(k)$. This leads to (\ref{item:prop:T_1_1}) with $\sh (k) = s$.

We have the following properties.

\begin{lemma}
	\label{prop:properties_quadratic_transform_1}
The transformation $T$ in \eqref{item:prop:T_1_1} satisfies that
	\begin{align} \label{item:prop:T_1_2}
		T^* (\mathcal N+1)^j T \leq C_j  (1+\|s\|_{\rm HS}^2)^{j} (\mathcal N+1)^j, \quad \forall j\geq1.
	\end{align}
	Here the constant $C_j$ is independent of $s$. 
\end{lemma}

\begin{proof}

Let us begin with the case $j=1$. We have, using (\ref{item:prop:T_1_1}) and the CCR,
\begin{align*}
T^* \mathcal N T 
	&= \int (a^*(c_{x}) +  a(s_{x}))(a(c_{x}) +  a^*(s_{x})) \nn\\
	&= \int \Big(  a^*(c_{x})a(c_{x}) + a^*(s_{x}) a(s_{x})  + a^*(c_{x})a^*(s_{x}) +a(c_{x}) a(s_{x}) + \|s_{x}\|^2_{L^2} \Big) \nn \\
	&= \|s\|_{\rm HS}^2 + \dd\Gamma(c^2 + s^2) + \Big[ \int (c s)(x,y) a^*_x a^*_y + \hc  \Big].
\end{align*}
By the Cauchy--Schwarz inequality 
\begin{align} \label{eq:CS-AB}
\pm (AB+B^* A^*) \leq \eps AA^* + \eps^{-1}B^*B, \quad \forall \eps>0
\end{align}
we have, for every Hilbert--Schmidt operator $K$ on $L^2(\R^3)$ with kernel $K(x,y)$,
\begin{align} \label{eq:Kxy-N}
&\pm \Big[ \int K(x,y) a^*_x a^*_y \dd x \dd y + \hc  \Big] = \pm  \int \Big[ (\cN+1)^{-1/2} a^*_x a^*_y K(x,y) (\cN+3)^{1/2}  + \hc   \Big] \dd x \dd y  \nn\\
&\leq \int  \Big[ \|K\|_{\rm HS}  (\cN+1)^{-1/2}a_x^* a_y^*  a_y a_x (\cN+1)^{-1/2} + \|K\|_{\rm HS}^{-1}  |K(x,y)|^2 (\cN+3)  \Big] \dd x \dd y  \nn\\
&\leq 3\|K\|_{\rm HS} (\cN+1).
\end{align}
Using $c^2 =  1 + s^2$, $\|c\|_{op} \leq C (1+ \|s\|_{op})$, $\|s\|_{op} \leq \|s\|_{\rm HS}$ we obtain 
\begin{align}
T^* \mathcal N T&= \|s\|_{\rm HS}^2 + \dd\Gamma(c^2 + s^2) + \Big[ \int (c s)(x,y) a^*_x a^*_y + \hc  \Big] \nn \\
	&\leq  \|s\|_{\rm HS}^2  + \mathcal N \|c^2 + s^2\|_{op} + 3\|c s\|_{\rm HS} (\mathcal N+1) \nn\\
	&\leq C_1 (1+\|s\|_{\rm HS}^2) (\mathcal N+1). \label{eq:lemma_second_quadra_transfo_3}
\end{align}
%
%
%where we used $c^2 =  1 + s^2$, $\|c\|_{op} \leq C (1+ \|s\|_{op})$, $\|s\|_{op} \leq \|s\|_{\rm HS}$ and the Cauchy--Schwarz inequality 
%\begin{align}
%&\pm \Big[ \int K(x,y) a^*_x a^*_y \dd x \dd y + \hc  \Big] =  \int \Big[ K(x,y) (\cN+1)^{1/2} (\cN+1)^{-1/2} a^*_x a^*_y + \hc   \Big] \dd x \dd y  \nn\\
%&\leq \int  \Big[ \|K\|_{\rm HS}^{-1}  |K(x,y)|^2 ( \cN+1) + \|K\|_{\rm HS}  (\cN+1)^{-1/2}a_x^* a_y^*  a_y a_x (\cN+1)^{-1/2} \Big] \dd x \dd y  \nn\\
%&\leq 2\|K\|_{\rm HS} (\cN+1).
%\end{align}
For $j\geq 2$, we use an iterative argument, assuming the inequality is known for $j-2$, we have
\begin{align*}
T^* (\mathcal N+1)^j T 
	&= (T^* (\mathcal N+1) T) T^* (\mathcal N+1)^{j-2} T (T^* (\mathcal N+1) T) \\
	&\leq C_{j-2} (1+\|s\|_{\rm HS}^2)^{j-2} (T^* (\mathcal N+1) T) (\mathcal N+1)^{j-2} (T^* (\mathcal N+1) T) \\
	&\leq C_{j}  (1+\|s\|_{\rm HS}^2)^{j} (\mathcal N+1)^{j}  
\end{align*}
where we used  (\ref{eq:lemma_second_quadra_transfo_3}), the Cauchy--Schwarz inequality and the bounds
\begin{align*}
\left\|\dd\Gamma(s^2) (\mathcal N+1)^{-1} \right\|_{op} 
	&\leq \|s^2\|_{op}  \leq \|s\|_{\rm HS}^2, \\
\left\|\int (c s)(x,y) a^*_x a^*_y (\mathcal N+1)^{-1} \right\|_{op} 
	&\leq C \|c s\|_{\rm HS} \leq C \|c\|_{op} \|s\|_{\rm HS} \leq C (1+ \|s\|_{op}) \|s\|_{\rm HS} .
\end{align*}
The proof of Lemma \ref{prop:properties_quadratic_transform_1} is complete.
\end{proof}

\section{The excitation Hamiltonian} \label{sec:U}

In this section, we will factor out the condensate mode $\varphi$ by using a unitary transformation introduced in   \cite{LewNamSerSol-15}. 
Denote 
\[ Q=\1-|\varphi \rangle\langle \varphi|, \quad \gH_+ = QL^2(\R^3), \quad \cF_+ = \cF(\gH_+), \quad \1_+^{\leq N}= \1(\cN\leq N) \1_{\cF_+}.\] 
As explained in \cite[Section 4]{LewNamSerSol-15}, we can define a unitary map $U=U(\varphi)$ from $\gH^N$ to $\cF_+^{\leq N}$ by 
\begin{equation} \label{eq:def-UN}
	U \left(  \sum_{k} \varphi^{\otimes N-k} \otimes_s \xi_k \right) = \bigoplus_{k=0}^N \xi_k, \quad \xi_k \in \gH_+^k. 
\end{equation}
Moreover, the actions of $U$ can be computed using the rules 
\begin{equation} \label{eq:actions-U}
	U a^*(\varphi)a(g_1) U^* =  \sqrt{N-\cN} a(g_1),  \quad U a^*(g_1)a(g_2) U^* = a^*(g_1)a(g_2), \quad \forall g_1,g_2\in \gH_+. 
\end{equation}
Roughly speaking, the transformation $U$ replaces $a^*(\varphi)$ and $a(\varphi)$ by $\sqrt{N-\cN}$, thus rigorously implementing Bogoliubov's c-number substitution. We have %The total actions of $U$ on the Hamiltonian $H_N$ are given by the following 

\begin{lemma}[Operator bound on Fock space] \label{lem:UHU*} The following operator identity holds on $\cF_+^{\leq N}$
\begin{align} \label{eq:UHU}
U H_N U^* = \1_+^{\leq N} \mathcal H \1_+^{\leq N} =  \1_+^{\leq N} \left(  \sum_{i=0}^4 \cL_i  +  \cE^{(U)} \right) \1_+^{\leq N}  
\end{align}
where $\cH=\sum_{i=0}^4 \cL_i  +  \cE^{(U)}$ is an operator on the full Fock space $\cF(\gH)$ with 
\begin{align*}
	\cL_0&=  (N-\cN) \int \left(|\nabla\varphi|^2 + \Vext |\varphi|^2 + \frac{1}{2}(N^3 V_N \ast \varphi^2)  \varphi^2 \right) - \frac{1}{2} (\cN +1) \int \left(N^3 V_N \ast \varphi^2\right)  \varphi^2 \\
	&\qquad+ \dd\Gamma \Big(-\Delta+V_{\rm ext} + N^3 V_N *\varphi^2 + N^3V_N(x-y) \varphi(x) \varphi(y)\Big),\\
	\cL_1&=\sqrt{N}   a  \left(  \left(-\Delta + V_{\rm ext} + N^3 V_N*\varphi^2\right)\varphi\right) + \hc , \\
	\cL_2&=\frac{1}{2} \left( 1 - \frac{\cN}{N} - \frac{1}{2N} \right)  \int N^3 V_N (x-y) \varphi(x) \varphi(y) a_x a_y  \dd x\, \dd y + \hc  ,\\
	\cL_3&= \sqrt{(1-\cN/N)_+} \int N^{5/2}V_N(x-y) \varphi(x) a^*_y a_{x} a_{y}  \dd x \dd y + \hc  , \\
	\cL_4&=  \int N^2 V_N(x-y)  a_x^* a^*_y a_{x} a_{y}   \dd x \dd y
\end{align*}
and $\cE^{(U)}$ satisfies the  quadratic form estimate  on $\cF_+$ 
\begin{align} \label{eq:EU}
\pm \cE^{(U)} \leq \eps \cL_4 + C\eps^{-1} \frac{(\cN+1)^2}{N^3}   + C \frac{(\cN+1)^{3/2}}{N^{1/2}}, \quad \forall \eps>0. 
\end{align}
\end{lemma}

Here we define $\cH$ on the full Fock space $\cF$ to avoid putting $Q$ everywhere (or equivalently to avoid using the annihilation and creation operators $c_x,c_x^*$ on $\cF_+$ as in the introduction). Nevertheless, for our application we are only interested in the restriction of $\cH$ on $\cF_+$. In particular, although  $\cE^{(U)}$ does not leave  $\cF_+$ invariant, the quadratic form estimate \eqref{eq:EU} is interpreted as $\pm \langle \xi, \mathcal E^{(U)} \xi\rangle \leq \langle \xi, A \xi\rangle$ for all $\xi \in \cF_+$, with $A$ the right side of \eqref{eq:EU}.

\begin{proof} From a straightforward computation using \eqref{eq:actions-U} as in \cite[Section 4]{LewNamSerSol-15} we obtain the operator identity \eqref{eq:UHU} with $\cE^{(U)}=\1_+^{\leq N} \widetilde{\cE}^{(U)} \1_+^{\leq N}$ where
\begin{align} \label{eq:EU-full}
&\widetilde{\cE}^{(U)}= \left[ \frac{\cN (\cN +1)}{2 N}  \int \left(N^3 V_N \ast \varphi^2\right)  \varphi^2 -  \frac{\cN}{N}  \dd\Gamma \Big(N^3 V_N *\varphi^2 + N^3V_N(x-y) \varphi(x) \varphi(y)\Big) \right] \nn \\
&+ \left[ N^{1/2} \left( \sqrt{1-\frac{\mathcal N}{N}} -1 \right)  a\Big( (-\Delta + V_{\textrm{ext}}) \varphi\Big) + \hc \right] \nn\\
	&+ \left[ N^{1/2} \left( \left (1-\frac{\mathcal N+1}{N} \right) \sqrt{1-\frac{\mathcal N}{N}} - 1 \right) a\Big((N^3V_N \ast \varphi^2) \varphi\Big) + \hc  \right]\nn\\
 &+ \left[ \frac{1}{2} \left( \sqrt{ \left( 1-\frac{\mathcal{N}+1}{N} \right) \left( 1-\frac{\mathcal{N}}{N} \right)} - 1 + \frac{\cN}{N} + \frac{1}{2N}\right)  \iint N^3V_N(x-y) \varphi(x) \varphi(y) a_x a_y  + \hc  \right] \nn \\
 &=: 	\cE^{(U,0)}+ \cE^{(U,1a)} + \cE^{(U,1b)} + \cE^{(U,2)}.
\end{align}

Now let us estimate $\cE^{(U,i)}$ term by term. 

 \subsubsection*{Analysis of $\cE^{(U,0)}$.} Using $\varphi \in L^2(\R^3) \cap L^\infty(\R^3)$ and $\int N^3 V_N = \int V$, we find that 
 $$
 \|N^3 V_N \ast \varphi^2 \|_{L^\infty} \leq C, \quad \|N^3V_N(x-y) \varphi(x) \varphi(y)\|_{op} \leq C, \quad 0\leq\int \left(N^3 V_N \ast \varphi^2\right)  \varphi^2 \leq C. 
 $$
Hence, 
\begin{align} \label{eq:EU-0-pre}
\pm \cE^{(U,0)} \leq \frac{C \cN^2}{N} \quad \text{ on }\cF. 
\end{align}
Restricting \eqref{eq:EU-0-pre} to the subspace $\cF_+^{\leq N}\subset \cF$ and using $\cN \le N$ on $\cF_+^{\leq N}$, we obtain 
\begin{align} \label{eq:EU-0}
\pm \cE^{(U,0)} \leq \frac{C \cN^2}{N} \leq C \frac{(\mathcal N+1)^{3/2}}{N^{1/2}}\quad \text{ on }\cF_+^{\leq N}. 
\end{align}

 \subsubsection*{Analysis of $\cE^{(U,1a)}$.} By the Cauchy--Schwarz inequality as in \eqref{eq:CS-AB} we get for all $\varepsilon>0$, on $\mathcal F^{\leq N}$
\begin{align*}
&\pm \cE^{(U,1a)} %= \pm \left[ N^{1/2} \left( \sqrt{1-\frac{\mathcal N}{N}} -1 \right)  a\Big((-\Delta + V_{\textrm{ext}}) \varphi)\Big) + \hc \right]\\
= \pm \left[ N^{1/2} \left( \sqrt{1-\frac{\mathcal N}{N}} -1 \right) (\cN+1)^{-1/4}   a\Big((-\Delta + V_{\textrm{ext}}) \varphi)\Big) \cN^{1/4} + \hc \right]\\
&\leq \eps \cN^{1/4} a^*\Big((-\Delta + V_{\textrm{ext}}) \varphi)\Big) a\Big((-\Delta + V_{\textrm{ext}}) \varphi)\Big) \cN^{1/4} + \eps^{-1} N \left( \sqrt{1-\frac{\mathcal N}{N}} -1 \right)^2 (\cN+1)^{-1/2}  \\
&\leq \eps \cN^{3/2} \| (-\Delta + V_{\textrm{ext}}) \varphi \|_{L^2}^2 +  \eps^{-1} N \left( \frac{\mathcal N} {N}\right)^2 (\cN+1)^{-1/2}  \leq \left( \eps + C \eps^{-1} N^{-1}\right) \cN^{3/2}.
\end{align*}
Here we used $a^*(g)a(g) \leq \cN \|g\|_{L^2}^2$ with $g=(-\Delta + V_{\textrm{ext}}) \varphi \in L^2(\R^3)$ and 
$$
|\sqrt{1- t} -1 | \leq C t, \quad \forall t\in [0,1].
$$
Taking $\eps=N^{-1/2}$ we obtain the quadratic form estimate on $\cF^{\leq N}$
\begin{align} \label{eq:EU-1a}
 \pm \cE^{(U,1a)}  \leq C \frac{\cN^{3/2}}{N^{1/2}}. 
\end{align}

 \subsubsection*{Analysis of $\cE^{(U,1b)}$.} Similarly to $\cE^{(U,1a)}$, we can bound for all $\varepsilon>0$ on $\cF^{\leq N}$
\begin{align*}
&\pm \cE^{(U,1b)} = \pm \left[ N^{1/2} \left( \left (1-\frac{\mathcal N+1}{N} \right) \sqrt{1-\frac{\mathcal N}{N}} - 1 \right) a\Big((N^3V_N \ast \varphi^2) \varphi)\Big) + \hc  \right]\\
%&= \pm \left[ N^{1/2} \left( \left (1-\frac{\mathcal N-1}{N} \right) \sqrt{1-\frac{\mathcal N-1}{N}} - 1 \right) (\cN+1)^{-1/4}a^*\Big((N^3V_N \ast \varphi^2) \varphi)\Big) (\cN+2)^{1/4} + \hc  \right]\\
&= \pm \left[ N^{1/2} \left( \left (1-\frac{\mathcal N+1}{N} \right) \sqrt{1-\frac{\mathcal N}{N}} - 1 \right) (\cN+1)^{-1/4} a\Big((N^3V_N \ast \varphi^2) \varphi)\Big) \cN^{1/4}+ \hc  \right]\\
&\leq \eps \cN^{1/4}  a^*\Big((N^3V_N \ast \varphi^2) \varphi)\Big) a\Big((N^3V_N \ast \varphi^2) \varphi)\Big) \cN^{1/4} \\
&\quad  + \eps^{-1} N  \left( \left (1-\frac{\mathcal N+1}{N} \right) \sqrt{1-\frac{\mathcal N}{N}} - 1 \right)^2  (\cN+1)^{-1/2}\\
&\leq  \eps (\cN+1)^{3/2}  \|(N^3V_N \ast \varphi^2) \varphi \|_{L^2}^2 +  \eps^{-1} N \left( \frac{\mathcal N +1} {N}\right)^2 (\cN+1)^{-1/2}\\
&\leq \left( \eps + C \eps^{-1} N^{-1}\right) (\cN+1)^{3/2}.
\end{align*}
Here we have used 
$$\| (N^3V_N \ast \varphi^2) \varphi \|_{L^2} \leq \| N^3V_N\|_{L^1} \|\varphi\|_{L^\infty}^2 \| \varphi \|_{L^2} \leq C$$
 and the elementary inequality 
$$
| (1-t-N^{-1})\sqrt{1- t} -1 | \leq C (t+N^{-1}), \quad \forall t\in [0,1].
$$
 Taking $\eps=N^{-1/2}$ we obtain the inequality on $\cF^{\leq N}$
\begin{align} \label{eq:EU-1b}
\pm \cE^{(U,1b)} \leq C \frac{(\cN+1)^{3/2}}{N^{1/2}}. 
\end{align}

 \subsubsection*{Analysis of $\cE^{(U,2)}$.} By the Cauchy--Schwarz inequality as in \eqref{eq:CS-AB}, we have for all $\varepsilon>0$, on $\cF^{\leq N}$
 \begin{align}\label{eq:EU-2}
 &\pm \cE^{(U,2)} 
% &= \pm \left[ \frac{1}{2} \left( \sqrt{ \left( 1-\frac{\mathcal{N}+1}{N} \right) \left( 1-\frac{\mathcal{N}}{N} \right)}  - 1 + \frac{\cN}{N} + \frac{1}{2N}\right)  \iint N^3V_N(x-y) \varphi(x) \varphi(y) a_x a_y   + \hc  \right] \nn\\
 \leq \eps \iint N^2 V_N(x-y)  a^*_x a^*_y a_x a_ y \dd x \dd y \nn \\
 &\quad + \eps^{-1} \iint N^4 V_N(x-y) \varphi(x)^2 \varphi(y)^2 \dd x \dd y   \left( \sqrt{ \left( 1-\frac{\mathcal{N}+1}{N} \right) \left( 1-\frac{\mathcal{N}}{N} \right)}  - 1 + \frac{\cN}{N} + \frac{1}{2N}\right)^2  \nn \\
 &\leq  \eps \cL_4 + \eps^{-1} CN  \left( \frac{\cN+1}{N^2}\right) ^2 , \quad \forall \eps>0. 
  \end{align}
 Here we have used the uniform bound $0\leq\int N^3 (V_N*\varphi^2)\varphi^2 \leq C$ and 
% $$
% 0\leq  \iint N^3 V_N(x-y) \varphi(x)^2 \varphi(y)^2 \dd x \dd y \leq \| \varphi\|_{L^\infty}^2 \|\varphi\|_{L_2}^2 \|N^3 V_N \|_{L^1} \leq C 
% $$ and 
 the elementary inequality 
 $$
 \left| \sqrt{ \left( 1-\frac{\mathcal{N}+1}{N} \right) \left( 1-\frac{\mathcal{N}}{N} \right)}  - 1 + \frac{\cN}{N} + \frac{1}{2N}\right| \leq C\frac{\cN+1}{N^2}, \quad \forall \cN \in \{0,1,...,N\}. 
 $$

Inserting \eqref{eq:EU-0}, \eqref{eq:EU-1a}, \eqref{eq:EU-1b} and \eqref{eq:EU-2} in \eqref{eq:EU-full}, we obtain the desired estimate \eqref{eq:EU}. 
\end{proof}

\section{The first quadratic transform} \label{sec:T1}

Let us recall  the notations in Section \ref{sec:scattering}. Let $0\leq f\leq 1$ be the minimizer for \eqref{eq:def-a} and denote $\omega = 1 - f$. Let $1\geq \ell \gg N^{-1}$ and denote 
$$
\omega_{\ell,N}(x)= \chi(\ell^{-1} x)\omega(Nx), \quad \eps_{\ell,N}(x)= 2\Delta (\omega_{\ell,N}(x) - \omega(Nx))
$$
where $\chi$ is a  smooth radial function satisfying $\chi(x) =1$ if $|x|\leq 1/2$ and $\chi(x) = 0$ if $|x| \geq 1$. Let  
$$\widetilde s_1 (x,y) = - N \omega_{\ell,N}(x-y) \varphi(x) \varphi(y) \in L^2(\R^6), \quad s_1 = Q^{\otimes 2} \widetilde s_1 \in \gH_+^2 \subset L^2(\R^6).$$
Since the function $s_1(x,y)$ is real, symmetric and in $L^2(\mathbb{R}^6)$, it is the kernel of a real, symmetric, Hilbert-Schmidt operator on $L^2(\R^3)$ that we also denote by $s_1$. Then as discussed in Section \ref{sec:quadratic-transforms}, we can find a unitary transform $T_1$ on Fock space $\cF(\gH)$ such that
\begin{align}  \label{eq:T1-def-tec-sec}
T_1^* a^*(g) T_1 = a^*(c_1(g)) + a(s_1(g)), \quad \forall g\in \gH,
\end{align}
where $c_1 = \sqrt{1 + s_1^2}$. Note that $T_1$ is also a unitary transformation on $\cF_+$ since  $s_1$ is an operator on $\gH_+$. The role of this transformation is to extract the contribution of order $N$ from the Hamiltonian $\cH$ in Lemma \ref{lem:UHU*}. More precisely, we have

%\begin{align*}
%T_1 = \exp \left(\frac{1}{2} \int k_1(x,y) a^*_x a^*_y - \hc  \right).
%\end{align*}
%
%
% $$k_1 = \sh^{-1} s_1$$
% is also a real, symmetric, Hilbert-Schmidt operator on $L^2(\R^3)$, and we denote by $k_1(x,y)$ its kernel. Note that both $s_1$ and $k_1$ can be also interpreted as operators on $\gH_+$.  We can now define 
%Recall that the action of $T_1$ on the Fock space $\cF(\gH)$ is given by (see Lemma \ref{prop:properties_quadratic_transform_1})

\begin{lemma}
	\label{lemma:main_lemma_quadratic_transform}
We have the operator equality on $\cF$
$$
T^*_1 \mathcal H T_1 
	= \mathcal H^{(T_1)} + \mathcal E^{(T_1)} =  \cL_0^{(T_1)} + \cL_2^{(T_1)} + \cL_3^{(T_1)} + \cL_4 + \mathcal E^{(T_1)}
	$$
	where
	\begin{align*}
	\cL_0^{(T_1)} &= N \cE_{\rm GP}(\varphi) - 4\pi \ao \int \varphi^4   - \frac{1}{2}N \int N^3 ((\omega_{\ell,N} \varepsilon_{\ell,N}) \ast \varphi^2) \varphi^2 \\
	&\qquad+ \dd\Gamma \Big(-\Delta+V_{\rm ext} + N^3 V_N *\varphi^2 + \varphi(x) N^3V_N(x-y)  \varphi(y) - \mu\Big),\\
	\cL_2^{(T_1)} &=   \frac{1}{2}\int N^{3}\varepsilon_{\ell,N}(x-y) \varphi(x) \varphi(y) a^*_x a^*_y + \hc, \\
	\cL_3^{(T_1)} &= \int N^{5/2} V_N (x-y) \varphi(x) a^*_x a^*_y a_y + \hc,\\
	\cL_4 &= \frac{1}{2}\int N^{2} V_N (x-y) \varphi(x) a^*_x a^*_y a_x a_y,
\end{align*}
and  $\mathcal E^{(T_1)}$ satisfies the following quadratic form bound on $\cF_+$
\begin{align} \label{eq:cET1}
\pm \mathcal E^{(T_1)} 
	&\leq C  \ell^{1/2} (\mathcal N+1) +   C \ell^{3/2} \mathcal K+ C \frac{(\mathcal N+1)^{3/2}}{N^{1/2}} + C \frac{(\mathcal N+1)^{5/2}}{N^{3/2}} \nn\\
	&\qquad + \eps \left( \cL_4 + N^{-1} + \ell^{1/2} \frac{(\mathcal N+1)}{N^2} + \frac{(\mathcal N+1)^2}{N^3} \right)\nn \\
	&\qquad + \eps^{-1}C \left(\ell(\mathcal N+1) + \frac{(\mathcal N+1)^2}{N}+ \frac{(\mathcal N+1)^5}{N^4} \right)
\end{align}
for all $\eps>0$. Here $\cK=\dd\Gamma(Q(-\Delta+1)Q) +1$ and $\mu$ was defined in (\ref{eq:GPe}).
\end{lemma}
Again, although $\cE^{(T_1)}$ does not leave  $\cF_+$ invariant,  the quadratic form estimate \eqref{eq:cET1} is interpreted as $\pm \langle \xi, \mathcal E^{(T_1)} \xi\rangle \leq \langle \xi, A \xi\rangle$ for all $\xi \in \cF_+$, with $A$ the right side of \eqref{eq:cET1}.

The proof of Lemma \ref{lemma:main_lemma_quadratic_transform}  occupies the rest of the section. 

%\textcolor{red}{For this lemma, the value of $\mathcal H$ is given by what follows. The difference is in $\cL_0$, the $-\frac{\mathcal N}{N} \dd\Gamma(...)$ should be controlled by $\mathcal N^2 / N$, which could go to the error term from the beginning ?}
%\begin{align*}
%\mathcal H 
%	&= (N-\mathcal N) (\int |\nabla \varphi| ^2 + V_{\rm{ext}}|\varphi|^2) + \frac{1}{2} (N-\mathcal N)(N-\mathcal N-1) \int N^2 V_N \ast \varphi ^2 \varphi^2 \\
%	&\quad  + N^{1/2} a^*((-\Delta + V_{\textrm{ext}} + N^3V_N \ast \varphi^2) \varphi))) + \hc \\
%	&\quad + \dd\Gamma(-\Delta + V_{\rm ext} + N^3 V_N \ast \varphi^2 + \varphi \hat{V}(p/N) \varphi) +\frac{1}{2} (1 - \frac{\mathcal N-1}{N}) \int N^3V_N(x-y) \varphi(x) \varphi(y) a^*_x a^*_y + \hc \\
%	&\quad + \mathds{1}^{\{\mathcal N\leq N\}}\sqrt{1-\frac{\mathcal N}{N}} \int N^{5/2} V_N (x-y) \varphi(x) a^*_x a^*_y a_y + \hc \\
%	&\quad +  \frac{1}{2}\int N^2 V_N (x-y) a^*_x a^*_y a_{x} a_{y}.
%\end{align*}
%

\subsection{Properties of $s_1$} %Let us start by collecting some key estimates for the operator $s_1$. 	

\begin{lemma}[Properties of $s_1$] \label{lem:s1} We have 
\begin{align} \label{eq:estimate_b_without_proj} 
	|s_1(x,y) & + N \varphi(x) \varphi(y) \omega_{\ell,N}(x-y)|  \leq C \ell^2 \varphi(x) \varphi(y), \\		
%	\label{eq:estimate_b_without_proj_2}
%	|s_1(x,y)| &\leq C \varphi(x) \varphi(y) \left(C  \ell^2 + N \omega_{\ell,N} (x-y) \right)  \leq \nn\\
%	&\leq \varphi(x) \varphi(y) \left(C  \ell^2 + \frac{\mathds{1}_{|x-y| \leq C \ell N^{-1}}}{|x-y|} \right). 
		\label{item:prop_s2_3-sa}
	\|s_1\|_{op} &\leq C \ell^2, \qquad \|s_1\|_{L^\infty L^2} + \|s_1\|_{ L^2} \leq C \ell^{1/2}, \\
 \|\nabla_1 s_1\|_{op} &\leq C \ell , \qquad \|\nabla_1 s_1\|_{L^\infty L^2} + \|\nabla_1 s_1\|_{L^2} \leq C N^{1/2}, \label{item:prop_s2_3-sb}
\end{align}
Moreover, the operator $p_1:=c_1-1=\sqrt{1+s_1^2} -1 =s_1^2 (1+ \sqrt{1+s_1^2})^{-1}$ satisfies
\begin{align}	
\|p_1\|_{op} &\leq C \ell^4, \qquad \|p_1\|_{L^\infty L^2} + \|p_1\|_{ L^2} \leq C \ell^{5/2}, \label{item:prop_s2_3-pa} \\ 
\|\nabla_1 p_1\|_{op} &\leq C \ell^3 , \qquad \|\nabla_1 p_1\|_{L^\infty L^2} + \|\nabla_1 p_1\|_{L^2} \leq C \ell^{3/2}.\label{item:prop_s2_3-pb}
\end{align}
Here the constant $C>0$ is independent of $N$ and $\ell$. 

%	\begin{align}	
%	\|s_1\|_{op} &\leq C \ell^2, \qquad \|\nabla_1 s_1\|_{op} \leq C \ell \\
%	
%	\label{item:prop_s2_3}
%\begin{align*}
%\|s_1\|_{op} &\leq C \ell^2, \qquad \|\nabla_1 s_1\|_{op} \leq C \ell \\
%\|s_1\|_{L^\infty L^2} + \|s_1\|_{ L^2} &\leq C \ell^{1/2}, \qquad \|\nabla_1 s_1\|_{L^\infty_2 L^2_1} + \|\nabla_1 s_1\|_{L^2} \leq C N^{1/2}, \\
%\|p_1\|_{op} &\leq C \ell^4, \qquad \|\nabla_1 p_1\|_{op} \leq C \ell^3, \\ 
%\|p_1\|_{L^\infty L^2} + \|p_1\|_{ L^2} &\leq C \ell^{5/2}, \qquad \|\nabla_1 p_1\|_{L^\infty_2 L^2_1} + \|\nabla_1 p_1\|_{L^2} \leq C \ell^{3/2}, \\
%\end{align*}
%\end{enumerate}
\end{lemma}
For shortness we have denoted $\|s_1\|_{L^\infty L^2} = \max(\|s_1\|_{L^\infty_2 L^2_1},\|s_1\|_{L^\infty_1 L^2_2})$, etc.

\begin{proof}
By writing $s_1 = \widetilde s_1 - (P \otimes 1) \widetilde s_1 - (1 \otimes P) \widetilde s_1 + (P \otimes P) \widetilde s_1$, we obtain
\begin{align}
	\label{eq:s_1_expanded}
s_1(x,y)
	&= -N \varphi(x) \varphi(y) \left( \omega_{\ell,N}(x-y) -  \omega_{\ell,N} \ast \varphi^2(y) -  \omega_{\ell,N} \ast \varphi^2(x) - \int   \varphi^2 ( \omega_{\ell,N} \ast \varphi^2) \right).
\end{align}
This implies (\ref{eq:estimate_b_without_proj}) immediately
\begin{align*}
|s_1(x,y) + N \varphi(x) \varphi(y) \omega_{\ell,N}(x-y)| 
	&\leq \varphi(x) \varphi(y) N \|\omega_{\ell,N}\|_{L^1} \|\varphi\|_{L^\infty}^2 (1 + \|\varphi\|_{L^2}^2) \\
	&\leq C \ell^2 \varphi(x) \varphi(y).
\end{align*}

Next, we turn to (\ref{item:prop_s2_3-sa}). We have
\begin{align*}
\|s_1\|_{op} = \|Q\varphi(x) N \widehat{\omega}_{\ell,N}(p)\varphi(x)Q\|_{op} \leq \|\varphi\|_{L^\infty}^2 N \|\omega_{\ell,N}\|_{L^1} \leq C \ell^2.
\end{align*}
Here the last inequality is taken from Lemma \ref{prop:prop_epsilon}. Moreover, using $\|s_1\|_{L^2} = \|s_1\|_{\rm HS}$ and the H\"older inequality in Schatten space we get 
\begin{align*}
\|s_1\|_{L^2} \leq \|\varphi N^{1/2} |\omega_{\ell,N}|^{1/2}\|_{L^4}^2 \leq \|\varphi\|_{L^4}^2 N \|\omega_{\ell,N}\|_{L^2} \leq C \ell^{1/2}.
\end{align*}
Here the last inequality  is taken from  Lemma \ref{prop:prop_epsilon} again. The bound of $\|s_1\|_{L^\infty L^2}$ follows easily from (\ref{eq:s_1_expanded}) and Lemma \ref{prop:prop_epsilon}.
 Thus (\ref{item:prop_s2_3-sa}) holds true. 

Now consider 
\begin{align*}
\nabla_1 s_1 
	&= p Q\varphi(x) N \widehat{\omega}_{\ell,N}(p)\varphi(x)Q \\
	&= [p, Q] \varphi(x) N \widehat{\omega}_{\ell,N}(p)\varphi(x)Q + Q[p,\varphi(x)] N \widehat{\omega}_{\ell,N}(p)\varphi(x)Q \\
	&\quad + Q\varphi(x) N p \widehat{\omega}_{\ell,N}(p)\varphi(x)Q.
\end{align*}
Using that $\|[p,Q]\|_{op} \leq 2 \|\varphi\|_{L^2} \|\nabla \varphi\|_{L^2}$, that $[p,\varphi(x)] = -i\nabla \varphi(x)$ and the H\"older inequality in Schatten space, we obtain
\begin{align*}
\|\nabla_1 s_1 \|_{op} 
	&\leq 2 \|\varphi\|_{L^2} \|\nabla \varphi\|_{L^2} \|\varphi\|_{L^\infty} \|\omega_{\ell,N}\|_{L^1} + \|\nabla \varphi \|_{L^6} \|N \omega_{\ell,N}\|_{L^{6/5}} \|\varphi\|_{L^\infty}  +  \|\varphi\|_{L^\infty}^2 \|N \nabla \omega_{\ell,N}\|_{L^{1}} \\
	&\leq C (\ell^2 + \ell^{3/2} +\ell)  \leq C \ell,
\end{align*}
where we used Lemma \ref{prop:prop_epsilon} and the fact that $\|\omega_{\ell,N}\|_{L^{6/5}} \leq \| \omega_{\ell,N}\|_{L^{1}}^{2/3} \| \omega_{\ell,N}\|_{L^{2}}^{1/3}$. On the other hand, by applying $\nabla_1$ to the pointwise formula  (\ref{eq:s_1_expanded}) we easily obtain
\begin{align*}
\|\nabla_1 s\|_{L^\infty L^2} + \|\nabla_1 s\|_{L^2} &\leq C N (\|\varphi\|_{H^2}^4 +1)( \|\omega_{\ell,N}\|_{L^2} + \|\omega_{\ell,N}\|_{L^1} + \| \nabla \omega_{\ell,N}\|_{L^2} + \|\nabla \omega_{\ell,N}\|_{L^1})\\
	&\leq C N^{1/2}.
\end{align*}
Here we used Lemma \ref{prop:prop_epsilon} again. Thus \eqref{item:prop_s2_3-sb} holds true. 

It remains to prove the bounds on $p_1 = s_1^2 (1+\sqrt{1+s_1^2})^{-1}$. Since $0\le  (1+\sqrt{1+s_1^2})^{-1} \le 1$, clearly we have 
\begin{align*}
\|p_1\|_{op} &\leq C \|s_1\|^2_{op} \leq C \ell^4, \\
\| \nabla p_1\|_{op} &\leq C \|\nabla s_1\|_{op} \|s_1\|_{op} \leq C \ell^3, \\
\|p_1\|_{L^2} &\leq C \|s_1\|_{\rm HS} \|s_1\|_{op} \leq C \ell^{5/2}, \\
\|\nabla p_1\|_{L^2} &\leq C \|\nabla s_1 \|_{op} \|s_1\|_{\rm HS}  \leq C \ell^{3/2}.
\end{align*}
 Denoting $B = (1+\sqrt{1+s_1^2})^{-1}$, we have, for all $f\in L^2(\mathbb{R}^3)$ and $x\in \mathbb{R}^3$
\begin{align*}
\left| \int_{\mathbb{R}^3} p_1(x,y) f(y) \dd y \right| 
	&= \left|\int_{\mathbb{R}^3} s_1(x,u) (s_1B)(u,y) f(y) \dd u \dd y \right| \\
	&\leq \|s_1(x,\cdot)\|_{L^2} \|s_1\|_{op} \|B\|_{op} \|f\|_{L^2} \\
	&\leq C \|s_1\|_{L^\infty L^2} \|s_1\|_{op} \|f\|_{L^2} \\
	&\leq C \ell^{5/2} \|f\|_{L^2}.
\end{align*}
Hence $\|p_1\|_{L^\infty L^2} \leq C \ell^{5/2}$. Thus   \eqref{item:prop_s2_3-pa} holds true. 

Finally, we will use that if $A$ and $AB$ have kernels $\widetilde{a}$ and $\widetilde c$, then $\|\widetilde c\|_{L^\infty_1 L^2_2} \leq \|\widetilde a\|_{L^\infty_1 L^2_2} \|B\|_{op}$. Let us write
\begin{align*}
\nabla_1 p_1 
	= (\nabla_1 s_1) s_1 B &= [p, Q] \varphi(x) N \widehat{\omega}_{\ell,N}(p)\varphi(x)Q s_1 B + Q[p,\varphi(x)] N \widehat{\omega}_{\ell,N}(p)\varphi(x)Q s_1 B \\
	&\quad + Q\varphi(x) N \widehat{\omega}_{\ell,N}(p)[p,\varphi(x)]Q  s_1 B + Q\varphi(x) N \widehat{\omega}_{\ell,N}(p)\varphi(x) [p,Q]  s_1 B \\
	&\quad + s_1  (\nabla_1s_1) B.
\end{align*}
Using that 
\begin{align*}
|([p, Q])(x,y)| \leq C |\nabla \varphi(x) | \varphi(y) + \varphi(x) |\nabla \varphi(y)|, \\
([p,\varphi] N \widehat{\omega}_{\ell,N}(p)\varphi) (x,y) \leq C |\nabla \varphi(x)| \varphi(y) \frac{1}{|x-y|},
\end{align*}
and when necessary that $Q = 1 - P$ with $P(x,y) = \varphi(x) \varphi(y)$, we obtain
\begin{align*}
\|\nabla_1 p_1 \|_{L^\infty_1 L^2_2} 
	&\leq C \|s_1\|_{op} + C  \|s_1\|_{L^\infty L^2} \|\nabla_1 s_1\|_{op}  \leq C \ell^{3/2}, \\
\|\nabla_1 p_1 \|_{L^\infty_2 L^2_1}
    &=\|(\nabla_1 p_1)^\dagger \|_{L^\infty_1 L^2_2} \leq \|s_1\|_{L^\infty L^2} \|B\|_{op} \|\nabla_1 s_1\|_{op} \leq \ell^{3/2}.
\end{align*}
Thus \eqref{item:prop_s2_3-pb} holds true. The proof of Lemma \ref{lem:s1} is complete. 
\end{proof}

\subsection{Estimating $T^*_1 \cL_0 T_1$} 
	\label{sec:TL_0T} 
Recall 
\begin{align*}
	\cL_0&=  (N-\cN) \int \left(|\nabla\varphi|^2 + \Vext |\varphi|^2 + \frac{1}{2}(N^3 V_N \ast \varphi^2)  \varphi^2 \right) - \frac{1}{2} (\cN +1) \int \left(N^3 V_N \ast \varphi^2\right)  \varphi^2 \\
	&\qquad+ \dd\Gamma (-\Delta+V_{\rm ext} + N^3 V_N *\varphi^2 + N^3V_N(x-y) \varphi(x) \varphi(y)). 
\end{align*}

\begin{lemma}
	\label{lemma:TL0T} We have the operator equality on $\cF$
\begin{align*}
T^*_1 \cL_0 T_1 &= \cL_0 - \frac{N^4}{2}  \int  (\omega_{\ell,N}\varepsilon_{\ell,N} - V_N (1-\omega_N) \omega_{N} )(x-y)\varphi(x)^2 \varphi(y)^2  \\
&\quad +  \frac{N^3}{2}   \int \varphi(x) \varphi(y) (\varepsilon_{\ell,N} - V_N (1-\omega_N) )(x-y) (a^*_x a^*_y + a_x a_y) + \cE^{(T_1)}_{\mathcal L_0}
\end{align*}
where $ \cE^{(T_1)}_{\mathcal L_0}$ satisfies the quadratic form estimate on $\cF_+$
\begin{align*}
\pm \cE^{(T_1)}_{\mathcal L_0}
&\leq  C \ell^{3/2}\mathcal K + C \ell^{1/2}(\cN +1).
\end{align*}
\end{lemma}

\begin{proof} The constant of $\cL_0$ is unchanged by the conjugation with $T_1$. The  other terms of $\cL_0$ are of the form $\dd\Gamma(A)$ where  $A \in \{1, N^3 V_N \ast \varphi^2, N^3V_N(x-y)\varphi(x)\varphi(y),\Vext,-\Delta\}$.  

\bigskip

\noindent{\bf Case 1: } $A \in \{1, N^3 V_N \ast \varphi^2, N^3V_N(x-y)\varphi(x)\varphi(y),\Vext\}$. Let us prove that 
\begin{align} \label{eq:TL0-a}
\pm (T_1^* \dd\Gamma(A)T_1 -\dd\Gamma(A) ) \leq C \ell^{1/2}(\cN +1). 
\end{align}

If $A \in \{1, N^3 V_N \ast \varphi^2, N^3V_N(x-y)\varphi(x)\varphi(y)\}$, then $A$ is a bounded operator and $\|A\|_{op} \leq C$ uniformly in $N$.  Using \eqref{eq:T1-def-tec-sec} and the decomposition $c_1 = 1 + p_1$, we obtain, with $A(x,y)$ the kernel of $A$, 
\begin{align}	\label{eq:expansion_quadratic_1}
&T_1^* \dd\Gamma(A) T_1 
	= \iint A(x,y) (a^*(c_{1,x}) + a(s_{1,x}))(a(c_{1,y}) + a^*(s_{1,y})) \nn \\
	&= \tr (s_1 A s_1) + \dd\Gamma( c_1 A c_1 + s_1 A s_1) + \left[ \iint A(x,y) a^*(c_{1,x}) a^*(s_{1,y}) + \hc \right] \\
	&= \dd\Gamma(A) +  \tr (s_1 A s_1) + \dd\Gamma( p_1A p_1 + p_1 A+ Ap_1  + s_1 A s_1) + \left[ \iint (c_1 As_1) (x,y) a^*_x a^*_y + \hc \right]. \nn
\end{align}
From Lemma \ref{lem:s1}, it is straightforward to estimate the error on the right side of \eqref{eq:expansion_quadratic_1}. First,  using $\|s_1\|_{\rm HS} = \|s_1\|_{L^2(\mathbb{R}^6)} \leq C \ell^{1/2}$ we have
$$
|\tr (s_1 A s_1)| \leq \|A\|_{op} \|s_1\|_{\rm HS}^2 \leq C  \|A\|_{op} \ell. 
$$
Moreover, since $\|s_1\|_{op} \leq C \ell^{2}$ and $\|p_1\|_{op} \leq C \ell^4$, 
\begin{align*}
\pm \dd\Gamma (p_1A p_1 + p_1 A+ Ap_1  + s_1 A s_1) &\leq \cN \|p_1A p_1 + p_1 A+ Ap_1  + s_1 A s_1\|_{op} \\
&\leq  \cN \|A\|_{op} (2\|p_1\|_{op}+ \|p_1\|_{op}^2 + \|s_1\|_{op}^2) \leq C \cN \|A\|_{op}  \ell^{4}. 
\end{align*}
Finally, using $$
\|c_1 A s_1\|_{\rm HS} \leq \|c_1\|_{op} \|A\|_{op} \|s_1\|_{\rm HS} \leq C \ell^{1/2}
$$ and the Cauchy--Schwarz inequality we get 
\begin{align*}
&\pm \left[ \iint (c_1 A s_1) (x,y) a^*_x a^*_y + \hc \right] \leq \| c_1 A s_1\|_{\rm HS} (\cN +1) \leq C \ell^{1/2}(\cN +1). 
\end{align*}
% 
%
%Using that $c_1 = 1 + p_1$, with $\|p\|_{L^2(\mathbb{R}^6)}\leq C \ell^{5/2}, \|s_1\|_{L^2(\mathbb{R}^6)} \leq C \ell^{1/2},  \|s_1\|_{op} \leq C \ell^{2}$, we obtain using \textcolor{red}{[general Lemma on bounding operator on Fock space]} easily 
%\begin{align*}
%T_1^*\dd\Gamma(A) T_1 &= \dd\Gamma(A) + \mathcal E,
%\end{align*}
%with $\mathcal E$ satisfying for all $\xi,\xi' \in \FM$,
%\begin{align*}
%|\langle \xi',\mathcal E  \xi \rangle| 
%	&\leq C \|A\|_{op}(\|p_1\|_{op} + \|s_1\|_{op}^2 + \|s_1\|_{L^2} (1+ \|p_1\|_{L^2}) ) \|(\mathcal N+1)^{1/2} \xi'\| \|(\mathcal N+1)^{1/2} \xi\| \\
%	&\leq C \ell^{1/2} \|(\mathcal N+1)^{1/2} \xi'\| \|(\mathcal N+1)^{1/2} \xi\|.
%\end{align*}
Thus \eqref{eq:TL0-a} holds true when $A \in \{1, N^3 V_N \ast \varphi^2, N^3V_N(x-y)\varphi(x)\varphi(y)\}$. 

When $A = \Vext$, although $A$ is unbounded, the  decomposition (\ref{eq:expansion_quadratic_1}) also holds. We can adapt the proof of \eqref{eq:TL0-a} by using $p_1 = s_1 (1+\sqrt{1+s_1^2})^{-1} s_1 = s_1 B s_1$ where $\|B\|_{op}\leq C$ and that 
$$
\|\Vext s_1 \|_{op} \leq \|\Vext s_1 \|_{\rm HS} \leq C (N^{-1} \ell)^{1/2}.
$$
To verify the latter bounds, let us decompose
$$s_1 = Q \widetilde s_1 Q = \widetilde s_1 Q - P \widetilde s_1 Q$$
with $\widetilde s_1(x,y) = \varphi(x) \omega_{\ell,N}(x-y) \varphi(y)$. Using $\varphi,\Vext \varphi \in L^2\cap L^\infty$ we have
\begin{align*}
\| V_{\rm ext}\widetilde s_1 Q\|_{\rm HS}  &\leq \| V_{\rm ext}\widetilde s_1\|_{\rm HS} = \|\Vext(x) \varphi(x) N \omega_{\ell,N}(x-y) \varphi(y)\|_{L^2(\dd x \dd y)}\\
	&\leq \| V_{\rm ext} \varphi \|_{L^2} \|N \omega_{\ell,N}\|_{L^2}\|\varphi\|_{L^\infty} \leq C \ell^{1/2},\\
\| V_{\rm ext} P \widetilde s_1 Q\|_{\rm HS} &\leq \| V_{\rm ext} P\|_{op} \|P\|_{\rm HS} \|\widetilde s_1\|_{op} \leq \| V_{\rm ext} \varphi \|_{L^2} \|\varphi\|_{L^2}^2 \|\widetilde s_1\|_{op} \leq C \ell^2.
\end{align*}
Thus  \eqref{eq:TL0-a} also holds true with $A=\Vext$. 

\bigskip
\noindent
{\bf Case 2:} $A=-\Delta$.  For the kinetic term, we will prove that 
%\begin{align} \label{eq:T1K-main}
%T^*_1 \dd\Gamma(-\Delta) T_1
%	&= \dd\Gamma(-\Delta) - \frac{1}{2}  N^3  \int \varphi(x) \varphi(y) (V_N f_N)(x-y) (a^*_x a^*_y + a_x a_y) \nn \\
%	& \quad  +  \frac{N^4}{2}  \int  (V_N f_N \omega_{N})(x-y)\varphi(x)^2 \varphi(y)^2 - \frac{N^4}{2} \int (\omega_{\ell,N}\varepsilon_{\ell,N})(x-y) \varphi(x)^2 \varphi(y)^2 \nn \\
%	&\quad +   \frac{1}{2}\int N^3 \varepsilon_{\ell,N} (x-y) \varphi(x) \varphi(y) (a^*_x a^*_y + a_x a_y)  + \mathcal E 
%\end{align}
\begin{align} \label{eq:T1K-main}
T^*_1 \dd\Gamma(-\Delta) T_1
	&= \dd\Gamma(-\Delta) - \frac{N^4}{2}  \int  (\omega_{\ell,N}\varepsilon_{\ell,N} - V_N (1-\omega_N) \omega_{N} )(x-y)\varphi(x)^2 \varphi(y)^2  \nn \\
	& \quad  +  \frac{N^3}{2}   \int \varphi(x) \varphi(y) (\varepsilon_{\ell,N} - V_N (1-\omega_N) )(x-y) (a^*_x a^*_y + a_x a_y)  + \mathcal E 
\end{align}
on $\cF$, with $\mathcal E$ satisfying the quadratic form estimate on $\cF_+$
\begin{align*}
\pm \mathcal E  \leq C \ell^{3/2} \mathcal K + C \ell^{1/2} (\mathcal N+1).
\end{align*}

Indeed, using $T_1^* a^*_x T_1 = a^*(c_{1,x}) + a(s_{1,x})$, expanding and using the CCR, we obtain
\begin{align}
T^*_1 \dd\Gamma(-\Delta) T _1
	&= \int \nabla_x (a^*(c_{1,x}) + a(s_{1,x})) \nabla_x (a(c_{1,x}) + a^*(s_{1,x})) \nn \\
	&= \int \nabla_x a^*(c_{1,x}) \nabla_x a(c_{1,x})  + \int \nabla_x a^*(s_{1,x}) \nabla_x a(s_{1,x}) \nn \\
	&\quad + \int (\nabla_x a^*(c_{1,x})  \nabla_x a^*(s_{1,x}) + \nabla_x a(c_{1,x})  \nabla_x a(s_{1,x}))  + \| \nabla_x s_1\|_{L^2}^2\nn \\
	&= \int \nabla_x a^*_x \nabla_x a_x  + \int  (-\Delta_x s_1)(x,y) (a^*_x a^*_y + a_x a_y)  + \| \nabla_x s_1\|_{L^2}^2 \label{eq:exp_kine_quadra_transfor} \\
	&\quad  + \int \nabla_x a^*_x a(\nabla_x p_{1,x}) +  a^*(\nabla_x p_{1,x}) \nabla a_x + \int a^*(\nabla_x p_{1,x}) a(\nabla_x p_{1,x}) \nn \\
	&\quad + \int  a^*(\nabla_x s_{1,x})  a( \nabla_xs_{1,x}) + \int a^*(\nabla_x p_{1,x}) a^*(\nabla_x s_{1,x}) + a(\nabla_x p_{1,x}) a(\nabla_x s_{1,x}). \nn
\end{align}
The main contribution comes from the terms in the first line of (\ref{eq:exp_kine_quadra_transfor}). Let us first bound the other terms of (\ref{eq:exp_kine_quadra_transfor}). For all $\xi',\xi \in \mathcal F_+$,  by the Cauchy--Schwarz inequality we have 
\begin{align*}
|\langle \xi', \int \nabla_x a^*_x a(\nabla_x p_{1,x}) \xi \rangle|
	&\leq \int \| \nabla_x a_x \xi'\| \| a(\nabla_x p_{1,x}) \xi\|  \\
	& \leq \left( \int \| \nabla_x a_x \xi'\|^2 \right)^{1/2 }\left( \int \| a(\nabla_x p_{1,x}) \xi\|^2 \right)^{1/2 } \\	
&\leq 	 \|\mathcal K^{1/2} \xi'\| \|\nabla_1 p_1\|_{op} \|\mathcal N^{1/2} \xi\| \\
&\leq C \ell^{3} \|\mathcal K^{1/2} \xi'\| \|\mathcal N^{1/2} \xi\|.
\end{align*}
Here we have used Lemma \ref{lem:s1} to control $\|\nabla_1 p_1\|_{op}$. Note that for operators $A,B$ satisfying $\pm A \leq B$ is equivalent to $|\braket{\xi,A\xi'}| \leq C \|B^{1/2}\xi\|B^{1/2}\xi'\|$ for some constant $C>0$ and all $\xi,\xi' \in \mathcal F$. 

Similarly, for all $\xi',\xi \in \mathcal F_+$, we have
\begin{align*}
|\langle \xi', \int  a^*(\nabla_x p_{1,x}) a(\nabla_x p_{1,x}) \xi \rangle| 
	&\leq C \ell^{6} \|\mathcal N^{1/2}\xi'\| \|\mathcal N^{1/2}\xi\|, \\
|\langle \xi', \int  a^*(\nabla_x s_{1,x}) a(\nabla_x s_{1,x}) \xi \rangle|
	&\leq C \|\nabla_1 s_1\|_{op}^2 \|\mathcal N^{1/2}\xi'\| \|\mathcal N^{1/2}\xi\| \leq C \ell^{2} \|\mathcal N^{1/2}\xi'\| \|\mathcal N^{1/2}\xi\|,\\
|\langle \xi', \int a^*(\nabla_x p_{1,x}) a^*(\nabla_x s_{1,x}) \xi \rangle|
	&\leq \int \| a(\nabla_x s_{1,x})  \xi'\| \| a^*(\nabla_x p_{1,x}) \xi\| \\
	&\leq \|\nabla_1 s_1\|_{op} \|\nabla_1 p_1\|_{L^\infty_2 L^2_1} \|\mathcal N^{1/2}\xi'\| \|\mathcal N^{1/2}\xi\| \\
	&\leq \ell^{5/2} \|\mathcal N^{1/2}\xi'\| \|\mathcal N^{1/2}\xi\|.
\end{align*}

Now let us extract the main contribution of the terms in the first line of (\ref{eq:exp_kine_quadra_transfor}). The first term in (\ref{eq:exp_kine_quadra_transfor}) is $\dd\Gamma(-\Delta)$. For the second term, we decompose 
\begin{align}\label{eq:D-s1-dec}
-\Delta_1 s_1(x,y) = -\Delta_1 (Q^{\otimes 2}\widetilde{s_1})  = Q^{\otimes 2}  (-\Delta_1 \widetilde{s_1}) + ([\Delta, Q] \otimes Q) \widetilde s_1.
\end{align}
Since the error term $\mathcal E$ appearing in (\ref{eq:T1K-main}) is only estimated on $\mathcal F_+$, we can replace for free $Q^2$ by $1$ in the quadratic terms and we only to estimate the second term on the right-hand side above. Note that $[\Delta, Q] = |\varphi \rangle \langle \Delta \varphi| - |\Delta \varphi \rangle \langle \varphi|$ and $\varphi \in H^2$. Hence, 
$$
\|([\Delta, Q] \otimes Q) \widetilde s_1\|_{L^2(\R^6)}  = \| [\Delta, Q] \widetilde s_1 Q\|_{\rm HS}  \leq \|[\Delta, Q] \|_{\rm HS} \|\widetilde s_1\|_{op} \leq C \ell^2. 
$$
Consequently, for all $\xi',\xi \in \mathcal F_+$,
\begin{align*}
\Big|\Big\langle \xi', \int  \Big( ([\Delta, Q] \otimes Q) \widetilde s_1  \Big) (x,y) a^*_x a^*_y \xi \Big\rangle\Big| 
	&\leq C \ell^2 \|(\mathcal N+1)^{1/2}\xi'\| \|(\mathcal N+1)^{1/2}\xi\|.
\end{align*}
It remains to consider   
\begin{align} \label{eq:Del-1-s-dec}
-\Delta_1 \widetilde s_1 (x,y) &=  - N \varphi(x) (\Delta \omega_{\ell,N}) (x-y) \varphi(y)  - 2 N \nabla \varphi(x) (\nabla \omega_{\ell,N}) (x-y) \varphi(y) \nn\\
&\quad- N(\Delta \varphi)(x) \omega_{\ell,N}(x-y) \varphi(y).
\end{align}
The first term on the right side of \eqref{eq:Del-1-s-dec} is the one appearing in our claim \eqref{eq:T1K-main}. The second term is almost antisymmetric because of the $(\nabla \omega_{\ell,N})(x-y)$ factor and because $x \simeq y$, namely %. We have
%\begin{align*}
%& N \int \nabla \varphi (x) \varphi(y) \cdot (\nabla \omega_{\ell,N})(x-y) a^*_x a^*_y =  N \int \nabla \varphi (y) \varphi(x) \cdot (\nabla \omega_{\ell,N})(y-x) a^*_x a^*_y \\
%	&= - N \int \nabla \varphi (y) \varphi(x) \cdot (\nabla \omega_{\ell,N})(x-y) a^*_x a^*_y  \\
%	&= \frac{1}{2} N \int (\nabla \varphi (x) \varphi(y) - \nabla \varphi (y) \varphi(x)) \cdot (\nabla \omega_{\ell,N})(x-y) a^*_x a^*_y
%\end{align*}
\begin{align*}
N \int \nabla \varphi (x) \varphi(y) \cdot (\nabla \omega_{\ell,N})(x-y) a^*_x a^*_y = \frac{N}{2} \int (\nabla \varphi (x) \varphi(y) - \nabla \varphi (y) \varphi(x)) \cdot (\nabla \omega_{\ell,N})(x-y) a^*_x a^*_y.
\end{align*}
Using $|\nabla\omega_{\ell,N}(x-y)| \leq N^{-1}|x-y|^2 \mathds{1}_{\{|x-y|\leq C\ell\}}$ from Lemma \ref{prop:prop_epsilon}, we have for all $\xi',\xi \in \mathcal F_+$, 
\begin{align*}
&\Big|\Big\langle \xi',  N \int \nabla \varphi (x) \varphi(y) \cdot (\nabla \omega_{\ell,N})(x-y) a^*_x a^*_y \xi \Big\rangle\Big| \\
&= \Big|\Big\langle \xi',  \frac{N}{2} \int (\nabla \varphi (x) \varphi(y) - \nabla \varphi (y) \varphi(x)) \cdot (\nabla \omega_{\ell,N})(x-y) a^*_x a^*_y  \xi \Big\rangle\Big| \\
	&\leq  \eta \int \Bigg(  \frac{|\nabla \varphi(x) -\nabla \varphi(y)|^2}{|x-y|^4} \varphi(y)^2 + |\nabla \varphi(y)|^2 \frac{|\varphi(x) - \varphi(y)|^2}{|x-y|^4}\Bigg) C\ell^2  \\
	&\qquad + \eta^{-1} \int \frac{1}{|x-y|^2} \| (\mathcal N+1)^{-1/2} a_x a_y \xi'\|^2 \| (\mathcal N+1)^{1/2}\xi\|^2 \\
	&\leq C \eta \ell^2\big(\|\nabla \varphi\|_{H^{3/2}}^2 \|\varphi\|_{L^\infty}^2 + \|\varphi\|_{H^{1/2}}^2 \|\nabla \varphi\|_{L^\infty}^2 \big)   + \eta^{-1} C \int \frac{1}{|x-y|^2}  \| (\mathcal N+1)^{-1/2}a_x a_y \xi'\|^2 \| (\mathcal N+1)^{1/2}\xi\|^2 \\
	&\leq C\ell  \|\mathcal K^{1/2} \xi' \| \|(\mathcal N+1)^{1/2}\xi\|
\end{align*}
with an appropriate choice of $\eta>0$. Here we have used Hardy's inequality $|x-y|^{-2} \leq 4 (-\Delta_x)$, the fundamental inequality  
$$
\int_{\mathbb{R}^{3}\times \mathbb{R}^{3}} \frac{|u(x)-u(y)|^2}{|x-y|^{4}}\dd x \dd y \leq C \|u\|_{H^{1/2}}^2, \quad \forall u\in H^{1/2}(\R^3),
$$
%and that for any $u \in H^{1/2}(\mathbb{R}^{3})$, $\iint_{\mathbb{R}^{3}\times \mathbb{R}^{3}} \frac{|u(x)-u(y)|^2}{|x-y|^{4}} = C \|u\|_{H^{1/2}(\mathbb{R}^{3})}^2$. 
and the fact that $\nabla \varphi \in L^\infty(\mathbb{R}^{3})$ which is an easy consequence of Lemma \ref{lemma:GPmin}. For the third term in \eqref{eq:Del-1-s-dec}, we have, for all $\xi',\xi \in \mathcal F_+$,
\begin{align*}
&|\langle \xi',  \int N \Delta \varphi(x) \omega_{\ell,N}(x-y) \varphi(y) a^*_x a^*_y \xi \rangle|\\
&\leq  \|N \Delta \varphi(x) \omega_{\ell,N}(x-y) \varphi(y)  \|_{L^2(\R^6)} \|(\mathcal N+1)^{1/2} \xi' \| \| (\mathcal N+1)^{1/2}\xi\|\\
&\leq \| \Delta \varphi\|_{L^2} \|N \omega_{\ell,N}\|_{L^2} \|\varphi\|_{L^\infty} \|(\mathcal N+1)^{1/2} \xi' \| \| (\mathcal N+1)^{1/2}\xi\| \\
	&\leq C \ell^{1/2} \| (\mathcal N+1)^{1/2} \xi' \| \| (\mathcal N+1)^{1/2}\xi\|.
\end{align*}

For the constant term in (\ref{eq:exp_kine_quadra_transfor}), using \eqref{eq:D-s1-dec}, $Q s_1 = s_1 Q = s_1$ and the cyclicity of the trace, we have 
\begin{align}
	\label{eq:constant_term_kin_quart_1}
&\int  \| \nabla_x s_1\|_{L^2}^2 
	= \tr (s_1 (-\Delta) Q \widetilde s_1 Q) = \tr (s_1 (-\Delta) \widetilde s_1) + \tr (s_1 [-\Delta, Q] \widetilde s_1) \nn \\
	&= \int s_1(x,y) (-\Delta_1 \widetilde s_1)(x,y) + \int s_1(x,y) ([-\Delta_1,Q] \widetilde s_1)(x,y) \nn \\
	&= - \int N (\Delta \omega_{\ell,N}(x-y)) \varphi(x) \varphi(y)s_1(x,y) - 2 \int N \nabla \varphi(x) \nabla \omega_{\ell,N}(x-y) \varphi(y) s_1(x,y) \nn \\
	&\quad - N \int \Delta \varphi(x) \omega_{\ell,N}(x-y) \varphi(y) s_1(x,y) +\tr (s_1 [-\Delta, Q] \widetilde s_1)
\end{align}
The last three terms in \eqref{eq:constant_term_kin_quart_1}  are error terms. Let us estimate them. Using 
$$|s_1(x,y) + N \omega_{\ell,N}(x-y) \varphi(x) \varphi(y)| \leq C\ell^{2} \varphi(x) \varphi(y)$$
 and doing an integration by part, we obtain
\begin{align*}
&\Big|\int N \nabla \varphi(x) \nabla \omega_{\ell,N}(x-y) \varphi(y) s_1(x,y) \Big| \\
&\leq C \left| \int N^2 \nabla (\varphi^2)(x) \nabla(\varphi^2)(y) \omega_{\ell,N}^2(x-y) \right| + C  \ell^{2} \int N |\nabla \varphi(x)| \varphi(x) \varphi(y)^2  |\nabla \omega_{\ell,N}(x-y)| \\
	&\leq C \|\varphi\|_{H^2}^4 (N^2 \|\omega_{\ell,N}\|_{L^2}^2 + \ell^{2} N\|\nabla \omega_{\ell,N}\|_{L^1})  \leq C \ell.
\end{align*}
We have
\begin{align*}
\Big|N \int \Delta \varphi(x) \omega_{\ell,N}(x-y) \varphi(y) s_1(x,y)\Big| 
	\leq C \|s_1\|_{L^2} \|\varphi\|_{H^2}^2 N \|\omega_{\ell,N}\|_{L^2}  \leq C \ell,
\end{align*}
and
\begin{align*}
|\tr (s_1 [-\Delta, Q] \widetilde s_1)| 
	\leq C \|[-\Delta, Q]\|_{\rm HS} \|\widetilde s_1\|_{\rm HS} \|s_1\|_{op} \leq C \ell^{5/2}.
\end{align*}
To finish we only need to estimate the difference of replacing $s_1$ by $\widetilde s_1$ in the first term in (\ref{eq:constant_term_kin_quart_1}). Using again that 
$$|s_1(x,y) + N \omega_{\ell,N}(x-y) \varphi(x) \varphi(y)| \leq \ell^{-2} C \varphi(x) \varphi(y),$$
we obtain
\begin{align*}
\Big| \int N &(\Delta \omega_{\ell,N}(x-y))\varphi(x) \varphi(y) s_1(x,y) + \int N (\Delta \omega_{\ell,N}(x-y)) \omega_{\ell,N}(x-y)\varphi(x)^2\varphi(y)^2\Big| \\
	&\leq C \ell^2 \int N |\Delta \omega_{\ell,N}(x-y)| \varphi(x)^2 \varphi(y)^2 \\
	&\leq C \ell^2 \int N^3 (V_N f_N + |\varepsilon_{\ell,N}|)(x-y) \varphi(x)^2 \varphi(y)^2  \\ 
	&\leq C \ell^2 (\|Vf\|_{L^1} + N^3\|\varepsilon_{\ell,N}\|_{L^1}) \|\varphi\|_{L^4}^4  \leq C \ell^2,
\end{align*}
where we used (\ref{eq:w-pointwise}) to infer that $N^3\|\varepsilon_{\ell,N}\|_{L^1} \leq C$.  Thus using the equation (\ref{eq:scattering_equation_truncated}), we see that \eqref{eq:T1K-main} holds true. Combining with \eqref{eq:TL0-a}, we conclude the proof of   Lemma \ref{lemma:TL0T}.

\end{proof}

\subsection{Estimating $T^*_1 \cL_1 T_1$}
Recall 
\begin{align*}
	\cL_1=\sqrt{N}   a  \left(  \left(-\Delta + V_{\rm ext} + N^3 V_N*\varphi^2\right)\varphi\right) + \hc .
\end{align*}
\begin{lemma}
	\label{lemma:T1L1T_1} We have the operator equality on $\cF$
$$
T^*_1 \cL_1 T _1 = \sqrt N  \int N^{3} ((V_N \omega_N) \ast \varphi^2)(x) \varphi (x) (a(c_{1,x}) + a^*(s_{1,x})) + \hc  + \cE^{(T_1)}_{\cL_1} 
$$
where satisfies the quadratic form estimate on $\cF_+$
$$
\pm \cE^{(T_1)}_{\cL_1}  \leq C N^{-3/2}(\cN +1)^{1/2}. 
$$	
\end{lemma}

\begin{proof} Using $T^*_1 a^*_x T_1 = a^*(c_{1,x}) + a(s_{1,x})$   we have 
\begin{align*}
T_1 \cE^{(T_1)}_{\cL_1}T_1^*   &= \cL_1   - \Big( \sqrt N  \int N^{3} ((V_N \omega_N) \ast \varphi^2)(x) \varphi (x) a_x + \hc  \Big) \\
&=  \sqrt{N}   a  \left(  \left(-\Delta + V_{\rm ext} + N^3 (V_Nf_N)* \varphi^2\right)\varphi\right) + \hc  
\end{align*}
with $f_N=1-\omega_N$. Hence, combining with the Gross--Pitaevskii equation (\ref{eq:GPe})  we can bound
\begin{align*}
\pm  \1_{\cF_+} T_1 \cE^{(T_1)}_{\cL_1}T_1^* \1_{\cF_+}  &= \pm \sqrt{N}   a  \left( Q \left(-\Delta + V_{\rm ext} + N^3 (V_Nf_N)* \varphi^2\right)\varphi\right) + \hc  \\
&\leq \sqrt{N} \| Q \left(-\Delta + V_{\rm ext} + N^3 (V_Nf_N)* \varphi^2\right)\varphi\|_{L^2} (\cN +1)^{1/2}\\
&=  \sqrt{N} \|Q (N^3 (V_N f_N) \ast \varphi^2 -8\pi \ao \varphi^2)\varphi \|_{L^2} (\cN +1)^{1/2}\\
&\leq \sqrt{N} \|N^3 (V_N f_N) \ast \varphi^2 -8\pi \ao \varphi^2\|_{L^2 }\|\varphi \|_{L^\infty} (\cN +1)^{1/2}. 
\end{align*}
Since $Vf$ is integrable and compactly supported, its Fourier transform $\widehat{Vf}$ is $C^{\infty}$.  Moreover, $\widehat{Vf}(0)=\int Vf =8\pi \ao$ and 
$\nabla \widehat{Vf}(0)  = \int_{\mathbb{R}^{3}} -i x (Vf)(x) \dd x = 0$ since $Vf$ is even. Thus
\begin{align}\label{eq:VNfN-pN-0}
\|N^3 (V_N f_N) \ast \varphi^2 -8\pi \ao \varphi^2\|_{L^2 } &= \| ( \widehat {V f} (p/N) - \widehat {V f} (0))  \widehat{\varphi^2}\|_{L^2 }\nn\\
&\leq N^{-2} \| \nabla^2_q \widehat {V f} (q)\|_{L^\infty} \|  q^2 \widehat{\varphi^2}\|_{L^2 } \leq CN^{-2}. 
\end{align}
Hence, we conclude that 
\begin{align*}
\pm  \1_{\cF_+} T_1 \cE^{(T_1)}_{\cL_1}T_1^* \1_{\cF_+}  \leq C N^{-3/2}(\cN_+ +1)^{1/2}. 
\end{align*}
The desired estimate thus follows from Lemma \ref{prop:properties_quadratic_transform_1} and the fact that $T_1$ leaves $\cF_+$ invariant. 
\end{proof}

\subsection{Estimating $T_1^* \cL_2 T_1$}
Recall from Lemma \ref{lem:UHU*} that
\begin{align*}
\cL_2 =  \frac{1}{2} \int N^3V_N(x-y) \varphi(x) \varphi(y) a^*_x a^*_y \left( 1 - \frac{\cN}{N} - \frac{1}{2N} \right) + \hc
\end{align*}

\begin{lemma}
	\label{lemma:quadratic_term_quadratic_transfo}
We have
\begin{align*}
T^*_1 \cL_2 T _1
	&= \frac{1}{2} \int N^3V_N(x-y) \varphi(x) \varphi(y) a^*_x a^*_y + \hc \\
	&\quad + \left(\mathcal N + \frac{1}{2} - N \right) \int N^3(V_N \omega_N)(x-y) \varphi(x)^2 \varphi(y)^2 + \mathcal E^{(T_1)}_{\mathcal L_2},
\end{align*}
where $\mathcal E^{(T_1)}_{\mathcal L_2}$ satisfies the following quadratic form estimate on $\cF_+$, for all $\eta >0$,
\begin{align*}
\pm \mathcal E^{(T_1)}_{\mathcal L_2}
	&\leq C \ell^{1/2} (\mathcal N+1)+ C N^{-1} (\mathcal N+1)^2 + (\eta \cL_4 + \eta^{-1} N^{-1} (\mathcal N+1)^2).
%	& \leq C \ell^{1/2} (\mathcal N+1)+ C N^{-1/2}  (\mathcal N+1) \mathcal K.
\end{align*}
\end{lemma}

\begin{proof}[Proof of Lemma \ref{lemma:quadratic_term_quadratic_transfo}] 
Let us denote $T_1^*\mathcal N T_1 = \mathcal N_{T_1}$. Using 
\begin{align*}
T_1^* a_x^* a_y^* T_1&= \Big(a^*(c_{1,x}) + a(s_{1,x}) \Big) \Big( a^*(c_{1,y}) + a(s_{1,y})\Big)\nn\\
&=\Big( a^*(c_{1,x}) a^*(c_{1,y}) + \braket{s_{1,x},c_{1,y}}  + a^*(c_{1,x}) a(s_{1,y}) + a^*(c_{1,y}) a(s_{1,x})+ a(s_{1,x}) a(s_{1,y}) \Big)
\end{align*}
we can write 
\begin{align}
T_1^* \mathcal L_2^{(1)} T_1 &=\frac{1}{2}  \int N^3V_N(x-y) \varphi(x) \varphi(y) \Big( a^*(c_{1,x}) a^*(c_{1,y}) + \braket{s_{1,x},c_{1,y}}  + a^*(c_{1,x}) a(s_{1,y}) + \nn \\
	&\qquad \qquad  + a^*(c_{1,y}) a(s_{1,x}) + a(s_{1,x}) a(s_{1,y}) \Big) \dd x \dd y \times \left(1 - \frac{\mathcal N_{T_1}+1/2}{N}\right)  + {\rm h.c}.
	\label{eq:T1L_2T_1}
\end{align}
As we will see, the factor $(\mathcal N_{T_1}+1/2)/N$ will contribute only to the constant term containing $\braket{s_{1,x},c_{1,y}}$. 
%. As we will derive bounds using expectations against $\ket{\xi} \bra{\xi'}$, we can forget about it in the first place and add it later by changing state $\xi \to (1 - \frac{\mathcal N_{T_1}+1/2}{N}) \xi$. By using $\| (\mathcal N+1)^{-1/2} \mathcal N_{T_1} (\mathcal N+1)^{-1/2} \|_{op} \leq C$, we will show that its contribution in front of the quadratic terms (without the constant $\braket{s_{1,x},c_{1,y}}$) goes to  the error term.
Let us extract the main contribution from the quadratic terms. 

\begin{lemma}
	\label{lemma:estimate_quadratic_quadratic transform}
For all $g,h \in L^2(\mathbb{R}^6)$, $\widetilde g, \widetilde h \in L^2(\mathbb{R}^3)$, $\sharp_1,\sharp_2 \in \{*,\cdot\}$ (namely $a^\sharp$ is either creation or annihilation operator), $\xi',\xi \in \mathcal F$, we have
\begin{align}
& \Big | \Big\langle \xi', \int N^3V_N(x-y) \varphi(x) \varphi(y) a^{\sharp_1}(g_x) a^{\sharp_2}(h_y) \xi \Big\rangle \Big| \nn \\
	&\qquad \qquad \qquad \qquad \leq C \|V\|_{L^1} \|g\|_{L^2} \|h\|_{L^2} \| \varphi\|_{L^\infty}^2 \|(\mathcal N+1)^{1/2} \xi'\| \|(\mathcal N+1)^{1/2} \xi\|, \label{eq:lemma:estimate_quadratic_quadratic transform_1} \\
&\Big| \Big\langle \xi', \int N^3V_N(x-y) \varphi(x) \varphi(y) a^*_x a^{\sharp_1}(g_y) \xi \Big\rangle \Big| \nn \\
	&\qquad \qquad \qquad \qquad  \leq C \|V\|_{L^1} \|g\|_{L^2} \|\varphi\|_{L^\infty}^2 \|(\mathcal N+1)^{1/2} \xi'\| \|(\mathcal N+1)^{1/2} \xi\|, \label{eq:lemma:estimate_quadratic_quadratic transform_2} \\
&\Big| \Big\langle \xi',\int N^2 V_N(x-y) \varphi(x) \varphi(y) a^*_x a^*_y \xi \Big\rangle \Big| 
	\leq C N^{-1/2} \|V\|_{L^1}^{1/2}\|\varphi\|_{L^\infty} \| \cL_4^{1/2} \xi'\|\|\xi\|,\label{eq:lemma:estimate_quadratic_quadratic transform_3}  \\
& \Big|\int N^3 V_N(x-y) \varphi(x) \varphi(y) \widetilde g(x) \widetilde h(y)\Big| 
 	\leq C \|V\|_{L^1} \|\widetilde g\|_{L^2} \|\widetilde h\|_{L^2} \| \varphi\|_{L^\infty}^2. \label{eq:lemma:estimate_quadratic_quadratic transform_4}
\end{align}
\end{lemma}

\begin{proof}[Proof of Lemma \ref{lemma:estimate_quadratic_quadratic transform}] 
We will denote $a^{\overline \sharp} = (a^\sharp)^\dagger$. Let us prove (\ref{eq:lemma:estimate_quadratic_quadratic transform_1}), for all $\xi,\xi' \in \mathcal F_+$, we have
\begin{align*}
& \Big|\Big\langle \xi',\int N^3 V_N(x-y)\varphi(x) \varphi(y) a^{\sharp_1}(g_{x}) a^{\sharp_2} (h_{y})    \xi \Big\rangle \Big| \\
	&\leq \int N^3 V_N(x-y)\varphi(x) \varphi(y) \|a^{\overline \sharp_1}(g_{x}) \xi'\| \|a^{\sharp_2} (h_{y})\xi\| \\
	&\leq \int N^3 V_N(x-y)\varphi(x) \varphi(y) \|g_{x}\|_{L^2} \|(\mathcal N+1)^{1/2}\xi'\| \|h_{y}\|_{L^2} \|(\mathcal N+1)^{1/2}\xi\| \\
	&\leq  C \|V\|_{L^1} \|\varphi\|_{L^\infty}^2 \|g\|_{L^2} \|h\|_{L^2} \|(\mathcal N+1)^{1/2}\xi'\|  \|(\mathcal N+1)^{1/2}\xi\|.
\end{align*}

The proof of (\ref{eq:lemma:estimate_quadratic_quadratic transform_2}) is similar. Let us now deal with (\ref{eq:lemma:estimate_quadratic_quadratic transform_3}), for all $\xi,\xi' \in \mathcal F_+$ and $\eta > 0$, we have
\begin{align*}
&  \Big| \Big\langle \xi',\int N^2 V_N(x-y)\varphi(x) \varphi(y) a^{*}_x a^*_y \xi  \Big\rangle  \Big| \\
	&\leq C  \int N^2 V_N(x-y)\varphi(x) \varphi(y) \|a_x a_y \xi'\| \|\xi\| \\
	&\leq C \eta  \int N^2 V_N(x-y)  \|a_x a_y \xi'\|^2 + C\eta^{-1} \int N^2 V_N(x-y)\varphi(x)^2 \varphi(y)^2 \|\xi\|^2 \\
	&\leq C \eta  \|\mathcal L_4^{1/2} \xi'\|^2 + C \eta^{-1} N^{-1} \|V\|_{L^1} \|\varphi\|_{L^\infty}^2 \|\varphi\|_{L^2}^2 \|\xi\|^2,
\end{align*}
optimizing in $\eta$ proves the claim. The estimate (\ref{eq:lemma:estimate_quadratic_quadratic transform_4}) is a simple consequence of the H\"older inequality.
\end{proof}

Now come back to the proof of Lemma \ref{lemma:quadratic_term_quadratic_transfo}. Recall that $c_1=\sqrt{1+s_1^2}=1+p_1$, developing (\ref{eq:T1L_2T_1}) and using the Lemma \ref{lemma:estimate_quadratic_quadratic transform} we obtain 	
\begin{align}
T_1^* \mathcal L_2 T_1
	&= \frac{1}{2} \int N^3V_N(x-y) \varphi(x) \varphi(y) (a^*_x a^*_y + a_x a_y) \nn \\
	&\quad  + \int N^3V_N(x-y) \varphi(x) \varphi(y) s_1(x,y) \nn \\
	&\quad -(T^*_1 \mathcal N T_1+1/2) \int N^2V_N(x-y) \varphi(x) \varphi(y) s_1(x,y) +  \mathcal E,
		\label{eq:L_2_quadra_transfo}
\end{align}
with for all $\xi',\xi \in \mathcal F_+$,
\begin{align*}
| \langle \xi', \mathcal E  \xi \rangle | 
	&\leq C (\|p_1\|_{L^2} + \|p_1\|_{L^2}^2 + \|s_1\|_{L^2} + \|s_1\|_{L^2}^2 + \|s_1\|_{L^2} \|p_1\|_{L^2}) \times \\
	&\qquad \times (\| (\mathcal N+1)^{1/2} \xi'\| \| (\mathcal N+1)^{1/2} (1 - \frac{T^*_1 \mathcal N T_1+1/2}{N}) \xi\|) \\
	&\qquad + CN^{-1/2}\| \cL_4^{1/2} \xi'\|\|(T^*_1 \mathcal N T_1+1/2)\xi\|\\
	&\leq C \ell^{1/2} \| (\mathcal N+1)^{1/2} \xi'\| \| (\mathcal N+1)^{1/2} \xi\| +  C \ell^{1/2}N^{-1} \| (\mathcal N+1)^{1/2} \xi'\| \| (\mathcal N+1)^{3/2} \xi\| \\
	&\qquad  + C N^{-1/2} \| \cL_4^{1/2} \xi'\| \|(\mathcal N +1) \xi\|.
\end{align*}
Here, we used that $T_1^* (\mathcal N+1) T_1 \leq (\mathcal N+1)$. Let us now deal with the second term in (\ref{eq:L_2_quadra_transfo}), using that 
$$|s_1(x,y) + N \omega_{\ell,N}(x-y) \varphi(x) \varphi(y)| \leq C \ell^{2} \varphi(x) \varphi(y),$$
we obtain
\begin{align*}
\int N^3V_N(x-y) \varphi(x) \varphi(y)s_1(x,y) = - \int N^4(V_N \omega_N) (x-y) \varphi(x)^2 \varphi(y)^2 + \mathcal E,
\end{align*}
where
$$| \langle \xi', \mathcal E  \xi \rangle |  \leq C \ell^2 \|\xi'\| \|\xi\|, \quad \forall \xi',\xi \in \mathcal F_+.$$
Similarly, and using also that $T_1^* (\mathcal N+1)T_1 \leq C (\mathcal N+1)$, we have
\begin{align*}
&- (T^*_1 \mathcal N T_1+1/2) \int N^2V_N(x-y) \varphi(x) \varphi(y) s_1(x,y) \\
&= (T^*_1 \mathcal N T_1+1/2) \int N^3(V_N\omega_N)(x-y) \varphi(x)^2 \varphi(y)^2 +  \mathcal E
\end{align*}
where 
$$| \langle \xi' ,\mathcal E  \xi \rangle |  \leq C N^{-1} \| (\mathcal N+1)^{1/2} \xi'\| \| (\mathcal N+1)^{1/2} \xi\|, \quad \forall \xi',\xi \in \mathcal F_+.$$

Finally, expanding 
$$T^*_1 \mathcal N T_1 = \mathcal N + 2 \dd \Gamma(s_1^2) + \int a^*(c_{1,x}) a^*(s_{1,x}) + \hc + \|s_1\|_{L^2}^2,$$
we obtain
\begin{align*}
 &(T^*_1 \mathcal N T_1+1/2) \int N^3(V_N\omega_N)(x-y) \varphi(x)^2 \varphi(y)^2 \\
 &= (\mathcal N + 1/2) \int N^3(V_N\omega_N)(x-y) \varphi(x)^2 \varphi(y)^2 + \mathcal E,
\end{align*}
where
\begin{align*} | \langle \xi', \mathcal E  \xi \rangle |   &\leq C ((1+ \|p_1\|_{L^2})\|s_1\|_{L^2} + \|s_1\|_{L^2}^2 + \|s_1\|_{op}^2)  \| (\mathcal N+1)^{1/2} \xi'\| \| (\mathcal N+1)^{1/2} \xi\| \\
&\leq C \ell^{1/2} \| (\mathcal N+1)^{1/2} \xi'\| \| (\mathcal N+1)^{1/2} \xi\|, \quad \forall \xi',\xi \in \mathcal F_+.
\end{align*}
The proof of  Lemma \ref{lemma:quadratic_term_quadratic_transfo} is complete.
\end{proof}

\subsection{Estimating $T^*_1 \cL_3 T_1$} Recall from Lemma \ref{lem:UHU*} 
\begin{align*}
\cL_3 %&=  \sqrt{\left( 1-\frac{\mathcal N}{N} \right)_+} \int N^{5/2} V_N (x-y) \varphi(x) a^*_x a^*_y a_y + \hc  \\
	=  \int N^{5/2} V_N (x-y) \varphi(x) a^*_x a^*_y a_y  \sqrt{\left( 1-\frac{\mathcal N}{N} \right)_+} + \hc  
 \end{align*}

\begin{lemma}
	\label{lemma:cubic_term_quadratic_transfo}
We have
\begin{align*}
T^*_1\mathcal L_3 T_1 
	&=   \int N^{5/2} V_N (x-y) \varphi(x) (a^*_x a^*_y a_y + \hc) \\
	&\quad - \sqrt{N} \left(\int N^{3} ((V_N \omega_N) \ast \varphi^2)(x) \varphi (x) (a(c_{1,x}) + a^*(s_{1,x})) + \hc \right) + \mathcal E^{(T_1)}_{\mathcal L_3},
\end{align*}
with $\mathcal E^{(T_1)}_{\mathcal L_3}$ satisfying the following quadratic form estimate on $\cF_+$, with for all $\varepsilon > 0$, 
\begin{align*}
\pm \mathcal E^{(T_1)}_{\mathcal L_3} 
	&\leq C \frac{(\mathcal N+1)^{3/2}}{N^{1/2}} +  C  \frac{(\mathcal N+1)^{5/2}}{N^{3/2}} \\
	&\quad + \eps \cL_4 + \eps^{-1} C \Big\{ \ell (\mathcal N+1) + \frac{(\mathcal N+1)^{3/2}}{N^{1/2}} + \frac{(\mathcal N+1)^2}{N}\Big\}.
\end{align*}
\end{lemma}

\begin{proof} Let us denote $\mathcal N_{T_1} = T^*_1 \mathcal N T_1$. Using $T^*_1 a^*_x T_1 = a^*(c_{1,x}) + a(s_{1,x})$, and the CCR, we have
\begin{align} \label{eq:T1L3-oo}
T^*_1 \mathcal L_3 T _1
	&=   \int N^{5/2} V_N (x-y) \varphi(x) (a^*(c_{1,x}) + a(s_{1,x})) ( a^*(c_{1,y}) + a(s_{1,y}))\times \\ 
	&\qquad\qquad\qquad\times (a(c_{1,y}) + a^*(s_{1,y}))  \sqrt{\left( 1-\frac{\mathcal N_{T_1}}{N} \right)_+}  + \hc  \nn\\
	&= \int N^{5/2} V_N (x-y) \varphi(x) \Big( a^*(c_{1,x}) a^*(c_{1,y}) + \braket{s_{1,x},c_{1,y}}  + a^*(c_{1,x}) a(s_{1,y}) + a^*(c_{1,y}) a(s_{1,x})\nn\\
	&\qquad \qquad \qquad \qquad + a(s_{1,x}) a(s_{1,y}) \Big) (a(c_{1,y}) + a^*(s_{1,y}))   \sqrt{\left( 1-\frac{\mathcal N_{T_1}}{N} \right)_+}+ \hc   \nn
\end{align}

Before expanding \eqref{eq:T1L3-oo}, let us gather some simple estimates.
\begin{lemma}
	\label{lemma:estimate_cubic_quadratic transform}
For all $g,h \in L^2(\mathbb{R}^6)$, $\sharp_1,\sharp_2 \in \{*,\cdot\}$, $\xi,\xi'\in \mathcal F$, we have
\begin{align}
& \Big| \Big \langle \xi', \int  N^{5/2}V_N(x-y) \varphi(x)  a^*(g_x) a^*_y a(c_{1,y})  \xi  \Big\rangle  \Big| \nn \\ 
& \qquad \leq C N^{-1/2} \|V\|_{L^1} \|g\|_{L^\infty L^2}  \|\varphi\|_{L^\infty} \| (\mathcal N+1)^{3/4}\xi'\| \| (\mathcal N+1)^{3/4}\xi\|,  \label{eq:lemma:estimate_cubic_quadratic transform_1} \\
& \Big|  \Big\langle \xi', \int N^{5/2}V_N(x-y) \varphi(x)  a^*_x a^{\sharp_1}(h_y) a(c_{1,y})  \xi  \Big\rangle  \Big| \nn \\
&\qquad \leq C N^{-1/2} \|V\|_{L^1} \|h\|_{L^\infty L^2} \|\varphi\|_{L^\infty} \| (\mathcal N+1)^{3/4}\xi'\| \| (\mathcal N+1)^{3/4}\xi\|, \label{eq:lemma:estimate_cubic_quadratic transform_2} \\
& \Big|  \Big\langle \xi', \int N^{5/2}V_N(x-y) \varphi(x)  a^{\sharp_1}(g_x) a^{\sharp_2}(h_y) a(c_{1,y})  \xi  \Big\rangle \Big | \nn  \\
&\qquad \leq C N^{-1/2} \|V\|_{L^1} \|g\|_{L^\infty L^2} \|h\|_{L^2} \|\varphi\|_{L^\infty} \| (\mathcal N+1)^{3/4}\xi'\| \| (\mathcal N+1)^{3/4}\xi\|, \label{eq:lemma:estimate_cubic_quadratic transform_3} \\
& \Big|  \Big\langle \xi' ,\int N^{5/2}V_N(x-y) \varphi(x)  a^{*}_x a^{\sharp_2}(h_y) a^*(s_{1,y})  \xi  \Big\rangle  \Big| \nn \\ 
&\qquad \leq C N^{-1/2} \|V\|_{L^1} \|h\|_{L^2} \| \varphi\|_{L^\infty}  \| (\mathcal N+1)^{3/4}\xi'\| \| (\mathcal N+1)^{3/4}\xi\|, \label{eq:lemma:estimate_cubic_quadratic transform_4} \\
& \Big|  \Big\langle \xi',\int N^{5/2}V_N(x-y) \varphi(x)  a^{*}(g_x) a^{*}_y a^*(s_{1,y}) \xi  \Big\rangle \Big | \nn \\ 
&\qquad \leq C N^{-1/2} \|V\|_{L^1} \|g\|_{L^2} \| \varphi\|_{L^\infty}  \| (\mathcal N+1)^{3/4}\xi'\| \| (\mathcal N+1)^{3/4}\xi\|,\label{eq:lemma:estimate_cubic_quadratic transform_5} \\
& \Big|  \Big\langle \xi', \int N^{5/2}V_N(x-y) \varphi(x)  a^{\sharp_1}(g_x) a^{\sharp_2}(h_y) a^*(s_{1,y})  \xi  \Big\rangle  \Big| \nn \\ 
&\qquad \leq C N^{-1/2} \|V\|_{L^1} \|g\|_{L^2} \|h\|_{L^2} \| \varphi\|_{L^\infty}  \| (\mathcal N+1)^{3/4}\xi'\| \| (\mathcal N+1)^{3/4}\xi\|, \label{eq:lemma:estimate_cubic_quadratic transform_6} \\
& \Big|  \Big\langle \xi', \int N^{5/2}V_N(x-y) \varphi(x)  a^{*}_x a^{*}_y a^{\sharp_1}(g_y)  \xi  \Big\rangle  \Big| \nn \\ 
&\qquad \leq C \|g\|_{L^2} (\|V\|_{L^1})^{1/2} \| \varphi\|_{H^2} \|\cL_4^{1/2} \xi'\| \|(\mathcal N+1)^{1/2}\xi\|, 
\label{eq:lemma:estimate_cubic_quadratic transform_7}\\
&  \Big| \Big \langle \xi', \int N^{5/2}V_N(x-y) \varphi(x) \braket{s_{1,x},p_{1,y}}  (a(c_{1,y}) + a^*(s_{1,y}))  \xi  \Big\rangle  \Big|  \nn \\
&\qquad \leq C N^{-1/2} \|V\|_{L^1} \| \varphi\|_{H^2} \|(\mathcal N+1)^{1/2}\xi'\| \|\xi\|.
	\label{eq:lemma:estimate_cubic_quadratic transform_8}
\end{align}
\end{lemma}
\begin{proof}[Proof of Lemma \ref{lemma:estimate_cubic_quadratic transform}] Let us prove (\ref{eq:lemma:estimate_cubic_quadratic transform_1}), for all $\xi,\xi' \in \mathcal F$ and $\eta >0$, we have
\begin{align*}
&  \Big| \Big\langle \xi', \int  N^{5/2}V_N(x-y) \varphi(x)  a^*(g_x) a^*_y a(c_{1,y}) \xi  \Big\rangle \Big| \\
	&\qquad \leq \int  N^{5/2}V_N(x-y) \varphi(x)  \| a(g_x) a_y \xi' \| \| a(c_{1,y}) \xi\| \\
	&\qquad \leq \int  N^{5/2}V_N(x-y) \varphi(x) \|g_x\|_{L^2}  \| a_y \mathcal N^{1/2}\xi' \| \| a(c_{1,y}) \xi\| \\
	&\qquad \leq \eta  \|\varphi\|_{L^\infty}^2 \|g\|_{L^\infty L^2}^2 \int  N^{5/2}V_N(x-y)  \| a_y \mathcal N^{1/2}\xi' \|^2 + \eta^{-1}  \int  N^{5/2}V_N(x-y) \| a(c_{1,y}) \xi\|^2 \\
	&\qquad \leq \eta N^{-1/2}  \|\varphi\|_{L^\infty}^2 \|g\|_{L^\infty L^2}^2 \|V\|_{L^1} \|\mathcal N \xi' \|^2 + \eta^{-1}   N^{-1/2} \|V\|_{L^1} \|c_1\|_{op}^2 \|\mathcal N^{1/2}\xi\|^2 \\
	&\qquad \leq N^{-1/2}  \|\varphi\|_{L^\infty} \|g\|_{L^\infty L^2} \|V\|_{L^1} \|\mathcal N \xi' \| \|\mathcal N^{1/2}\xi\|,
\end{align*}
where we optimized over $\eta$ to obtain the last inequality. Finally, changing state $\xi \to (\mathcal N+1)^{1/4} \xi$ and $\xi' \to (\mathcal N+1)^{-1/4} \xi'$ and using that $(\mathcal N+1) a^*_x = a^*_x \mathcal N$ proves (\ref{eq:lemma:estimate_cubic_quadratic transform_1}).

The proofs of (\ref{eq:lemma:estimate_cubic_quadratic transform_2})--(\ref{eq:lemma:estimate_cubic_quadratic transform_6}) are similar. For (\ref{eq:lemma:estimate_cubic_quadratic transform_7}), we have for all  $\xi,\xi' \in \mathcal F$ and all $\eta>0$,
\begin{align*}
& \Big |  \Big\langle \xi', \int N^{5/2}V_N(x-y) \varphi(x)  a^{*}_x a^{*}_y a^{\sharp_1}(g_y)  \xi  \Big\rangle  \Big| \nn \\ 
&\qquad \leq  \eta \int N^{2}V_N(x-y)  \| a_x a_y \xi' \|^2 + \eta^{-1} \int N^{3}V_N(x-y) \varphi(x)^2 \| a^{\sharp_1}(g_y)\xi\|^2 \\
&\qquad \leq \eta \|\mathcal L_4^{1/2} \xi'\|^2 + \eta^{-1} \|\varphi\|_{L^\infty}^2 \int N^{3}V_N(x-y) \|g_y\|_{L^2}^2 \| (\mathcal N+1)^{1/2}\xi\|^2 \\
&\qquad \leq \eta \|\mathcal L_4^{1/2} \xi'\|^2 + \eta^{-1} \|\varphi\|_{H^2}^2 \|V\|_{L^1} \|g\|_{L^2}^2 \| (\mathcal N+1)^{1/2}\xi\|^2.
\end{align*}
The proof of (\ref{eq:lemma:estimate_cubic_quadratic transform_7}) is finished by optimizing over $\eta >0$. Finally, (\ref{eq:lemma:estimate_cubic_quadratic transform_8}) is an easy consequence of the pointwise inequality from Lemma \ref{lem:s1}:
$$|\braket{s_{1,x},p_{1,y}}  | \leq \|s_1\|_{L^\infty L^2} \|p_1\|_{L^\infty L^2} \leq C.$$ 
\end{proof}

Now let us come back to the proof of Lemma \ref{lemma:cubic_term_quadratic_transfo}. From \eqref{eq:T1L3-oo}, we decompose $c_1(x,y) = \delta_{x,y} + p_1(x,y)$, then we use that 
$$\|p_1\|_{L^\infty L^2}, \|s_1\|_{L^\infty L^2} \leq C, \quad \|s_1\|_{L^2}, \|p_1\|_{L^2} \leq \ell^{1/2}, \quad |p_1(x,y)| \leq C |\varphi(x)| |\varphi(y)|,$$
from Lemma \ref{lem:s1} and we can control the error by Lemma \ref{lemma:estimate_cubic_quadratic transform}. We obtain 
\begin{align}\label{eq:L_3_quadratic_transform}
&T^*_1 \mathcal L_3 T_1
	= \int N^{5/2} V_N (x-y) \varphi(x) \left(a^*_x a^*_y a_y   \sqrt{\left( 1-\frac{\mathcal N_{T_1}}{N} \right)_+}  + \hc\right) \\
	&+ \left(\int N^{5/2} V_N (x-y) \varphi(x) s_1(x,y) (a(c_{1,y}) + a^*(s_{1,y}))   \sqrt{\left( 1-\frac{\mathcal N_{T_1}}{N} \right)_+}   + \hc\right) +\mathcal E, \nn
\end{align}
where for all $\xi,\xi'\in \mathcal F$, by  Lemma \ref{prop:properties_quadratic_transform_1}, 
\begin{align*}
| \langle \xi',  \mathcal E \xi \rangle | 
	& \leq C N^{-1/2}  \| (\mathcal N+1)^{3/4}\xi'\| \Big \| (\mathcal N+1)^{3/4}    \sqrt{\left( 1-\frac{\mathcal N_{T_1}}{N}\right)_+} \xi \Big\| \\
	&\quad +  C \ell^{1/2} \|\cL_4^{1/2} \xi'\| \Big\|(\mathcal N+1)^{1/2}   \sqrt{\left( 1-\frac{\mathcal N_{T_1}}{N} \right)_+} \xi \Big\| \\
	& \leq C N^{-1/2}  \| (\mathcal N+1)^{3/4}\xi'\| \| (\mathcal N+1)^{3/4} \xi\|   \\
	&\quad +  C \ell^{1/2} \|\cL_4^{1/2} \xi'\| \|(\mathcal N+1)^{1/2} \xi\|.
\end{align*}
%Here we have used Lemma \ref{prop:properties_quadratic_transform_1}. 
%where we used that 
%$$(\mathcal N+1)^k \leq C_k T^*_1 (\mathcal N+1)^k T_1 \leq C_k' (\mathcal N+1)^k.$$
To extract the main contribution from the first term in (\ref{eq:L_3_quadratic_transform}), we decompose
\begin{align} \label{eq:1-sqrt-abc}
1 -   \sqrt{\left( 1-\frac{\mathcal N_{T_1}}{N}\right)_+} = \mathds{1}^{\{\mathcal N_{T_1} > N\}} + \mathds{1}^{\{\mathcal N_{T_1} \leq N\}} \left(1-\sqrt{\left( 1-\frac{\mathcal N_{T_1}}{N} \right)_+} \right).
\end{align}
The corresponding error terms can be bounded as, for all $\xi,\xi'\in \mathcal F$,  
\begin{align*}
\Big|\Big\langle \xi', \int N^{5/2} &V_N (x-y) \varphi(x) a^*_x a^*_y a_y \mathds{1}^{\{\mathcal N_{T_1} \leq N\}} \left(1-\sqrt{\left( 1-\frac{\mathcal N_{T_1}}{N}\right)_+} \right) \xi \Big\rangle\Big| \\
	&\leq C \|V\|_{L^1}^{1/2} \|\varphi\|_{L^\infty} \|\cL_4^{1/2} \xi'\| \Big\|\mathcal N^{1/2} \mathds{1}^{\{\mathcal N_{T_1} \leq N\}} \left(1-\sqrt{\left( 1-\frac{\mathcal N_{T_1}}{N}\right)_+} \right) \xi \Big\| \\
	&\leq C N^{-1/2}  \|\cL_4^{1/2} \xi'\| \|(\mathcal N_{T_1}+1) \xi\|  \leq C N^{-1/2}  \|\cL_4^{1/2} \xi'\| \|(\mathcal N+1) \xi\|,
\end{align*}
and
\begin{align*}
&|\langle \xi', \int N^{5/2} V_N (x-y) \varphi(x) a^*_x a^*_y a_y \mathds{1}^{\{\mathcal N_{T_1} > N\}}  \xi \rangle| \\
	&\leq C \|\cL_4^{1/2} \xi'\| \|\mathcal N^{1/2} \mathds{1}^{\{\mathcal N_{T_1} > N\}} \xi\|  \leq C N^{-1/2}  \|\cL_4^{1/2} \xi'\| \|(\mathcal N+1) \xi\|,
\end{align*}
where we used that 
$$\mathds{1}^{\{\mathcal N_{T_1} > N\}} \mathcal N \mathds{1}^{\{\mathcal N_{T_1} > N\}} \leq C \mathds{1}^{\{\mathcal N_{T_1} > N\}} (\mathcal N_{T_1}+1) \leq  C N^{-1} (\mathcal N_{T_1}+1)^2 \leq  C N^{-1} (\mathcal N+1)^2.$$

Finally we turn to the second term in (\ref{eq:L_3_quadratic_transform}). Using that 
$$|s_1(x,y) + N \omega_{\ell,N} (x-y) \varphi(x) \varphi(y)| \leq C\ell^{2} \varphi(x) \varphi(y) $$ 
we obtain that, for all $\xi,\xi'\in \mathcal F$,
\begin{align*}
&\Big|\Big\langle \xi',  \int N^{5/2} V_N (x-y) \varphi(x) (s_1(x,y) + N \omega_{\ell,N} (x-y) \varphi(x) \varphi(y))  (a(c_{1,y}) + a^*(s_{1,y})) \times  \\
&\qquad\qquad\qquad\qquad\qquad\qquad\qquad\qquad\qquad\qquad\qquad\qquad \times \sqrt{\left( 1-\frac{\mathcal N_{T_1}}{N} \right)_+} \xi \Big\rangle \Big| \\
	&\leq C N^{-1/2} \|V\|_{L^1} \|\varphi\|_2 \|\varphi\|_{L^\infty}^2 (1+ \|s_1\|_{L^\infty L^2})\|\xi'\| \Big\|(\mathcal N+1)^{1/2}  \sqrt{\left( 1-\frac{\mathcal N_{T_1}}{N} \right)_+}\xi \Big\| \\
	&\leq  C N^{-1/2} \|\xi'\| \|(\mathcal N+1)^{1/2} \xi\|.
\end{align*}
Thus it remains to consider
\begin{align*}
&- \int N^{5/2} V_N (x-y)  N \omega_{\ell,N} (x-y) \varphi^2(x) \varphi(y)  (a(c_{1,y}) + a^*(s_{1,y})) \sqrt{\left( 1-\frac{\mathcal N_{T_1}}{N} \right)_+} \\
&= - \int N^{7/2} ( (V_N \omega_N) * \varphi^2) (x) \varphi(x)  (a(c_{1,x}) + a^*(s_{1,x})) \sqrt{\left( 1-\frac{\mathcal N_{T_1}}{N} \right)_+}. 
\end{align*}
To extract the main contribution from the latter expression, we use the decomposition \eqref{eq:1-sqrt-abc} again. The corresponding error terms can be bounded as, for all  $\xi,\xi'\in \mathcal F$,
\begin{align*}
&\Big|\Big\langle \xi', \int N^{7/2} (V_N\omega_N) \ast \varphi(y)^2 \varphi(y) (a(c_{1,y}) + a^*(s_{1,y}))  \mathds{1}^{\{\mathcal N_{T_1} \leq N\}} \left(1-\sqrt{1-\frac{\mathcal N_{T_1}}{N}}\right)  \xi \Big\rangle \Big| \\
	&\leq C N^{1/2} \|V\|_{L^1}\|\varphi\|_{L^\infty}^2 \|\varphi\|_{L^2} \|\xi'\| \|\mathcal N_{T_1}^{1/2}\mathds{1}^{\{\mathcal N_{T_1} \leq N\}} \left(1-\sqrt{1-\frac{\mathcal N_{T_1}}{N}}\right)\xi \| \\
	&\leq CN^{-1/2} \|\xi'\| \|(\mathcal N_{T_1}+1)^{3/2}\xi\|  \leq CN^{-1/2} \|\xi'\| \|(\mathcal N+1)^{3/2}\xi\|.
\end{align*}
and 
\begin{align*}
&|\langle \xi', \int N^{7/2} (V_N\omega_N) \ast \varphi(y)^2 \varphi(y) (a(c_{1,y}) + a^*(s_{1,y}))  \mathds{1}^{\{\mathcal N_{T_1} > N\}}  \xi \rangle| \\
	&\leq C N^{1/2} \|\xi'\| \|\mathcal N_{T_1}^{1/2}\mathds{1}^{\{\mathcal N_{T_1} > N\}}  \xi \| \leq C N^{-1/2}\|\xi'\| \|\mathcal N_{T_1}^{3/2}\xi \| \leq C N^{-1/2}\|\xi'\| \|(\mathcal N+1)^{3/2}\xi \|.
\end{align*}
Here we used again that $\mathds{1}^{\{\mathcal N_{T_1} > N\}} \leq N^{-1}\mathcal N_{T_1}$. The proof of Lemma \ref{lemma:cubic_term_quadratic_transfo} is complete. 
\end{proof}

\subsection{Estimating $T^*_1 \cL_4 T_1$}

Recall  
\begin{align*}
\mathcal L_4  = \frac{1}{2}\int N^2 V_N (x-y) a^*_x a^*_y a_{x} a_{y}.
\end{align*}

\begin{lemma}
	\label{lemma:quartic_term_quadratic_transfo}
We have the operator identity on $\cF$
\begin{align*}
T^*_1 \mathcal L_4 T _1
	&= \mathcal L_4 -\frac{1}{2} \int N^3 (V_N\omega_N)(x-y) \varphi(x) \varphi(y) (a^*_x a^*_y + a_x a_y)\\
	&\qquad + \frac{N^4}{2} \int (V_N \omega_N^2)(x-y) \varphi(x)^2 \varphi(y)^2  + \mathcal E^{(T_1)}_{\mathcal L_4},
\end{align*}
where $ \mathcal E^{(T_1)}_{\mathcal L_4}$ satisfies the quadratic form estimate on $\cF_+$, for all $\eta>0$, 
\begin{align*}
\pm \mathcal E^{(T_1)}_{\mathcal L_4} \leq  C \ell^{1/2} (\mathcal N+1) + C \frac{(\mathcal N+1)^2}{N} + C \Big (\eta \cL_4 + \eta^{-1}\frac{(\mathcal N+1)^2}{N} \Big).
\end{align*}
\end{lemma}

\begin{proof}

Using the relation $T^*_1 a^*_x T_1 = a^*(c_{1,x}) + a(s_{1,x})$, and the CCR, we have
\begin{align}
&T^*_1 \mathcal L_4 T_1
	= \frac{1}{2}\int N^2 V_N(x-y)  (a^*(c_{1,x}) + a(s_{1,x}))( a^*(c_{1,y}) + a(s_{1,y})) \times \nn \\
	&\qquad\qquad\qquad\qquad\qquad\qquad\qquad \times (a(c_{1,x}) + a^*(s_{1,x})) (a(c_{1,y}) + a^*(s_{1,y})) \nn \\
	&= \frac{1}{2} \int N^2 V_N(x-y) \Big( a^*(c_{1,x}) a^*(c_{1,y}) + \braket{s_{1,x},c_{1,y}}  + a^*(c_{1,x}) a(s_{1,y}) + \nn\\
	&\qquad\qquad\qquad\qquad\qquad\qquad\qquad\qquad\qquad\qquad\qquad  + a^*(c_{1,y}) a(s_{1,x})  + a(s_{1,x}) a(s_{1,y}) \Big) \nn \\
	&\qquad \times \Big( a(c_{1,x}) a(c_{1,y}) + \braket{c_{1,x},s_{1,y}}  + a^*(s_{1,y})a(c_{1,x}) + a^*(s_{1,x})a(c_{1,y}) + a^*(s_{1,x}) a^*(s_{1,y}) \Big) .\label{eq:dev_quartic_quadratic_transfo}
\end{align}

Let us write $c_1(x,y) = \delta_{x,y} + p_1(x,y)$ and use that $\|p_1\|_{L^\infty L^2}, \|s_1\|_{L^\infty L^2} \leq C$ and that $|p_1(x,y)| \leq C |\varphi(x)| |\varphi(y)|$. After expanding, we obtain a sum of quartic, quadratic and constant terms. The quartic terms are of the form
\begin{align*}
\frac{1}{2} \int N^2 V_N(x-y) a^{\sharp_1} (g_{1,x}) a^{\sharp_2} (g_{2,y}) a^{\sharp_3} (g_{3,x}) a^{\sharp_1} (g_{4,y})
\end{align*}
with $g_{i} \in \{\delta_{x,y}\} \cup L^2(\mathbb{R}^{6})$ and $\sharp_i \in \{*,\cdot\}$ with the condition that if one of the $g_i$ is $\delta_{x,y}$ then $(\sharp_1,\sharp_2,\sharp_3,\sharp_4)$ is in normal order. Among the quartic terms, only the term with $g_i = \delta_{x,y}$ for all $i \in \{1,\dots,4\}$ contributes, the other terms are errors. We classify them in function of the number of $g_i$ equal to $\delta_{x,y}$. Up to symmetries coming for hermitian conjugation, it is enough to bound the following terms. 

\begin{lemma}
		\label{prop:estimate_L_4_quartic_terms}
Let $g_i \in L^2(\mathbb{R}^6)$, we have for all $\eta>0$,
\begin{align}
\pm & \Big(\int N^2 V_N(x-y) a^{\sharp_1}(g_{1,x}) a^{\sharp_2} (g_{2,y}) a^{\sharp_3} (g_{3,x}) a^{\sharp_4} (g_{4,y}) + \hc  \Big) \nn \\
& \qquad \qquad \qquad\qquad\qquad \leq C N^{-1}\|V\|_{L^1} \|g_1\|_{L^2} \|g_2\|_{L^\infty L^2} \|g_3\|_{L^2} \|g_4\|_{L^\infty L^2} (\mathcal N+1)^2, \label{eq:prop:estimate_L_4_quartic_terms_1} \\
\pm & \Big(\int N^2 V_N(x-y) a^{*}_x a^{\sharp_2} (g_{2,y}) a^{\sharp_3} (g_{3,x}) a^{\sharp_4} (g_{4,y}) + \hc  \Big) \nn\\
&\qquad \qquad \qquad \qquad\qquad\leq C N^{-1}\|V\|_{L^1} \|g_2\|_{L^\infty L^2} \|g_3\|_{L^2} \|g_4\|_{L^\infty L^2} (\mathcal N+1)^2, \label{eq:prop:estimate_L_4_quartic_terms_2}\\
\pm & \Big(\int N^2 V_N(x-y) a^{*}_x a^{*}_y a^{\sharp_3} (g_{3,x}) a^{\sharp_4} (g_{4,y}) + \hc  \Big) \nn \\
&\qquad \qquad \qquad \qquad\qquad\leq C N^{-1/2}(\|V\|_{L^1})^{1/2} \|g_3\|_{L^2} \|g_4\|_{L^\infty L^2} (\eta \cL_4 + \eta^{-1} (\mathcal N+1)^2), \label{eq:prop:estimate_L_4_quartic_terms_3} \\
\pm & \Big(\int N^2 V_N(x-y) a^{*}_x a^{\sharp_2} (g_{2,y}) a^{\sharp_3} (g_{3,x}) a_y + \hc  \Big) \nn \\ 
&\qquad \qquad \qquad\qquad\qquad  \leq C N^{-1}\|V\|_{L^1} \|g_2\|_{L^\infty L^2} \|g_3\|_{L^\infty L^2} (\mathcal N+1)^2, \label{eq:prop:estimate_L_4_quartic_terms_4}\\
\pm & \Big(\int N^2 V_N(x-y) a^{*}_x a^{\sharp_2} (g_{2,y}) a^{\sharp_4} (g_{4,y}) a_x + \hc  \Big) \nn \\ 
&\qquad \qquad \qquad \qquad\qquad\leq C N^{-1}\|V\|_{L^1} \|g_2\|_{L^\infty L^2} \|g_4\|_{L^\infty L^2} (\mathcal N+1)^2, \label{eq:prop:estimate_L_4_quartic_terms_5}\\
\pm & \Big(\int N^2 V_N(x-y) a^{*}_x a^{*}_y a^{\sharp_4} (g_{4,y}) a_x + \hc  \Big)   \nn \\ 
&\qquad \qquad \qquad\qquad\qquad \leq C N^{-1/2}(\|V\|_{L^1})^{1/2} \|g_4\|_{L^\infty L^2} (\eta \cL_4 + \eta^{-1} (\mathcal N+1)^2).\label{eq:prop:estimate_L_4_quartic_terms_6}
\end{align}
\end{lemma}

\begin{proof}[Proof of Lemma \ref{prop:estimate_L_4_quartic_terms}] Recall that we use the notation $a^{\overline \sharp} = (a^\sharp)^\dagger$. Let us prove (\ref{eq:prop:estimate_L_4_quartic_terms_1}). For all $\xi,\xi' \in \mathcal F_+$, we have
\begin{align*}
&  \Big| \Big\langle \xi',\int N^2 V_N(x-y) a^{\sharp_1}(g_{1,x}) a^{\sharp_2} (g_{2,y}) a^{\sharp_3} (g_{3,x}) a^{\sharp_4} (g_{4,y})  \xi  \Big\rangle  \Big| \\
	&\leq \int N^2 V_N(x-y) \| a^{\overline \sharp_2} (g_{2,y})  a^{\overline \sharp_1}(g_{1,x}) \xi'\|  \|a^{\sharp_3} (g_{3,x}) a^{\sharp_4} (g_{4,y}) \xi\| \\
	&\leq C \int N^2 V_N(x-y) \| g_{1,x}\|_{L^2} \| g_{2,y}\|_{L^2} \| (\mathcal N+1) \xi'\|  \| g_{3,x}\|_{L^2} \| g_{4,y}\|_{L^2} \|(\mathcal N+1)\xi\|  \\
	&\leq C N^{-1}\|V\|_{L^1} \|g_1\|_{L^2} \|g_2\|_{L^\infty L^2} \|g_3\|_{L^2} \|g_4\|_{L^\infty L^2} \|(\mathcal N+1)\xi'\| \|(\mathcal N+1)\xi\|.
\end{align*}

Similarly for all $\xi,\xi' \in \mathcal F_+$
\begin{align*}
&  \Big| \Big\langle \xi',\int N^2 V_N(x-y) a^{*}_x a^{\sharp_2} (g_{2,y}) a^{\sharp_3} (g_{3,x}) a^{\sharp_4} (g_{4,y})  \xi \Big \rangle \Big | \\
	&\leq \int N^2 V_N(x-y) \| a^{\overline \sharp_2}(g_{2,y}) a_x \xi'\|  \|a^{\sharp_3} (g_{3,x}) a^{\sharp_4} (g_{4,y}) \xi\| \\
	&\leq C \int N^2 V_N(x-y) \| g_{2,y}\|_{L^2} \| (\mathcal N+1)^{1/2} a_x \xi'\| \| g_{3,x}\|_{L^2} \| g_{4,y}\|_{L^2} \|(\mathcal N+1)\xi\|  \\
	&\leq C N^{-1}\|V\|_{L^1} \|g_2\|_{L^\infty L^2} \|g_3\|_{L^2} \|g_4\|_{L^\infty L^2} \|(\mathcal N+1)\xi'\| \|(\mathcal N+1)\xi\|,
\end{align*}
this proves (\ref{eq:prop:estimate_L_4_quartic_terms_2}). Let us now deal with (\ref{eq:prop:estimate_L_4_quartic_terms_3}), for all $\xi,\xi' \in \mathcal F_+$ and all $\eta>0$ we have
\begin{align*}
&  \Big| \Big\langle \xi',\int N^2 V_N(x-y) a^{*}_x a^{*}_y a^{\sharp_3} (g_{3,x}) a^{\sharp_4} (g_{4,y})   \xi  \Big\rangle  \Big| \\
	&\leq C \eta  \int N^2 V_N(x-y)\| a_x a_y \xi'\|^2 + \eta^{-1} \int N^2 V_N(x-y) \|a^{\sharp_3} (g_{3,x}) a^{\sharp_4} (g_{4,y}) \xi\|^2 \\
	&\leq C \eta \| \mathcal L_4^{1/2}\xi'\|^2 + C\eta^{-1}  \int N^2 V_N(x-y) \| g_{3,x}\|_{L^2}^2 \| g_{4,y}\|_{L^2}^2 \|(\mathcal N+1)\xi\|^2 \\
	&\leq C\eta  \| \mathcal L_4^{1/2}\xi'\|^2 + C \eta^{-1} N^{-1}\|V\|_{L^1} \|g_3\|_{L^2}^2 \|g_4\|_{L^\infty L^2}^2 \|(\mathcal N+1)\xi\|^2,
\end{align*}
optimizing in $\eta$ proves the claim. We now prove (\ref{eq:prop:estimate_L_4_quartic_terms_4}), for all $\xi,\xi' \in \mathcal F_+$
\begin{align*}
&  \Big| \Big\langle \xi',\int N^2 V_N(x-y) a^{*}_x a^{\sharp_2} (g_{2,y}) a^{\sharp_3} (g_{3,x}) a_y   \xi  \Big\rangle \Big | \\
	&\leq C  \int N^2 V_N(x-y)\| a^{\overline \sharp_2} (g_{2,y}) a_x \xi'\| \|a^{\sharp_3} (g_{3,x}) a_y\xi\| \\
	&\leq C  \int N^2 V_N(x-y) \|g_{2,y}\|_{L^2}\|g_{3,x}\|_{L^2} \| (\mathcal N+1)^{1/2}a_x \xi'\| \|(\mathcal N+1)^{1/2} a_y\xi\| \\
	&\leq  C N^{-1}\|V\|_{L^1} \|g_2\|_{L^\infty L^2} \|g_3\|_{L^\infty L^2} \|(\mathcal N+1)\xi'\| \|(\mathcal N+1)\xi\|.
\end{align*}
The estimate (\ref{eq:prop:estimate_L_4_quartic_terms_5}) is obtained in a similar way as  (\ref{eq:prop:estimate_L_4_quartic_terms_4}). Finally, we prove (\ref{eq:prop:estimate_L_4_quartic_terms_6}), 
for all $\xi,\xi' \in \mathcal F_+$ and $\eta>0$, we have
\begin{align*}
&  \Big| \Big\langle \xi',\int N^2 V_N(x-y) a^{*}_x a^{*}_y a^{\sharp_4} (g_{4,y}) a_x   \xi  \Big\rangle  \Big| \\
	&\leq C  \int N^2 V_N(x-y)\|a_x a_y \xi'\| \|a^{\sharp_3} (g_{4,y}) a_x\xi\| \\
	&\leq C  \eta \int N^2 V_N(x-y)\|a_x a_y \xi'\|^2 + C \eta^{-1} \int N^2 V_N(x-y) \|a^{\sharp_3} (g_{4,y}) a_x\xi\|^2  \\
	&\leq C \eta \|\mathcal L_4^{1/2} \xi'\|^2 + C \eta^{-1} \int N^2 V_N(x-y) \|g_{4,y}\|_{L^2} \|(\mathcal N+1)^{1/2} a_x\xi\|^2 \\
	&\leq C \eta \|\mathcal L_4^{1/2} \xi'\|^2 + C \eta^{-1} N^{-1} \|V\|_{L^1} \|g_{4}\|_{L^\infty L^2} \|(\mathcal N+1)\xi\|^2,
\end{align*}
optimizing in $\eta$ proves the claim.

\end{proof}

The quadratic terms are of the forms,
\begin{align*}
\int N^2 V_N(x-y) \widetilde s_1 (x,y) a^{\sharp_1} (g_{1,x}) a^{\sharp_2} (g_{2,y}), \\
\int N^2 V_N(x-y) \braket{p_{1,y},s_{1,x}} a^{\sharp_1} (g_{1,x}) a^{\sharp_2} (g_{2,y}),
\end{align*}
with some $g_{i} \in \{\delta_{x,y}\} \cup L^2(\mathbb{R}^{6})$ and $\sharp_i \in \{*,\cdot\}$ with the condition that if one of the $g_i$ is $\delta_{x,y}$ then $(\sharp_1,\sharp_2)$ is in normal order. Only the terms of the first kind, will contribute to the first two orders, and they are dealt with as for $T^*_1 \widetilde {\mathcal L}_2 T_1$, that is as in Lemma \ref{lemma:quadratic_term_quadratic_transfo}. Let us bound the others, using the symmetries, and using that $|\braket{p_{1,y},s_{1,x}}| \leq C \varphi(x) \varphi(y)$ it is enough to bound terms of the following form.

\begin{lemma}
		\label{prop:estimate_L_4_quadratic_terms}
Let $g_i \in L^2(\mathbb{R}^6)$, for all $\eta>0$ we have
\begin{align}
&\pm \Big(\int N^2 V_N(x-y)\varphi(x)^2 \varphi(y)^2 a^{\sharp_1}(g_{1,x}) a^{\sharp_2} (g_{2,y}) + \hc \Big) \nn \\ 
&\qquad \qquad \qquad \qquad \qquad \qquad\qquad \qquad 
\leq C N^{-1} \|V\|_{L^1} \|\varphi\|_{L^\infty}^4 \|g_1\|_{L^2} \|g_2\|_{L^2} (\mathcal N+1), \label{eq:prop:estimate_L_4_quadratic_terms_1}\\
&\pm \Big(\int N^2 V_N(x-y)\varphi(x)^2 \varphi(y)^2 a^{*}_x a^{\sharp_2} (g_{2,y}) + \hc \Big)  \nn \\ 
&\qquad \qquad \qquad \qquad \qquad \qquad \qquad \qquad
\leq C N^{-1} \|V\|_{L^1} \|\varphi\|_{L^\infty}^4 \|g_2\|_{L^2} (\mathcal N+1), \label{eq:prop:estimate_L_4_quadratic_terms_2}\\
&\pm \Big(\int N^2 V_N(x-y)\varphi(x)^2 \varphi(y)^2 a^{*}_x a^*_y + \hc \Big)  \nn \\ 
&\qquad \qquad \qquad \qquad \qquad \qquad \qquad \qquad
\leq  C N^{-1/2}(\|V\|_{L^1})^{1/2} \|\varphi\|_{L^\infty}^2 \|\varphi\|_{L^4}^2 (\eta \cL_4 + \eta^{-1}) \label{eq:prop:estimate_L_4_quadratic_terms_3}.
\end{align}
\end{lemma}
The proof of Lermma \ref{prop:estimate_L_4_quadratic_terms} is similar to the one of Lemma \ref{lemma:estimate_quadratic_quadratic transform} and we skip the details. Using Lemmas \ref{prop:estimate_L_4_quartic_terms} and \ref{prop:estimate_L_4_quadratic_terms} we have
\begin{align*}
T^*_1 \mathcal L_4 T_1
	&= - \frac{1}{2} \int N^3 (V_N \omega_N( (x-y) \varphi(x) \varphi(y)( a^*_x a^*_y + a_x a_y) \\
	&\quad+ \frac{1}{2} \int N^4(V_N \omega_N^2) (x-y) \varphi(x)^2 \varphi(y)^2 + \mathcal E,
\end{align*}
where the error term $\mathcal E $ satisfies the quadratic form estimate on $\cF_+$
\begin{align*}
\pm \mathcal E \leq C \ell^{1/2} (\mathcal N+1) + C N^{-1} (\mathcal N+1)^2+ C N^{-1/2} (\eta \cL_4 + \eta^{-1} (\mathcal N+1)^2). 
\end{align*}
The proof of Lemma \ref{lemma:quartic_term_quadratic_transfo} is complete. 
\end{proof}

\subsection{Conclusion of Lemma \ref{lemma:main_lemma_quadratic_transform}}

\begin{proof}
From Lemmas \ref{lemma:TL0T}, \ref{lemma:T1L1T_1}, \ref{lemma:quadratic_term_quadratic_transfo}, \ref{lemma:cubic_term_quadratic_transfo} and \ref{lemma:quartic_term_quadratic_transfo},
we obtain 
\begin{align*}
T^*_1 \mathcal H T_1 
	&= \sum_{i=0}^4 T^*_1 \cL_i  T_1 + T^*_1 \cE^{(U)} T_1 =  \sum_{\substack{i=0 \\ i \neq 1}}^4 \cL_i^{(T_1)} + \cE^{(T_1)}_1 + T^*_1 \cE^{(U)} T _1 \\
	&\quad + \frac{N-1- 2 \mathcal N}{2}\int_{\R^3}  ( N^3(V_Nf_N)\ast \varphi^2 - 8\pi \ao \varphi^2) \varphi^2\end{align*}
with $\cE^{(T_1)}_1 $ satisfying the quadratic form estimate on $\cF_+$, for all $\varepsilon>0$,
\begin{align*}
\pm \cE^{(T_1)}_1 
	&\leq \ell^{1/2} (\mathcal N+1) + C \ell^{3/2} \mathcal K +  C \frac{(\mathcal N+1)^{3/2}}{N^{1/2}} + C \frac{(\mathcal N+1)^2}{N} + \frac{(\mathcal N+1)^{5/2}}{N^{3/2}} \\
	&\quad + \eps \cL_4 + \eps^{-1} C \Big\{  \ell^{1/2} (\mathcal N+1) + \frac{(\mathcal N+1)^{3/2}}{N^{1/2}} + C \frac{(\mathcal N+1)^2}{N} \Big\}.
\end{align*}
Here we used again $f_N=1-\omega_N$. By  \eqref{eq:VNfN-pN-0} we can bound
\begin{align*}
\bigg|\int ( N^3(V_Nf_N)\ast \varphi^2 - 8\pi \ao \varphi^2) \varphi^2\bigg| \leq \| N^3(V_Nf_N)\ast \varphi^2 - 8\pi \ao \varphi^2\|_{L^2} \|\varphi^2 \|_{L^2}\leq C N^{-2}.
\end{align*}

The error term $\cE^{(U)}$ was estimated in Lemma \ref{lem:UHU*}. To bound $T^*_1 \cE^{(U)} T _1$, we use Lemma \ref{prop:properties_quadratic_transform_1} and
\begin{align*}
T^*_1 \mathcal L_4 T_1 \leq C \Big(N + \mathcal L_4 +  \ell^{1/2} (\mathcal N+1) + \frac{(\mathcal N+1)^2}{N}\Big)
\end{align*}
which follows from  Lemma \ref{lemma:quartic_term_quadratic_transfo}. Thus we obtain for all $\eps >0$,
\begin{align*}
\pm T^*_1 \cE^{(U)} T _1
	\leq \eps \Big\{(N + \mathcal L_4 +  \ell^{1/2} (\mathcal N+1) + \frac{(\mathcal N+1)^2}{N} \Big\} + C\eps^{-1} \frac{(\cN+1)^2}{N^3}  + C \frac{(\cN+1)^{3/2}}{N^{1/2}}. 
	\end{align*}
	Equivalently, replacing $\eps\mapsto \eps N^{-2}$, we have for all $\eps >0$,
	$$
\pm T^*_1 \cE^{(U)} T _1	\leq \eps \Big\{(N^{-1} + N^{-2} \mathcal L_4 + \ell^{1/2} \frac{(\mathcal N+1)}{N^2} + \frac{(\mathcal N+1)^2}{N^3} \Big\} + C\eps^{-1} \frac{(\cN+1)^2}{N}  + C \frac{(\cN+1)^{3/2}}{N^{1/2}}.
$$
%Denoting $\cE^{(T_1)} = \cE^{(T_1)}_1 + \cE^{(U)} + CN^{-1} + C\mathcal N/N^{2}$, this finishes the proof.
The proof of Lemma \ref{lemma:main_lemma_quadratic_transform} is complete. 
%Denoting $\cE^{(T_1)} = \cE^{(T_1)}_1 + \cE^{(U)} + CN^{-1} + C\mathcal N/N^{2}$, this finishes the proof.
\end{proof}

%prop:properties_quadratic_transform_1

\section{The cubic transform}\label{sec:Tc}

Recall that $\ell \gg N^{-1}$ and let us define the operator $\widetilde k_{c} : L^2(\mathbb{R}^3) \to L^2(\mathbb{R}^3)^{\otimes 2}$ by the kernel
\begin{align*}
\widetilde k_c (x,y,y') = - N^{1/2}\varphi(x) \omega_{\ell,N}(x-y)\delta_{y,y'},
\end{align*}
let us also define $k_{c} = Q^{\otimes 2} \widetilde k_{c} Q$ and for $M\geq 50$, the function $\theta_M : \mathbb{R}_+ \to \mathbb{R}_+$ by
\begin{align*}
\theta_M (x) = 
\left\{
\begin{array}{lll}
1 &\qquad x \leq M/2+10, \\
\frac{1}{2}(\frac{4x - 3M}{40-M} +1)&\qquad \frac{M}{2}+10 \leq x \leq M-10, \\
0 &\qquad x \geq M - 10.
\end{array}
\right.
\end{align*}
It satisfies, for all $j \in [0,10]$, $\theta_{M} (x\pm j) = 1$ when $x \leq M/2$ and $\theta_{M} (x\pm j) = 0$ when $x \geq M$, and
\begin{align*}
\|\theta_M - \theta_{M}(\cdot\pm j)\|_{L^\infty} \leq C M^{-1},
\end{align*}
for some constant $C>0$. We also define the notation $\theta_{M,+j} = \theta_{M}(\cdot \pm j)$ for $j\geq 0$. We will use the abuse of notation $\theta_M = \theta_M(\mathcal N)$, and similarly $\theta_{M,+j} = \theta_{M,+j} (\mathcal N)$.

Let us define 
\begin{align*}
K_{c}^* = \int k_{c}(x,y,y') a^*_x a^*_y a_{y'} , \qquad K_{c} = \int  \overline{k_{c}(x,y,y')}  a^*_{y'} a_{y} a_x,
\end{align*}
and
\begin{align*}
T_c = \exp \left(\theta_M \int k_{c}(x,y,y') a^*_x a^*_y a_{y'}  - \hc \right) = \exp \left(\theta_M K_{c}^* - K_c \theta_M\right)
\end{align*}

The main result of the section is the following Lemma.

\begin{lemma}
	\label{prop:recap_cubic_transform}
We have the operator equality on $\cF$
\begin{align*}
T_{c}^*T^*_1\mathcal H T_1 T_{c} 
	&= \mathcal H^{(T_{c})} + \mathcal E^{(\theta_M)}+ \mathcal E^{(T_c)},
\end{align*}
where
\begin{align}
	\mathcal H^{(T_{c})} 
	&= N \cE_{\rm GP}(\varphi)  - 4\pi \ao \int \varphi^4 - \frac{1}{2}N^4 \int ((\omega_{\ell,N} \varepsilon_{\ell,N}) \ast \varphi^2) \varphi^2  \nn\\
	&\quad +\dd\Gamma(D) + \frac{1}{2}  \int N^{3}\varepsilon_{\ell,N} (x-y) \varphi(x) \varphi(y)  (2a^*_x a_y + a^*_x a^*_y + a_x a_y )  \dd x \dd y +  \cL_4,\\
 \mathcal E^{(\theta_M)} 
 	&=  (1-\theta_M)\int N^{5/2} V_N (x-y) \varphi(x) a^*_x a^*_y a_y + \hc \nn \\
 	&\qquad + (1-\theta_{M})^2 \dd\Gamma((N^3V_N\omega_N)\ast \varphi^2 + \varphi (x) \widehat{(V\omega_N)}(p/N))\label{eq:E_theta_M}
\end{align}
and $\mathcal E^{(T_c)}$ satisfies the quadratic form estimate on $\cF_+$ 
\begin{align}
\pm \mathcal E^{(T_c)} 
	&\leq C (\ell^{1/2} + M^{1/2} N^{-1/2} \ell^{-1}) (\mathcal N+1) \nn \\
	&\qquad + C \big( M^{1/2} N^{-1/2}  +  \ell^{3/2} \big)  (\mathcal K + \cL_4 + 1)  \nn \\
	&\qquad + C \frac{(\mathcal N+1)^{3/2}}{N^{1/2}} + C \frac{(\mathcal N+1)^{5/2}}{N^{3/2}}\nn \\
	&\qquad + \eps \left( \cL_4 +  MN^{-1} \mathcal K + (\mathcal N+1) + \ell^{1/2} \frac{(\mathcal N+1)}{N^2} + \frac{(\mathcal N+1)^2}{N^3} \right) \nn \\
	&\qquad + \eps^{-1}C \left(\ell(\mathcal N+1) + \frac{(\mathcal N+1)^2}{N}+ \frac{(\mathcal N+1)^5}{N^4} \right) \label{eq:error_cubic}
\end{align}
for all $\eps>0$ and $MN^{-1} \lesssim \ell \lesssim 1$. Recall that $\mathcal K$ and $\mathcal{L}_4$ are given in Lemma \ref{lemma:main_lemma_quadratic_transform}. 
\end{lemma}

The rest of the section is devoted to the proof of Lemma \ref{prop:recap_cubic_transform}. In subsection \ref{sec:prop_kc} we prove some properties of the kernel $k_c$ and that $T_{c}^* (\mathcal N+1)^k  T_{c} \leq C_k (\mathcal N+1)^k$. Then in the subsection \ref{sec:Tc_K_L_4Tc} we estimate $T^*_c \dd\Gamma(-\Delta) T_c$ and $ T^*_c \cL_4 T_c$, while in the subsections \ref{sec:L0T_1}, \ref{sec:L2T_1}, \ref{sec:L3T_1}, we estimate respectively $T^*_c \cL_i^{(T_1)} T_c$, $i \in \{0,2,3\}$, where we used the notation $ \cL_i^{(T_1)}$ from Lemma \ref{lemma:main_lemma_quadratic_transform}. 
%
%where,
%
%are so that $\mathcal H^{(T_1)} =  \cL_0^{(T_1)} + \cL_2^{(T_1)} + \cL_3^{(T_1)} + \cL_4^{(T_1)}$. 
Finally, in subsection \ref{sec:proof_cubic} we collect all the terms and finish the proof of Lemma \ref{prop:recap_cubic_transform}.

\subsection{Properties of the cubic kernel}
	\label{sec:prop_kc}
\begin{lemma}[Properties of the cubic kernel $k_c$]
	\label{prop:cubic_kernel_est} We have, the following decomposition 
	$$k_{c}(x,y,y') = k_{c,1}(x,y) \delta_{y,y'} + k_{c,2}(x,y,y'),\quad \forall x,y,y' \in \mathbb{R}^3$$
	with the following pointwise bounds
\begin{align*}
|k_{c,1}(x,y)| &\leq C \varphi(x)  (N^{-1/2} \ell^2 + N^{1/2}\omega_{\ell,N}(x-y)), \\
|k_{c,2}(x,y,y')| &\leq C \varphi(x) \varphi(y) \varphi(y')(N^{-1/2} \ell^2 + N^{1/2}\omega_{\ell,N}(x-y) +N^{1/2} \omega_{\ell,N}(x-y')).
\end{align*}
Moreover, 
\begin{align}
\|k_c\|_{op} \leq C N^{-1/2} \ell^{1/2}, \label{item:prop:cubic_kernel_est_2}\\
\|k_{c,1}\|_{L^2} + \|k_{c,2}\|_{L^2} \leq C N^{-1/2} \ell^{1/2}, 	\label{eq:prop_cub_1} \\ 
\|k_{c,1}\|_{L^\infty_2 L^2_1} + \|k_{c,1}\|_{L^\infty_1 L^2_2}  \leq C N^{-1/2} \ell^{1/2} , 	\label{eq:prop_cub_2} \\
\|k_{c,2}\|_{L^\infty_{2,3} L^2_1} + \|k_{c,2}\|_{L^\infty_{1} L^2_{2,3}} \leq C N^{-1/2} \ell^{1/2} , 	\label{eq:prop_cub_3} \\
\|k_{c,2}\|_{L^\infty_{3} L^2_{1,2}} + \|k_{c,2}\|_{L^\infty_{2} L^2_{1,3}}  \leq C N^{-1/2} \ell^{1/2}. 	\label{eq:prop_cub_4}
\end{align}

\end{lemma}

\begin{proof} 
Let us expand
\begin{align*}
 Q^{\otimes 2} \widetilde k_{c} Q 
 	&=  (1-P)^{\otimes 2} \widetilde k_{c} (1-P) \\
 	&= \widetilde k_{c} - P\otimes 1 \widetilde k_{c} - 1\otimes P \widetilde k_{c}  - \widetilde k_{c} P + P^{\otimes 2} \widetilde k_{c} + P\otimes 1 \widetilde k_{c} P + 1 \otimes P \widetilde k_{c} P  - P^{\otimes 2} \widetilde k_{c} P.
\end{align*}
In terms of kernel, this reads
\begin{align}
&k_{c}(x,y,y') 
	= N^{1/2}\varphi(x) \omega_{\ell,N}(x-y) \delta_{y,y'} - N^{1/2}\varphi(x)  \int \dd t \varphi(t)^2 \omega_{\ell,N}(t-y)  \delta_{y,y'}\nn  \\
	&\quad - N^{1/2}\varphi(x) \varphi(y) \varphi(y') \omega_{\ell,N}(x-y') - N^{1/2}\varphi(x) \varphi(y) \varphi(y') \omega_{\ell,N}(x-y)\nn  \\
	&\quad + \varphi(x) \varphi(y) \varphi(y') \int \dd t \varphi(t)^2 N^{1/2} \omega_{\ell,N}(t-y') + \varphi(x) \varphi(y) \varphi(y') \int \dd t \varphi(t)^2 N^{1/2}\omega_{\ell,N}(t-y) \nn \\
	&\quad + \varphi(x) \varphi(y) \varphi(y') \int \dd t \varphi(t)^2 N^{1/2} \omega_{\ell,N} (x-t) + \varphi(x)\varphi(y) \varphi(y') \int \dd t\dd u \varphi(t)^2 \varphi(u)^2 N^{1/2} \omega_{\ell,N} (u-t) \nn \\
	&=: k_{c,1}(x,y) \delta_{y,y'} + k_{c,2}(x,y,y'). \label{eq:full_decompo_cubic_kernel}
\end{align}
Using this decomposition and that $N^{1/2} \|\omega_{\ell,N}\|_{L^1} \leq CN^{-1/2} \ell^2$, the first two estimates follow easily. To prove (\ref{item:prop:cubic_kernel_est_2}), it is enough to prove it for $\widetilde k_{c}$. For any $f\in L^2(\mathbb{R}^3), g\in L^2(\mathbb{R}^3)^{\otimes 2}$, we have
\begin{align*}
|\langle g, \widetilde k_{c} f \rangle| 
	&\leq N^{1/2} \int \varphi(x) \omega_{\ell,N}(x-y) |g(x,y)| |f(y)| \nn \\
	&\leq N^{1/2} \|\varphi\|_{L^\infty} \|\omega_{\ell,N}\|_{L^2} \|g\|_{L^2} \|f\|_{L^2} \leq C N^{-1/2} \ell^{1/2} \|g\|_{L^2} \|f\|_{L^2}.
\end{align*}
Hence $\|\widetilde k_{c}\|_{op} \leq C N^{-1/2} \ell^{1/2}$. For (\ref{eq:prop_cub_1}), from (\ref{eq:full_decompo_cubic_kernel}), we have
\begin{align*}
\|k_{c,1}\|_{L^2} 
	&\leq \|\varphi\|_{L^2} N^{1/2}\|\omega_{\ell,N}\|_{L^2} + \|\varphi\|_{L^2}^3 N^{1/2} \|\omega_{\ell,N}\|_{L^2} \\
	&\leq C N^{-1/2} \ell^{1/2}, \\
\|k_{c,2}\|_{L^2} 
	&\leq 2 \|\varphi\|_{L^\infty}\|\varphi\|_{L^2}^2N^{1/2} \|\omega_{\ell,N}\|_{L^2} + 3 \|\varphi\|_{L^\infty}\|\varphi\|_{L^2}^4 N^{1/2}\|\omega_{\ell,N}\|_{L^2} + \|\varphi\|_{L^\infty}^2 \|\varphi\|_{L^2}^5 N^{1/2}\|\omega_{\ell,N}\|_{L^1} \\
	&\leq C N^{-1/2} \ell^{1/2}.
\end{align*}

The inequalities in (\ref{eq:prop_cub_2}), (\ref{eq:prop_cub_3}) and (\ref{eq:prop_cub_4}) follow easily from  above pointwise bound, except the estimate on $\|k_{c,1}\|_{L^\infty L^2}$ that is derived similarly as (\ref{eq:prop_cub_1}) from (\ref{eq:full_decompo_cubic_kernel}).
\end{proof}

For every $t\in \mathbb{R}$, we define the unitary operator on $\mathcal{F}$:
\begin{align*}
T_c^t = \exp \left(t \left(\theta_M K_{c}^* - K_c \theta_M\right)\right).
\end{align*}

We have following analogue of Lemma \ref{prop:properties_quadratic_transform_1}.  

\begin{lemma}
	\label{lemma:number_part_cubic}
For all $k\geq1$ and $N^{-1}\ll \ell \lesssim 1$, $ M \leq N$ and $\lambda \in [0,1]$, there is a constant $C_k$ independent of $N$, such that on $\mathcal F$,
\begin{align*}
T_{c}^{-\lambda} (\mathcal N+1)^k T_{c}^{\lambda} \leq C_k (\mathcal N+1)^k.
\end{align*}
\end{lemma}
\begin{proof} First note that the cut-off $\theta_M$ makes the kernel $\theta_MK_c^*-K_c \theta_M$ bounded on $\mathcal{F}$, for instance 
$$\|\theta_M K_c^* - K_c \theta_M\|_{op} \le C \|\theta_M (\mathcal N + 1)^{3/2}\|_{op} \|(\mathcal N + 1)^{-3/2}K_c^*\|_{op}  \leq C M^{3/2} \|k_c\|_{op},$$
and also satisfies that $\mathds{1}^{\{\mathcal N> M\}} (\theta_MK_c^*-K_c \theta_M)=0$. Therefore, $T_c^\lambda$ leaves invariant the domain of $(\mathcal{N}+1)^s$ for every $s>0$, namely if $\|  (\mathcal{N}+1)^s \xi \|<\infty$ for some $\xi \in \mathcal{F}$, then 
\begin{align*}
&\| (\cN+1)^s T_c^\lambda \xi\| = \left\| \sum_{m=0}^\infty  (\cN+1)^s  \frac{\lambda^m (\theta_M K_c^* - K_c \theta_M)^m}{m!} \xi \right\| \\
&= \left\| (\cN+1)^s \mathds{1}^{\{\mathcal N> M\}} \xi +  \sum_{m=0}^\infty  (\cN+1)^s \mathds{1}^{\{\mathcal N\leq M\}}  \frac{\lambda^m (\theta_M K_c^* - K_c \theta_M)^m}{m!} \xi \right\|\\
&\le \left\| (\cN+1)^s \mathds{1}^{\{\mathcal N\leq M\}} \xi \right\| +  \sum_{m=0}^\infty  \left\|(\cN+1)^s \mathds{1}^{\{\mathcal N\leq M\}} \right\|_{op} \left\| \frac{\lambda^m (\theta_M K_c^* - K_c \theta_M)^m}{m!} \right\|_{op} \|\xi\|\\
&\le \left\| (\cN+1)^s \xi \right\| + (M+1)^s e^{C \lambda M^{3/2} \|k_c\|_{op}} \|\xi\| <\infty. 
\end{align*}
%acts like the identity on the sector $\{\mathcal N \geq M\}$ and therefore the domain of $(\mathcal N+1)^k$ for all $k\geq 0$. To conclude the proof, it is in fact clearer to invoke the Grönwall lemma. 

However the bound obtained as above is far from optimal. To improve the bound, we will use a Gr\"onwall argument. Let $k \geq 1$, we have by the Duhamel formula, for all $ 0 \leq \lambda \leq 1$,
\begin{align*}
T_{c}^{-\lambda}  (\mathcal N+1)^k T_{c}^{\lambda}
	&=  (\mathcal N+1)^k -  \int_0^\lambda T_{c}^{-t} [\theta_M K_{c}^* - K_c \theta_M, (\mathcal N+1)^k] T_{c}^{t} \dd t \\
	& = (\mathcal N+1)^k -  \int_0^\lambda T_{c}^{- t} \Bigg( \theta_M K_c^*((\mathcal N+1)^k - (\mathcal N+2)^k) + \hc \Bigg) T_{c}^{t} \dd t 
\end{align*}
where we used $\mathcal N K_{c}^* = K_{c}^* (\mathcal N +1)$ and $\mathcal N \theta_M= \theta_M \mathcal{N}$. Combining with the bound $\|(\mathcal N+1)^{-3/2} K_{c}^*\|_{op} \leq CN^{-1/2} \ell^{1/2}$, following from (\ref{eq:prop_cub_2}), the elementary estimate $(\mathcal N+2)^k - \mathcal (N+1)^k \leq C_k (\mathcal N+1)^{k-1}$, and the fact that $\theta_M=0$ outside the sector $\{\mathcal{N} \le M\}$, we find that 
\begin{align*}
T_{c}^{-\lambda}  (\mathcal N+1)^k T_{c}^{\lambda} &\leq  (\mathcal N+1)^k + C_kN^{-1/2} \ell^{1/2}   \int_0^\lambda T_{c}^{-t} \mathds{1}^{\{\mathcal N\leq M\}}(\mathcal N+1)^{k+1/2} T_{c}^{ t} \dd t \\
	&\leq  (\mathcal N+1)^k + C_k M^{1/2} N^{-1/2}  \int_0^\lambda T_{c}^{- t}(\mathcal N+1)^{k} T_{c}^{t} \dd t,
\end{align*}

We now take any $\xi \in \mathcal F$ such that $\braket{\xi, (\mathcal N + 1)^k \xi} < \infty$. Then $u(t)=\langle \xi, T_{c}^{-t}  (\mathcal N+1)^k T_{c}^{t} \xi\rangle$ is finite (due to the above non-optimal estimate) and it satisfies 
$$
u(\lambda) \le \braket{\xi, (\mathcal N + 1)^k \xi}  + C_k  M^{1/2} N^{-1/2} \int_0^\lambda u(t) \dd t. 
$$
The desired claim follows from the obvious bound $M^{1/2} N^{-1/2}\le 1$ and the Gr\"onwall lemma. 
%Because of the cut-off $\theta_M$, $T_{c}^{\lambda}$ acts trivially on the sectors $\{\mathcal N \geq M\}$ and $\theta_M K_c^*$ is a bounded operator, therefore $u(\lambda) :=  \braket{\xi, T_{c}^{-\lambda}  (\mathcal N+1)^k T_{c}^{\lambda} \xi}$ is finite for any $\lambda$ and defines a continuous function. Now, using the obvious bound $\lambda M^{1/2} N^{-1/2} \leq 1$ and the Grönwall Lemma for $u(\lambda)$, we obtain our claim.
\end{proof}

\subsection{Estimates involving $\dd\Gamma(-\Delta)$ and $\cL_4$}
	\label{sec:Tc_K_L_4Tc}

Here we estimate $T_c^* \dd\Gamma(-\Delta) T_c$ and $T^*_c \cL_4 T_c$, those bounds are important to estimate the other terms $T^*_c \cL_i^{(T_1)} T_c$.
\begin{lemma}
	\label{prop:estimate_kinetic_L4_cubictransform}
We have, for all $\lambda \in [0,1]$, and $MN^{-1} \lesssim \ell \lesssim 1$,
\begin{align}
	\label{eq:estimate_K_L_4_1}
&e^{-\lambda (\theta_M K_{c}^* - K_c \theta_M)} \mathcal K e^{\lambda (\theta_M K_{c}^* - K_c \theta_M)}  \leq C  (\mathcal K + \cL_4), \\
&e^{-\lambda (\theta_M K_{c}^* - K_c \theta_M)} \mathcal L_4 e^{\lambda (\theta_M K_{c}^* - K_c \theta_M)} \leq C( \cL_4 + \mathcal N + 1 + MN^{-1} \mathcal K).
		\label{eq:estimate_K_L_4_2}
\end{align}
We also have
\begin{align}
T_{c}^* \dd\Gamma(-\Delta) T_{c}
	&= \dd\Gamma(-\Delta) - \theta_M \int \varphi(x) N^{5/2} (V_N f_N)(x-y) a^*_x a^*_y a_y + \hc \nn \\
	&\quad + \theta_{M,+1}^2 \int N^{3} (V_Nf_N\omega_N)(x-y) \varphi(x)^2 a^*_y a_y \nn \\
	&\quad + \theta_{M,+1}^2\int N^{3} (V_Nf_N\omega_N)(x-y) \varphi(x)\varphi(y) a^*_y a_y + \mathcal E^{(T_c)}_{-\Delta},
		\label{eq:TcKTc}
\end{align}
with $\mathcal E$ satisfying, on $\mathcal F_+$,
\begin{align} \label{eq:estimate_error_kinetic_cubic_transfo}
\pm \mathcal  E^{(T_c)}_{-\Delta}  \leq C M^{1/2}N^{-1/2} (\mathcal K + \cL_4 + \ell^{-1}\mathcal N ),
\end{align}
and
\begin{align}
T_{c}^* \mathcal L_4 T_{c}
	&=\mathcal L_4   - \theta_M \int N^{5/2} (V_N\omega_N) (x-y) \varphi(x)  a^*_x a^*_y a_y + \hc \nn \\
	&+ \theta_{M,+1}^2 \int N^{3} (V_N\omega_N^2)(x-y) \varphi(x)^2 a^*_y a_y \nn \\
	& + \theta_{M,+1}^2\int N^{3} (V_N\omega_N^2)(x-y) \varphi(x)\varphi(y) a^*_x a_y + \mathcal E_{\mathcal L_4}^{(T_c)}, \label{eq:prop:bound_L4_cubic_transfo_1}
\end{align}
with $\mathcal E_{\mathcal L_4}^{(T_c)}$ satisfying, on $\F_+$,
\begin{align}
	\label{eq:estimate_error_L_4_cubic_transfo}
\pm \mathcal E_{\mathcal L_4}^{(T_c)} \leq  C M^{1/2} N^{-1/2} (\mathcal K + \cL_4).
\end{align}
\end{lemma}

To prove Lemma \ref{prop:estimate_kinetic_L4_cubictransform} we first need some a priori estimates on the commutators $\ad_{K_c-K_{c}^*}(\dd\Gamma(-\Delta))$, $\ad_{K_{c}-K_c^*}(\mathcal L_4)$ and $\ad^{(2)}_{K_{c}-K_c^*}(\dd\Gamma(-\Delta)) \simeq \ad^{(2)}_{K_{c}-K_c^*}(\mathcal L_4) \simeq \ad_{K_{c}-K_c^*} (\mathcal L_3^{(T_1)})$ which are given by the following Lemmas.

\begin{lemma}[Estimate on $\ad_{K_{c}^*}(\dd\Gamma(-\Delta))$]
	\label{lemma:adjoint_kinetic_cubic}
We have on $\FM$,
\begin{align*}
[\dd\Gamma(-\Delta), K_{c}^*] = - \int N^{5/2} (V_N f_N)(x-y) \varphi(x)a^*_x a^*_y a_y + \mathcal E, 
\end{align*}
where $\mathcal E$ satisfies on $\FM$
\begin{align*}
 \pm \mathcal E + \hc \leq C M^{1/2} N^{-1/2}  (\mathcal K + \ell^{-1} \mathcal N).
\end{align*}
\end{lemma}

\begin{lemma}[Estimate on $\ad_{K_{c}^*}(\cL_4)$]
	\label{lemma:adjoint_1_L_4_cubic}
There is some constant independent of $N$, such that on $\FM$,
\begin{align*}
[\mathcal L_4,K_{c}^*] =  -\int N^{5/2} (V_N \omega_N) (x-y) \varphi(x)a^*_x a^*_y a_y  + \mathcal E_{\ad_{c}(\mathcal L_4)},
\end{align*}
with for all $\xi,\xi' \in \FM$,
\begin{align*}
|\langle \xi',  \mathcal E_{\ad_{c}(\mathcal L_4)}  \xi \rangle | &\leq C \|\cL_4^{1/2} \xi'\| \Big( MN^{-1} \ell^{1/2} \|(\mathcal N+1)^{1/2} \xi\| \\ 
	&\qquad\qquad\qquad\qquad\qquad+ M^{1/2} N^{-1/2}  \ell^{1/2}  \|\cL_4^{1/2} \xi\|  + M N^{-1} \|\mathcal K^{1/2}\xi\| \Big)\\
	 &\leq C M^{1/2} N^{-1/2}\|(\mathcal K + \cL_4 +\mathcal N+1)^{1/2}\xi'\|\|(\mathcal K + \cL_4 +\mathcal N)^{1/2}\xi\|.
\end{align*}
\end{lemma}

The main contributions coming from $[\dd\Gamma(-\Delta), K_{c}^*]$ and $[\mathcal L_4,K_{c}^*]$ are, as expected, similar to $\cL_3^{(T_1)}$ defined in Lemma \ref{lemma:main_lemma_quadratic_transform}, which we recall is given by
\begin{align*}
 \cL_3^{(T_1)} 
 	= \int N^{5/2} V_N (x-y) \varphi(x) a^*_x a^*_y a_y + \hc   :=  \cL_3^{(T_1,*)} + \cL_3^{(T_1,\circ)}.
\end{align*}
This amounts to replacing $V_N$ by $V_N f_N$ or $V_N \omega_N$. To compute the next order, that is to compute the commutator of these three quantities with $K^*_c $ or $K_c$, it is enough to do it for $ \cL_3^{(T_1)}$ only. Therefore, we only deal with the latter and will refer to the same Lemma below for the others. 

\begin{lemma}[Estimate on $\ad_{K_c-K_{c}^*}(\cL_3^{(T_1)})$]
\label{lemma:adjoint_L3_cubic}
We have
\begin{align*}
[\cL_3^{(T_1,*)},K_{c}] =  \int N^{3} (V_N\omega_N)(x-y) \varphi(x)^2 a^*_y a_y + \int N^{3} (V_N\omega_N)(x-y) \varphi(x)\varphi(y) a^*_x a_y + \mathcal E
\end{align*}
with for all $ \xi' ,\xi \in \FM$,
\begin{align*}
	&|\langle \xi', [\cL_3^{(T_1,*)},K_{c}^*] \xi \rangle| + |\langle \xi' , \mathcal E \xi \rangle| \\
	&\leq C M N^{-1}  \| (\mathcal N+1)^{1/2}\xi'\| \| \mathcal K^{1/2}\xi\| + C M^{1/2} N^{-1/2} \|\cL_4^{1/2}\xi'\| \|\mathcal K^{1/2}\xi\| .
\end{align*}
\end{lemma}

In the rest of this subsection, we first prove Lemma \ref{prop:estimate_kinetic_L4_cubictransform} assuming Lemmas \ref{lemma:adjoint_kinetic_cubic}, \ref{lemma:adjoint_1_L_4_cubic} and \ref{lemma:adjoint_L3_cubic} and then give the proofs of the latter three Lemmas.

\begin{proof}[Proof of Lemma \ref{prop:estimate_kinetic_L4_cubictransform}]
Let us first prove (\ref{eq:estimate_K_L_4_1}), for this we will use and iterate the Duhamel formula. Let us recall the notation for $\lambda \in \mathbb{R}$, $T_c^\lambda = e^{\lambda(\theta_M K_{c}^* - K_c \theta_M)}$ and let us compute for $\lambda \in [0,1]$,
\begin{align}
	\label{eq:duhamel_kinetic_2}
&e^{-\lambda (\theta_M K_{c}^* - K_c \theta_M)}  \dd \Gamma(-\Delta) e^{\lambda (\theta_M K_{c}^* - K_c \theta_M)} = T_c^{-\lambda}\dd \Gamma(-\Delta) T_c^\lambda \nn \\
	&=  \dd \Gamma(-\Delta ) + \lambda \int_0^1 T_c^{-\lambda t} [\dd \Gamma(-\Delta),\theta_{M}K_{c}^* - K_{c}\theta_{M}] T_c^{\lambda t} \dd t \nn \\
	&=  \dd \Gamma(-\Delta ) + \lambda \int_0^1  T_c^{-\lambda t} (\theta_{M} [\dd \Gamma(-\Delta),K_{c}^*] + \hc )  T_c^{\lambda t} \dd t \nn \\
	&=  \dd \Gamma(-\Delta ) - \lambda \bigg(\theta_{M} \int \varphi(x) N^{5/2} (V_N f_N)(x-y) a^*_x a^*_y a_y + \hc  \bigg)\nn \\
	&\quad + \lambda^2 \theta_{M,+1}^2 \int N^{3} (V_Nf_N\omega_N)(x-y) \varphi(x)^2 a^*_y a_y + \lambda^2 \theta_{M,+1}^2  \int N^{3} (V_Nf_N\omega_N)(x-y) \varphi(x)\varphi(y) a^*_x a_y \nn\\
	&\quad + \lambda \int_0^1  T_c^{- \lambda t} (\theta_{M}[\dd \Gamma(-\Delta),K_{c}^*] +  \theta_{M}\int \varphi(x) N^{5/2} (V_N f_N)(x-y) a^*_x a^*_y a_y)  T_c^{\lambda t} \dd t  + \hc  \nn \\
	&\quad - \lambda^2 \int_0^1 \int_0^t  T_c^{-\lambda u} \Bigg\{ [\theta_{M} \int \varphi(x) N^{5/2} (V_N f_N)(x-y) a^*_x a^*_y a_y ,\theta_M K_{c}^* - K_c \theta_M] \nn \\
	&\qquad \qquad \qquad\qquad\qquad + \theta_{M,+1}^2 \int N^{3} (V_Nf_N\omega_N)(x-y) \varphi(x)^2 a^*_y a_y \nn  \\
		&\qquad \qquad \qquad\qquad\qquad + \theta_{M,+1}^2  \int N^{3} (V_Nf_N\omega_N)(x-y) \varphi(x)\varphi(y) a^*_x a_y\Bigg\} T_c^{\lambda u}  \dd u \dd t + \hc  \nn \\
	&\quad - 2 \lambda^3 \int_0^1 \int_0^t \int_0^u T_c^{-\lambda v}  [\theta_{M,+1}^2 \int N^{3} (V_Nf_N\omega_N)(x-y) \varphi(x)^2 a^*_y a_y,\theta_M K_{c}^* - K_c \theta_M]  T_c^{\lambda v} \dd v \dd u \dd t  \nn \\
	&\quad - 2 \lambda^3 \int_0^1 \int_0^t \int_0^u  T_c^{\lambda v} [\theta_{M,+1}^2  \int N^{3} (V_Nf_N\omega_N)(x-y) \varphi(x)\varphi(y) a^*_x a_y,\theta_M K_{c}^* - K_c \theta_M]   T_c^{\lambda v} \dd v \dd u \dd t.
\end{align}
The second term on the right hand side of \eqref{eq:duhamel_kinetic_2} is bounded by the Cauchy--Schwarz inequality
\begin{align}
		\label{eq:cor:estimate_kinetic_cubic_2}
\pm (\theta_{M}\int \varphi(x) N^{5/2} (V_N f_N)(x-y) a^*_x a^*_y a_y  + \hc ) 
	& \leq C \cL_4 +  C \int N^{3} (V_N f_N)\ast\varphi(y)^2  a^*_y a_y  \nn \\
	&\leq C (\cL_4 +  (\mathcal N+1)),
\end{align}
where we used that $\|N^3 V_Nf_N \ast \varphi^2\|_{L^\infty} \leq \|V\|_{L^1} \|\varphi\|_{L^\infty}^2$. The third and fourth terms in (\ref{eq:duhamel_kinetic_2}) are easily bounded by $C\|V\|_{L^1}\|\varphi\|_{L^\infty}^2 (\mathcal N+1)$. Let us now deal with the remaining integral terms and bound the terms between the exponentials. The fifth term in (\ref{eq:duhamel_kinetic_2}) is bounded using Lemma \ref{lemma:adjoint_kinetic_cubic} and that $M\leq N$,
\begin{multline*}
\pm(\theta_{M}[\dd \Gamma(-\Delta),K_{c}^*] +  \theta_{M}\int \varphi(x) N^{5/2} (V_N f_N)(x-y) a^*_x a^*_y a_y) + \hc  \\
 \leq C M^{1/2} N^{-1/2} (\mathcal K + \ell^{-1}(\mathcal N+1)) .
\end{multline*}
For the sixth term in (\ref{eq:duhamel_kinetic_2})  we can write  
\begin{align}
	\label{eq:cor:estimate_kinetic_cubic_3_2}
&[\theta_{M} \int \varphi(x) N^{5/2} (V_N f_N)(x-y) a^*_x a^*_y a_y,\theta_M K_{c}^* - K_c \theta_M]\nn \\
	%&= 2 \theta_{M}\left(\theta_{M} - \theta_{M,-1}\right) \int \varphi(x) N^{5/2} (V_N f_N)(x-y) a^*_x a^*_y a_y K_c^* \nn \\
	&= \theta_{M} \theta_{M,-1} \Big[\int \varphi(x) N^{5/2} (V_N f_N)(x-y) a^*_x a^*_y a_y, K_{c}^*\Big] \nn\\
	&\quad  - \theta_{M,+1}^2 \Big [\int \varphi(x) N^{5/2} (V_N f_N)(x-y) a^*_x a^*_y a_y, K_{c}\Big] \nn \\
	&\quad  + (\theta_{M,+1}^2 -\theta_{M}^2) \int \varphi(x) N^{5/2} (V_N f_N)(x-y) a^*_x a^*_y a_y K_{c}. 
\end{align}
Let us explain \eqref{eq:cor:estimate_kinetic_cubic_3_2} in detail. Noting $L = \int \varphi(x) N^{5/2} (V_N f_N)(x-y) a^*_x a^*_y a_y$, we have
\begin{align*}
[L,\theta_{M}] &= (\theta_{M,-1}-\theta_M) L, \\ 
[K^*_c,\theta_{M}] &= (\theta_{M,-1}-\theta_M) K^*_c, \\
 [K_c,\theta_M] &= (\theta_{M,+1}-\theta_M)K_c.
\end{align*}
Then we can compute 
\begin{align*}
[\theta_{M}L,& \theta_{M} K_c^* - K_c \theta_{M}] = [\theta_{M}L, \theta_{M} K_c^*] - [\theta_{M}L,K_c \theta_{M}] = {\rm (I)} - {\rm (II)}
\end{align*}
with 
\begin{align*}
	{\rm (I)}
	&=  [\theta_{M}L, \theta_{M}]K_c^* + \theta_{M} [\theta_{M}L, K_c^*] \\
	&= \theta_{M} [L, \theta_{M}]K_c^* + \theta_{M}^2 [L, K_c^*] + \theta_{M} [\theta_{M}, K_c^*] L \\
	&=  \theta_{M}(\theta_{M,-1}-\theta_M) L K_c^* + \theta_{M}^2 [L, K_c^*] - \theta_{M} (\theta_{M,-1}-\theta_M) K^*_c L \\
	&=\theta_{M}(\theta_{M,-1}-\theta_M) L K_c^* + \theta_{M}^2 [L, K_c^*] - \theta_{M} (\theta_{M,-1}-\theta_M) ([K^*_c, L] + L K_c^*) \\
	&= \theta_{M} \theta_{M,-1}[L,K^*_c]
\end{align*}
and 
\begin{align*}
	{\rm (II)}
	&= [\theta_{M}L,K_c] \theta_{M} +  K_c [\theta_{M}L, \theta_{M}] \\
	& = \theta_{M} [L,K_c] \theta_{M} + [\theta_{M},K_c] L \theta_{M} + K_c\theta_{M} [L, \theta_{M}]  \\
	& = \theta_{M}^2 [L,K_c]  -\theta_{M}(\theta_{M,+1}-\theta_M)K_c L  +\theta_{M,+1} (\theta_{M}-\theta_{M,+1})  K_cL  \\
	& =  \theta_{M}^2 [L,K_c]  + (\theta_{M}^2 - \theta_{M,+1}^2) ([K_c, L] + L K_c) \\
	&= \theta_{M,+1}^2 [L,K_c] +  (\theta_{M}^2 - \theta_{M,+1}^2) L K_c.
\end{align*}

%The first term is bounded in the following way, for all $\xi,\xi' \in \FM$,
%\begin{align*}
%&2 \theta_{M}\left(\theta_{M} - \theta_{M,-1}\right) \Big|\Big\langle \xi', \int \varphi(x) N^{5/2} (V_N f_N)(x-y) a^*_x a^*_y a_y K_{c}^* \xi \Big\rangle\Big| \\
%	&\leq C \|V\|_{L^1} M^{-1} \|\cL_4^{1/2} \xi'\|  \|\cN^{1/2} K_c^* \xi\|^2 \\
%	&\leq C M^{1/2} N^{-1/2} \ell^{1/2} \|\cL_4^{1/2} \xi'\| \|\cN^{1/2} \xi\|,
%\end{align*}
%where we used that $\|\theta_{M} -\theta_{M,-1}\|_{op} \leq C M^{-1} $ and that $\|k_c\|_{op} \leq C N^{-1/2} \ell^{1/2}$. 
Now we focus on the right hand side of \eqref{eq:cor:estimate_kinetic_cubic_3_2}. Applying Lemma \ref{lemma:adjoint_L3_cubic} with the potential $V_N f_N$ instead of $V_N$ in the definition of $\cL_3^{(T_1)}$ allows to deal with the first two terms (note that $0\leq f_N \leq 1$), we obtain that for all $\xi,\xi' \in \FM$,
\begin{align*}
&\Big|\Big\langle \xi',\Big[\int \varphi(x) N^{5/2} (V_N f_N)(x-y) a^*_x a^*_y a_y, K_{c}^*\Big] \xi \Big\rangle\Big| \\
	&\leq C M N^{-1} \| (\mathcal N+1)^{1/2}\xi'\| \| \mathcal K^{1/2}\xi\|  + CM^{1/2} N^{-1/2} \|\cL_4^{1/2}\xi'\| \|\mathcal K^{1/2}\xi\|,
\end{align*}
and
\begin{align*}
&\Big[\int \varphi(x) N^{5/2} (V_N f_N)(x-y) a^*_x a^*_y a_y, K_{c}\Big]  \\
	&= \int N^{3} (V_Nf_N\omega_N)(x-y) \varphi(x)^2 a^*_y a_y  + \int N^{3} (V_Nf_N\omega_N)(x-y) \varphi(x)\varphi(y) a^*_x a_y  + \mathcal E,
\end{align*}
where for all $\xi,\xi' \in \FM$,
\begin{align*}
|\langle \xi',  \mathcal E \xi \rangle|  \leq  CM N^{-1}  \| (\mathcal N+1)^{1/2}\xi'\| \| \mathcal K^{1/2}\xi\| + C M^{1/2} N^{-1/2} \|\cL_4^{1/2}\xi'\| \|\mathcal K^{1/2}\xi\|.
\end{align*}
Moreover we have, for all $\xi,\xi' \in \FM$
\begin{align}
&\left|\langle \xi', \int \varphi(x) N^{5/2} (V_N f_N)(x-y) (\theta_{M,+1}^2 -\theta_{M}^2) a^*_x a^*_y a_y K_{c} \xi \rangle \right| \nn \\
	&\leq \eta  \|\cL_4^{1/2}  \xi'\|^2 + \eta ^{-1} \int N^{3} V_N\ast \varphi^2 (y)  \|a_y K_{c} (\theta_{M,+1}^2 -\theta_{M}^2) \xi\|^2 \nn \\
	&\leq C N^{-1/2} \ell^{1/2} \|\varphi\|_{L^\infty} \|V\|_{L^1}^{1/2} \|\cL_4^{1/2} \xi'\| \|(\theta_{M,+3}^2 -\theta_{M,+2}^2)  (\mathcal N+1)^{2} \xi\| \nn\\
	&\leq C M^{1/2} N^{-1/2} \ell^{1/2} \|\cL_4^{1/2} \xi'\| \|(\mathcal N+1)^{1/2} \xi\|\label{eq:cor:estimate_kinetic_cubic_4},
\end{align}
where we used again that $\|\theta_{M,+3}^2 -\theta_{M,+2}^2\|_{op} \leq C M^{-1} $ and that $\|k_c\|_{op} \leq C N^{-1/2} \ell^{1/2}$. This
concludes the analysis of the terms on the right-hand side of (\ref{eq:cor:estimate_kinetic_cubic_3_2}).

Finally, the last two terms in (\ref{eq:duhamel_kinetic_2}) can be bounded by
\begin{align*}
C \|V\|_{L^1}\|\varphi\|_{L^\infty}^2 M^{1/2} N^{-1/2} \ell^{1/2} (\mathcal N+1)
\end{align*}
in a similar way as in the proof Lemma \ref{lemma:TcL0Tc} (Case 1, which is independent of Lemma \ref{prop:estimate_kinetic_L4_cubictransform}). Hence, using all the above estimates and that $M\leq N$ we obtain
\begin{align}
T_c^{-\lambda}  \mathcal K T_c^{\lambda}
	\leq  C ( \mathcal K + \cL_4)  + C \lambda \int_0^1 T_c^{-\lambda t} \mathcal K T_c^{\lambda t} \dd t  + C\lambda^2 \int_0^1 \int_0^t T_c^{-\lambda u}\big( \mathcal K + \cL_4 \big)T_c^{-\lambda u} \dd u \dd t.	\label{eq:estimate_K_iteration}
\end{align}
Let us now estimate in a similar fashion
\begin{align}
&T_c^{-\lambda} \mathcal L_4 T_c^{\lambda}
	= \mathcal L_4 + \lambda \int_0^1T_c^{-\lambda t} [\mathcal L_4,\theta_M K_{c}^* - K_{c} \theta_M] T_c^{\lambda t}\dd t \nn \\
	&= \mathcal L_4 + \lambda \int_0^1 T_c^{-\lambda} \theta_M [\mathcal L_4,K_{c}^*] T_c^{\lambda} \dd t + \hc  \nn \\
	&=  \mathcal L_4 -\lambda \bigg(\theta_M  \int N^{5/2} (V_N\omega_N) (x-y) \varphi(x) a^*_x a^*_y a_y + \hc  \bigg) \nn \\
	&\quad +\lambda^2  \theta_{M,+1}^2 \int N^{3} (V_N\omega_N^2)(x-y) \varphi(x)^2 a^*_y a_y + \lambda^2 \theta_{M,+1}^2\int N^{3} (V_N\omega_N^2)(x-y) \varphi(x)\varphi(y) a^*_x a_y \nn \\
	&\quad + \lambda \int_0^1  T_c^{- \lambda t} (\theta_{M}[\cL_4,K_{c}^*] +  \theta_{M}\int \varphi(x) N^{5/2} (V_N \omega_N)(x-y) a^*_x a^*_y a_y)  T_c^{\lambda t} \dd t  + \hc  \nn \\
	&\quad - \lambda^2 \int_0^1 \int_0^t  T_c^{-\lambda u} \Bigg\{ [\theta_{M} \int \varphi(x) N^{5/2} (V_N \omega_N)(x-y) a^*_x a^*_y a_y ,\theta_M K_{c}^* - K_c \theta_M] \nn \\
	&\qquad \qquad \qquad\qquad\qquad + \theta_{M,+1}^2 \int N^{3} (V_N\omega_N^2)(x-y) \varphi(x)^2 a^*_y a_y \nn  \\
		&\qquad \qquad \qquad\qquad\qquad + \theta_{M,+1}^2  \int N^{3} (V_N\omega_N^2)(x-y) \varphi(x)\varphi(y) a^*_x a_y\Bigg\} T_c^{\lambda u}  \dd u \dd t + \hc  \nn \\
	&\quad - 2 \lambda^3 \int_0^1 \int_0^t \int_0^u T_c^{-\lambda v} [\theta_{M,+1}^2 \int N^{3} (V_N\omega_N^2)(x-y) \varphi(x)^2 a^*_y a_y,\theta_M K_{c}^* - K_c \theta_M] T_c^{\lambda v}\dd v \dd u \dd t  \nn \\
	&\quad - 2 \lambda^3 \int_0^1 \int_0^t \int_0^u T_c^{-\lambda v} [\theta_{M,+1}^2  \int N^{3} (V_N\omega_N^2)(x-y) \varphi(x)\varphi(y) a^*_x a_y,\theta_M K_{c}^* - K_c \theta_M] T_c^{\lambda v}\dd v \dd u \dd t .\label{eq:bound_L4_cubic_transfo_2}
\end{align}
The four first terms are the main contributions, we nevertheless need to bound them to first obtain (\ref{eq:estimate_K_L_4_2}). The second term in (\ref{eq:bound_L4_cubic_transfo_2}) is bounded by the Cauchy-Schwarz inequality by $\cL_4 + \mathcal N+1$ in the same way as in (\ref{eq:cor:estimate_kinetic_cubic_2}). The third and fourth term are easily bounded by $C \|V\|_{L^1}\|\varphi\|_{L^\infty}^2(\mathcal N+1)$. To bound the fifth and the sixth term, we use Lemmas \ref{lemma:adjoint_1_L_4_cubic} and \ref{lemma:adjoint_L3_cubic} in a similar way as for estimating the fifth and the sixth terms in (\ref{eq:duhamel_kinetic_2}), they are bounded by $ \mathcal K + \cL_4$.
 And the last two terms are bounded similarly as in Lemma \ref{lemma:TcL0Tc} (Case 1), by
\begin{align*}
C \|V\|_{L^1}\|\varphi\|_{L^\infty}^2 M^{1/2} N^{-1/2} \ell^{1/2} (\mathcal N+1).
\end{align*}
Hence, we obtain
\begin{align}
T_c^{-\lambda} \mathcal L_4 T_c^{\lambda}
	&\leq C (\mathcal L_4 + \mathcal N  + 1) + C \lambda  \int_0^1 T_c^{-\lambda t} (\mathcal K + \cL_4 )  T_c^{\lambda t} \dd t   + C \lambda ^2 \int_0^1 \int_0^t  T_c^{-\lambda u} (\mathcal K + \cL_4 ) T_c^{\lambda u} \dd u \dd t. \label{eq:estimate_L_4_iteration}
\end{align}
Combing (\ref{eq:estimate_K_iteration}) and (\ref{eq:estimate_L_4_iteration}), using the bound 
$$ T_c^{-t}(\mathcal N+1) T_c^{t} \leq C (\mathcal N+1),\quad \forall t\in [0,1],$$
from Lemma \ref{lemma:number_part_cubic} and iterating the Duhamel formula, we obtain 
 \begin{equation*}
e^{-\lambda (\theta_M K_{c}^* - K_c \theta_M)}  (\mathcal K + \cL_4) e^{\lambda (\theta_M K_{c}^* - K_c \theta_M)}  \leq C  (\mathcal K + \cL_4)
 \end{equation*}
from which (\ref{eq:estimate_K_L_4_1}) follows. Thanks to (\ref{eq:estimate_K_L_4_1}), we can estimate more precisely the error terms in the Duhamel expansions, i.e. the terms with the integrals in (\ref{eq:duhamel_kinetic_2}) and (\ref{eq:bound_L4_cubic_transfo_2}), to obtain (\ref{eq:estimate_error_kinetic_cubic_transfo}) and (\ref{eq:estimate_error_L_4_cubic_transfo}), in particular allowing to also prove (\ref{eq:estimate_K_L_4_2}). 
\end{proof}

\begin{proof}[Proof of Lemma \ref{lemma:adjoint_kinetic_cubic}]
We have
\begin{align*}
[\dd\Gamma(-\Delta)&, K_{c}^*]
	= \sum_{i=1}^3 \mathcal K^{(i)}, \\
\mathcal K^{(1)}
	&=- \int (\Delta_1 \circ k_{c})(x,y,y') a^*_{x} a^*_y a_{y'}, \\
\mathcal K^{(2)}
	&= - \int (\Delta_2 \circ k_c)(x,y,y') a^*_x a^*_y a_{y'}, \\
\mathcal K^{(3)}
	&=  \int (k_c \circ \Delta)(x,y,y') a^*_{x} a^*_y a_{y'},
\end{align*}
where $\circ$ denotes the composition between two applications. Let us estimate $\mathcal K^{(i)}$. On $\FM$, we have
\begin{align}
\mathcal K^{(1)} 
	&= \int - \Delta_1 (Q^{\otimes 2} \widetilde k_{c} Q)(x,y,y') a^*_{x} a^*_y a_{y'}, \nn \\
	&= \int - \Delta_1 (Q \otimes 1 \widetilde k_{c})(x,y,y') a^*_{x} a^*_y a_{y'}, \nn \\
	&= \int - \Delta_1 ((\varphi \otimes 1 ) N^{1/2}\omega_N) (x,y) a^*_{x} a^*_y a_{y} + \int  \Delta_1 \varphi(x) N^{1/2} \varphi(t)^2 \omega_N (t-y) a^*_{x} a^*_y a_{y} \nn \\
	&= N^{1/2} \int \varphi(x) (-\Delta \omega_{\ell,N})(x-y) a^*_x a^*_y a_y - 2N^{1/2} \int \nabla \varphi (x) \cdot \nabla \omega_{\ell,N} (x-y) a^*_x a^*_y a_y\nn  \\
	&\quad + N^{1/2} \int (-\Delta \varphi)(x) \omega_{\ell,N}(x-y) a^*_x a^*_y a_y + \int  \Delta_1 \varphi(x) N^{1/2} \varphi(t)^2 \omega_{\ell,N} (t-y) a^*_{x} a^*_y a_{y}. \label{eq:est_kin_cubic_1}
\end{align}
Let us show that the last three terms above are error terms. Recall from Lemma \ref{prop:prop_epsilon} that $|\nabla\omega_{\ell,N}(x-y)| \leq N^{-1}|x-y|^{-2} \mathds{1}_{\{|x-y|\leq C\ell\}}$. Hence, for all $\xi,\xi'\in \FM$, and all $\eta>0$,
\begin{align*}
| \langle \xi', N^{1/2} &\int \nabla \varphi (x) \cdot \nabla \omega_{\ell,N} (x-z)  a^*_x a^*_z a_z \xi \rangle| \\
	&\leq \eta  \int N^{1/2} |\nabla \omega_{\ell,N} (x-z)|   \|a_x a_z \xi'\|^2  + \eta^{-1} \int N^{1/2}  |\nabla \varphi (x)|^2 |\nabla \omega_{\ell,N} (x-z)|  \|a_z \xi \|^2 \\
	&\leq \eta CN^{-1/2}  \int \frac{1}{|x-z|^2}   \|a_x a_z \xi'\|^2 +  \eta^{-1} N^{-1/2} \int  |\nabla \varphi (x)|^2  \frac{1}{|x-z|^2} \|a_z \xi \|^2 \\
	&\leq C \eta  N^{-1/2} \| (\mathcal N+1)^{1/2}\mathcal K^{1/2} \xi'\|^2 + C \eta^{-1} N^{-1/2} \|\varphi\|_{H^1}^2 \|\mathcal K^{1/2} \xi\|^2 \\
	&\leq C M^{1/2} N^{-1/2} \| \mathcal K^{1/2} \xi'\| \|\mathcal K^{1/2} \xi\|,
\end{align*}
where we used that $\xi'\in \FM$ and optimized over $\eta>0$ to obtain the last inequality. Similarly, for all $\eta>0$, we have on $\FM$,
\begin{align*}
\pm N^{1/2} \int (-\Delta \varphi)(x) \omega_{\ell,N}(x-z) a^*_x a^*_z a_z + \hc
	&\leq \eta \mathcal N^2 + \eta^{-1} \|-\Delta \varphi\|_{L^\infty}^2 \|N^{1/2} \omega_{\ell,N}\|_{L^2}^2 \mathcal N \\
	&\leq  \eta C \mathcal N^2 + \eta^{-1} C N^{-1} \mathcal N \\
	&\leq C M^{1/2} N^{-1/2} \mathcal N,
\end{align*}
where we optimized over $\eta>0$. Again, for all $\eta>0$, we have on $\FM$, 
\begin{align*}
\pm \int  \Delta_1 \varphi(x) N^{1/2} \varphi(t)^2 \omega_{\ell,N} (t-z) a^*_{x} a^*_z a_{z} + \hc
	&\leq \eta \mathcal N^2 + \eta^{-1} \|\Delta \varphi\|_{L^\infty}^2 \| N \omega_{\ell,N}^2 \ast |\varphi|^2 \|_{L^\infty} \mathcal N \\
	&\leq \eta \mathcal N^2 + \eta^{-1} C \|N^{1/2} \omega_N\|_{L^2}^2 \mathcal N \\
	&\leq \eta \mathcal N^2 + \eta^{-1} CN^{-1} \mathcal N \\
	&\leq C M^{1/2} N^{-1/2} \mathcal N,
\end{align*}
where we optimized over $\eta>0$.
We now deal with the first term on the right-hand side of (\ref{eq:est_kin_cubic_1}). Since $-2\Delta \omega_{\ell,N} = N^2 V_N f_N - N^2 \varepsilon_{\ell,N}$, we only need to estimate, for all $\xi, \xi' \in \FM$,
\begin{align*}
\Big|\Big\langle \xi', N^{5/2} &\int \varepsilon_{\ell,N}(x-y) \varphi(x) a^*_x a^*_y a_y \xi \Big\rangle\Big| \\
	&\leq \eta N^{5/2}  \|\varphi\|_{L^\infty} \int|\varepsilon_{\ell,N}(x-y)| \|a_x a_y \xi'\|^2 + \eta^{-1}  N^{5/2} \|\varphi\|_{L^\infty} \int \varepsilon_{\ell,N}(x-y) \|a_y \xi\|^2 \\
	&\leq \eta C N^{5/2}  \|\varepsilon_{\ell,N} \|_{L^{3/2}} M \|\mathcal K^{1/2} \xi'\|^2 + \eta^{-1} C N^{5/2} \|\varepsilon_{\ell,N}\|_{L^{1}} \|(\mathcal N+1)^{1/2} \xi\|^2 \\
	&\leq \eta C N^{-1/2} {\ell}^{-1} M \|\mathcal K^{1/2} \xi'\|^2  + \eta^{-1} C N^{-1/2}  \|(\mathcal N+1)^{1/2} \xi\|^2 \\
	&\leq C M^{1/2}N^{-1/2} {\ell}^{-1/2} \|\mathcal K^{1/2} \xi' \| \|(\mathcal N+1)^{1/2} \xi\|,
\end{align*}
where we used \eqref{eq:w-pointwise} and optimized over $\eta >0$. We now will estimate the sum $\mathcal K^{(2)} + \mathcal K^{(3)}$. Let us first note that, on $\{\varphi\}^\perp$, we have
\begin{align*}
-\Delta_2 k_c + k_c \Delta 
	&= 1 \otimes (-\Delta Q) \widetilde k_{c} + \widetilde k_{c} Q \Delta \\
	&= -1 \otimes \Delta \widetilde k_{c} + \widetilde k_{c} \Delta + 1\otimes \ket{ \Delta \varphi} \bra{\varphi} \widetilde k_{c} - \widetilde k_{c} \ket{\varphi} \bra {\Delta \varphi} \\
	&= (-\Delta_2 (\widetilde k_{c})) - 2 \nabla_2(\widetilde k_{c}) \cdot \nabla  + 1\otimes \ket{ \Delta \varphi} \bra{\varphi} \widetilde k_{c} - \widetilde k_{c} \ket{\varphi} \bra {\Delta \varphi}.
\end{align*}
 we have, on $\FM$,
\begin{align*}
&(\mathcal K^{(2)} + \mathcal K^{(3)})
	=  N^{1/2} \int \varphi(x) (-\Delta \omega_{\ell,N})(x-y) a^*_x a^*_y a_y + 2 N^{1/2} \int \varphi(x) \nabla \omega_{\ell,N}(x-y) a^*_x a^*_y \nabla a_y \\
	&\quad + N^{1/2} \int (\Delta \varphi)(y) \varphi (x) \varphi(z) \omega_{\ell,N}(x-z) a^*_x a^*_y a_z - N^{1/2} \int \varphi(x) \omega_{\ell,N}(x-y) \varphi(y) (\Delta \varphi)(z) a^*_x a^*_y a_z.
\end{align*}
The first term is equal to the main contribution of $\mathcal K^{(1)}$, let us show that the last three terms are error terms. We have for all $\eta>0$,  on $\FM$,
\begin{align*}
&\pm \Big(N^{1/2} \int \varphi(x) \nabla \omega_{\ell,N}(x-y) a^*_x a^*_y \nabla a_y + \hc \Big) \\
	&\leq \eta \int  N^{1/2} |\nabla \omega_{\ell,N}| (x-y) a^*_x a^*_y a_x a_y 
		+ \eta^{-1} \int\varphi(x)^2  N^{1/2} |\nabla \omega_{\ell,N}| (x-y) \nabla a^*_y \nabla a_y \\
	&\leq \eta C \int  N^{-1/2}\frac{1}{|x-y|^2} a^*_x a^*_y a_x a_y + \eta^{-1} N^{1/2}\|\nabla \omega_{\ell,N}\|_{L^1} \mathcal K \\
	&\leq C M^{1/2} N^{-1/2} \mathcal K,
\end{align*}
where we used Hardy's inequality $|x|^{-2} \leq  4(-\Delta)$, that $N^{1/2} \|\nabla \omega_{\ell,N}\|_{L^1} \leq C N^{-1/2}$ and where we chose $\eta = M^{-1/2}$. Similarly, for all $\eta>0$, we have on $\FM$,
\begin{align*}
\pm (N^{1/2} \int (\Delta \varphi)(y) \varphi (x) \varphi(z) &\omega_N(x-z) a^*_x a^*_y a_z + \hc ) \\
	&\leq \eta \int \varphi(z)^2 a^*_x a^*_y a_x a_y + \eta^{-1} \int N (\Delta \varphi)(y)^2\varphi(x)^2 \omega_N(x-z)^2 a^*_z a_z \\
	&\leq \eta \|\varphi\|_{L^2}^2 \mathcal N^2 + \eta^{-1} \|\Delta \varphi\|_{L^2}^2 \|\varphi\|_{L^\infty}^2 \|N^{1/2} \omega_N\|_{L^2}^2 \mathcal N \\
	&\leq C M^{1/2} N^{-1/2} \mathcal N,
\end{align*}
where we optimized over $\eta>0$. Again, we have for all $\eta>0$,  on $\FM$,
\begin{align*}
\pm (N^{1/2} \int \varphi(x) \omega_N(x-y) &\varphi(y) (\Delta \varphi)(z) a^*_x a^*_y a_z + \hc ) \\
	&\leq \eta \int (\Delta \varphi(z))^2 a^*_x a^*_y a_x a_y + \eta^{-1} \int N \varphi (y)^2\varphi(x)^2 \omega_N(x-y)^2 a^*_z a_z \\
	&\leq \eta \|\varphi\|_{L^2}^2 \mathcal N^2 + \eta^{-1} \|\varphi\|_{L^2}^2 \|\varphi\|_{L^\infty}^2 \|N^{1/2} \omega_N\|_{L^2}^2 \mathcal N \\
	&\leq C M^{1/2} N^{-1/2} \mathcal N,
\end{align*}
where we optimized over $\eta>0$.
\end{proof}

\begin{proof}[Proof of Lemma \ref{lemma:adjoint_1_L_4_cubic}]
We have on $\mathcal F_+$
\begin{align*}
[\cL_4,K_{c}^*]
	&=  \dd\Gamma(L_4^{(1)}) + \sum_{i=2}^4 \dd\Gamma(L_4^{(i)}), \\
 \dd\Gamma(L_4^{(1)}) 
 	&= \int N^2 V_N (x-y) k_{c}(x,y,y') a^*_x a^*_y a_{y'}, \\
 \dd\Gamma(L_4^{(2)}) 
 	&= \int N^2 V_N (x-y) k_{c}(x,z,z') a^*_x a^*_y a^*_z a_{y} a_{z'}, \\
 \dd\Gamma(L_4^{(3)}) 
 	&= \int N^2 V_N (x-y) k_{c}(z,y,y') a^*_x a^*_y a^*_z a_{x} a_{y'}, \\
 \dd\Gamma(L_4^{(4)}) 
 	&= - \int N^2 V_N (x-y') k_{c}(z,y,y') a^*_x a^*_y a^*_z a_{x} a_{y'}.
\end{align*}
Let us estimate $\dd\Gamma(L_4^{(i)})$. For $\dd\Gamma(L_4^{(1)})$, note that 
\begin{align*} Q^{\otimes 2} \widetilde k_{c}  =  \widetilde k_{c} - P\otimes 1 \widetilde k_{c} - 1\otimes P \widetilde k_{c}+ P^{\otimes 2} \widetilde k_{c}.
\end{align*} 
Therefore, it is enough to bound, for all $\xi,\xi' \in \FM$
\begin{align*}
\Big|\Big\langle \xi',  &(\dd\Gamma(L_4^{(1)}) -  \int N^{5/2} (V_N\omega_N) (x-y) \varphi(x)  a^*_x a^*_y a_y)) \xi\Big\rangle\Big| \\
&\leq \Big|\Big\langle \xi',\int N^2 V_N(x-y) (P\otimes 1 \widetilde k_{c}+  1\otimes P \widetilde k_{c} - P^{\otimes 2} \widetilde k_{c})(x,y,y')a^*_x a^*_y a_{y'})) \xi \Big\rangle\Big|\\
 	&\leq \int N^2 V_N (x-y) \varphi(x) \varphi(t)^2 N^{1/2} \omega_{\ell,N}(t-y) \|a_x a_y \xi'\| \| a_y \xi \| \\
	&\qquad + \int N^2 V_N (x-y) \varphi(x) \varphi(y) \varphi(y') N^{1/2} \omega_{\ell,N}(x-y') \|a_x a_y \xi'\| \| a_{y'} \xi \|  \\
	&\qquad + \int N^2 V_N (x-y) \varphi(x) \varphi(y) \varphi(t)^2 \varphi(y') N^{1/2} \omega_{\ell,N}(t-y') \|a_x a_y \xi'\| \| a_{y'} \xi \| \\
	&\leq \eta \|\cL_4^{1/2} \xi'\|^2 + \eta^{-1} \|V\|_{L^1} ( \|\varphi\|_{L^\infty}^3+  \|\varphi\|_{L^\infty}^2 \|\varphi\|_{L^2} + \|\varphi\|_{L^\infty}^3 \|\varphi\|_{L^2}^2) \|\omega_{\ell,N}\|_{L^2}^2 \|\mathcal N^{1/2}\xi\|^2 \\
	&\leq  \eta \|\cL_4^{1/2} \xi'\|^2  + \eta^{-1} C N^{-2} \ell \|\mathcal N^{1/2}\xi\|^2 \\
	&\leq C N^{-1}\ell^{1/2} \|\cL_4^{1/2} \xi'\| \|\mathcal N^{1/2}\xi\|
\end{align*}
where $\eta$ has been chosen to be proportional to $ N^{-1} \ell^{1/2}\|\mathcal N^{1/2}\xi\| \|\cL_4^{1/2} \xi'\|^{-1}$. For  $\dd\Gamma(L_4^{(2)})$, using Lemma \ref{prop:cubic_kernel_est}, we have, for all $\xi,\xi' \in \FM$,
\begin{align*}
&|\langle \xi',  \dd\Gamma(L_4^{(2)})  \xi \rangle|  = \Big| \Big\langle \xi', \int N^2 V_N (x-y) k_{c}(x,z,z') a^*_x a^*_y a^*_z a_{y} a_{z'} \xi \Big\rangle \Big| \\
	&\leq \eta \int N^2 V_N (x-y) \|(\mathcal N+1)^{-1/2}a_x a_y a_z \xi' \|^2 \\
	&\qquad+ \eta^{-1} \int N^2 V_N (x-y) k_{c,2}(x,z,z')^2 \|(\mathcal N+1)^{1/2} a_y a_{z'} \xi \|^2 \\
	&\qquad +\eta^{-1} \int N^2 V_N (x-y) k_{c,1}(x,z)^2 \|(\mathcal N+1)^{1/2}a_y a_{z} \xi \|^2 \\
	&\leq \eta \| \cL_4^{1/2}\xi'\|^2 + \eta^{-1} \int N^2 V_N (x-y) \varphi(x)^2 (N^{-1} \ell^4 + N \omega_{\ell,N}(x-z)^2)  \|(\mathcal N+1)^{1/2}a_y a_z \xi\|^2 \\
	&\quad +  \eta^{-1} \int N^2 V_N (x-y) \varphi(x)^2 \varphi(z)^2 \varphi(z')^2 \times \\
	&\qquad\qquad\qquad\qquad\qquad \times(N^{-1} \ell^4 + N \omega_{\ell,N}(x-z)^2 +  N \omega_{\ell,N}(x-z')^2) \| (\mathcal N+1)^{1/2} a_y a_{z'} \xi\|^2 \\
	&\leq \eta \| \cL_4^{1/2}\xi'\|^2 + \eta^{-1} \Bigg(\|\varphi\|_{L^\infty}^2 \|V\|_{L^1}  N^{-2} \ell^4 \|(\mathcal N+1)^{3/2} \xi\|^2 \\
	&\qquad \qquad \qquad \qquad \qquad \qquad  + \|\varphi\|_{L^\infty}^2 \int N V_N(x-y)\frac{1}{|x-z|^2} \| (\mathcal N+1)^{1/2}a_y a_z \xi\|^2  \Bigg) \\
	&\quad + \eta^{-1} \Bigg( \|\varphi\|_{L^\infty}^4 \|\varphi\|_{L^2}^2 \|V\|_{L^1} N^{-2} \ell^4  \|(\mathcal N+1)^{3/2} \xi\|^2 + \|\varphi\|_{L^\infty}^6 \| \omega_N\|_{L^2}^2 \|V\|_{L^1}  \|(\mathcal N+1)^{3/2} \xi\|^2  \\
	&\qquad \qquad \qquad +\|\varphi\|_{L^2}^2 \|\varphi\|_{L^\infty}^4 \int N V_N(x-y)\frac{1}{|x-z'|^2} \|(\mathcal N+1)^{1/2} a_y a_{z'} \xi \|^2   \Bigg) \\
	&\leq \eta \| \cL_4^{1/2}\xi'\|^2 + \eta^{-1} C N^{-2}\ell M^2 \|(\mathcal N+1)^{1/2} \xi\|^2 + \eta^{-1} C N^{-2} \|V\|_{L^1} \|(\mathcal N+1) \mathcal K^{1/2} \xi\|^2) \\
	&\leq C M N^{-1} \|\cL_4^{1/2}\xi'\| \|\mathcal K^{1/2} \xi\|
\end{align*}
where $\eta$ has been chosen to be proportional to $M N^{-1} \|\mathcal K^{1/2}\xi\| \|\cL_4^{1/2} \xi'\|^{-1}$. %
%\begin{align*}
%|\langle g,  &\widetilde L_4^{(2)}  f\rangle | \\
%	&\leq \int N^2 V_N (x-y) |k_{c,2}(x,z,z')|  |g(x,y,z)| |f(y,z')| + \int N^2 V_N (x-y) |k_{c,1}(x,z)|  |g(x,y,z)| |f(y,z)|  \\
%	&\leq \eta \int N^2 V_N (x-y) |g(x,y,z)|^2 + \eta^{-1} \int N^2 V_N (x-y) k_{c,2}(x,z,z')^2 |f(y,z')|^2 \\
%	&\qquad \qquad +\eta^{-1} \int N^2 V_N (x-y) k_{c,1}(x,z)^2 |f(y,z)|^2  \\
%	&\leq \eta \|V\|_{L^{3/2}} \|\nabla_1 g\|_{L^2}^2 + \eta^{-1} \int N^2 V_N (x-y) \varphi(x)^2 (N^{-7/3} + N \omega_N(x-z)^2) |f(y,z)|^2 \\
%	&\quad +  \eta^{-1} \int N^2 V_N (x-y) \varphi(x)^2 \varphi(z)^2 \varphi(z')^2 (N^{-7/3} + N \omega_N(x-z)^2 +  N \omega_N(x-z')^2) |f(y,z')|^2  \\
%	&\leq \eta \|V\|_{L^{3/2}} \|\nabla_1 g\|_{L^2}^2 + \eta^{-1} \Bigg(\|\varphi\|_{L^\infty}^2 \|V\|_{L^1} N^{-10/3} \|\mathcal N \xi\|^2 + \|V\|_{L^1} \|\varphi\|_{L^\infty}^2 \int \mathds{1}_{|y-z| \leq C \ell N^{-1}} \| a_y a_z \xi\|^2  \Bigg) \\
%	&\quad + \eta^{-1} \Bigg( \|\varphi\|_{L^\infty}^4 \|\varphi\|_{L^2}^2 \|V\|_{L^1} N^{-10/3} \|f\|_{L^2}^2 + \|\varphi\|_{L^\infty}^6 \| \omega_N\|_{L^2}^2 \|V\|_{L^1} \|f\|_{L^2}^2 \\
%	&\qquad \qquad \qquad + \|V\|_{L^1} \|\varphi\|_{L^\infty}^4 \|\varphi\|_{L^2}^2 \int \mathds{1}_{|y-z'| \leq C \ell N^{-1}} |f(y,z')|^2   \Bigg) \\
%	&\leq \eta \|V\|_{L^{3/2}} \|\nabla_1 g\|_{L^2}^2  + \eta^{-1} (C N^{-7/3} \|f\|_{L^2}^2 + N^{-2/3} \|\nabla_1 f\|_{L^2}^2) \\
%	&\leq C N^{-1/3} (\|\nabla_1 f\|_{L^2}^2 + \|f\|_{L^2}^2)  (\|\nabla_1 g\|_{L^2}^2 + \|g\|_{L^2}^2).
%\end{align*}
Now we turn to $\dd\Gamma(L_4^{(3)})$, using again Lemma \ref{prop:cubic_kernel_est}, we have, for all $\xi,\xi' \in \FM$,
\begin{align*}
&|\langle \xi',  \dd\Gamma(L_4^{(3)})  \xi \rangle|  = \left| \langle \xi', \int N^2 V_N (x-y) k_{c}(z,y,y'') a^*_x a^*_y a^*_z a_{x} a_{y''} \xi \rangle \right| \\
	&\leq \eta \|\mathcal N^{1/2} \cL_4^{1/2}\xi'\|^2 + \eta^{-1} \int N^2 V_N (x-y) k_{c}(z,y,y'')^2 \|a_{x} a_{y''} \xi\|^2 \\
	&\leq \eta \|\mathcal N^{1/2} \cL_4^{1/2}\xi'\|^2 + \eta^{-1} \int N^2 V_N (x-y) \varphi(z)^2 (N^{-1} \ell^4 + N \omega_{\ell,N}(z-y)^2) \|a_x a_y \xi\|^2 \\
	&\quad + \eta^{-1} \int N^2 V_N (x-y) \varphi(z)^2 \varphi(y)^2 \varphi(y')^2 (N^{-1} \ell^4 + N \omega_{\ell,N}(z-y)^2 + N \omega_{\ell,N}(z-y')^2 ) \|a_x a_{y'} \xi\|^2 \\
	&\leq \eta \|\mathcal N^{1/2} \cL_4^{1/2}\xi'\|^2 + \eta^{-1} \|\varphi\|_{L^2} N^{-1} \ell^4 \|\cL_4^{1/2}\xi \|^2 + \eta^{-1} \|\varphi\|_{L^\infty}^2 N \| \omega_{\ell,N}\|_{L^2}^2 \|\cL_4^{1/2}\xi \|^2 \\
	&\quad + \eta^{-1}N^{-2} \ell^4 \|V\|_{L^1} \|\varphi\|_{L^\infty}^2 \|\varphi\|_{L^2}^4 \|\mathcal N \xi\|^2 + 2 \eta^{-1} \|\varphi\|_{L^\infty}^6 \| \omega_{\ell,N}\|_{L^2}^2 \|V\|_{L^1} \|\mathcal N \xi\|^2 \\
	&\leq \eta \|\mathcal N^{1/2} \cL_4^{1/2}\xi'\|^2 + \eta^{-1} CN^{-1} \ell \|\cL_4^{1/2}\xi\|^2 + \eta^{-1} CN^{-2} \ell\|\mathcal N \xi\|^2 \\
	&\leq C M^{1/2} N^{-1/2} \ell^{1/2} \| \cL_4^{1/2}\xi'\| \| \cL_4^{1/2}\xi\| +  C M N^{-1} \ell^{1/2}  \| \cL_4^{1/2}\xi'\| \|\mathcal N^{1/2} \xi\|
	\end{align*}
	with an appropriate choice of $\eta>0$.  Finally, similar computations show that for all $\xi,\xi' \in \FM$,
\begin{align*}
|\langle \xi',  \dd\Gamma(L_4^{(4)})  \xi \rangle| 	&\leq C M^{1/2} N^{-1/2} \ell^{1/2} \| \cL_4^{1/2}\xi'\| \| \cL_4^{1/2}\xi\| +  C M N^{-1} \ell^{1/2}  \| \cL_4^{1/2}\xi'\| \|\mathcal N^{1/2} \xi\|.
\end{align*}
Collecting all the above estimates proves Lemma \ref{lemma:adjoint_1_L_4_cubic}.
\end{proof}

\begin{proof}[Proof of Lemma \ref{lemma:adjoint_L3_cubic}]

Let us compute and estimate 
\begin{align*}
- [\cL_3^{(T_1,*)}, K_{c}] &= \sum_{i=1}^{7} \cL_3^{(T_1,i)} ,\\
\cL_3^{(T_1,1)}
	&=\int N^{5/2} V_N(x-y) \varphi(x) \overline{k_{c}(x,y,y')} a^*_{y'} a_y , \\
\cL_3^{(T_1,2)}
	&=\int N^{5/2} V_N(x-y) \varphi(x) \overline{k_{c}(y,x,x')} a^*_{x'} a_y , \\
\cL_3^{(T_1,3)}
	&=-\int N^{5/2} V_N(x-y) \varphi(x) \overline{k_{c}(z,y',y)} a^*_{x} a^*_y a_{y'} a_{z} , \\
\cL_3^{(T_1,4)}
	&=\int N^{5/2} V_N(x-y) \varphi(x) \overline{k_{c}(y,z,z')} a^*_{x} a^*_{z'} a_{y} a_{z} , \\
\cL_3^{(T_1,5)}
	&=\int N^{5/2} V_N(x-y) \varphi(x) \overline{k_{c}(z,y,y')} a^*_{x} a^*_{y'} a_{y} a_{z} , \\
\cL_3^{(T_1,6)}
	&=\int N^{5/2} V_N(x-y) \varphi(x) \overline{k_{c}(x,z,z')} a^*_{z'} a^*_{y} a_{y} a_{z} , \\
\cL_3^{(T_1,7)}
	&=\int N^{5/2} V_N(x-y) \varphi(x) \overline{k_{c}(z,x',x)} a^*_{x'} a^*_{y} a_{y} a_{z} ,
\end{align*}
and
\begin{align*}
[\cL_3^{(T_1,*)}, K_{c}^*] &= \sum_{i=8}^{11} \cL_3^{(T_1,i)} ,\\
\cL_3^{(T_1,8)}
	&=\int N^{5/2} V_N(x-y) \varphi(x) k_{c}(y,z,z') a^*_{x} a^*_y a^*_z a_{z'} , \\
\cL_3^{(T_1,9)}
	&=\int N^{5/2} V_N(x-y) \varphi(x) k_{c}(z,y,y') a^*_{x} a^*_y a^*_{z} a_{y'} , \\
\cL_3^{(T_1,10)}
	&=-\int N^{5/2} V_N(x-y) \varphi(x) k_{c}(z,y',y) a^*_{x} a^*_{y'} a^*_{z} a_{y} , \\
\cL_3^{(T_1,11)}
	&=-\int N^{5/2} V_N(x-y) \varphi(x) k_{c}(z,x',x) a^*_{z} a^*_{x'} a^*_{y} a_{y}.
\end{align*}
Now let us estimate $\cL_3^{(T_1,i)}$. We begin with $\cL_3^{(T_1,1)}$, on $\mathcal F_+$ we have, 
\begin{align*}
\cL_3^{(T_1,1)}
	&= - \int N^{3} V_N(x-y) \varphi(x) \Bigg(\varphi(x) \omega_{\ell,N}(x-y) \delta_{y,y'} - \varphi(x) \int \dd z \varphi(z)^2 \omega_{\ell,N}(z-y) \delta_{y,y'} \\
	&\qquad - \varphi(x) \varphi(y) \varphi(y') \omega_{\ell,N}(x-y') + \varphi(x) \varphi(y) \int \dd z \varphi(z)^2 \varphi(y') \omega_{\ell,N}(z-y')\Bigg) a^*_{y'} a_{y}.
\end{align*}
Let us prove that the last three terms above are error terms, we have for all $\xi,\xi'\in \FM$
\begin{align*}
&\Big|\Big\langle \xi', \int N^{3} V_N(x-y) \varphi(x)^2 \omega_{\ell,N}(z-y) a^*_{y} a_y \xi \Big\rangle \Big| \\
&\leq \|\varphi\|_{L^\infty}^2\|V\|_{L^1} \|\omega_{\ell,N}\|_{L^1} \| \mathcal N^{1/2} \xi'\|  \| \mathcal N^{1/2} \xi\| \leq C N^{-1} \ell^2 \| \mathcal N^{1/2} \xi'\|  \| \mathcal N^{1/2} \xi\|.
\end{align*}
Similarly for all $\xi,\xi'\in \FM$,
\begin{align*}
&\Big|\Big\langle \xi',  \int N^{3} V_N(x-y) \varphi(x)^2 \varphi(y) \varphi(y') \omega_{\ell,N}(x-y') a^*_{y'} a_y\xi \Big\rangle \Big| \\
	&\leq \|\varphi\|_{L^\infty}^4 \|V\|_{L^1} \|\omega_{\ell,N}\|_{L^1} \| \mathcal N^{1/2} \xi'\|  \| \mathcal N^{1/2} \xi\|  \leq CN^{-1} \ell^2 \| \mathcal N^{1/2} \xi'\|  \| \mathcal N^{1/2} \xi\|,
\end{align*}
and for all $\xi,\xi'\in \FM$
\begin{align*}
&\Big|\Big\langle \xi', \int N^{3} V_N(x-y)\varphi(x)^2 \varphi(y)\varphi(z)^2 \varphi(y') \omega_{\ell,N}(z-y') a^*_{y'} a_y y\xi \Big\rangle \Big| \\
	&\leq \|\varphi\|_{L^2}^4 \|\varphi\|_{L^\infty}^2 \|V\|_{L^1} \|\omega_{\ell,N}\|_{L^1} \| \mathcal N^{1/2} \xi'\|  \| \mathcal N^{1/2} \xi\|  \leq C N^{-1}\ell^2  \| \mathcal N^{1/2} \xi'\|  \| \mathcal N^{1/2} \xi\|.
\end{align*}
Therefore
\begin{align*}
\cL_3^{(T_1,1)} 
	&=-  \int N^{3} (V_N\omega_N)(x-y) \varphi(x)^2 a^*_y a_y + \mathcal E,
\end{align*}
with for all $\xi,\xi'\in \FM$  $|\langle \xi', \mathcal E \xi \rangle | \leq CN^{-1}\ell^2 \| \mathcal N^{1/2} \xi'\|  \| \mathcal N^{1/2} \xi\|$.

Similarly as for $\cL_3^{(T_1,1)}$, one shows that
\begin{align*}
\cL_3^{(T_1,2)} 
	&= - \int N^{3} (V_N\omega_N)(x-y) \varphi(x)\varphi(y) a^*_y a_y + \mathcal E,
\end{align*}
with $\mathcal E$ satisfying 
$$|\langle \xi', \mathcal E \xi \rangle | \leq CN^{-1}\ell^2 \| \mathcal N^{1/2} \xi'\|  \| \mathcal N^{1/2} \xi\|,\quad \xi,\xi'\in \FM.$$
Next, we turn to $\cL_3^{(T_1,3)}$, we have, for all $\xi,\xi' \in \FM$, 
\begin{align*}
&|\langle \xi', \cL_3^{(T_1,3)} \xi \rangle|
	\leq \int N^{5/2} V_N(x-y) \varphi(x) \varphi(z) \varphi(y') \varphi(y) \times \\
	&\qquad \qquad \times (N^{-1/2}\ell^2 + N^{1/2} \omega_{\ell,N}(z-y') + N^{1/2}\omega_{\ell,N}(z-y)) \| a_{x} a_y \xi'\| \| a_{y'} a_{z} \xi \| \\
	&\qquad + \int N^{5/2} V_N(x-y) \varphi(x) \varphi(z) (N^{-1/2}\ell^2 + N^{1/2}\omega_{\ell,N}(z-y)) \| a_{x} a_y \xi'\| \| a_{y} a_{z} \xi \| \\
	&\leq \eta \int N^2 V_N(x-y) \varphi(z)^2 \varphi(y')^2  \| a_{x} a_y \xi'\|^2 + \eta \int N^2 V_N(x-y) \varphi(z)^2 \| a_{x} a_y \xi'\|^2 \\
	&\qquad + \eta^{-1}  C \int N^3 V_N(x-y) \varphi(x)^2 \varphi(y)^2  (N^{-1}\ell^4 + N \omega_{\ell,N}(z-y')^2 \\
	&\qquad\qquad\qquad\qquad\qquad\qquad\qquad\qquad\qquad\qquad\qquad+ N\omega_{\ell,N}(z-y)^2) \| a_{y'} a_z \xi' \|^2 \\
	&\qquad + \eta^{-1} C \int N^3 V_N(x-y) \varphi(x)^2 N \omega_{\ell,N}(z-y)^2 \|a_{y} a_z \xi\|^2 \\
	&\leq \eta C  \|\cL_4^{1/2} \xi'\|^2 + \eta^{-1} C \|V\|_{L^1} \Bigg(\|\varphi\|_{L^\infty}^2 \|\varphi\|_{L^2}^2 N^{-1}\ell^4 \| \mathcal N \xi\|^2 \\
	&\qquad \qquad + \|\varphi\|_{L^\infty}^4 \|N^{1/2}\omega_{\ell,N}\|_{L^2}^2 \|\mathcal N\xi\|^2+ \|\varphi\|_{L^\infty}^2 (1 +  \|\varphi\|_{L^2}^2)  N^{-1}  \| \mathcal K^{1/2} \mathcal N^{1/2} \xi\|^2   \Bigg) \\
	&\leq  \eta C  \|\cL_4^{1/2} \xi'\|^2 + C \eta^{-1} N^{-1} (\ell^4  \|\mathcal N\xi\|^2  + \ell \|\mathcal N\xi\|^2 +   \| \mathcal K^{1/2} \mathcal N^{1/2} \xi\|^2  ) \\
	&\leq C N^{-1/2}   \|\cL_4^{1/2} \xi'\| \| \mathcal K^{1/2}\mathcal N^{1/2}  \xi\| \\
	&\leq C M^{1/2} N^{-1/2}  \| \cL_4^{1/2}\xi'\| \| \mathcal K^{1/2}  \xi\|,
\end{align*}
where we used that 
\begin{align} \label{eq:wlN2-bb}
N\omega_{\ell,N}(z-y')^2 \leq C N^{-1} |y'-z|^2 \leq C N^{-1} (-\Delta_z)
\end{align}
which follows from Lemma \ref{prop:prop_epsilon} and Hardy's inequality. 

We now estimate $\cL_3^{(T_1,4)}$, we have, for all $\xi,\xi' \in \FM$, 
\begin{align*}
&|\langle \xi', \cL_3^{(T_1,4)} \xi \rangle|
	\leq \int N^{5/2} V_N(x-y) \varphi(x) \varphi(y) \varphi(z) \varphi(z') \times \\
	&\qquad \qquad \times (N^{-1/2} \ell^2 + N^{1/2} \omega_{\ell,N}(y-z) + N^{1/2}\omega_{\ell,N}(y-z')) \| a_{x} a_{z'} \xi'\| \| a_{y} a_{z} \xi \| \\
	&\qquad + \int N^{5/2} V_N(x-y) \varphi(x) \varphi(y) (N^{-1/2} \ell^2+ N^{1/2}\omega_{\ell,N}(y-z)) \| a_{x} a_{z} \xi'\| \| a_{y} a_{z} \xi \| \\
	&\leq \eta \int N^2 V_N(x-y) \varphi(z)^2  \| a_{x} a_{z'} \xi'\|^2 + \eta \int N^2 V_N(x-y) \| a_{x} a_{z} \xi'\|^2 \\
	&\qquad + \eta^{-1}\int N^3 V_N(x-y) \varphi(x)^2 \varphi(y)^2 \varphi(z')^2 \times \\
	&\qquad \qquad\qquad \times (N^{-1} \ell^4 + N \omega_{\ell,N}(y-z)^2 + N\omega_{\ell,N}(y-z')^2)\| a_{y} a_{z} \xi \|^2 \\
	&\qquad +  \eta^{-1}\int N^3 V_N(x-y) \varphi(x)^2 \varphi(y)^2 (N^{-1} \ell^4 + N \omega_{\ell,N}(y-z)^2) \| a_{y} a_{z} \xi \|^2 \\
	&\leq \eta N^{-1}\|V\|_{L^1} (\|\varphi\|_{L^2}^2 +1) \|\mathcal N \xi'\|^2 \\
	&\qquad + \eta^{-1} \|V\|_{L^1} \|\varphi\|_{L^\infty}^2(1 + \|\varphi\|_{L^2}^2) ( N^{-1} \ell^4 \|\mathcal N\xi\|^2 + N^{-1} \|\mathcal K^{1/2} \mathcal N^{1/2} \xi\|^2) \\
	&\leq C M N^{-1} \|(\mathcal N+1)^{1/2} \xi'\| \| \mathcal K^{1/2} \xi\|
\end{align*}
with an appropriate choice of $\eta>0$. Similarly, as for $\cL_3^{(T_1,3)}$ we obtain, for all $\xi,\xi' \in \FM$,
\begin{align*}
|\langle \xi',\cL_3^{(T_1,5)} \xi \rangle| 
	&\leq C M^{1/2} N^{-1/2}  \| \cL_4^{1/2}\xi'\| \| \mathcal K^{1/2}  \xi\|,
\end{align*}
Let us estimate $\cL_3^{(T_1,6)}$, we have, for all $\xi,\xi' \in \FM$, 
\begin{align*}
&|\langle \xi', \cL_3^{(T_1,6)} \xi \rangle| \leq \int N^{5/2} V_N(x-y) \varphi(x)^2 \varphi(z) \varphi(z') \times \\
	&\qquad \qquad \times (N^{-1/2} \ell^2 + N^{1/2} \omega_{\ell,N}(x-z) + N^{1/2}\omega_{\ell,N}(x-z')) \| a_{y} a_{z'} \xi'\| \| a_{y} a_{z} \xi \| \\
	&\qquad + \int N^{5/2} V_N(x-y) \varphi(x)^2 (N^{-1/2} \ell^2 + N^{1/2}\omega_{\ell,N}(x-z)) \| a_{y} a_{z} \xi'\| \| a_{y} a_{z} \xi \| \\
	&\leq \eta \int N^2 V_N(x-y) (\varphi(z)^2 + \delta_{z,z'})  \| a_{y} a_{z'} \xi'\|^2 \\
	&\qquad + \eta^{-1} \int N^3 V_N(x-y)\varphi(x)^4 \varphi(z')^2(N^{-1} \ell^4 + N \omega_{\ell,N}(x-z)^2 + N\omega_{\ell,N}(x-z')^2)\| a_{y} a_{z} \xi \|^2 \\
	&\qquad + \eta^{-1} \int N^3 V_N(x-y)\varphi(x)^4 (N^{-1} \ell^4 + N \omega_{\ell,N}(x-z)^2)\| a_{y} a_{z} \xi \|^2 \\
	&\leq \eta (1+\|\varphi\|_{L^2}^2) \|V\|_{L^1} N^{-1} \|\mathcal N\xi'\|^2 \\
	&\qquad  + \eta^{-1} (\|V\|_{L^1} \|\varphi\|_{L^\infty}^4  (\|\varphi\|_{L^2}^2 N^{-1} \ell^4 + \|\varphi\|_{L^\infty}^2 \|N^{1/2} \omega_{\ell,N}\|_{L^2}^2 ) \|\mathcal N \xi\|^2 ) \\
	&\qquad + \eta^{-1} C (1+\|\varphi\|_{L^2}^2) \|\varphi\|_{L^\infty}^4 \int N^3 V_N(x-y) N^{-1} \frac{1}{|x-z|^2}\| a_{y} a_{z} \xi \|^2 \\
	&\leq C \eta N^{-1} \|\mathcal N \xi'\|^2 + C \eta^{-1}(N^{-1} \ell^4 + N^{-1} \ell)\|\mathcal N\xi\|^2 + C \eta^{-1} \|V\|_{L^1} N^{-1} \| \mathcal K^{1/2} \mathcal N^{1/2} \xi\|^2 \\
	&\leq C \eta  N^{-1} \|\mathcal N \xi'\|^2 + C \eta^{-1} N^{-1} \|\mathcal K^{1/2} \mathcal N^{1/2} \xi\|^2 \\
	&\leq C N^{-1} \|\mathcal N \xi'\| \|\mathcal K^{1/2} \mathcal N^{1/2} \xi\|  \leq C M N^{-1} \|\mathcal N^{1/2} \xi'\| \|\mathcal K^{1/2} \xi\|,
\end{align*}
with an appropriate choice of $\eta>0$, where we have used \eqref{eq:wlN2-bb}. Similarly as for $\cL_3^{(T_1,3)}$, we can show that for all $\xi,\xi \in \FM$,
\begin{align*}
|\langle \xi', \cL_3^{(T_1,7)} \xi \rangle| 
	&\leq C M^{1/2} N^{-1/2}  \| \cL_4^{1/2}\xi'\| \| \mathcal K^{1/2}  \xi\|.
\end{align*}
We now estimate $\cL_3^{(T_1,8)}$, we have, for all $\xi,\xi' \in \FM$,
\begin{align*}
&|\langle \xi', \cL_3^{(T_1,8)} \xi \rangle|
	\leq \int N^{5/2} V_N(x-y) \varphi(x) \varphi(y) \varphi(z) \varphi(z') \times \\
	&\qquad \qquad \times (N^{-1/2} \ell^2 + N^{1/2} \omega_{\ell,N}(y-z) + N^{1/2}\omega_{\ell,N}(y-z')) \| a_{x} a_y a_z \xi'\| \|a_{z'} \xi\| \\
	&\qquad + \int N^{5/2} V_N(x-y) \varphi(x) \varphi(y) (N^{-1/2} \ell^2+ N^{1/2} \omega_{\ell,N}(y-z)) \| a_{x} a_y a_z \xi'\| \|a_{z} \xi\| \\
	&\leq \eta \int N^2 V_N(x-y) \varphi(z')^2  \| a_{x} a_y a_z \xi'\|^2 + \eta \int N^2 V_N(x-y) \| a_{x} a_y a_z \xi'\|^2 \\
	&\qquad + \eta^{-1} \int N^3 V_N(x-y) \varphi(x)^2 \varphi(z)^2 ((N^{-1} \ell^4 + N \omega_{\ell,N}(y-z)^2 \\
	&\qquad \qquad\qquad\qquad\qquad\qquad\qquad\qquad\qquad\qquad\qquad\qquad + N\omega_{\ell,N}(y-z')^2) \|a_{z'} \xi\|^2 \\
	&\qquad + \eta^{-1} \int N^3 V_N(x-y) \varphi(x)^2 (N^{-1} \ell^4 + N \omega_{\ell,N}(y-z)^2) \|a_{z} \xi\|^2 \\
	&\leq \eta (\|\varphi\|_{L^2}^2 + 1)  \| (\mathcal N+1)^{1/2}\cL_4^{1/2} \xi'\|^2 \\
	&\qquad + \eta^{-1}\Bigg( \|V\|_{L^1} \|\varphi\|_{L^2}^4 N^{-1} \ell^4 + 2\|V\|_{L^1} \|\varphi\|_{L^\infty}^2 \|\varphi\|_{L^2}^2 \|N^{1/2} \omega_{\ell,N}\|_{L^2}^2\Bigg)\|\mathcal N^{1/2} \xi\|^2 \\
	&\leq \eta C  M \|\cL_4^{1/2} \xi'\|^2 + \eta^{-1} C (N^{-1} \ell^4 + N^{-1} \ell) \|\mathcal N^{1/2} \xi\|^2 \\
	&\leq C M^{1/2} N^{-1/2} \ell^{1/2} \|\cL_4^{1/2} \xi'\| \|\mathcal N^{1/2} \xi\|
\end{align*}
with an appropriate choice of $\eta>0$. Similarly, we can prove that , for all $\xi,\xi' \in \FM$,
\begin{align*}
|\langle \xi', \cL_3^{(T_1,9)}\xi \rangle| 
	&\leq C  M^{1/2} N^{-1/2} \ell^{1/2} \|\cL_4^{1/2} \xi'\| \|\mathcal N^{1/2} \xi\|.
\end{align*}
Next, we estimate $\cL_3^{(T_1,10)}$, we have, for all $\xi,\xi' \in \FM$
\begin{align*}
&|\langle \xi', \cL_3^{(T_1,10)} \xi \rangle| 
	\leq  \int N^{5/2} V_N(x-y) \varphi(x) ( k_{c,2}(z,y',y) + k_{c,1}(z,y) \delta_{y,y'}) \| a_{x} a_{y'} a_{z} \xi' \| \| a_{y}\xi\| \\
	&\leq C \eta \int N^2 V_N(x-y) (1+\delta_{y,y'}) \| a_{x} a_{y'} a_{z} \xi' \|^2   \\
	&\qquad + C \eta^{-1} \int N^3 V_N(x-y) \varphi(x)^2 (|k_{c,2}(z,y',y)|^2 + |k_{c,1}(z,y)|^2 \delta_{y,y'})  \|a_{y}\xi\|^2 \\
	&\leq \eta \|V\|_{L^1} N^{-1} \| (\mathcal N+1)^{3/2} \xi'\|^2+ \eta \|( \mathcal N +1)^{1/2} \cL_4^{1/2} \xi'\|^2
		\\
	&\quad + \eta^{-1} \|\varphi\|_{L^\infty}^2 \|V\|_{L^1} (\|k_{c,2}\|_{L^\infty_3 L^2_{1,2}}^2 + \|k_{c,1}\|_{L^\infty_2 L^2_{1}}^2) \|(\mathcal N+1)^{1/2} \xi \|^2 \\
	&\leq C \eta (M^2 N^{-1} \|(\mathcal N+1)^{1/2}\xi'\|^2 + M \|\cL_4^{1/2} \xi'\|^2) + C \eta^{-1} N^{-1} \ell \|(\mathcal N+1)^{1/2} \xi \|^2 \\
	&\leq C M^{1/2} N^{-1/2} \ell^{1/2}  \|\cL_4^{1/2}\xi'\| \|(\mathcal N+1)^{1/2} \xi \| \\
	&\quad  + C M N^{-1} \ell^{1/2}  \|(\mathcal N+1)^{1/2}\xi'\| \|(\mathcal N+1)^{1/2} \xi \|
\end{align*}
with an appropriate choice of $\eta>0$. Finally, similarly as for $\cL_3^{(T_1,8)}$, we can show that for all $\xi,\xi' \in \FM$,
\begin{align*}
|\langle \xi', \cL_3^{(T_1,11)} \xi \rangle| &\leq C M^{1/2} N^{-1/2} \ell^{1/2} \|\cL_4^{1/2} \xi'\| \|\mathcal N^{1/2} \xi\|.
\end{align*}
\end{proof}

\subsection{Estimating $T_{c}^* \cL_0^{(T_1)} T_{c}$}
\label{sec:L0T_1}
Recall
\begin{align}
	\label{eq:recall_L0T_1}
	\cL_0^{(T_1)} &= N \cE_{\rm GP}(\varphi) - 4\pi \ao \int \varphi^4   - \frac{1}{2}N \int N^3 ((\omega_{\ell,N} \varepsilon_{\ell,N}) \ast \varphi^2) \varphi^2 \nn\\
	&\qquad+ \dd\Gamma \Big(-\Delta+V_{\rm ext} + N^3 V_N *\varphi^2 + N^3V_N(x-y) \varphi(x) \varphi(y) - \mu\Big).
\end{align}
	
	\begin{lemma}
		\label{lemma:TcL0Tc}
		 We have 
\begin{align*}
	T_{c}^* \cL_0^{(T_1)} T_{c} 
	&= \cL_0^{(T_1)} - \theta_M \int \varphi(x) N^{5/2} (V_N f_N)(x-y) a^*_x a^*_y a_y + \hc   \\
	&\quad + \theta_{M,+1}^2 \int N^{3} (V_Nf_N\omega_N)(x-y) \varphi(x)^2 a^*_y a_y  \\
	&\quad + \theta_{M,+1}^2\int N^{3} (V_Nf_N\omega_N)(x-y) \varphi(x)\varphi(y) a^*_y a_y + \mathcal E,
\end{align*}
where on $\F_+$
\begin{align*}
\pm \mathcal E \leq C M^{1/2}N^{-1/2} ( 1+ \ell^{-1/2}) (\mathcal K + \mathcal L_4),
\end{align*}
	\end{lemma}
	
\begin{proof}
	The constant in $\cL_0^{(T_1)}$ is left unchanged by the conjugation with $T_c$. The term in the second line  in (\ref{eq:recall_L0T_1}) can be decomposed into  a sum $\dd\Gamma(A)$ where $A$ is either a bounded operator with $\|A\|_{op}\leq C$, in our case $A\in\{1,N^3 V_N *\varphi^2, N^3V_N(x-y) \varphi(x) \varphi(y)\}$, or $A$ is unbounded, in our case $A \in\{ -\Delta, V_{\rm ext} \}$.

\bigskip

\noindent{\bf Case 1: } $A \in \{1, N^3 V_N \ast \varphi^2, N^3V_N(x-y)\varphi(x)\varphi(y)\}$. Let us prove that if $A$ is a bounded operator then, we have on $\F$
\begin{align*}
\pm (T_{c}^* \dd\Gamma(A) T_{c} - \dd\Gamma(A)) \leq C \|A\|_{op} M^{1/2}N^{-1/2} \ell^{1/2} (\mathcal N+1).
\end{align*}
By the Duhamel formula we have
\begin{align}
	\label{eq:duhamel_quad_terms_cubic_transfo}
T_{c}^* \dd\Gamma(A) T_{c} = \dd\Gamma(A) + \int_0^1 T_c^{- t} [\dd\Gamma(A),\theta_M K_{c}^* - K_c \theta_M]  T_c^{ t} \dd t
\end{align}
and
\begin{align*}
[\dd\Gamma(A),\theta_M K_{c}^* - K_{c} \theta_M] &= \theta_M \sum_{i=1}^3 \dd\Gamma(A_i) + \hc ,\\
\dd\Gamma(A_1)
	&= \int A(x,y) k_{c}(y,z,z') a^*_x a^*_z a_{z'}, \\
\dd\Gamma(A_2)
	&= \int A(x,y) k_{c}(t,y,z') a^*_x a^*_t a_{z'}, \\	
\dd\Gamma(A_3)
	&= - \int A(x,y) k_{c}(t,z,x) a^*_t a^*_z a_y.
\end{align*}
We will use that for any bounded operator $T : \gH_+ \to \gH_+^2$ and for all $\xi,\xi' \in \mathcal F$, we have the estimate
\begin{align}
	\label{eq:bound_general_quad_cubic_transfo}
|\langle \xi', \int T(x,y,z) a^*_x a^*_y a_z\xi \rangle | \leq C \|T\|_{op} \|(\mathcal N+1)^{3/4}\xi'\|\|(\mathcal N+1)^{3/4}\xi\|
\end{align}
Let us show that $\|A_i\|_{op} \leq C \|A\|_{op} N^{-1/2}  \ell^{1/2}$ for all $1\leq i \leq 3$.  For all $f \in L^2(\mathbb{R}^3), g \in L^2(\mathbb{R}^6)$, we have
\begin{align*}
&\Big| \int A(x,y) k_{c}(y,z,z') \overline{g(x,z)} f(z') \Big|\\
	&\leq\Big | \int A(x,y) k_{c,1}(y,z) \overline{g(x,z)} f(z) \Big| +\Big | \int A(x,y) k_{c,2}(y,z,z') \overline{g(x,z)} f(z') \Big| \\
	&\leq \Big| \int k_{c,1}(y,z) (A_1\overline{g})(y,z) f(z) \Big| + \Big| \int k_{c,2}(y,z,z') (A_1\overline{g}(y,z) f(z') \Big| \\
	&\leq \int \|k_{c,1}(\cdot,z)\|_{L^2} \|A_1\overline{g}(\cdot,z)\|_{L^2} |f(z)| + \|k_{c,2}\|_{L^2} \|A_1 \overline{g}\|_{L^2} \|f\|_{L^2} \\
	&\leq  \|A\|_{op} (\|k_{c,1}\|_{L^\infty_2 L^2_1} + \|k_{c,2}\|_{L^2}) \|f\|_{L^2} \|g\|_{L^2}\\
	&\leq C \|A\|_{op} N^{-1/2} \ell^{1/2}  \|f\|_{L^2} \|g\|_{L^2}.
\end{align*}
For all $f \in L^2(\mathbb{R}^3), g \in L^2(\mathbb{R}^6)$, we have
\begin{align*}
&\Big| \int A(x,y) k_{c}(t,y,z') \overline{g(x,t)} f(z') \Big| \\
	&\leq\Big | \int A(x,y) k_{c,1}(t,y) \overline{g(x,t)} f(y)\Big | +\Big | \int A(x,y) k_{c,2}(t,y,z') \overline{g(x,t)} f(z')\Big| \\
	&\leq \Big| \int  k_{c,1}(t,y) (A_1\overline{g})(y,t) f(y)\Big| +\Big | \int  k_{c,2}(t,y,z') (A_1\overline{g})(y,t) f(z') \Big| \\
	&\leq \int \|k_{c,1}(\cdot,y)\|_{L^2} \|A_1 \overline g(y,\cdot)\|_{L^2} |f(y)| + \|A\|_{op} \int \|g(\cdot,t)\|_{L^2} \|k_{c,2}(t,\cdot,z')\|_{L^2} |f(z')| \\
	&\leq \|A\|_{op} (\|k_{c,1}\|_{L^\infty_2 L^2_1} + \|k_{c,2}\|_{L^2}) \|f\|_{L^2} \|g\|_{L^2} \\
	&\leq C \|A\|_{op} N^{-1/2} \ell^{1/2}  \|f\|_{L^2} \|g\|_{L^2}.
\end{align*}
For all $f \in L^2(\mathbb{R}^3), g \in L^2(\mathbb{R}^6)$, we have
\begin{align*}
&\Big| \int A(x,y) k_{c}(t,z,x) \overline{g(t,z)} f(y) \Big| \\
	&\leq\Big | \int A(x,y) k_{c,1}(t,x) \overline{g(t,x)} f(y) \Big| + \Big| \int A(x,y) k_{c,2}(t,z,x) \overline{g(t,z)} f(y) \Big| \\
	&\leq \int |k_{c,1}(t,x)| |g(t,x)| |A(f)(x)| +  \int |k_{c,2}(t,z,x)| |g(t,z)| |A(f)(x)| \\
	&\leq \int \|k_{c,1}(\cdot,x)\|_{L^2} \|g(\cdot,x)\|_{L^2} |A(f)(x)| +  \|k_{c,2}\|_{L^2} \|g\|_{L^2} \|A(f)\|_{L^2} \\
	&\leq \|A\|_{op} (\|k_{c,1}\|_{L^\infty_2 L^2_1} + \|k_{c,2}\|_{L^2}) \|f\|_{L^2} \|g\|_{L^2} \\
	&\leq C \|A\|_{op} N^{-1/2} \ell^{1/2}  \|f\|_{L^2} \|g\|_{L^2}.
\end{align*}

Therefore, using (\ref{eq:bound_general_quad_cubic_transfo}) we have
\begin{align*}
\pm [\dd\Gamma(A),\theta_M K_{c}^* - K_c \theta_M] 
	&\leq C \|A\|_{op} N^{-1/2} \ell^{1/2} (\mathcal N+1)^{3/2} \1^{\{\cN \leq M\}} \\
	&\leq C \|A\|_{op} M^{1/2} N^{-1/2} \ell^{1/2} (\mathcal N+1)
\end{align*}
and using Lemma \ref{lemma:number_part_cubic}, we obtain from (\ref{eq:duhamel_quad_terms_cubic_transfo}) that
\begin{align*}
\pm (T_{c}^* \dd \Gamma(A) T_{c} - \dd\Gamma(A)) \leq C \|A\|_{op}M^{1/2} N^{-1/2} \ell^{1/2} (\mathcal N+1).
\end{align*}

\bigskip

\noindent{\bf Case 2: } $A  = V_{\rm ext}$. We will show that on $\F_+$, for all $\eta>0$,
\begin{align*}
\pm (T_{c}^* \dd\Gamma(V_{\rm ext}) T_{c} - \dd\Gamma(V_{\rm ext})) \leq C M^{1/2} N^{-1/2} (\eta (\mathcal K + \cL_4 + \mathcal N+1) + \eta^{-1} (\mathcal N+1)).
\end{align*}
Following the Case 1 and using Lemma \ref{prop:estimate_kinetic_L4_cubictransform}, we only need to estimate
\begin{align*}
&[\dd\Gamma(V_{\rm ext}),\theta_M K_{c}^* - K_{c} \theta_M] = \theta_M \sum_{i=1}^3 \dd\Gamma(V_{\rm ext,i}) + \hc ,\\
\dd\Gamma(V_{\rm ext,1})
	&= \int V_{\rm ext}(x) k_{c}(x,y,y') a^*_x a^*_y a_{y'}, \\
\dd\Gamma(V_{\rm ext,2})
	&= \int V_{\rm ext}(x) k_{c}(y,x,x') a^*_x a^*_y a_{x'} \\
	&= \int V_{\rm ext}(x) k_{c,1}(y,x) a^*_x a^*_y a_{x} + \int V_{\rm ext}(x) k_{c,2}(y,x,x') a^*_x a^*_y a_{x'}, \\	
\dd\Gamma(V_{\rm ext,3})
	&= - \int V_{\rm ext}(x) k_{c}(y,x',x) a^*_y a^*_{x'} a_x \\
	&= - \int V_{\rm ext}(x) k_{c,1}(y,x) a^*_y a^*_{x} a_x - \int V_{\rm ext}(x) k_{c,2}(y,x',x) a^*_y a^*_{x'} a_x.
\end{align*}
Where we wrote $k_{c}(x,y,y') = k_{c,1}(x,y) \delta_{y,y'} + k_{c,2}(x,y,y')$. We can see that the first terms in $\dd\Gamma(V_{\rm ext,2})$ and $\dd\Gamma(V_{\rm ext,3})$ cancels and it remains to estimate the rest. We will use Lemma \ref{prop:cubic_kernel_est},
\begin{align*}
|k_{c,1}(x,y)| &\leq C |\varphi(x) | (N^{-1/2} \ell^2 + N^{1/2}\omega_{\ell,N}(x-y)), \\
|k_{c,2}(x,y,y')| &\leq C |\varphi(x) \varphi(y) \varphi(y')|(N^{-1/2} \ell^2 + N^{1/2}\omega_{\ell,N}(x-y) +N^{1/2} \omega_{\ell,N}(x-y')).
\end{align*}
For all $\xi,\xi' \in \FM$, we have, for all $\eta_1,\eta_2,\eta_3>0$,
\begin{align*}
&|\langle \xi', \dd\Gamma(V_{\rm ext,1})\xi \rangle | \leq \int V_{\rm ext}(x) \varphi(x)  (N^{-1/2} \ell^2 + N^{1/2}\omega_{\ell,N}(x-y)) \|a_x a_y \xi'\| \|a_y \xi\| \\
	&\quad +\int V_{\rm ext}(x) \varphi(x) \varphi(y) \varphi(y')  (N^{-1/2} \ell^2 + N^{1/2}\omega_{\ell,N}(x-y) +N^{1/2} \omega_{\ell,N}(x-y')) \|a_x a_y \xi'\| \|a_{y'} \xi\| \\
	&\leq  C N^{-1/2} \ell^2 (1+\|\varphi\|_{L^2}^2)\| V_{\rm ext} \varphi\|_{L^2} \|(\mathcal N+1)\xi'\| \|\mathcal N^{1/2} \xi\| \\
	&\qquad + \eta_1 \int N \omega_{\ell,N}(x-y)^2 \|a_x a_y \xi'\|^2 + \eta_1^{-1} \int |V_{\rm ext} \varphi(x)|^2 \|a_y \xi\|^2 \\
	&\qquad +  \eta_2 \int \varphi(y')^2 \|a_x a_y \xi'\|^2 + \eta_2^{-1} \int N \omega_{\ell,N}(x-y)^2 \varphi(y)^2  |V_{\rm ext} \varphi(x)|^2 \|a_{y'} \xi\|^2 \\
	&\qquad +  \eta_3 \int  N \omega_{\ell,N}(x-y')^2 \varphi(y')^2  \|a_x a_y \xi'\|^2 + \eta_3^{-1} \int \varphi(y)^2  |V_{\rm ext} \varphi(x)|^2 \|a_{y'} \xi\|^2 \\
	&\leq \eta_1 \int N^{-1} \frac{1}{|x-y|^2} \|a_x a_y \xi'\|^2 +  \eta_1^{-1} \| V_{\rm ext} \varphi\|_{L^2}^2 \|\mathcal N^{1/2} \xi\|^2  \\
	&\qquad + C ( N^{-1/2} \ell^2 + N^{1/2} \|\omega_{\ell,N}\|_{L^2}) \| V_{\rm ext} \varphi\|_{L^2}(1 + \|\varphi\|_{H^2}^2)\|(\mathcal N+1)\xi'\| \|\mathcal N^{1/2} \xi\| \\
	&\leq C M^{1/2} N^{-1/2} \ell^{1/2} \|(\mathcal N+1)^{1/2}\xi'\|\|(\mathcal N+1)^{1/2}\xi\| + C M^{1/2} N^{-1/2} \|\mathcal K^{1/2}\xi'\|\|(\mathcal N+1)^{1/2}\xi\| \\
	&\leq  C M^{1/2} N^{-1/2} \|\mathcal K^{1/2}\xi'\|\|(\mathcal N+1)^{1/2}\xi\|
\end{align*}
with an appropriate choice of $\eta_1,\eta_2,\eta_3>0$ and where we used  Hardy's inequality $|x|^{-2} \leq 4 (-\Delta)$.
For all $\xi,\xi' \in \FM$, we have
\begin{align*}
\Big|\Big\langle \xi', &\int V_{\rm ext}(x) k_{c,2}(y,x,x') a^*_x a^*_y a_{x'} \xi \Big\rangle\Big |  \\
	&\leq \int V_{\rm ext}(x) \varphi(x) \varphi(y) \varphi(x') \times \\
	&\qquad \times (N^{-1/2} \ell^2 + N^{1/2}\omega_{\ell,N}(y-x) +N^{1/2} \omega_{\ell,N}(y-x')) \|a_x a_y \xi'\| \|a_{x'} \xi\| \\
	&\leq C (N^{-1/2} \ell^2 \|V_{\rm ext}\varphi\|_{L^2} \|\varphi\|_{L^2}^2 \|(\mathcal N+1) \xi'\| \|(\mathcal N+1)^{1/2} \xi\| \\
	&\qquad + 2 C ( \|V_{\rm ext}\varphi\|_{L^2} \|\varphi\|_{L^2} \|\varphi\|_{L^\infty} N^{1/2} \|\omega_{\ell,N}\|_{L^2})\|(\mathcal N+1) \xi'\| \|(\mathcal N+1)^{1/2} \xi\| \\
	&\leq C M^{1/2} N^{-1/2} \ell^{1/2} \|(\mathcal N+1)^{1/2} \xi'\| \|(\mathcal N+1)^{1/2} \xi\|,
\end{align*}
where we used that $V_{\rm ext} \geq 1$.
Similarly, for all $\xi,\xi' \in \FM$, we have
\begin{align*}
|\langle \xi', \int V_{\rm ext}(x) k_{c,2}(y,x',x) a^*_y a^*_{x'} a_x \xi \rangle |  \leq C M^{1/2} N^{-1/2} \ell^{1/2} \|(\mathcal N+1)^{1/2} \xi'\| \|(\mathcal N+1)^{1/2} \xi\|.
\end{align*}

\bigskip

\noindent{\bf Case 3: } $A  = -\Delta$. This case is dealt with in Lemma \ref{prop:estimate_kinetic_L4_cubictransform} where (\ref{eq:TcKTc}) and (\ref{eq:estimate_error_kinetic_cubic_transfo}) are proven and are enough for our claim.

\end{proof}

\subsection{Estimating $T_{c}^* \cL_2^{(T_1)} T_{c}$}
	\label{sec:L2T_1}
Recall 
\begin{align*}
\cL_2^{(T_1)} =  \frac{1}{2}\int N^3 \varepsilon_{\ell,N} (x-y) \varphi(x) \varphi(y) a^*_x a^*_y + \hc 
\end{align*}
Let us recall that we have the pointwise inequality $N^3 |\varepsilon_{\ell,N}(x)| \leq C \ell^{-3} \mathds{1}_{\{|x|\leq \ell \}}$.
\begin{lemma}
	\label{lemma:quadra_epsilon_cubic_transfo}
Assume that $ MN^{-1} \lesssim \ell \lesssim 1$, then we have on $\mathcal F_+$,
\begin{align*}
T_{c}^*\cL_2^{(T_1)} T_{c} = \cL_2^{(T_1)}+ \mathcal E,
\end{align*}
with $\mathcal E$ satisfying, on $\mathcal F_+$, 
\begin{align*}
\pm \mathcal E \leq C M^{1/2} N^{-1/2}   \ell^{-1}  (\mathcal N+1).
\end{align*}
\end{lemma}

\begin{proof}
By the Duhamel formula we have
\begin{align*}
e^{-(\theta_M K_{c}^* - K_{c} \theta_M)} \cL_2^{(T_1)} e^{ \theta_M K_{c}^* - K_{c} \theta_M} 
	&=  \cL_2^{(T_1)} +  \int_0^1 T_c^{-t}[\cL_2^{(T_1)},\theta_M K_{c}^* - K_{c} \theta_M] T_c^{t} \dd t.
\end{align*}
In view of Lemma \ref{lemma:number_part_cubic} it is enough to show that 
\begin{align*}
\pm [\cL_2^{(T_1)},\theta_M K_{c}^* - K_{c} \theta_M] \leq C M^{1/2} N^{-1/2}   \ell^{-1} (\mathcal N+1).
\end{align*}

Let us then bound $ [\cL_2^{(T_1)},\theta_M K_{c}^* - K_{c} \theta_M]$, we have on $\mathcal F_+$
\begin{align}
	\label{eq:quadratic_cubic_transfo}
&[\cL_2^{(T_1)},\theta_M K_{c}^* - K_{c} \theta_M]  = \frac{1}{2}[\int N^3\varepsilon_{\ell,N}(x-y) \varphi(x) \varphi(y) a^*_x a^*_y , \theta_MK_{c}^*-K_{c}\theta_M] + \hc  \nn \\
	&=  \frac{1}{2} \theta_M [\int N^3\varepsilon_{\ell,N}(x-y) \varphi(x) \varphi(y) a^*_x a^*_y, K_{c}^*]   - \frac{1}{2}\theta_{M,+2} [\int N^3\varepsilon_{\ell,N}(x-y) \varphi(x) \varphi(y) a^*_x a^*_y,K_{c}]\nn \\
	&\quad + \frac{1}{2}(\theta_{M,-2} -\theta_M) \int N^3\varepsilon_{\ell,N}(x-y) \varphi(x) \varphi(y) a^*_x a^*_y K_{c}^*\nn \\
	&\quad  - \frac{1}{2}\int N^3\varepsilon_{\ell,N}(x-y) \varphi(x) \varphi(y) a^*_x a^*_yK_{c} (\theta_M-\theta_{M,+2}).
\end{align}

Let us first estimate the first two terms of (\ref{eq:quadratic_cubic_transfo}), disregarding the prefactor $\frac{1}{2}\theta_M$ for the moment, we have

\begin{align*}
- \frac{1}{2}\Big[\int N^3\varepsilon_{\ell,N}(x-y) &\varphi(x) \varphi(y) a^*_x a^*_y ,K_{c}\Big]  = \sum_{i=1}^3 \widetilde{\mathcal L}_2^{(i)}, \\
\widetilde{\mathcal L}_2^{(1)}
	&=   \int N^3 \varepsilon_{\ell,N} (x-y) \varphi(x) \varphi(y) \overline{k_{c}(x,y,y')} a_{y'}, \\
\widetilde{\mathcal L}_2^{(2)}
	&=    \int N^3\varepsilon_{\ell,N} (x-y) \varphi(x) \varphi(y) \overline{k_{c}(x,z,z')} a^*_{y}   a^*_{z'}a_{z}, \\
\widetilde{\mathcal L}_2^{(3)}
	&=    \int N^3 \varepsilon_{\ell,N} (x-y) \varphi(x) \varphi(y) \overline{k_{c}(z,x,z')}  a^*_{y} a^*_{z'} a_{z},
\end{align*}
and
\begin{align*}
	&\frac{1}{2}\Big[\int N^3\varepsilon_{\ell,N}(x-y) \varphi(x) \varphi(y) a^*_x a^*_y ,K_{c}^*\Big] \\
	& = \widetilde{\mathcal L}_2^{(4)} = - \int N^3 \varepsilon_{\ell,N} (x-y) \varphi(x) \varphi(y) k_{c}(z,t,x) a^*_y a^*_z a^*_t.
\end{align*}

Let us now estimate $\widetilde{\mathcal L}_2^{(i)}$. Thanks to the prefactor $\theta_M$ it is enough to estimate the above terms on $\FM$.
Let us write $k_{c}(x,y,y') = k_{c,1}(x,y)\delta_{y,y'} + k_{c,2}(x,y,y')$. We have for all $\xi,\xi' \in \FM$, using Lemma \ref{prop:cubic_kernel_est}
\begin{align*}
&| \langle \xi', \widetilde{\mathcal L}_2^{(1)} \xi \rangle|
	\leq  \int N^3 |\varepsilon_{\ell,N}(x-y) \varphi(x) \varphi(y) k_{c,1}(x,y)| \|\xi'\| \|a_y \xi\| \\
	&\quad +  \int N^3 |\varepsilon_{\ell,N}(x-y) \varphi(x) \varphi(y) k_{c,2}(x,y,y')| \|\xi'\| \|a_{y'} \xi\| \\
	&\leq C \|\varphi\|_{L^\infty}^2 \int N^3 \varepsilon_{\ell,N}(x-y) ( N^{-1} \ell^{1/2} \varphi(x) + N^{1/2} \omega_{\ell,N}(x-y) \varphi(x)) \|\xi'\| \|a_y \xi\| \\
	&\quad + C \|\varphi\|^2_{L^\infty} \int N^3 |\varepsilon_{\ell,N}(x-y)|  \varphi(x) \varphi(y) \varphi(y')( N^{-1} \ell^{1/2} + N^{1/2}\omega_{\ell,N}(x-y) \\
	&\qquad\qquad\qquad\qquad\qquad\qquad\qquad\qquad\qquad\qquad\qquad+N^{1/2} \omega_{\ell,N}(x-y')) \|\xi'\| \|a_{y'} \xi\| \\
	&\leq C  (\|\varphi\|_{L^2} + \|\varphi\|_{L^2}^3) (N^{-1} \ell^{1/2} \|N^3 \varepsilon_{\ell,N}\|_{L^1}  + N^{1/2} \|N^3 \varepsilon_{\ell,N} \omega_{\ell,N}\|_{L^1}) \|\xi'\| \|(\mathcal N+1)^{1/2} \xi\| \\
	&\quad + C \|\varphi\|_{L^2}^2  \|\varphi\|_{L^\infty}  N^{1/2} \|\omega_{\ell,N}\|_{L^1} \| N^3\varepsilon_{\ell,N}\|_{L^1} \|\xi'\| \|(\mathcal N+1)^{1/2} \xi\| \\
	&\leq  C ((N^{-1} \ell^{1/2} +  N^{-1/2} \ell^{-1} + N^{-1/2} \ell^2 ) \|\xi'\| \|(\mathcal N+1)^{1/2} \xi\| \\
	&\leq C ((N^{-1} \ell^{1/2}+ N^{-1/2} \ell^{-1}) \|\xi'\| \|(\mathcal N+1)^{1/2} \xi\|,
\end{align*}
where we used that 
\begin{align*}
N^{1/2} \int N^3 |\varepsilon_{\ell,N} \omega_{\ell,N} (x)| \leq N^{1/2} \ell^{-3}\int \frac{\mathds{1}_{\{|x|\leq \ell\}}}{N |x|} \leq  N^{-1/2} \ell^{-1}.
\end{align*}
Let us write $k_{c}(x,y,y')=k_{c,1}(x,y) \delta_{y,y'} + k_{c,2}(x,y,y')$ and estimate for all  $\xi,\xi' \in \FM$,
\begin{align*}
| \langle \xi', \widetilde{\mathcal L}_2^{(2)} \xi \rangle |
	&\leq \int N^3 |\varepsilon_{\ell,N}(x-y)\varphi(x) \varphi(y) k_{c,2}(x,z,z')|  \|a_{y} a_{z'} \xi'\| \| a_z \xi \| \\
	&\quad + \int N^3 |\varepsilon_{\ell,N}(x-y) \varphi(x) \varphi(y) k_{c,1}(x,z)| \|a_{y} a_{z} \xi'\| \| a_z \xi \| \\
	&\leq \|N^3 \varepsilon_{\ell,N}\|_{L^1} \|\varphi\|_{L^\infty}^2 \int \|k_{c,2}(\cdot,z,z')\|_{L^2}   \|(\mathcal N+1)^{1/2} a_{z'} \xi'\| \| a_z \xi \|  \\
	&\quad + \|N^3 \varepsilon_{\ell,N}\|_{L^1} \|\varphi\|_{L^\infty}^2 \int \|k_{c,1}(\cdot,z)\|_{L^2}  \|(\mathcal N+1)^{1/2} a_{z} \xi'\| \| a_z \xi \| \\
	&\leq C (\|k_{c,2}\|_{L^\infty_{3} L^2_{1,2}} + \|k_{c,2}\|_{L^\infty_{2} L^2_{1,3}} + \|k_{c,1}\|_{L^\infty_2 L^2_{1}})  \|(\mathcal N+1) \xi'\|  \|(\mathcal N+1)^{1/2}\xi\| \\
	&\leq C M^{1/2} N^{-1/2} \ell^{1/2} \|(\mathcal N+1)^{1/2}\xi\| \|(\mathcal N+1)^{1/2} \xi\|.
\end{align*}
We have for all  $\xi,\xi' \in \FM$,
\begin{align*}
| \langle \xi',  \widetilde{\mathcal L}_2^{(3)} \xi \rangle |
	&\leq \int N^3 |\varepsilon_{\ell,N}(x-y) \varphi(x) \varphi(y) k_{c,2}(z,x,z')|\|a_{y} a_{z'} \xi'\|  \| a_z \xi \|   \\
	&\quad+ \int N^3 |\varepsilon_{\ell,N}(x-y) \varphi(x) \varphi(y) k_{c,1}(z,x)|  \|a_{y} a_{z} \xi'\| \| a_z \xi \|  \\
	&\leq \|N^3 \varepsilon_{\ell,N}\|_{L^1} \|\varphi\|_{L^\infty}^2 \int \|k_{c,2}(z,\cdot,z')\|_{L^2}   \|(\mathcal N+1)^{1/2} a_{z'} \xi'\| \| a_z \xi \|  \\
	&\quad + \|N^3 \varepsilon_{\ell,N}\|_{L^1} \|\varphi\|_{L^\infty}^2 \int \|k_{c,1}(z,\cdot)\|_{L^2} \|(\mathcal N+1)^{1/2} a_{z} \xi'\| \| a_z \xi \|  \\
	&\leq C (\|k_{c,2}\|_{L^\infty_{3} L^2_{1,2}} + \|k_{c,2}\|_{L^\infty_{2} L^2_{2,3}} + \|k_{c,1}\|_{L^\infty_1 L^2_{2}})  \|(\mathcal N+1) \xi'\| \|(\mathcal N+1)^{1/2}\xi\| \\
	&\leq C M^{1/2} N^{-1/2} \ell^{1/2} \|(\mathcal N+1)^{1/2}\xi'\| \|(\mathcal N+1)^{1/2} \xi\|.
\end{align*}
Let us estimate $\widetilde{\cL}_2^{(4)}$, for this we write $k_{c}(x,y,y')=k_{c,1}(x,y) \delta_{y,y'} + k_{c,2}(x,y,y')$ and estimate for all  $\xi,\xi' \in \FM$,
\begin{align*}
\Big| \Big\langle \xi', &\int N^3 \varepsilon_{\ell,N}(x-y) \varphi(x) \varphi(y) k_{c,2}(z,t,x) a^*_y a^*_z a^*_t \xi \Big\rangle \Big| \\
	&\leq \int N^3 |\varepsilon_{\ell,N}(x-y)| \varphi(x) \varphi(y) \| a_y \xi' \| \|\int \dd z \dd t k_{c,2}(z,t,x) a^*_z a^*_t \xi \| \\
	&\leq \|k_{c,2}\|_{L^\infty_3 L^2_{1,2}} \int N^3 |\varepsilon_{\ell,N}(x-y)| \varphi(x) \varphi(y) \| a_y \xi' \| \| (\mathcal N+1) \xi \| \\
	&\leq C N^{-2/3} \|\varphi\|_{L^2} \|N^3 \varepsilon_{\ell,N}\|_{L^1} \|\varphi\|_{L^\infty} \|(\mathcal N+1)^{1/2}\| \| (\mathcal N+1) \xi \| \\
	&\leq C M^{1/2}  N^{-1/2} \ell^{1/2}   \|(\mathcal N+1)^{1/2}\| \| (\mathcal N+1)^{1/2} \xi \|,
\end{align*}
and
\begin{align*}
\Big| \Big\langle \xi', &\int N^3 \varepsilon_{\ell,N} (x-y) \varphi(x) \varphi(y) k_{c,1}(z,x) a^*_y a^*_z a^*_x \xi\Big \rangle \Big| \\
	&\leq \|k_{c,1}\|_{L^\infty_2 L^2_{1}} \int N^3 |\varepsilon_{\ell,N}(x-y)| \varphi(x) \varphi(y) \| a_x a_y \xi' \| \| (\mathcal N+1)^{1/2} \xi \| \\
	&\leq C \|\varphi\|_{L^2}  \|\varphi\|_{L^\infty}  \|k_{c,1}\|_{L^\infty_1 L^2_{2}} N^3 \|\varepsilon_{\ell,N}\|_{L^2} \|(\mathcal N+1) \xi' \|  \|(\mathcal N+1)^{1/2} \xi \| \\
	&\leq C M^{1/2} N^{-1/2} \ell^{-1} \|(\mathcal N+1)^{1/2} \xi' \|  \|(\mathcal N+1)^{1/2} \xi \|.
\end{align*}
Therefore, for all  $\xi,\xi' \in \FM$,
\begin{align*}
| \langle \xi',  \widetilde{\mathcal L}_2^{(4)} \xi \rangle | &\leq C M^{1/2} N^{-1/2} \ell^{-1} \|(\mathcal N+1)^{1/2} \xi' \|  \|(\mathcal N+1)^{1/2} \xi \|.
\end{align*}

The last two terms of (\ref{eq:quadratic_cubic_transfo}) are bounded in the following way, for all $\xi,\xi' \in \mathcal F_+$,
\begin{align*}
\Big|\Big\langle \xi',  (\theta_{M,-2} &-\theta_M) \int N^3\varepsilon_{\ell,N}(x-y) \varphi(x) \varphi(y) a^*_x a^*_y K_{c}^* \xi\Big \rangle \Big| \\
	&\leq \|N^3\varepsilon_{\ell,N}\|_{L^2(\mathbb{R}^3)} \|\varphi\|_{L^2} \|\varphi\|_{L^\infty} \| (\mathcal N+1) (\theta_{M} -\theta_{M,+2}) \xi'\| \|K_{c}^* \xi\| \\
	&\leq C \ell^{-3/2} \|(\theta_{M,-2} -\theta_{M}) \|_{op} N^{-1/2} \ell^{1/2} \| \mathds{1}^{\leq M+2}(\mathcal N+1)\xi'\| \|\mathds{1}^{\leq M-1}(\mathcal N+1)^{3/2}\xi\| \\
	&\leq C M^{1/2} N^{-1/2}   \ell^{-1} \|(\mathcal N+1)^{1/2}\xi'\| \|(\mathcal N+1)^{1/2}\xi\|.
\end{align*}
Similarly, we  have for all $\xi,\xi' \in \mathcal F_+$,
\begin{multline*}
\Big|\Big\langle \xi', \int N^3\varepsilon_{\ell,N}(x-y) \varphi(x) \varphi(y) a^*_x a^*_yK_{c} (\theta_M-\theta_{M,+2}) \xi \Big\rangle \Big| \\
\leq C  M^{1/2} N^{-1/2}   \ell^{-1} \|(\mathcal N+1)^{1/2}\xi'\| \|(\mathcal N+1)^{1/2}\xi\|.
\end{multline*}

\end{proof}
\subsection{Estimating $T_{c}^* \cL_3^{(T_1)} T_{c}$}
	\label{sec:L3T_1}
Let us recall from Lemma \ref{lemma:main_lemma_quadratic_transform} that 
\begin{align*}
 \cL_3^{(T_1)} 
 	= \int N^{5/2} V_N (x-y) \varphi(x) a^*_x a^*_y a_y + \hc   =  \cL_3^{(T_1,*)} + \cL_3^{(T_1,\circ)}.
\end{align*}
\begin{lemma}
	\label{lemma:cubic_cubic_transfo}
Assume that $ MN^{-1} \lesssim \ell \lesssim 1$, then we have on $\mathcal F_+$,
\begin{align*}
T_{c}^*  \cL_3^{(T_1)} T_{c} &=  \cL_3^{(T_1)}  - 2 \theta_{M,+1} \int N^{3} (V_N\omega_N)(x-y) \varphi(x)^2 a^*_y a_y \\
&\quad- 2 \theta_{M,+1}\int N^{3} (V_N\omega_N)(x-y) \varphi(x)\varphi(y) a^*_x a_y  + \mathcal E,
\end{align*}
with $\mathcal E$, satisfying for all $ \xi' ,\xi \in \F_+$,
\begin{align*}
	|\langle \xi',  \mathcal E  \xi \rangle| 
	&\leq  C M^{1/2} N^{-1/2} \|(\mathcal K+\cL_4 )^{1/2} \xi'\|\|(\mathcal K+\cL_4  )^{1/2} \xi\|.
\end{align*}
\end{lemma}

\begin{proof}
By the Duhamel formula we have
\begin{align*}
T_{c}^* \cL_3^{(T_1,*)} T_{c} &= e^{-(\theta_M K_{c}^* - K_{c} \theta_M))} \cL_3^{(T_1,*)} e^{\theta_M K_{c}^* - K_{c} \theta_M} \\
	&=  \cL_3^{(T_1,*)} +  \int_0^1 T_c^{-t} [\cL_3^{(T_1,*)},\theta_M K_{c}^* - K_{c} \theta_M] T_c^{t}\dd t \\
	&=  \cL_3^{(T_1,*)} +  \int_0^1 T_c^{-t} (\theta_M [\cL_3^{(T_1,*)},K_{c}^*]T_c^{t} \dd t -  \int_0^1 T_c^{-t}[\cL_3^{(T_1,*)},K_{c}] \theta_{M,+1})T_c^{t} \dd t  \\
	&\qquad + \int_0^1 T_c^{-t} (\cL_3^{(T_1,*)} K_{c}^*(\theta_{M,+1}-\theta_{M,+2})   -\cL_3^{(T_1,*)} K_{c} (\theta_M-\theta_{M,+1}))T_c^{t} \dd t
\end{align*}
The second term is an error term bounded using Lemma \ref{lemma:adjoint_L3_cubic} and the last two are also error terms and we bound them using Lemma \ref{prop:estimate_kinetic_L4_cubictransform} and the following estimates. We have, for all all $ \xi' ,\xi \in \mathcal F_+$,
\begin{align*}
|\langle \xi', \cL_3^{(T_1,*)} K_{c}^*(\theta_{M,+1}-\theta_{M,+2}) \xi \rangle|
	&\leq \int N^{5/2} V_N (x-y) \varphi(x) \| a_x a_y \xi'\| \|a_y K_{c}^* (\theta_{M,+1}-\theta_{M,+2})\xi\| \\
	&\leq C \|\varphi\|_{L^\infty} \|V\|_{L^1}^{1/2} \|\cL_4^{1/2} \xi'\| \|\mathcal N^{1/2} K_{c}^*(\theta_{M,+1}-\theta_{M,+2}) \xi\| \\
	&\leq C M^{-1} N^{-1/2} \ell^{1/2} \|\cL_4^{1/2}\xi'\| \|(\mathcal N+1)^{2} \xi\| \\
	&\leq C M^{1/2} N^{-1/2} \ell^{1/2}\|\cL_4^{1/2}\xi'\|  \|(\mathcal N+1)^{1/2} \xi\|.
\end{align*}
And similarly,
\begin{align*}
|\langle \xi', \cL_3^{(T_1,*)} K_{c} (\theta_M-\theta_{M,+1}) K_{c}^* \xi \rangle| 
	\leq  C M^{1/2} N^{-1/2} \ell^{1/2}\|\cL_4^{1/2}\xi'\|  \|(\mathcal N+1)^{1/2} \xi\|.
\end{align*}
Hence, using Lemma \ref{prop:estimate_kinetic_L4_cubictransform}, we obtain
\begin{align}
T_{c}^* \cL_3^{(T_1,*)} T_{c}
	&=  \cL_3^{(T_1)} - \theta_{M,+1} \int N^{3} (V_N\omega_N)(x-y) \varphi(x)^2 a^*_y a_y \nn \\
	&\quad-  \theta_{M,+1}  \int N^{3} (V_N\omega_N)(x-y) \varphi(x)\varphi(y) a^*_x a_y \nn  \\
	&\quad -\int_0^1 \int_0^{t} T_c^{-u} \Big[\theta_{M,+1}\int N^{3} (V_N\omega_N)(x-y) \varphi(x)^2 a^*_y a_y,K_{c}^* \theta_M - K_{c} \theta_M]\Big]T_c^{u} \dd u \dd t \nn  \\
	&\quad -\int_0^1 \int_0^{t} T_c^{-u} \Big[\theta_{M,+1}\int N^{3} (V_N\omega_N)(x-y) \varphi(x)\varphi(y)a^*_x a_y,K_{c}^* \theta_M - K_{c} \theta_M]\Big]T_c^{u} \dd u \dd t \nn \\
	&\quad + \mathcal E, \label{eq:cubic_cubic_transfo_1}
\end{align}
with $\mathcal E$ satisfying 
\begin{align*}
	|\langle \xi' , \mathcal E \xi \rangle| \leq C M^{1/2} N^{-1/2} \|(\mathcal K+\cL_4 )^{1/2} \xi'\|\|(\mathcal K+\cL_4)^{1/2} \xi\|, \quad \forall  \xi,\xi' \in \mathcal F_+.
\end{align*}

The last two terms in (\ref{eq:cubic_cubic_transfo_1}) are bounded by
\begin{align*}
C\|V\|_{L^1} \|\varphi\|_{L^\infty}^2 M^{1/2} N^{-1/2} \ell^{1/2} (\mathcal N+1)
\end{align*}
as in the proof of Lemma \ref{lemma:TcL0Tc} (Case 1). This concludes the proof of Lemma \ref{lemma:cubic_cubic_transfo}. 
\end{proof}

\subsection{Conclusion of Lemma \ref{prop:recap_cubic_transform} }
	\label{sec:proof_cubic}

\begin{proof}[Proof of Lemma \ref{prop:recap_cubic_transform}]
From Lemmas \ref{lemma:main_lemma_quadratic_transform}, \ref{lemma:TcL0Tc}, \ref{lemma:quadra_epsilon_cubic_transfo}, \ref{lemma:cubic_cubic_transfo} and Lemma \ref{prop:estimate_kinetic_L4_cubictransform} Eqs. (\ref{eq:prop:bound_L4_cubic_transfo_1}--\ref{eq:estimate_error_L_4_cubic_transfo}), we have
\begin{align}\label{eq:proof_cubic}
T_{c}^* T^*_1\mathcal H T_1T_{c} 
	&= T_{c}^* \mathcal H^{(T_1)} T_{c} + T_{c}^*\mathcal E^{(T_1)}T_{c} = \mathcal H^{(T_c)} + \cE^{(T_c)}_1+ T_{c}^*\mathcal E^{(T_1)}T_{c} +\mathcal E^{(\theta_M)}   \\
	&+ \dd\Gamma\left((N^3V_Nf_N)\ast \varphi^2 - 8 \pi \ao \varphi^2 + \varphi (x) N^3 (V_N f_N - \eps_{\ell,N})(x-y) \varphi(y)  \right),\nn
\end{align}
where $\mathcal E^{(T_1)}$ was defined in Lemma \ref{lemma:main_lemma_quadratic_transform}, $\cE^{(\theta_M)}$ in Lemma \ref{prop:recap_cubic_transform} and $\cE^{(T_c)}_1 $ satisfies
\begin{align*}
\pm  \cE^{(T_c)}_1 
	&\leq  C M^{1/2}N^{-1/2} (\mathcal K + \cL_4) + M^{1/2} N^{-1/2}\ell^{-1} (\mathcal N+1).
\end{align*}
Using Lemmas \ref{lemma:number_part_cubic} and \ref{prop:estimate_kinetic_L4_cubictransform}, we have
\begin{align*}
\pm T_{c}^* \mathcal E^{(T_1)} T_{c} 
	&\leq  C  \ell^{1/2} (\mathcal N+1) +   C \ell^{3/2} (\mathcal K + \cL_4) \\
	&\quad + C \frac{(\mathcal N+1)^{3/2}}{N^{1/2}} + C \frac{(\mathcal N+1)^{5/2}}{N^{3/2}} \\
	&\quad + \eps \left( \cL_4 + M^{1/2}N^{-1/2} \mathcal K + (\mathcal N +1 ) +  \ell^{1/2} \frac{(\mathcal N+1)}{N^2} + \frac{(\mathcal N+1)^2}{N^3} \right) \\
	&\quad + \eps^{-1}C \left(\ell(\mathcal N+1) + \frac{(\mathcal N+1)^2}{N}+ \frac{(\mathcal N+1)^5}{N^4} \right).
\end{align*}
It remains to estimate the last term in (\ref{eq:proof_cubic}). By the same argument leading to \eqref{eq:VNfN-pN-0}, since $Vf$ is even, compactly supported and $\int Vf = 8\pi \ao$, we have
\begin{align*}
\| (N^3V_Nf_N)\ast \varphi^2 - 8 \pi \ao \varphi^2 \|_{L^2} &= \| (\widehat {Vf} (p/N) - \widehat {Vf}(0)) \widehat{\varphi^2} (p) \|_{L^2} \\
&\leq\| |p|^{-2}(\widehat {Vf} (p/N) - \widehat {Vf}(0))\|_{L^\infty}  \|p^2\widehat{\varphi^2} (p) \|_{L^2}  \leq CN^{-2}. 
\end{align*}
Therefore, we have the operator inequality
\begin{align*}
\pm \left((N^3V_Nf_N)\ast \varphi^2 - 8 \pi \ao \varphi^2\right) \leq  C N^{-2} (1-\Delta). 
\end{align*}
Next, consider $\Phi=N^3(V_Nf_N - \eps_{\ell,N})$. Recall from Lemma \ref{prop:prop_epsilon} that $\eps_{\ell,N}$ is even,  $\int N^3 \eps_{\ell,N}=8\pi \ao$ and $N^3 |\eps_{\ell,N}|\leq C \ell^3 \1_{\{|x| \leq \ell\}}$. Thus $\widehat  \Phi$ is smooth, satisfies $\widehat  \Phi(0)=0$, $\nabla \widehat\Phi(0)=0$, and hence 
\begin{align*}
| \widehat \Phi(p)| = | \widehat \Phi(p) -\widehat  \Phi(0) - p \cdot \nabla \widehat\Phi(0)| \leq C\|(|x|^2+1) \Phi\|_{L^1} |p|^2 \leq C \ell^2  |p|^2, \quad \forall p\in \R^3.  
\end{align*}
Therefore, for all $f \in H^1(\mathbb{R}^3)$,
\begin{align} \label{eq:Phi-delta-Tc}
&\Big| \int \overline{f(x)} \varphi(x)  \Phi(x-y) \varphi(y) f(y) \dd x \dd y\Big| = \Big| \int \overline{\widehat {f\varphi}}(p) \widehat{\Phi}(p) \widehat{f\varphi}(p) \dd p \Big|\nn\\
& \leq C\ell^2 \| |p| \widehat {f\varphi}(p)\|_{L^2}^2 \leq C \ell^2 \|\varphi\|_{H^1}^2 \|f\|_{H^1}^2.
\end{align}
Hence,
$$
\pm \varphi(x) \Phi(x-y) \varphi(y) \leq C\ell^2 (1-\Delta). 
$$
We conclude that
$$
\pm \dd\Gamma\left((N^3V_Nf_N)\ast \varphi^2 - 8 \pi \ao \varphi^2 + \varphi (x) N^3 (V_Nf_N -\eps_{\ell,N}) (x-y)\varphi(y) \right)  \leq C \ell^2 \dd\Gamma(1-\Delta).
$$
Defining $\mathcal E^{(T_c)} := T_{c}^* T^*_1\mathcal H T_1T_{c}  - \mathcal H^{(T_c)} - \mathcal E^{(\theta_M)} $ finishes the proof of Lemma \ref{prop:recap_cubic_transform}.

%For $\Phi \in \{Vf, \varepsilon_{\ell,N}\}$, $\Phi$ is even and have the following expansion
%\begin{align*}
%\Big| \widehat \Phi(p/N) - \int \Phi \Big| \leq C\|(|x|^2+1) \Phi\|_{L^1} N^{-2} |p|^2,
%\end{align*}
%so that for all $f,g \in L^2(\mathbb{R}^3)$,
%\begin{align*}
%\Big| \int \overline{g(x)} \varphi(x) N^3 \Phi(N(x-y)) &\varphi(y) f(y) - 8 \pi \int\overline{g(x)} \varphi(x)^4 f(x)\Big| \\
%	&\leq C\|(|x|^2+1) \Phi\|_{L^1}  \|\nabla (\varphi f)\|_{L^2}\|\nabla(\varphi g)\|_{L^2} \\
%	&\leq C \|(|x|^2+1) \Phi\|_{L^1} \|\varphi\|_{H^2}^2 \|f\|_{H^1} \|g\|_{H^1}
%\end{align*}
%Now using that 
%\begin{align*}
%&\int \varepsilon_{\ell,N} = \int V f = 8\pi \ao, \\
%&\int |x|^2 |\varepsilon_{N \ell}| 
%	\leq C (N\ell)^{-3} \int_{|x|\leq N \ell} |x|^2 \leq C N^2 \ell^2,
%\end{align*}
%and that $\||x|^2 Vf\|_{L^1}\leq C$, we obtain
%\begin{align*}
%|(N^3V_Nf_N)\ast \varphi^2 - 8\pi a \varphi^4| 
%	&\leq C N^{-2} \|\varphi\|_{H^2}^4 \\
%|(N^3 \varepsilon_{N,\ell})\ast \varphi^2 - 8\pi a \varphi^4|
%	&\leq C \ell^2 \|\varphi\|_{H^2}^4,
%\end{align*}
%and therefore 
%\begin{multline*}
%\pm \dd\Gamma\left((N^3V_Nf_N)\ast \varphi^2 + \varphi \widehat{(Vf)}(p/N) \varphi) - 8 \pi a \varphi^2 - \varphi \widehat {\eps}_{\ell}(p/N) \varphi\right)  \\ \leq C \ell^2 \mathcal K + C N^{-1}+ C\mathcal N/N^2.
%\end{multline*}
%Defining $\mathcal E^{(T_c)} := T_{c}^* T^*_1\mathcal H T_1T_{c}  - \mathcal H^{(T_c)} - \mathcal E^{(\theta_M)} $ finishes the proof of Lemma \ref{prop:recap_cubic_transform}.
\end{proof}

\section{The last quadratic transform} \label{sec:T2}

\subsection{Diagonalisation of $\mathcal H$}

Following \cite{GreSei-13}, let us define
\begin{align*}
K &= Q \widetilde K Q, \qquad \widetilde K(x,y) = N^3 \varphi(x) \varepsilon_{\ell,N} (x-y) \varphi(y),
\end{align*}
and
\begin{align*}
E &= (D^{1/2} (D+2K)D^{1/2})^{1/2}.
\end{align*}
Note that $D$ and $E$ leaves $\gH_+$ invariant, and they are strictly positive on $\gH_+$. We can now define the operators on $\gH_+$
\begin{align*}
A &= D^{1/2} E^{-1/2}, \quad k_2 = - \log |A|.
\end{align*}
It can equivalently be define through the polar decomposition $A = W e^{k_2}$, where $W$ is a unitary operator on $\gH_+$ which we extend by $\1$ in direction $\mathbb C \varphi$, so that $W$ now denotes a unitary operator of $\gH$. The kernel of $k_2$ is real because $D$ and $E$ are real and we will prove later that $k_2$ is Hilbert-Schmidt. Assuming this, we define the second quadratic transform
\begin{align*}
T_2 = \exp \Big(\frac{1}{2}\int k_2(x,y) a^*_x a^*_y - \frac{1}{2} \int k_2(x,y) a_x a_y \Big) \mathcal W.
\end{align*}
where $k_2 \in \gH^2_+$ and $\mathcal W$ implements $W$ on the Fock space $\mathcal F \simeq \mathcal F_+ \otimes \mathcal F(\mathbb C \varphi)$, that is $\mathcal W$ is unitary on $\mathcal F$ and satisfies $\mathcal W^* a^*(f) \mathcal W = a^*(W f)$ for all $f \in \mathcal H$. Note that $T_2$ preserves $\F_+$.
We can now state the main result of this section.
\begin{lemma}
	\label{prop:second_quadratic_transform}
Assume that $MN^{-1} \lesssim \ell \lesssim 1$, we have
\begin{align}
T_2^* T_{c}^* T_1^* \mathcal H T_1 T_{c} T_2 
	&= N \cE_{\rm GP}(\varphi) - 4\pi \ao \int \varphi^4   - \frac{N^4}{2} \int ((\omega_{\ell,N} \varepsilon_{\ell,N})\ast \varphi^2) \varphi^2 -\frac{1}{2} \tr (D + K - E)   \nn \\
	&\quad + \dd\Gamma(E)  + T^*_2 \mathcal L_4 T_2 + T^*_2\mathcal E^{(\theta_M)}T_2 +  \mathcal E^{(T_2)},
		\label{eq:second_quadratic_transform}
\end{align}
where $\mathcal E^{(\theta_M)}$ is given by (\ref{eq:E_theta_M}) and $\mathcal E^{(T_2)}$ satisfies the following quadratic form estimate on $\cF_+$
\begin{align}
\pm \mathcal E^{(T_2)}
	&\leq C (\ell^{1/2} +M^{1/2} N^{-1/2} \ell^{-1} ) (\mathcal N+1) \nn \\
	&\qquad + C \big(  M^{1/2}N^{-1/2} + \ell^{3/2} \big) ( \dd\Gamma(D) + \ell^{-1} + T^*_2\mathcal L_4 T_2)  \nn \\
	&\qquad + C\frac{(\mathcal N+1)^{3/2}}{N^{1/2}} + C \frac{(\mathcal N+1)^{5/2}}{N^{3/2}}\nn \\
	&\quad + \eps \left( T_2^*\cL_4T_2 +  MN^{-1} (\dd\Gamma(D) + \ell^{-1}) + (\mathcal N+1) + \ell^{1/2} \frac{(\mathcal N+1)}{N^2} + \frac{(\mathcal N+1)^2}{N^3} \right) \nn \\
	&\quad + \eps^{-1}C \left(\ell(\mathcal N+1) + \frac{(\mathcal N+1)^2}{N}+ \frac{(\mathcal N+1)^5}{N^4} \right), \label{eq:lemma_second_quad_transfo_error}
\end{align}
for all $\eps>0$.
Moreover, we have
\begin{align*}
- \frac{1}{2}N^4 \int (\omega_{\ell,N} \varepsilon_{\ell,N})\ast \varphi^2 \varphi^2  -\frac{1}{2} \tr (D + K - E)    \geq - C
\end{align*}
for some constant $C>0$.
\end{lemma}

\begin{remark}
The error term $T^*_2 \mathcal L_4 T_2$ in (\ref{eq:second_quadratic_transform}) is non-negative and will be used to absorb parts of the other error term $\mathcal E^{(T_2)}$ when we compute a lower bound for the eigenvalues of $\mathcal H$. To prove the matching upper bounds, we still need to estimate it, this is done in Lemma \ref{lemma:estimate_L4_second_quadratic_transfo}.
\end{remark}

\begin{lemma}[Estimation of $T_2^* \mathcal L_4 T_2$]
	\label{lemma:estimate_L4_second_quadratic_transfo}
We have, on $\F_+$,
\begin{align*}
\pm T_2^* \mathcal L_4 T_2
	&\leq  C N^{-1} \ell^{-7}  (\mathcal N+1)^2 + C \mathcal L_4.
\end{align*}
\end{lemma}

In the subsection \ref{sec:propT_2} we prove some properties about the transform $T_2$. Then in the subsection \ref{sec:proof_L4_T2} we prove Lemma \ref{lemma:estimate_L4_second_quadratic_transfo}, that will be useful for the proof of Lemma \ref{lem:excitation-pre} (convergence of the excitation spectrum). Finally, in subsection \ref{sec:proof_propT2} we give the proof of Lemma \ref{prop:second_quadratic_transform}.

\subsection{Properties of $T_2$}
	\label{sec:propT_2}

Again following \cite{GreSei-13}, let us define
\begin{align*}
c_2 &= \frac{1}{2}(D^{1/2} E^{-1/2} + D^{-1/2} E^{1/2}), \\
s_2 &= \frac{1}{2}( D^{1/2} E^{-1/2} - D^{-1/2} E^{1/2}).
\end{align*}
One can check that we have $c_2^* c_2 - s_2^* s_2 = 1$ and that we can rewrite $c_2$ and $s_2$ as
\begin{align*}
c_2 &= W \ch k_2 = \sum_{i\geq 1} \ch(\alpha_i) \ket{\psi_i} \bra{\varphi_i}  , \\
s_2 &= W \sh k_2 = \sum_{i\geq 1} \sh(\alpha_i) \ket{\psi_i} \bra{\varphi_i},
\end{align*}
where $k_2 = \sum_{i\geq 1} \alpha_i \ket{\varphi_i} \bra{\varphi_i}$ is a spectral decomposition of $k_2$, with $(\varphi_i)$ an orthonormal family and $\psi_i = W \varphi_i$. Finally, let us also set the notation
\begin{align*}
p_2  = c_2 -1.
\end{align*}

Assuming that $s_2$ is Hilbert-Schmidt, we can use Lemma \ref{prop:properties_quadratic_transform_1}, to obtain
\begin{align*}
T^*_2 a^*(f) T_2 = a^*(c_2(f)) + a(s_2(f)), \\
 T^*_2 a(f) T_2 = a(c_2(f)) + a^*(s_2(f)).
\end{align*}

The following Lemma gathers estimates on $c_2$, $s_2$ and more generally on the action of $T_2$ on the Fock space. These estimates are needed for the proof of Lemma \ref{prop:second_quadratic_transform}.

\begin{lemma}
	\label{lemma:bound_second quadratic transform}
We have, for $\ell$ small enough, 
\begin{align}
& \|s_2\|_{\rm HS} \leq C , \label{item:lemma_second_quadra_transfo_1}\\
&\|D^{3/2} s_2\|_{\rm HS} + \|D^{3/2} s_2^*\|_{\rm HS} + \|Ds_2 D^{1/2}\|_{\rm HS} + \|D s_2^* D^{1/2}\|_{\rm HS} \leq C \|K D\|_{\rm HS}, \label{item:lemma_second_quadra_transfo_2a}\\
&\|D c^*_2 s_2 D \|_{\mathfrak S_1} \leq C \|D K D\|_{\mathfrak S_1}, \label{item:lemma_second_quadra_transfo_2b}\\
&\|D p_2\|_{\rm HS} + \|D p_2^*\|_{\rm HS} \leq C \|K D\|_{\rm HS}, \label{item:lemma_second_quadra_transfo_2c}\\
&\|KD\|_{\rm HS} \leq C \ell^{-7/2}, \label{item:lemma_second_quadra_transfo_3}\\
&\|D K D\|_{\mathfrak S_1} \leq C \ell^{-7}, \label{item:lemma_second_quadra_transfo_4}\\
&\|\nabla_1 s_2\|_{L^2(\mathbb{R}^6)} + \|s_2\|_{L^\infty L^2} +\|\nabla_1 s_2^*\|_{L^2(\mathbb{R}^6)} + \|s_2^*\|_{L^\infty L^2} \leq C ( \|K D\|_{\rm HS} +1), \label{item:lemma_second_quadra_transfo_5a}\\
&\|\nabla_1 p_2\|_{L^2(\mathbb{R}^6)} + \|p_2\|_{L^\infty L^2} + \|\nabla_1 p_2^*\|_{L^2(\mathbb{R}^6)} + \|p_2^*\|_{L^\infty L^2}  \leq C( \|KD\|_{\rm HS}+1), \label{item:lemma_second_quadra_transfo_5b}\\
&T_2^* (\mathcal N+1)^k T_2 \leq C_k (\mathcal N+1)^k, \quad \forall k\geq 1 \label{item:lemma_second_quadra_transfo_6}\\
&- \tr (D + K - E) + \frac{1}{2}\tr D^{-1} K^2 \geq - C, \label{item:lemma_second_quadra_transfo_7}\\
&\frac{1}{2}\tr D^{-1} K^2 = - \int N^4 ( (\varepsilon_{\ell,N} \omega_{\ell,N} ) \ast \varphi^2) \varphi^2 + \mathcal O(1) = \mathcal O(\ell^{-1}), \label{item:lemma_second_quadra_transfo_8}\\
&T_2^* \dd\Gamma (D) T_2 \leq C (\dd\Gamma (D) + \ell^{-1} + 1). \label{item:lemma_second_quadra_transfo_9}
\end{align}

\end{lemma}

\begin{proof}[Proof of Lemma \ref{lemma:bound_second quadratic transform}]  
Let us first establish some relations between $E$ and $D$. Since $\varepsilon_{\ell,N}$ is even, supported in $\{|x|\leq \ell\}$ and $\int N^3 \varepsilon_{\ell,N}=8\pi \ao$, we have $|N^3 \widehat {\varepsilon}_{\ell,N} (p) - 8\pi \ao| \leq C \ell^2 p^2 $. Hence, 
%\begin{align}
%K - 8\pi\ao Q\varphi^2Q &=  Q  N^3 \varphi (\widehat {\varepsilon}_{\ell,N} (p) -8\pi \ao)\varphi Q \nn \\
%	&\leq  C\ell^2 Q  \varphi (-\Delta) \varphi Q \nn\\
%	&\leq C \|\varphi\|_{H^1}^2 \ell^2 Q (-\Delta) Q \nn \\
%	&\leq C \ell^2 D, \label{eq:bound_K_D}
%\end{align}
\begin{align}
\pm (K - 8\pi\ao Q\varphi^2Q) &=  \pm Q  N^3 \varphi (\widehat {\varepsilon}_{\ell,N} (p) -8\pi \ao)\varphi Q  \leq  C\ell^2 Q  \varphi (-\Delta) \varphi Q \nn\\
	&\leq C \|\varphi\|_{H^1}^2 \ell^2 Q (-\Delta) Q  \leq C \ell^2 D, \label{eq:bound_K_D}
\end{align}
where we used that $\inf \sigma (D) > 0$ (where $D$ is interpreted as an operator on $\gH_+$). Hence for $\ell$ small enough and using that $0 \leq 8\pi \ao Q\varphi^2 Q \leq C D$, we have
\begin{align*}
C^{-1} D^2 \leq E^2 = D^{1/2} (D+2K) D^{1/2} \leq C D^2.
\end{align*}
Since  $x \mapsto x^{\alpha}$ is operator monotone for $0 < \alpha < 1$, we also have $C^{-1} D^{2\alpha} \leq E^{2\alpha}\leq C D^{2\alpha}$ and therefore $\|E^{-\alpha} D^{\alpha}\|_{op} + \|D^{-\alpha} E^{\alpha}\|_{op} \leq C_{\alpha}$ for some $C_{\alpha}>0$ independent of $N$.

\bigskip
\noindent 
\emph{Proof of \eqref{item:lemma_second_quadra_transfo_1}.} Let us rewrite $s_2$, using the formula
\begin{align*}
x = \frac{2}{\pi} \int_0^\infty \frac{x^2}{x^2 + t^2} \dd t,
\end{align*}
we have
\begin{align}
s_2 
	&= \frac{-1}{\pi} \int_0^\infty  \frac{D^{-1/2}}{D^2+t^2} D^{1/2} 2 K D^{1/2}\frac{E^{-1/2}}{E^2 + t^2} t^2 \dd t \nn \\
	&= \frac{-2}{\pi} \int_0^\infty\frac{t^2}{D^2 + t^2} 2  K (D^{1/2}E^{-1/2})  \frac{1}{E^2+t^2}  \dd t. \label{eq:development_s_2}
\end{align}
We obtain, 
\begin{align*}
\|s_2 \|_{\rm HS} 
	&\leq C  \int_0^\infty \|\frac{D^{2/5}}{D^2 + t^2}Q\varphi\|_{\rm op} \| D^{-2/5} Q \varphi N^3\widehat {\varepsilon}_{\ell,N} (p) \varphi Q D^{-2/5}\|_{HS} \|D^{9/10}E^{-9/10}\|_{op}  \|\frac{t^2 E^{2/5}}{E^2+t^2}\|_{op} \\
	&\leq C   \|D^{-2/5} Q\varphi\|_{\rm \mathfrak{S}_4}^2 \|\varepsilon_{\ell,N}\|_{L^1} \int_0^\infty  \frac{1}{(\inf \sigma_{\varphi^\perp} (D)  + t)^{16/5}}  \\
	&\leq C \|(D+1)^{-2/5} (1+p^2)^{2/5} \|_{op}^2 \| (1+p^2)^{-2/5} \varphi\|_{\rm \mathfrak{S}_4}^2 \\
	&\leq C \|\varphi\|_{L^4}^2 \leq C,
\end{align*}
where we used that $-\Delta + 1 \leq C (D+C)$ for some $C >0$ and therefore $\| (D+C)^{-\alpha/2} (-\Delta + 1)^{\alpha/2}\|_{op} \leq C_{\alpha}$ for all $0< \alpha < 1$.

\bigskip
\noindent 
\emph{Proof of \eqref{item:lemma_second_quadra_transfo_2a}.}
We only prove the estimate for $\|D^{3/2} s_2\|_{\rm HS} + \|Ds_2 D^{1/2}\|_{\rm HS}$ since the part with $s_2^*$ can be dealt with similarly. 
Starting from (\ref{eq:development_s_2}), we have
\begin{align*}
D^{3/2} s_2
	&= \frac{-2}{\pi} \int_0^\infty \frac{D^{1/2}}{D^2 + t^2} (DK) (D^{1/2} E^{-1/2})  \frac{t^2}{E^2+t^2}  \dd t,
\end{align*}
so that
\begin{align*}
\|D^{3/2} s_2\|_{\rm HS}
	&\leq C \int_0^\infty   \frac{1}{(\inf \sigma(D)^2+t^2)^{3/4}} \|D K\|_{\rm HS} \|D^{1/2}E^{-1/2}\|_{op}   \|\frac{t^2}{E^2 + t^2}\|_{op} \dd t \
	&\leq C \|D K\|_{\rm HS}.
\end{align*}
Similarly, one proves that $\|D s_2D^{1/2}\|_{\rm HS} \leq C \|D K\|_{\rm HS}$.

\emph{Proof of \eqref{item:lemma_second_quadra_transfo_2b}.}
Since $\|DE^{-1}\|_{op} \leq C$ we rather estimate,
\begin{align*}
E c^*_2 s_2 E
	&= \frac{1}{4} E^{1/2}(D-E)E^{1/2} + \frac{1}{4} E^{1/2} (D-E) D^{-1} E^{3/2} \\
	&= \frac{1}{\pi} \int_0^\infty   \frac{E^{1/2}}{E^2+t^2} D^{1/2} K  D^{1/2} \frac{t^2}{D^2 + t^2} E^{1/2}  \dd t  \\
	&\quad + \frac{1}{\pi} \int_0^\infty  \frac{E^{1/2}}{E^2+t^2} D^{1/2} K  D^{1/2} D^{-1} \frac{t^2}{D^2 + t^2} E^{3/2}  \dd t \\
	&= \frac{1}{\pi} \int_0^\infty   \frac{1}{E^2+t^2} (E^{1/2}D^{-1/2}) (D K  D) \frac{t^2}{D^2 + t^2} (D^{-1/2}E^{1/2})  \dd t  \\
	&\quad + \frac{1}{\pi} \int_0^\infty  \frac{ t }{E^2+t^2} (E^{1/2}D^{-1/2})  (D K D) D^{-3/2} E \frac{E^{1/2}t}{E^2 + t^2} \dd t\\ 
%	&\quad - \frac{1}{2\pi} \int_0^\infty2  \frac{D^{1/2}}{D^2+t^2} D^{1/2} 2 K  D^{1/2} \frac{t^2}{E^2 + t^2} (D^{1/2} 2K D^{1/2}) \frac{1}{D^2 +t^2} D^{1/2}  \dd t  \\
	& \quad + \frac{2}{\pi} \int_0^\infty \frac{ 1}{E^2+t^2} (E^{1/2}D^{-1/2})  (D K D) D^{-3/2}\frac{t^2}{D^2 + t^2} D^{1/2}K D^{1/2} \frac{E^{3/2}}{E^2 +t^2}   \dd t.
\end{align*}
Hence,
\begin{align*}
&\|D c^*_2 s_2 D \|_{\mathfrak S _1} 
	\leq C  \int_0^\infty \frac{1}{\inf \sigma (E^2)  + t^2}  \|E^{1/2} D^{-1/2}\|_{op} \|DKD\|_{\mathfrak S_1} \|\frac{t^2}{D^2 + t^2}\|_{op} \|D^{-1/2}E^{1/2}\|_{op} \\
	&\quad + C  \int_0^\infty \frac{t}{\inf \sigma (E^2)  + t^2}  \|E^{1/2} D^{-1/2}\|_{op}  \|DKD\|_{\mathfrak S_1} \|D^{-1/2}\|_{op}\|D^{-1} E\|_{op}  \frac{1}{(\inf\sigma E^2 + t^2)^{1/4}} \dd t \\
	&\quad + C  \int_0^\infty \frac{1}{\inf \sigma (E^2)  + t^2} \|E^{1/2}D^{-1/2}\|_{op} \|DKD\|_{\mathfrak S_1}  \|D^{-1}\|_{op} \|\frac{t^2}{D^2 + t^2}\|_{op} \times \\
	&\qquad \qquad \qquad \qquad\qquad\qquad\qquad \times \|D^{-1}\|_{op} \|K\|_{op}\| \|D^{1/2}E^{-1/2}\|_{op} \|\frac{E^2}{E^2 + t^2}\|_{op} \dd t \\
	&\leq C (1+ \|K\|_{op}) \|D K D\|_{\mathfrak S_1}  \leq C \|D K D\|_{\mathfrak S_1}.
\end{align*}

\noindent
\emph{Proof of \eqref{item:lemma_second_quadra_transfo_2c}.} 
We only prove the inequality for $\|D p_2\|_{\rm HS}$ since the one for $p_2^*$ is proven in a similar way. Let us develop
\begin{align*}
p_2 = c_2 -1  
	&= \frac{1}{2}(D^{1/2}E^{-1/2} -1) + \frac{1}{2}(D^{-1/2}E^{1/2} -1) \\
	&= \frac{1}{2}(D^{1/2} -E^{1/2})E^{-1/2} + \frac{D^{-1/2}}{2}(E^{1/2} -D^{1/2}) \\
	&= \frac{1}{\pi} \int_0^\infty \frac{1}{D+t^2} (D-E) E^{-1/2} \frac{t^2}{E+t^2} \dd t + \frac{1}{\pi} \int_0^\infty \frac{1}{D+t^2} D^{-1/2}(E-D) \frac{t^2}{E+t^2} \dd t \\
	&= \frac{2}{\pi} \int_0^\infty \frac{1}{D+t^2} D^{1/2} s_2 \frac{t^2}{E+t^2} \dd t - \frac{2}{\pi} \int_0^\infty \frac{1}{D+t^2} s_2 E^{1/2} \frac{t^2}{E+t^2} \dd t.
\end{align*}
There we have
\begin{align*}
\|D p_2\|_{\rm HS} \leq C( \|D^{3/2} s_2\|_{\rm HS} + \| D s_2 D^{1/2}\|_{\rm HS}\|D^{-1/2}E^{1/2}\|_{op}) \int_0^\infty \frac{1}{\inf \sigma(D) + t^2} \dd t 
	\leq  C \|D K\|_{\rm HS}.
\end{align*}

\noindent
\emph{Proof of \eqref{item:lemma_second_quadra_transfo_3}.} Using that $[-\Delta + V_{\rm ext} + 8\pi \ao \varphi^2,Q]=0$, we have
\begin{align*}
\tr DK^2D
 	&= \tr (-\Delta + V_{\rm ext} + 8\pi a \varphi^2-\mu) (\varphi N^3 \widehat{\varepsilon}_{\ell,N} (p) \varphi)^2(-\Delta + V_{\rm ext} + 8\pi a \varphi^2 -\mu) \\
	&\leq C (\tr K^2 + \tr ((-\Delta) K^2(-\Delta)) + \tr ( V_{\rm ext} K^2 V_{\rm ext})) \\
	&\leq C \|\varphi\|_{L^4}^4 \|N^3 \varepsilon_{\ell,N}\|_{L^2}^2 + C \tr [-\Delta,\varphi] N^3 \widehat{\varepsilon}_{\ell,N} (p) \varphi^2 N^3 \widehat{\varepsilon}_{\ell,N} (p) [-\Delta,\varphi] \\
	&\quad + C \tr \Big( \varphi N^3 p^2 \widehat{\varepsilon}_{\ell,N} (p) \varphi^2 N^3 p^2 \widehat{\varepsilon}_{\ell,N} (p) \varphi \Big) + \|V_{\rm ext} \varphi\|_{L^2}^2  \|N^3 \varepsilon_{\ell,N}\|_{L^2}^2 \|\varphi\|_{L^\infty}^2 \\
	&\leq C  \|N^3 \varepsilon_{\ell,N}\|_{L^2}^2 + C \| (\Delta \varphi) N^3 \widehat{\varepsilon}_{\ell,N} (p) \varphi \|_{\rm HS} + C \| \nabla \varphi \cdot p N^3 \widehat{\varepsilon}_{\ell,N} (p) \varphi \|_{\rm HS} \\
	&\quad + C \| \varphi  p^2 N^3 \widehat{\varepsilon}_{\ell,N} (p) \varphi \|_{\rm HS} \\
	&\leq \|\varphi\|_{H^2}^2 \|N^3 \varepsilon_{\ell,N}\|_{H^2}  \leq C \ell^{-7/2}.
\end{align*}

\noindent
\emph{Proof of \eqref{item:lemma_second_quadra_transfo_4}.} To estimate $\|D K D\|_{\mathfrak S_1}$, using again $[-\Delta + V_{\rm ext} + 8\pi \ao \varphi^2,Q]=0$, we have
\begin{align*}
\|DKD\|_{\mathfrak S_1}  &=  \|Q  (-\Delta + V_{\rm ext} + 8\pi \ao \varphi^2-\mu) (\varphi N^3 \widehat{\varepsilon}_{\ell,N} (p) \varphi)(-\Delta + V_{\rm ext} + 8\pi a \varphi^2 -\mu) Q\|_{\mathfrak S_1} \nn\\
	&\leq  C(1 + \|\varphi\|_{L^\infty}^2) \|\varphi N^3 \widehat{\varepsilon}_{\ell,N} (p) \varphi\|_{\mathfrak S_1} + C\| V_{\rm ext} \varphi N^3 |\widehat{\varepsilon}_{\ell,N} (p)| \varphi  V_{\rm ext}\|_{\mathfrak S_1}  \nn\\
	&\quad + C \|(-\Delta)(\varphi N^3 |\widehat{\varepsilon}_{\ell,N} (p)| \varphi)(-\Delta)\|_{\mathfrak S_1}\\
	&\leq  C(\| \varphi\|_{L^2}^2 + \|V_{\rm ext} \varphi\|_{L^2}^2) \| N^3 \widehat{\varepsilon}_{\ell,N}(p)\|_{L^1} + C \| \varphi \|_{H^2}^2  \| (1+|p|^4)N^3 \widehat{\varepsilon}_{\ell,N}(p)\|_{L^1}. 
\end{align*}
Here we have estimated $\|(-\Delta)(\varphi N^3 |\widehat{\varepsilon}_{\ell,N} (p)| \varphi)(-\Delta)\|_{\mathfrak S_1}$ by using $[\Delta,\varphi]=\Delta \varphi + 2\nabla \varphi \cdot \nabla$ and the Cauchy--Schwarz inequality. Since $N^3 \eps_{\ell,N}$ is smooth and bounded by $C\ell^3 \1_{\{|x|\leq \ell\}}$, we have
\begin{align*}
\|DKD\|_{\mathfrak S_1}	&\leq C  \|  (1+p^4) N^3 \widehat{\varepsilon}_{\ell,N}(p) \|_{L^1} \\
& \leq C   \|  (1+p^4) (\ell^{-2}+|p|^2)^{-4} \|_{L^1} \| (\ell^{-2}+p^2)^4 N^3 \widehat{\varepsilon}_{\ell,N}(p)\|_{L^\infty}  \\
&\leq C \| (\ell^{-2} + |p|^{2})^{-2}  \|_{L^1}  \| (\ell^{-2}-\Delta)^4 N^3 \varepsilon_{\ell,N}\|_{L^1}\\
&\leq C \ell \cdot \ell^{-8}= C \ell^{-7}.
\end{align*}

\noindent
\emph{Proof of \eqref{item:lemma_second_quadra_transfo_5a}-\eqref{item:lemma_second_quadra_transfo_5b}.} Take a constant $C_0 \ge 1 - ( \Vext + 8\pi \ao \varphi^2 -\mu)$. Using $D+C_0\ge 1-\Delta$ we have
$$
\| \nabla_x s_2(x,y)\|_{L^2 L^2} \leq \| (D+C_0)_x^{1/2} s_2(x,y)\|_{L^2 L^2} = C\|(D+C_0)^{1/2} s_2\|_{\rm HS} \leq C (\|KD\|_{\rm HS}+1). 
$$ 
Here we have used \eqref{item:lemma_second_quadra_transfo_1} and \eqref{item:lemma_second_quadra_transfo_2a} in the last estimate. Moreover, using the kernel estimate 
$$
0\leq (D+C_0)^{-1}(x,y)= (-\Delta+ \Vext +  8\pi \ao \varphi^2 -\mu +C_0) ^{-1}(x,y) \leq (1-\Delta)^{-1}(x,y)
$$
and the Sobolev embedding $H^2(\R^3)\subset L^\infty(\R^3)$ we find that
$$
\|(D+C_0)^{-1} g\|_{L^\infty} \leq  \|(1-\Delta)^{-1} |g| \|_{L^\infty} \leq C \|g\|_{L^2}, \quad \forall g\in L^2(\R^3).
$$
Consequently, 
\begin{align} \label{eq:D-Sobolev}
\|g\|_{L^\infty} \leq  \|(D+C_0)g\|_{L^2} \le C \|(D+1) g\|_{L^2}
\end{align}
and hence
$$
\|s_2(x,y)\|_{L^\infty L^2} \leq C\| (D+1)_x s_2(x,y)\|_{L^2 L^2} = C\|(D+1) s_2\|_{\rm HS} \leq C (\|KD\|_{\rm HS}+1). 
$$ 
The bound for $s_2^*$ are obtained similarly. The bound  \eqref{item:lemma_second_quadra_transfo_5b} can be deduced from \eqref{item:lemma_second_quadra_transfo_2c} by the same argument. 
%
%
%Note the kernel equality $(D s_2)(x,y) = D_x s_2(x,y)$, moreover since the domain of $D$ is included in $H^2$ and since $D\geq 0$, there is some constant $C >0$ independent of $N$ such that {\bf Do we need some condition on $\Vext$ for that?}
%\begin{align*}
%(-\Delta)^2 \leq C (D + 1)^2.
%\end{align*}
%Hence (\ref{item:lemma_second_quadra_transfo_5a}) is a consequence of (\ref{item:lemma_second_quadra_transfo_2a}) and the Sobolev inequalities. The bound \eqref{item:lemma_second_quadra_transfo_5b} is obtained similarly. 

\noindent
\emph{Proof of \eqref{item:lemma_second_quadra_transfo_6}.}
This is a consequence of (\ref{item:lemma_second_quadra_transfo_1}), Lemma \ref{prop:properties_quadratic_transform_1} and the fact that 
\begin{align*}
\mathcal W \mathcal N \mathcal W^* = \mathcal N. 
\end{align*}

\noindent
\emph{Proof of \eqref{item:lemma_second_quadra_transfo_7}.} From \cite[Theorem 6]{NamNapRicTri-21}, we obtain
\begin{align*}
-  \tr (D + K - E) \geq - \frac{1}{2}\tr D^{-1} K^2 - C  \|K\|_{op} \tr_{\gH_+} \Big[ (D+K)^{-2} K^2\Big].
\end{align*}
If one had $(D+K)^2\geq C^{-2} (1-\Delta)^2 - C$ on $\gH$, then a simple adaptation of \cite[Lemma 8]{NamNapRicTri-21} would give
\begin{align}\label{eq:NNRT-simple}
\tr_{\gH_+} ( (D+K)^{-2} K^2 ) 
	&\leq C \tr_{\gH} \Big(  (1-\Delta)^{-2} \widetilde K^2 \Big) + C \nn \\
	&\leq  C \tr \Big( (1+p^2)^{-2} \Big(\varphi(x) N^3 \widehat{\varepsilon_{\ell,N}}(p) \varphi(x) \Big)^2  \Big)+ C\nn \\
	&\leq \|(1+p^2)^{-1} \varphi \|_{\rm HS}^2 \|\varphi\|_{L^\infty}^2 \|N^3 \widehat{\varepsilon_{\ell,N}}\|_{L^\infty}^2 \nn \\
	&\leq   \|(1+p^2)^{-1}\|_{L^2(\mathbb{R}^3)} \|\varphi\|_{L^2}^2  \|\varphi\|_{L^\infty}^2 \|N^3 \varepsilon_{\ell,N}\|_{L^1}^2 \leq C. 
\end{align}
However, in the present paper we are not assuming an additional condition on $\Vext$ as in \cite{NamNapRicTri-21}, and hence  the operator inequality $(D+K)^2\geq C^{-2} (1-\Delta)^2 - C$ is not available. To overcome this issue, we  write 
\begin{align*}
\tr_{\gH_+} ( (D+K)^{-2} K^2 ) 
	\leq C \tr_{\gH} \Big(  (-\Delta +\Vext+1)^{-2} \widetilde K^2 \Big) + C 
\end{align*}
and use the kernel estimate
$$
0\le (-\Delta +\Vext+1)^{-1}(x,y) \le (1-\Delta)^{-1}(x,y). 
$$
This gives
\begin{align*}
&\tr_{\gH} \Big(  (-\Delta +\Vext+1)^{-2} \widetilde K^2 \Big) = \|(-\Delta +\Vext+1)^{-1} \widetilde K\|_{\rm HS}^2\\
&= \iint_{\R^3} \Big| \int (-\Delta +\Vext+1)^{-1}(x,z)   \varphi(z) N^3 \eps_{\ell,N}(z,y) \varphi(y) \dd z \Big|^2 \dd x \dd y \\
&= \iint_{\R^3} \Big| \int (1-\Delta)^{-1}(x,z)  \varphi(z) N^3 |\eps_{\ell,N}(z,y)| \varphi(y) \dd z  \Big|^2 \dd x \dd y \\
&= \tr_{\gH} \Big( (1+p^2)^{-2} \Big( \varphi(x) N^3 \widehat{ |\varepsilon_{\ell,N}|}(p) \varphi(x) \Big)^2 \Big)\\
&\le  \|(1+p^2)^{-1}\|_{L^2}  \|\varphi\|_{L^2}^2  \|\varphi\|_{L^\infty}^2 \|N^3 |\varepsilon_{\ell,N}|\|_{L^1}^2 \le C. 
\end{align*}
Here the last estimate goes exactly like in \eqref{eq:NNRT-simple} where the replacement $\eps_{\ell,N}$ by its absolute value does not cause any problem. This completes the proof of \eqref{item:lemma_second_quadra_transfo_7}.

\noindent \emph{Proof of \eqref{item:lemma_second_quadra_transfo_8}.} Take a constant $C_0>0$ such that $D + K +C_0\ge 1-\Delta$. By \eqref{item:lemma_second_quadra_transfo_6},  
$$
0\le \tr_{\gH_+} \Big( (D+K)^{-1} -  (D + K +C_0)^{-1})  K^2 \Big)  \le C_0 \tr_{\gH_+} \Big( (D+K)^{-2}  K^2 \Big) \le C. 
$$
On the other hand, using  the operator inequality $(D+K+C_0)^{-1} \le (1-\Delta)^{-1}$, we can proceed similarly as in \cite[Lemma 8, Eq. (59) and (66)]{NamNapRicTri-21} to get
\begin{align*}
\tr_{\gH_+} (D+K)^{-1} K^2 
	&\leq \tr (1-\Delta)^{-1} \widetilde K^2 + C = 2 \int N^4 (\varepsilon_{\ell,N} (\omega_{N}- \omega_{\ell,N})) \ast \varphi^2) \varphi^2 + \mathcal O(1).
\end{align*}
Here we have used the equation $\frac{1}{2} N^2\varepsilon_{\ell,N} = \Delta (\omega_{\ell,N} -   \omega_N) = - \Delta [(1-\chi_{\ell}) \omega_{N}]$ from  \eqref{eq:def_omega_ell}. By an integration by part, and using that $(-\Delta \omega_N) (1-\chi_{\ell}) = \frac{1}{2} N^2V_N f_N(1-\chi_{\ell})  = 0 $, we obtain
\begin{align*}
&\frac{1}{2}\int N^4 (\varepsilon_{\ell,N} \omega_{N} \ast \varphi^2) \varphi^2 
	= N^2 \int [(1-\chi_{\ell}) \omega_{N}(-\Delta \omega_N)] (x-y) \varphi(x)^2 \varphi(y)^2 \\
	&\quad - 2 N^2 \int [(1-\chi_{\ell}) \omega_{N}(\nabla \omega_N)] (x-y) \nabla(\varphi^2)(x) \varphi(y)^2 \\
	&\quad + N^2 \int [(1-\chi_{\ell}) \omega_{N}^2] (x-y) (-\Delta\varphi^2)(x) \varphi(y)^2 \\
	&\leq C  N^2 \int \omega_{N}^2 (x-y) |\nabla \varphi^2(x)|^2 |\nabla \varphi^2(y)|  + N^2 \int  \omega_{N}^2(x-y) |(-\Delta \varphi^2)(x)| \varphi(y)^2\\
	&\quad +  C  N^2 \int \omega_{N}^2 (x-y) |\nabla \chi_\ell(x-y) | |\nabla \varphi^2(x)|^2 |\varphi^2(y)|  \leq C \|\varphi\|_{H^2(\mathbb{R}^3)}^4,
\end{align*}
where we used the Hardy-Littlewood-Sobolev inequality and that $\int N^2 \omega_{N}^2 |\nabla \chi_\ell| \leq C$. This finishes to prove  (\ref{item:lemma_second_quadra_transfo_8}).

\noindent
\emph{Proof of \eqref{item:lemma_second_quadra_transfo_9}.} In view of (\ref{eq:bound_K_D}), for $\ell$ small enough, we can use the standard lower bound on quadratic hamiltonians \cite[Lemma 9]{NamNapSol-16}, that is
\begin{equation*}
\dd \Gamma(\frac{1}{2}D + K) + \frac{1}{2} \int K(x,y) (a^*_x a^*_y + a_x a_y) \geq - C  \tr (K D^{-1} K).
\end{equation*}
Therefore, using the above inequality, (\ref{eq:bogoliubov_hamiltonian}) and $E \leq C D$, we obtain
\begin{align*}
-C \tr (K D^{-1} K)  + \frac{1}{2} T_2^* \dd\Gamma(D) T_2 
	&\leq T_2^* \left( \dd \Gamma(D + K) + \frac{1}{2} \int K(x,y) (a^*_x a^*_y + a_x a_y)\right)T_2 \\
	&= \dd\Gamma(E)   -\frac{1}{2} \tr (D + K - E) \leq C \dd\Gamma(D),
\end{align*}
where we also used that $\tr (D + K - E) = 2 \braket{\dd\Gamma(E)}_{T_2^* \Omega} \geq 0$, see (\ref{eq:bogoliubov_hamiltonian}). Therefore, it remains to prove the bound on $\tr (K D^{-1} K)$, that is (\ref{item:lemma_second_quadra_transfo_9}). From (\ref{item:lemma_second_quadra_transfo_8}), we have
\begin{align*}
\tr (K D^{-1} K) 
	&\leq - 2 \int N^4 \varepsilon_{\ell,N} \omega_{\ell,N} \ast \varphi^2 \varphi^2 + C\\
	&\leq C \frac{N}{\ell^3} \int \frac{\mathds{1}_{|x|\leq \ell }}{N |x|} \|\varphi\|_{L^2}^2 \|\varphi\|_{L^\infty}^2 + C \leq C (\ell^{-1} +1),
\end{align*}
where we used that $N^3 \varepsilon_{\ell,N}\leq C \ell^{-3}\mathds{1}_{\{|x|\leq \ell\}}$, see Lemma \ref{prop:prop_epsilon}.
The proof of  Lemma \ref{lemma:bound_second quadratic transform} is complete.
\end{proof}

\subsection{Proof of Lemma \ref{lemma:estimate_L4_second_quadratic_transfo}}
	\label{sec:proof_L4_T2}

\begin{proof}[Proof of Lemma \ref{lemma:estimate_L4_second_quadratic_transfo}]
Let us expand $T^*_2 \mathcal L_4 T_2$ exactly as in (\ref{eq:dev_quartic_quadratic_transfo}), that is we expand in an economical way where the CCR are only used to set in normal order only the terms where $g_i = \delta_{x,y}$ for some $1\leq i \leq 4$. We obtain quartic terms, quadratic terms and one constant term. The quartic terms are of the form
\begin{align*}
\int N^2 V_N(x-y) a^{\sharp_1} (g_{1,x}) a^{\sharp_2} (g_{2,y}) a^{\sharp_3} (g_{3,x}) a^{\sharp_1} (g_{4,y})
\end{align*}
with $g_{i} \in \{\delta_{x,y}\} \cup \{s_2(x,y),p_2(x,y)\}$ and $\sharp_i \in \{*,\cdot\}$ with the condition that if one of the $g_i$ is $\delta_{x,y}$ then $(\sharp_1,\sharp_2,\sharp_3,\sharp_4)$ is in normal order. By Lemma \ref{prop:estimate_L_4_quartic_terms} and Lemma \ref{lemma:bound_second quadratic transform}, we can bound all these quartic terms in the following way
\begin{align*}
&\pm \int N^2 V_N(x-y) a^{\sharp_1} (g_{1,x}) a^{\sharp_2} (g_{2,y}) a^{\sharp_3} (g_{3,x}) a^{\sharp_1} (g_{4,y}) + \hc \\
&\qquad\leq CN^{-1}(1+\|s_2\|_{\rm HS} + \|p_2\|_{\rm HS})^2 (1+\|D s_2\|_{\rm HS} + \|D p_2\|_{\rm HS})^2 (\mathcal N+1)^2 + C \mathcal L_4 \\
&\qquad\leq CN^{-1}\ell^{-7} (\mathcal N+1)^2 + C \mathcal L_4.
\end{align*}
It remains to bound the quadratic terms appearing in $T_2^* \mathcal L_4 T_2$, in view of (\ref{eq:dev_quartic_quadratic_transfo}), they are of the form
\begin{align*}
\int N^2 V_N(x-y) \braket{c_{2,x},s_{2,y}}  a^{\sharp_1} (g_{1,x}) a^{\sharp_2} (g_{2,y})
\end{align*}
with $g_{i} \in \{\delta_{x,y}\} \cup \{s_2(x,y),p_2(x,y)\}$ and $\sharp_i \in \{*,\cdot\}$ with the condition that if one of the $g_i$ is $\delta_{x,y}$ then $(\sharp_1,\sharp_2)$ is in normal order. Note that
\begin{align*}
\braket{c_{2,x},s_{2,y}} = s_2(x,y) + \braket{p_{2,x},s_{2,y}},
\end{align*}
so that from \eqref{eq:D-Sobolev}, we have $\|\braket{c_{2,x},s_{2,y}}\|_{L^2_x L^\infty_y} \leq \|s_2\|_{L^2L^\infty} (1+\|p_2\|_{L^2}) \leq C(1+ \|Ds_2\|_{HS})$. This estimate and a simple adaptation of the proof of Lemma \ref{lemma:estimate_quadratic_quadratic transform} leads to
\begin{align*}
\pm &\int N^2 V_N(x-y) \braket{c_{2,x},s_{2,y}}  a^{\sharp_1} (g_{1,x}) a^{\sharp_2} (g_{2,y}) + \hc \\
&\leq  CN^{-1}(1+\|s_2\|_{\rm HS} + \|p_2\|_{\rm HS})^2 (1+\|D s_2\|_{\rm HS} + \|D p_2\|_{\rm HS})^2 (\mathcal N+1)  +  C \mathcal L_4 \\
&\leq  CN^{-1} \ell^{-7} (\mathcal N+1)  +  C \mathcal L_4.
\end{align*}
It remains to estimate the constant term, that is
\begin{align*}
\left| \int N^2 V_N(x-y)\braket{c_{2,x},s_{2,y}}\braket{s_{2,x},c_{2,y}}\right|  
	&\leq \int N^2 V_N(x-y) |\braket{c_{2,x},s_{2,y}}|^2 \\
	&\leq  N^{-1}\|V\|_{L^1} \|\braket{c_{2,x},s_{2,y}}\|_{L^2_x L^\infty_y}^2 \leq C N^{-1} \ell^{-7}.
\end{align*}
Gathering the above estimates finishes to prove Lemma \ref{lemma:estimate_L4_second_quadratic_transfo}.

\end{proof}

\subsection{Conclusion of Lemma \ref{prop:second_quadratic_transform}}
\label{sec:proof_propT2}
Let us recall, from Lemma \ref{prop:recap_cubic_transform}, that 
\begin{align*}
\mathcal H^{(T_c)} &= N \cE_{\rm GP}(\varphi) - 4\pi a \int \varphi^4 + \frac{N^4}{2}\int (\omega_{\ell,N} \varepsilon_{\ell,N}) \ast \varphi^2 \varphi^2 \\
	&\quad + \dd\Gamma(D + K) + \frac{1}{2} \int K(x,y) (a^*_x a^*_y + a_x a_y) + \cL_4 .
	\end{align*}
From \cite[Eq. (28)]{GreSei-13}, we have 
\begin{align}
T_2^* \Big(  \dd \Gamma(D + K) + \frac{1}{2} \int K(x,y) (a^*_x a^*_y + a_x a_y) \Big) T_2 
 	=   \dd\Gamma(E)   -\frac{1}{2} \tr (D + K - E). \label{eq:bogoliubov_hamiltonian}
\end{align}
Hence,
\begin{align*}
T^*_2 T_{c}^*T_{1}^*\mathcal H T_{1} T_{c} T_2 
	&=N \cE_{\rm GP}(\varphi) - 4\pi a \int \varphi^4 + \frac{N^4}{2} \int ((\omega_{\ell,N} \varepsilon_{\ell,N})\ast \varphi^2) \varphi^2 \nn\\
	&  -\frac{1}{2} \tr (D + K - E) + 	\dd\Gamma(E)   + T^*_2 \mathcal L_4 T_2 + T^*_2\mathcal E^{(\theta_M)}T_2 + T_2^*\mathcal E^{(T_c)} T_2.
\end{align*}
To prove (\ref{eq:second_quadratic_transform}), in view of (\ref{eq:bogoliubov_hamiltonian}) we only need to estimate $T_2^*\mathcal E^{(T_c)} T_2$. Using Lemma \ref{lemma:bound_second quadratic transform} and that $\mathcal K \leq C (\dd\Gamma(D) +1)$, we obtain easily that $\mathcal E^{(T_2)} :=T_2^*\mathcal E^{(T_c)} T_2$ satisfies (\ref{eq:lemma_second_quad_transfo_error}). The proof of Lemma \ref{prop:second_quadratic_transform} is complete.

\section{Optimal BEC} \label{sec:BEC}

In this section we will prove the following result.

\begin{lemma}[Optimal rate]
	\label{lemma:optimal_rate}
	 On the truncated Fock space $\cF_+^{\leq N}$ we have
\begin{align*}
U H_N U^* =\1_+^{\leq N} \mathcal H \1_+^{\leq N} \geq N e_{GP} + C^{-1} \mathcal N - C
\end{align*}
for some constant $C>0$ independent of $N$. 
\end{lemma}

\begin{proof}
We use a localization technique from \cite{LewNamSerSol-15}. Let $f,g \in \mathbb{R} \to [0,1]$ be smooth such that $f^2(x) + g^2(x) =1$ for all $x\in \mathbb{R}$ such that $f(x) = 1$ for $x<1/2$ and $f(x) = 0$ for $x>1$. Now let $M_0 \geq 1$ and let us define $f_{M_0} = f(\mathcal N_+/M_0)$ and $g_{M_0} = g(\mathcal N_+ / M_0)$. We have
\begin{align}
\mathcal H &= f_{M_0} \mathcal H f_{M_0} + g_{M_0} \mathcal H g_{M_0} + \mathcal E_{M_0}, \label{eq:localization}\\
 \mathcal E_{M_0} &= \frac{1}{2} \left([f_{M_0},[f_{M_0},\mathcal H]]+ [g_{M_0},[g_{M_0},\mathcal H]]\right). \nn
\end{align}
One can easily show that 
\begin{align}
\pm \mathcal E_{M_0} \leq C M_0^{-2} (\mathcal L_4 + \dd\Gamma(E) +  N ) \mathds{1}^{\{M_0/2 \leq \mathcal N \leq M_0 \}}.
\label{eq:error_localization_0} 
\end{align}

Now let us choose $M_0 = \delta_{\alpha} N$, $M = \delta_{\beta} N$ with $\delta_{\alpha}, \delta_{\beta}$, small but fixed to be chosen later. Here $M,\ell$ are the parameters associated with $T_1,T_c$, satisfying $M N^{-1} \lesssim \ell \lesssim 1$. From Lemma \ref{prop:second_quadratic_transform} we have
\begin{align*}
T_{c}^* T_1^* \mathcal H T_1 T_{c} 
	&=  N \cE_{\rm GP}(\varphi) - 4\pi a \int \varphi^4 + \frac{1}{2}N^4 \int (\omega_{\ell,N} \varepsilon_{\ell,N})\ast \varphi^2 \varphi^2 -\frac{1}{2} \tr (D + K - E)   \nn \\
	&\quad  +T_2 \dd\Gamma(E) T_2^* + \mathcal L_4  + \mathcal E^{(\theta_M)} +  T_2 \mathcal E^{(T_2)} T_2^{*},
\end{align*}
and (\ref{eq:lemma_second_quad_transfo_error}) becomes
\begin{align*}
\pm T_2 \mathcal E^{(T_2)} T_2^{*}
	&\leq C (\ell^{1/2}  + \delta_{\beta}^{1/2}\ell^{-1}) (\mathcal N+1) \nn \\
	&\quad + C \big\{ \delta_{\beta}^{1/2} +\ell^{3/2}\big\} ( \dd\Gamma(D) + \ell^{-1}+ \mathcal L_4)  \nn \\
	&\quad + C \frac{(\mathcal N+1)^{3/2}}{N^{1/2}} + C \frac{(\mathcal N+1)^{5/2}}{N^{3/2}}\nn \\
	&\quad + \eps \left( \cL_4 +  \delta_{\beta} (\dd\Gamma(D) + \ell^{-1}) + (\mathcal N+1) +  \ell^{1/2} \frac{(\mathcal N+1)}{N^2} + \frac{(\mathcal N+1)^2}{N^3} \right) \\
	&\quad + \eps^{-1}C \left( \ell(\mathcal N+1) + \frac{(\mathcal N+1)^2}{N}+ \frac{(\mathcal N+1)^5}{N^4} \right),
\end{align*}
where we also used that $\dd\Gamma (D) \leq C  ( T_2 \dd \Gamma (E) T_2^* + C\ell^{-1})$, see Lemma \ref{lemma:bound_second quadratic transform}. Hence we see that if we choose $\eps=1/4$ and $\delta_{\beta} = \ell^{4}$, then for $\ell$ small enough, we can absorb in $T_2 \dd\Gamma(E) T_2^*$ and $\mathcal L_4$ all the terms except the ones proportional to $(\mathcal N+1)^{j}$ with $j > 1$ and $\mathcal E^{(\theta_M)}$. Using again Lemma \ref{lemma:bound_second quadratic transform} to estimate the constant terms, we obtain
\begin{align*}
T_{c}^* T_1^* \left(\mathcal H - N e_{\rm GP}\right) T_1 T_{c} 
	\geq \frac{1}{2} \left( T_2 \dd\Gamma(E) T_2^* + \mathcal L_4\right) + \mathcal E^{(\theta_M)} - C_{\delta_\alpha,\ell,\eps}( 1 + \frac{(\mathcal N+1)^{3/2}}{N^{1/2}} + \frac{(\mathcal N+1)^{5}}{N^{4}}).
\end{align*}
Using the expression of $\mathcal E^{(\theta_M)}$ given in (\ref{eq:E_theta_M}) and that $\theta_M \mathds{1}^{\{\mathcal N \leq M/2\}} = 1$ we easily prove that
\begin{align}
	\label{eq:estimate_E_theta}
\pm \mathcal E^{(\theta_M)} 
	&\leq \frac{1}{4} \mathcal L_4  + C \mathds{1}^{\{\mathcal N > M/2\}} \mathcal N \leq  \frac{1}{4} \mathcal L_4 +   C \frac{\mathcal N^{2}}{M}.
\end{align}
Therefore, denoting $\mathcal N_{T_1T_c} = T_c^* T_1^* \mathcal N T_1T_c$, and using it satisfies $(\mathcal N+1)^{j} \leq C_j (\mathcal N_{T_1T_c} +1)^{j} \leq C_j' (\mathcal N +1)^{j}$ we have
\begin{align*}
&T_{c}^* T_1^* f_{M_0}\left(\mathcal H - N e_{\rm GP}\right) f_{M_0}T_1 T_{c} \\
	&\quad =f(\mathcal N_{T_1T_c} / M_0) (T_{c}^* T_1^* \left(\mathcal H - N e_{\rm GP}\right) T_1 T_{c}) f(\mathcal N_{T_1T_c} / M_0) \\
	&\quad \geq \frac{1}{4} f(\mathcal N_{T_1T_c} / M_0) \left( T_2 \dd\Gamma(E) T_2^* + \mathcal L_4\right)f(\mathcal N_{T_1T_c} / M_0)  \\
	&\qquad - C_{\delta_\alpha,\ell,\eps}f(\mathcal N_{T_1T_c} / M_0)( 1 + \frac{(\mathcal N+1)^{3/2}}{N^{1/2}} + \frac{(\mathcal N+1)^{5}}{N^{4}} + \frac{\mathcal N^{2}}{M} )f(\mathcal N_{T_1T_c} / M_0) \\
	&\quad \geq \frac{\inf \sigma E}{4}  f(\mathcal N_{T_1T_c} / M_0) \mathcal N f(\mathcal N_{T_1T_c} / M_0)  \\
	&\qquad - C_{\delta_\alpha,\ell,\eps}f(\mathcal N_{T_1T_c} / M_0)( 1 + \frac{(\mathcal N_{T_1T_c}+1)^{3/2}}{N^{1/2}} + \frac{(\mathcal N_{T_1T_c}+1)^{5}}{N^{4}} +  \frac{\mathcal N_{T_1T_c}^{2}}{M} ) \\
	&\quad \geq C^{-1} f(\mathcal N_{T_1T_c} / M_0)^2 \left( \mathcal N_{T_1T_c} - C_{\delta_\alpha,\ell,\eps}( 1 +(M_0^{1/2}N^{-1/2} +  M_0^{4}N^{-4} +  M_0M^{-1} \mathcal N) \right) \\
	&\quad \geq  C^{-1} f(\mathcal N_{T_1T_c} / M_0)^2 \left( (1 - (\delta_{\alpha}^{1/2}+  \delta_{\alpha}^{4} +  \delta_{\alpha} \delta_{\beta}^{-1})\mathcal N_{T_1T_c} - C_{\delta_\alpha,\ell,\eps}\right).
\end{align*}
Taking $\delta_\beta$ sufficiently small but fixed and conjugating back by $T_1$ and $T_c$ we obtain
\begin{align}
f_{M_0}\left(\mathcal H - N e_{\rm GP}\right) f_{M_0} \geq C_{\delta_\alpha,\delta_\beta,\ell,\eps}^{-1} \mathcal N f_{M_0}^2 - C_{\delta_\alpha, \delta_\beta,\ell,\eps}f_{M_0}^2.
	\label{eq:error_localization_f}
\end{align}

Now by a standard argument, we know that 
\begin{align}
\1_+^{\leq N} g_{M_0} (\mathcal H - N e_{\rm GP}) g_{M_0}\1_+^{\leq N} \geq C \1_+^{\leq N} g_{M_0}^2 N \geq  C \1_+^{\leq N}  g_{M_0}^2 \mathcal N. \label{eq:error_localization_g}
\end{align}
for a constant $C$ independent of $N$. Putting together (\ref{eq:error_localization_0}),  (\ref{eq:error_localization_f}) and  (\ref{eq:error_localization_g}) we obtain
\begin{align*}
\1_+^{\leq N} \mathcal H \1_+^{\leq N}  \geq \1_+^{\leq N}  \left( N e_{GP} + C^{-1} \mathcal N - C - C \delta_{\alpha}^{-2} N^{-2}\mathcal L_4 \right) \1_+^{\leq N} .
\end{align*}
Combining this bound with the easy bound
\begin{align*}
\1_+^{\leq N} \mathcal H \1_+^{\leq N} \geq \1_+^{\leq N}  \left( \frac{1}{2} \mathcal L_4 - CN\right) \1_+^{\leq N} 
\end{align*}
(dividing by $N$ the second inequality and adding the first) we conclude the proof.
\end{proof}

\section{Excitation spectrum} \label{sec:conclusion}

We denote again $M_0 = \delta_{\alpha} N$, $M = \delta_{\beta} N$  and make the following choice for the parameters which is different from that in Section \ref{sec:BEC}. For computational purposes we will eventually take $\kappa = 1/8$ and 
\begin{align}
\ell \simeq N^{-\kappa}, \quad \delta_\alpha \simeq N^{\kappa/4 - 1/2}, \quad \delta_\beta = N^{-7\kappa/2}, \quad \eps \simeq N^{-\kappa/2}.
	\label{eq:choice_parameters}
\end{align}

Recall that $D = Q(-\Delta + V_{\textrm{ext}} + 8\pi a \varphi^2 - \mu)Q$ and $K=Q \widetilde K Q$ 
\begin{align*}
\mu = \int |\nabla \varphi|^2 + V_{\textrm{ext}} |\varphi|^2 + 8\pi a \varphi^4, \quad \widetilde K(x,y) =  N^3 \varphi(x) \varepsilon_{\ell,N} (x-y) \varphi(y).
\end{align*}
We also denote $E=(D^{1/2} (D+2K) D^{1/2})^{1/2}$ and
\begin{align*}
\widetilde \lambda_1(H_N) = N \mathcal{\cE}_{\rm GP}(\varphi) - 4\pi a \int \varphi^4 - \frac{N^4}{2} \int ((\omega_{\ell,N} \varepsilon_{\ell,N})\ast \varphi^2)  \varphi^2 -\frac{1}{2} \tr (D + K - E).
\end{align*}
From Section \ref{sec:BEC} we know that $\widetilde \lambda_1(H_N)= \lambda_1(H_N) + \mathcal{O}(1)$. In this section we prove

\begin{lemma}[Convergence of the excitation spectrum] \label{lem:excitation-pre} The eigenvalues $\lambda_1( \widetilde H_N)\leq \lambda_2( \widetilde H_N) \leq ...\leq \Lambda \leq N^{1/16} $ of $\widetilde H_N=H_N - \widetilde{\lambda}_1(H_N)$ on $\gH^N$ satisfy  
$$
\lambda_{L}(\widetilde H_N) =  \lambda_L ( \dd\Gamma(E) ) + \mathcal O(\Lambda N^{-1/16}) 
$$
where $\dd\Gamma(E)$ is interpreted as an operator on $\cF_+$. 
\end{lemma}

\begin{proof} We set the notations 
\begin{equation}
	\label{eq:H_tilde}
\widetilde{\mathcal H} = \mathcal H - \widetilde \lambda_1(H_N).
\end{equation} Recall that  
$$
\widetilde {\mathcal H} = T_1 T_c T_2 \left(\dd\Gamma (E) + T^*_2 \mathcal L_4 T_2 + T^*_2\mathcal E^{(\theta_M)}T_2 +  \mathcal E^{(T_2)}\right) T_2^* T_c^* T_1^*
$$
and on $\mathcal F_+^{\leq N}$
$$
U \widetilde H_N U^* = \mathds{1}^{\leq N}  \widetilde {\mathcal H} \mathds{1}^{\leq N} \geq C^{-1} \mathcal N \1^{\leq N} - C.
$$
%The statement in Lemma \ref{lem:excitation-pre} is equivalent to the fact that the eigenvalues of $\widetilde H_N$ satisfy 
%$$
%\lambda_L(\widetilde H_N) = \lambda_L ( \dd\Gamma(E) ) + o(N^{-1/32})_{N\to \infty}, \quad \forall L\in \mathbb{N}. 
%$$
We will use the variational principle, where the  lower and upper bounds are derived separately.

\subsection{Proof of the lower bound: $\lambda_L (\widetilde H_N) \geq \lambda_L(\dd\Gamma(E))+ \mathcal O(\Lambda N^{-\kappa/2})_{N\to \infty}$} \quad \\

Using again the localization (\ref{eq:localization}) and multiplying on both sides by $\mathds{1}_+^{\leq N}$, we obtain 
\begin{align}
	\label{eq:localization_upper_bound}
U \widetilde H_N U^* = \mathds{1}_+^{\leq N}\widetilde {\mathcal H}\mathds{1}_+^{\leq N} &= f_{M_0} \widetilde {\mathcal H} f_{M_0} + \mathds{1}_+^{\leq N} g_{M_0} \widetilde{\mathcal H} g_{M_0} \mathds{1}_+^{\leq N} + \mathcal E_{M_0}.
\end{align}
Let $Z$ be the space generated by the first $L$ eigenfunctions of $\widetilde  H_N$. Let $Y = UZ \subset \mathcal F_+^{\leq N}$ and let $P_Y$ be the orthogonal projection onto $Y$. Since $U$ is a unitary operator from $\gH^N$ to $\cF_+^{\leq N}$, we have $\dim (Y)=L$. By the min-max principle, we have the operator inequality 
\begin{align} \label{eq:low-low-0}
\lambda_L(\widetilde H_N) \geq P_Y U  \widetilde H_N U^* P_Y  = P_Y \widetilde {\mathcal H}   P_Y =  P_Y \left( f_{M_0} \widetilde {\mathcal H} f_{M_0}  +   g_{M_0} \widetilde {\mathcal H} g_{M_0}+  \mathcal E_{M_0}\right) P_Y .
\end{align}
It remains to bound the restriction of $f_{M_0} \widetilde {\mathcal H} f_{M_0}$, $g_{M_0} \widetilde {\mathcal H} g_{M_0}$ and $\mathcal E_{M_0}$ on $Y$. 

\bigskip
\noindent{\bf Estimating $f_{M_0} \widetilde {\mathcal H} f_{M_0}$.}  Considering Lemma \ref{prop:second_quadratic_transform} and the estimate on the error term $\mathcal E^{(T_2)} $ in (\ref{eq:lemma_second_quad_transfo_error}), using that ${\rm Ran}( f_{M_0}) \subset \mathcal F_+^{\leq M_0}$, that on $\mathcal F_+$
\begin{align*}
\mathcal N \leq C \dd \Gamma(E)
\end{align*}
and the choice of parameters (\ref{eq:choice_parameters}), we obtain on $ T_2^* T_c^* T_1^* \mathcal F_+^{\leq M_0}$,
\begin{align*}
\mathcal E^{(T_2)}  \geq -C N^{-\kappa/2} T_2^* \mathcal L_4 T_2 - C N^{-\kappa/2} (\dd\Gamma(E) + 1).
\end{align*}

We now turn to $\mathcal E^{(\theta_M)}$, similarly as in (\ref{eq:estimate_E_theta}) but using that $(1-\theta_M) \leq CM^{-2} \mathcal N^3$, we obtain on $ T_2^* T_c^* T_1^* \mathcal F_+^{\leq M_0}$,
\begin{align*}
 \mathcal E^{(\theta_M)} 
	\geq - \frac{1}{4} \mathcal L_4  -  C  \frac{M_0^2}{M^2} \dd\Gamma(E) \geq - \frac{1}{4}  \mathcal L_4  -  C N^{-\kappa/2}\dd\Gamma(E),
\end{align*}
where we used that $M_0^2 / M^2 = N^{\frac{15\kappa-2}{2}} = N^{-\kappa/2}$ for $\kappa = 1/8$.
In summary,  we have the quadratic form estimate  on $\cF_+$
\begin{align} \label{eq:low-low-1}
f_{M_0} \widetilde {\mathcal H} f_{M_0} \geq (1- CN^{-\kappa /2}) f_{M_0}  T_1 T_c T_2 \dd\Gamma(E)T_2^* T_c^* T_1^* f_{M_0} - N^{-\kappa/2}. 
\end{align}

%
%
%Recall that $M_0 = \delta_{\alpha} N$, $M = \delta_{\beta} N$ 
%
%\begin{align*}
%\ell \simeq N^{-\kappa}, \quad \delta_\alpha \simeq N^{\kappa - 1/2}, \quad \delta_\beta = N^{-5\kappa}, \quad \eps \simeq N^{-\kappa/2}.
%	%\label{eq:choice_parameters}
%\end{align*}

\bigskip

\noindent{\bf Estimating $g_{M_0} \widetilde {\mathcal H} g_{M_0}$}.  For every $\xi \in Y$, $\|\xi\|=1$, using that $g_{M_0} \leq \1^{\{\mathcal N \geq M_0/2\}}$ and the optimal rate of BEC in Lemma \ref{lemma:optimal_rate}, we obtain  
\begin{align} \label{eq:gM0-nn-a}
\|g_{M_0} \xi\|^2 
	&\leq C \langle \xi, (\cN/M_0)\xi\rangle \leq C  \langle \xi, (U \widetilde H_N U^* + C) \xi \rangle M_{0}^{-1} \nn \\
	& \leq C ( \lambda_L(\widetilde H_N) +C)  M_0^{-1}\leq C \Lambda N^{-1/2 + \kappa/4} \leq C N^{-\kappa/2},
\end{align}
where we used that $\Lambda \leq CN^{\kappa/2}$ by assumption.
%Here we have used also the upper bound $\lambda_L(\widetilde H_N)\leq \lambda_L(\dd\Gamma(E))  + \mathcal O(\Lambda^3 N^{-\kappa/2})$ derived earlier and that $\Lambda \leq C N^{\kappa/2}$. 
Hence, again from Lemma \ref{lemma:optimal_rate}, we obtain
\begin{align} \label{eq:low-low-2}
P_Y g_{M_0} \widetilde {\mathcal H} g_{M_0} P_Y \geq - C  P_Y g_{M_0}^2 P_Y \geq    - C N^{-\kappa/2}.
\end{align}

\bigskip

\noindent{\bf Estimating $\mathcal{E}_{M_0}$.} From (\ref{eq:error_localization_0}), we obtain
\begin{align} \label{eq:low-low-3}
P_Y \mathcal E_{M_0}P_Y & \geq  -C M_0^{-2} (P_Y\mathcal L_4P_Y + P_Y\dd\Gamma(E)P_Y +  N ) \nn \\
	&\geq -C M_0^{-2} (\lambda_1(H_N) + \lambda_L(\widetilde H_N) + N)  \geq-C N^{-1-\kappa/2} (N + \Lambda) \geq  -C N^{-\kappa/2} 
\end{align}
where we used that the expectation of $\dd\Gamma(E)$ and $\mathcal L_4$ against any normalized vector $\xi \in Y$ is bounded by $C (\lambda_1(H_N) + \lambda_L(\widetilde H_N)+ N)$, and also that  $\Lambda \leq CN^{\kappa/2}$.

\bigskip

\noindent{\bf Conclusion of the lower bound.}  Inserting \eqref{eq:low-low-1}, \eqref{eq:low-low-2}, and \eqref{eq:low-low-3} in \eqref{eq:low-low-0}, we find that 
 \begin{align*}  
\lambda_L(\widetilde H_N) &\geq  P_Y \left( f_{M_0} \widetilde {\mathcal H} f_{M_0}  +   g_{M_0} \widetilde {\mathcal H} g_{M_0}+  \mathcal E_{M_0}\right) P_Y \nn \\
&\geq (1- CN^{-\kappa/2}) P_Y f_{M_0} T_1 T_c T_2 \dd\Gamma(E)T_2^* T_c^* T_1^* f_{M_0}P_Y  - N^{-\kappa/2}.  
\end{align*}
Testing against $\xi \in Y$ and taking the supremum, we obtain
 \begin{align*}
\lambda_L(\widetilde H_N) \geq (1- CN^{-\kappa/2}) \sup_{\xi \in Y}  \frac{ \langle f_{M_0} \xi,   T_1 T_c T_2\dd\Gamma(E)T_2^* T_c^* T_1^* f_{M_0} \xi\rangle} {\|\xi\|^2} - N^{-\kappa/2}.  
\end{align*}
For all $\xi \in Y$, $\xi \neq 0$, by using a bound similar to \eqref{eq:gM0-nn-a}, we have
\begin{align*}
\frac{\| f_{M_0} \xi\|^2}{\|\xi\|^2} = 1 - \frac{\|g_{M_0} \xi \|^2}{\|\xi\|^2}  \geq 1 - \frac{\langle \xi, (2\cN/M_0)  \xi\rangle}{\|\xi\|^2}   \geq 1 - C \Lambda M_0^{-1} \geq 1 - C N^{-1/2}.
\end{align*}
Therefore, using that $T_2^* T_c^* T_1^*$ is unitary, we have
 \begin{align} \label{eq:low-low-4}
\lambda_L(\widetilde H_N) &\geq   (1- CN^{-\kappa/2}) \sup_{\xi \in Y}  \frac{ \langle f_{M_0} \xi,  T_1 T_c T_2\dd\Gamma(E)T_2^* T_c^* T_1^* f_{M_0} \xi\rangle} {\|f_{M_0}\xi\|^2} - N^{-\kappa/2}
  \nonumber\\
&=(1- CN^{-\kappa/2}) \sup_{\xi \in T_2^* T_c^* T_1^*f_{M_0}Y}  \frac{ \langle \xi,  \dd\Gamma(E) \xi\rangle} {\|\xi\|^2} - N^{-\kappa/2}.  
\end{align}
Now let us show that for $N$ large enough $\dim f_{M_0}Y = L$. It is enough to prove that $f_{M_0}$ is injective on $Y$. Let $\xi \in Y$ such that $\|\xi\| = 1$ and $ f_{M_0} \xi =0$, then using that $\xi = (f_{M_0}^2 + g_{M_0}^2) \xi =  g_{M_0}^2 \xi$ and that $g_{M_0}^2 \leq \1^{\leq M_0/2} \leq 2 M_0^{-1} \mathcal N$, we obtain
\begin{align*}
M_0 \leq 2 \langle \cN \rangle_{\xi} \leq C^{-1} \langle \widetilde H_N \rangle_{U^* \xi} + C \leq C^{-1} \Lambda + C,
\end{align*}
with a constant $C >0$ independent of $N$, which cannot be true for $N$ large enough since $M_0 \sim N^{1/2 + \kappa/4}$ and $\Lambda \leq N^{\kappa/2}$. Consequently by the min-max principle, we find that
$$
\sup_{\xi \in T_2^* T_c^* T_1^* f_{M_0}Y}  \frac{ \langle \xi,   \dd\Gamma(E) \xi\rangle} {\|\xi\|^2} \geq \lambda_L(\dd\Gamma(E)). 
$$
Inserting the latter bound in \eqref{eq:low-low-4} we get the desired lower bound 
\begin{align*}
\lambda_L(\widetilde H_N) \geq (1- CN^{-\kappa/2})\lambda_L(\dd\Gamma(E)) -C  N^{-\kappa/2}.
\end{align*}
Combining this with the assumption $\lambda_L(\widetilde H_N) \leq \Lambda$ we get $\lambda_L(\dd\Gamma(E)) \leq  \Lambda + \mathcal O(\Lambda N^{-\kappa/2})$ and hence
\begin{align*}
\lambda_L(\widetilde H_N)  \geq \lambda_L(\dd\Gamma(E)) + \mathcal O(\Lambda N^{-\kappa/2})_{N\to\infty}.
\end{align*}

\subsection{Proof of the upper bound: $\lambda_L (\widetilde H_N) \leq \lambda_L(\dd\Gamma(E)) + O(\Lambda N^{-\kappa/2})_{N\to \infty}$} \quad \\

We know that for every $L \geq 1$, the eigenvalue $\lambda_L(\dd\Gamma(E))$ is associated with an eigenvector of the form
%
%Let us denote by $(\xi_j)_{j\geq 1}$ the eigenvectors of $\dd\Gamma (E)$ associated to the eigenvalues $\lambda_1(\dd\Gamma(E)) \leq \lambda_2(\dd\Gamma(E)) \leq \dots$. Let $L \geq 1$, we know that $\xi_L$ belongs to some sector $\{\mathcal N=k\}$ with $k\leq L$ and that 
\begin{align}
	\label{eq:eigenvectors_E}
\xi_L = \prod_{k=1}^p \frac{(a^*(\varphi_{j_k}))^{n_{j_k}}}{\sqrt{n_{j_k} !}} \Omega
\end{align} 
for some $p \geq 1$, where $\varphi_{j_k}$ are normalized eigenfunctions of $E$ associated to $\lambda_{j_k}(E)$ (listed in increasing order and repeated in case of multiplicity) with $1 \leq j_1 < \dots  < j_p \leq L$. 
%Moreover, we have 
%$$\| \varphi_j \|_{H^2} \leq C \|(-\Delta + 1) \varphi_j\|_{L^2} \leq C \|E \varphi_j\|_{L^2} \leq C \lambda_j(E)$$
%for some constants $C$ independent of $j$. 
We also have, with the notations above, that $\mathcal N \xi_L = N_L \xi_L$ with
\begin{align*}
 N_L= \sum_{k=1}^p n_{j_k} \leq \frac{1}{\lambda_1(E)} \sum_{k=1}^p n_{j_k} \lambda_{j_k}(E) = \frac{\lambda_L(\dd\Gamma(E))}{\lambda_1(E)} \leq C \Lambda,
\end{align*}
where we used that $\lambda_L(\dd\Gamma(E)) \leq C \Lambda$ as proven earlier.
% (it suffices to consider functions in $C_c^\infty$ and supported in $[0,1]^3$) In fact, from  the assumption (\ref{eq:Vext-ass}) we have that $D(-\Delta_{\rm{Dir}, [0,1]^{3}}) \subset D(E)$ and therefore $E \leq C (-\Delta_{\rm{Dir}, [0,1]^{3}})$ for some constant $C$, so that by monotonicity of the eigenvalues $\lambda_L(E) \leq C \lambda_L(-\Delta_{\rm{Dir}, [0,1]^{3}}) \leq C L^{1/3}$.

Let $Y = \mathrm{span}\{\xi_j\}_{1 \leq j \leq L} \subset \cF_+^{\leq N}$ and $P_Y$ the orthogonal projection onto $Y$. 
Recall that $U\widetilde{H}_NU^*$ is defined on $\mathcal F_+^{\leq N}$, we extend it by $0$ on the whole $\mathcal F_+$ and we consider the extended Hamiltonian 
\begin{align*}
\mathcal H_{\rm ext} := U\widetilde{H}_NU^* + 2 \mathds{1}^{>N} \Lambda = \1^{\leq N}_+ \widetilde{\mathcal H} \1^{\leq N}_+ + 2\mathds{1}^{>N} \Lambda,
\end{align*}
where we recall that $\widetilde{\mathcal H}$ was defined in (\ref{eq:H_tilde}).
Then $\lambda_L(\widetilde{H}_N) = \lambda_L( \mathcal H_{\rm ext})$ since by assumption $\lambda_L(\widetilde{H}_N) \leq \Lambda $. Expanding $\1^{\leq N}_+ = 1 - \1^{> N}_+$, we obtain
\begin{align*}
\1^{\leq N}_+ \widetilde{\mathcal H} \1^{\leq N}_+ 
	&= \widetilde{\mathcal H} - \1^{> N}_+ \widetilde{\mathcal H} \1^{\leq N}_+ - \1^{\leq N}_+ \widetilde{\mathcal H} \1^{> N}_+ -\1^{> N}_+ \widetilde{\mathcal H} \1^{> N}_+ \\
	&\leq \widetilde{\mathcal H} -  \1^{> N}_+ \left(\cL_1 + \cL_2 + \cL_3 + \cE^{(U,1a)} + \cE^{(U,1b)} + \cE^{(U,2)}  \right) \1^{\leq N}_+ + \hc 
\end{align*}
where the $\cL_i$'s were defined in Lemma \ref{lem:UHU*} and $\cE^{(U,1a)},\cE^{(U,1b)}$ and $\cE^{(U,2)}$ in (\ref{eq:EU-full}). The terms appearing in the second summand above are the contributions of $\widetilde{\mathcal H}$ that do not commute with $\mathcal N$, contrarily to $\cL_0, \cL_4$ and $ \cE^{(U,0)}$. Using that $(\mathcal N-1) a^*_x = a^*_x \mathcal N$, one can see that these cross terms are localized on the sector of particles $\{\mathcal N \geq N-2\}$ and are therefore very small on $T_1  T_c T_2 Y$. To estimate them, we use Cauchy-Schwarz inequality as in the proof of Lemma \ref{eq:UHU}, we obtain for all $\eta>0$, on $\mathcal F_+$
\begin{align*}
 \1^{> N}_+ \left(\cL_1 +  \cE^{(U,1a)} + \cE^{(U,1b)}  \right) \1^{\leq N}_+ + \hc  
	&\leq \eta C N (\mathcal N+1) + \eta^{-1} \1^{> N}_+ \\
	&\leq  \eta C N (\mathcal N+1) + \eta^{-1} N^{-4} \mathcal N^{4}\1^{> N}_+ \\
	&\leq  C N^{-3/2} (\mathcal N+1)^4,
\end{align*}
where we used that $ \1^{> N}_+ \leq C N^{-2} \mathcal N^2$ and took $\eta = N^{-5/2}$.
Similarly, for all $\eta>0$, we have on $\mathcal F_+$
\begin{align*}
 \1^{> N}_+ \left( \cL_2 + \cE^{(U,2)} \right) \1^{\leq N}_+ + \hc  
	&= \pm \1^{\geq N+2}_+ \left( \cL_2 + \cE^{(U,2)} \right) \1^{> N-2}_+ \1^{\leq N}_+ + \hc  \\
	&\leq \eta \cL_4 + \eta^{-1} C N \1^{> N-2}_+ \\
	&\leq N^{-2} \cL_4 + C N^{-1}\mathcal N^4,
\end{align*}
where we used that $ N \1^{> N-2}_+ \leq C N^{-3} \mathcal N^4$ and took $\eta = N^{-2}$. Finally, for all $\eta>0$, we have on $\mathcal F_+$
\begin{align*}
 \1^{> N}_+ \cL_3 \1^{\leq N}_+ + \hc  
 	&=  \1^{> N}_+ \cL_3  \1^{> N-1}_+  \1^{\leq N}_+ + \hc  \\
 	&\leq \eta \cL_4 + C \eta^{-1} \mathcal N \1^{> N-1}_+ \\
	&\leq N^{-2} \cL_4 + C N^{-1} \mathcal N^4,
\end{align*}
where we used again that $ \1^{> N-1}_+ \leq C N^{-4} \mathcal N^4$ and took $\eta = N^{-1}$. Using the rough estimate
\begin{align*}
 T_2^*T_c^*T_1^* (\cL_4 + (\mathcal N+1)^4) T_1  T_c T_2
	&\leq C \left(\dd\Gamma(E) + T_2^*\cL_4T_2 + (\mathcal N+1)^4 + N \right)
\end{align*}
which follows from the Lemmas \ref{prop:properties_quadratic_transform_1}, \ref{lem:s1}, \ref{lemma:number_part_cubic}, \ref{prop:estimate_kinetic_L4_cubictransform} and \ref{lemma:bound_second quadratic transform}, we obtain that on $T_1  T_c T_2 Y$, we have
\begin{align*}
\1^{\leq N}_+ \widetilde{\mathcal H} \1^{\leq N}_+ 
	&\leq \widetilde{\mathcal H} + C T_1 T_c \cL_4 T_c^* T_1^* + C N^{-2}(\Lambda^4 + N)  \\
	&\leq \widetilde{\mathcal H} + C T_1 T_c \cL_4 T_c^* T_1^* + C \Lambda N^{-\kappa/2},
\end{align*}
where we used that $\Lambda^3 N^{-2} \leq C N^{-3\kappa/2 -2} \leq C N^{-\kappa/2}$ for our choice $\kappa = 1/8$.

Therefore, by the min-max principle we obtain that
\begin{align}
\lambda_L(\widetilde{H}_N) = \lambda_L( \mathcal H_{\rm ext})
	&\leq \sup_{ \xi  \in T_1  T_c T_2 Y} \frac{\langle \xi,  ( \1^{\leq N} \widetilde {\mathcal H} \1^{\leq N} + 2 \Lambda \1^{>N})  \xi \rangle }{\|\xi\|^2} \nn \\
	&\leq  \sup_{ \xi  \in T_1  T_c T_2 Y} \frac{\langle \xi,  (\widetilde {\mathcal H} + 2 \1^{>N}\Lambda  + C T_1 T_c \cL_4 T_c^* T_1^*)   \xi \rangle }{\|\xi\|^2} + C \Lambda N^{-\kappa/2}. \label{eq:var-prin-upp-0}
\end{align}
On $T_1  T_c T_2 Y$, we have $$2 \1^{>N}\Lambda \leq C \Lambda N^{-1}T_1  T_c T_2 (\cN_++1) T_1^*  T_c^* T_2^* \leq C \Lambda^2 N^{-1} \leq C \Lambda N^{-\kappa/2}.$$
It remains to bound $\widetilde {\mathcal H}$ and $T_1 T_c \cL_4 T_c^* T_1^*$ on $ T_1  T_c T_2 Y$.

\bigskip
\noindent {\bf Estimating $\cL_4$.} Let us now estimate $T_1 T_c \cL_4 T_c^* T_1^*$ on $ T_1  T_c T_2 Y$, in view of Lemma \ref{lemma:estimate_L4_second_quadratic_transfo} we have
\begin{align*}
P_Y T^*_2 \mathcal L_4 T_2  P_Y 
	&\leq P_Y \left(C N^{-1} \ell^{-7}  (\mathcal N+1)^2 + C \mathcal L_4\right)P_Y \\
	&\leq C N^{-1+7\kappa}  \Lambda^2 + C P_Y\mathcal L_4 P_Y.
\end{align*}
To estimate $P_Y \mathcal L_4 P_Y$ we \cite[Lemma 5.3]{ErdYau-01}, that is
\begin{equation}
	\label{eq:second_moment}
N^2V(N(x-y)) \leq C N^{-1} \|V\|_{L^1(\mathbb{R}^{3})} (1-\Delta_x) (1-\Delta_y) \leq CN^{-1} E_x E_y,
\end{equation}
we find
\begin{align*}
 P_Y \mathcal L_4 P_Y \leq C P_Y \dd\Gamma(E)^2 \leq C \Lambda^2 N^{-1}.
\end{align*}
Hence, on $ T_1  T_c T_2 Y$, we have
\begin{align}\label{eq:est_L4PY}
T_1 T_c \cL_4 T_c^* T_1^* \leq  C\Lambda^2 N^{7\kappa -1} \leq C \Lambda N^{-\kappa/2},
\end{align}
where we used that $\Lambda N^{7\kappa -1} \leq C N^{-\kappa/2}$ for our choice $\kappa=1/8$ and with $\Lambda \leq C N^{\kappa/2}$.

\bigskip
\noindent{\bf Estimating $\widetilde {\mathcal H}$.} Recall from Lemma \ref{prop:second_quadratic_transform} that
$$
\widetilde {\mathcal H} = T_1 T_c T_2 \left(\dd\Gamma (E) + T^*_2 \mathcal L_4 T_2 + T^*_2\mathcal E^{(\theta_M)}T_2 +  \mathcal E^{(T_2)}\right) T_2^* T_c^* T_1^*.
$$
On $ T_1  T_c T_2 Y$, the main contribution is 
$$
T_1 T_c T_2 \dd\Gamma (E) T_2^* T_c^* T_1^* \leq \lambda_L(E)
$$
which  follows from the definition of $Y$. The term $ T^*_2 \mathcal L_4 T_2$ was already estimated above in (\ref{eq:est_L4PY}), which we use to estimate $T^*_2\mathcal E^{(\theta_M)}T_2$. Using (\ref{eq:estimate_E_theta}) and that $\rm{Ran} \, P_Y \subset \mathcal F^{\leq C\Lambda}$, we find that
\begin{align*}
P_Y T^*_2 \mathcal E^{(\theta_M)} T_2 P_Y 
	&\leq P_Y T^*_2 \mathcal L_4 T_2 P_Y + C P_Y T_2^* \frac{\mathcal N^{2}}{M} T_2P_Y \\
	&\leq C\Lambda^2 N^{7\kappa -1} + C \frac{\Lambda^2}{M} \\
	&\leq C\Lambda^2 N^{7\kappa -1}  + C \Lambda^2 N^{7\kappa/2-1} \leq C \Lambda N^{-\kappa/2}.
\end{align*}
where we used that $T_2 ^* \mathcal N^{2} T_2 \leq C (\mathcal N+1)^2$, the bound (\ref{eq:est_L4PY}) and that $M\simeq N^{1-7\kappa/2}$ from (\ref{eq:choice_parameters}). Here, the last inequality holds as soon as $\kappa \leq 1/8$. Finally, in view of (\ref{eq:lemma_second_quad_transfo_error}) the $\mathcal E^{(T_2)}$ term is bounded by
\begin{align*}
P_Y \mathcal E^{(T_2)} P_Y
	&\leq C (\ell^{1/2}  + \delta_{\beta}^{1/2}\ell^{-1}) \Lambda \nn \\
	&\quad + C ( \delta_{\beta}^{1/2} +\ell^{3/2}) (\Lambda + \ell^{-1}+ \|P_Y T^*_2 \mathcal L_4 T_2 P_Y\|_{op})  \nn \\
	&\quad + C \frac{\Lambda^{3/2}}{N^{1/2}} + C \frac{\Lambda^{5/2}}{N^{3/2}}\nn \\
	&\quad + \eps \left( \|P_Y T^*_2 \mathcal L_4 T_2 P_Y\|_{op} +  \delta_{\beta} (\Lambda+ \ell^{-1}) + \Lambda +  \ell^{1/2} \frac{\Lambda}{N^2} + \frac{\Lambda^2}{N^3} \right) \\
	&\quad + \eps^{-1}C \left( \ell \Lambda + \frac{\Lambda^2}{N}+ \frac{\Lambda^5}{N^4} \right) \\
%	&\leq C \ell^{1/2} \Lambda + C \Lambda^3 N^{-1/2} + C \Lambda^5 N^{-3/2} \\
%	&\quad + C \ell^{1/2}(\Lambda^3 \ell^{-5/2}  N^{-1/2} + \ell^{5} (\Lambda + \ell^{-1}) + \Lambda + \ell^{1/2}\Lambda N^{-2} + \Lambda^2 N^{-3}) \\
%	&\quad + \ell^{-1/2}(\ell \Lambda +\Lambda N^{-1} + \Lambda^5 N^{-4}  )\\
	&\leq C\Lambda N^{-\kappa/2},
\end{align*}
where we used that $P_Y \dd\Gamma(D) P_Y \leq C\Lambda$, the estimate (\ref{eq:est_L4PY}), that $\ell^{1/2}, \delta_{\beta}^{1/2}\ell^{-1}, \varepsilon, \varepsilon^{-1}\ell \lesssim N^{-\kappa/2}$ with our choice of parameters (\ref{eq:choice_parameters}) and that $\Lambda \leq N^{\kappa/2}$. Finally, for $\kappa = 1/8$ and $\Lambda\leq N^{\kappa/2}$, we obtain that on $T_1  T_c T_2 Y$
 \begin{align} \label{eq:upp-upp-1}
\widetilde {\mathcal H} 
	\leq T_1 T_c T_2\dd \Gamma (E)T_2^* T_c^* T_1^* + CN^{-\kappa/2}\Lambda \leq \lambda_{L}(\dd\Gamma(E)) + C N^{-\kappa/2}\Lambda. 
 \end{align}

%Here we have used that 
%$$
%\langle \xi, 
%$$
%taking $M \simeq M_0$ is enough. The factor $ g_{M_0}^2$ above is important, since we will use that $\|g_{M_0} \xi\|^2 \leq CLM_0^{-1} \|\xi\|$ for all $\xi \in T_1  T_c T_2 Y$ 
  \bigskip
 \noindent {\bf Conclusion of the upper bound.} Putting \eqref{eq:est_L4PY} and \eqref{eq:upp-upp-1} together, we get the operator inequality on $ T_1  T_c T_2 Y$: 
\begin{align*}
 f_{M_0} U \widetilde{H}_N U^* f_{M_0} &= \widetilde {\mathcal H}- g_{M_0} \widetilde {\mathcal H} g_{M_0} - \mathcal E_{M_0} \leq \lambda_L(\dd\Gamma(E)) + C\Lambda N^{-\kappa/2}.
 \end{align*}
 Hence, from \eqref{eq:var-prin-upp-0} we find that 
 \begin{align*}
\lambda_L( \widetilde{H}_N )  \leq (1- C\Lambda N^{-1/2-\kappa})^{-1} \left[  \lambda_{L}(\dd\Gamma(E)) + C \Lambda N^{-\kappa/2} \right]. 
\end{align*}
Using that $\lambda_{L}(\dd\Gamma(E)) \leq \Lambda $ we can conclude that 
$$
\lambda_L( \widetilde{H}_N ) \leq \lambda_{L}(\dd\Gamma(E))  + \mathcal O(\Lambda N^{-\kappa/2})_{N\to \infty}. 
$$

%\begin{align*}
%\lambda_{L}(\dd\Gamma(E)) + C N^{-\delta_\star} \geq (1- CLM_0^{-1}) \lambda_L( \widetilde{H}_N ) -  CLM_0^{-1}(1 - L^2 M^{-1}).
%\end{align*}

The proof of Lemma \ref{lem:excitation-pre} is complete.
\end{proof}

\subsection{One-body eigenvalues} From Lemma \ref{lem:excitation-pre}, we know that the spectrum of $H_N -\lambda_1(H_N)$ below any energy level of order $\Lambda$ is equal to finite sums of the form  
\begin{align}
	\label{eq:spectrum_E}
\sum_{i \geq 1} n_i \widetilde{e_i} + \mathcal O(\Lambda N^{-1/16})_{N\to \infty},\quad n_i\in \{0,1,2,...\}.
\end{align}
where $\widetilde e_i$'s are eigenvalues of $E=(D^{1/2}(D+2K)D^{1/2})^{1/2}$ on $\gH_+$. Note that the operator $E$ still depends on $N$ via $K$. To finish the proof of Theorem \ref{thm:main}, we still need to compare the eigenvalues of $E$ and $E_\infty=(D^{1/2}(D+Q(16 \pi \ao \varphi^2)Q )D^{1/2})^{1/2}$. We have 

\begin{lemma}\label{lem:1-body-ev} For all $N,L\geq 1$, the eigenvalues of operators $E,E_\infty$ on $\gH_+$ satisfy
$$
 |\lambda_L(E) - \lambda_L( E_{\infty}) | \leq C \lambda_L(E) \ell^2
$$ 
where $\lambda_L$ denotes the $L$-th lowest eigenvalue. The constant $C$ is independent of $N,L$. 
\end{lemma}

\begin{proof}[Proof of Lemma \ref{lem:1-body-ev}] The operators $E$ and $E_\infty$ are positive and have compact resolvent, it suffices for our claim to show that $\|E^{-2} - E_\infty^{-2}\|_{op}^2 \leq C \ell^2$. By the resolvent formula, we can write 
\begin{align*}
E^{-2}  - E_\infty^{-2}&= D^{-1/2} \Big[ (D+2K)^{-1} - (D+Q(16 \pi \ao \varphi^2)Q )^{-1}\Big] D^{-1/2} \\
&= - D^{-1/2} (D+2K)^{-1} Q(2\widetilde K- 16 \pi \ao \varphi^2) Q (D+Q(16 \pi \ao \varphi^2)Q )^{-1} D^{-1/2}. 
\end{align*}
Since $D^{-1}$ has compact resolvent, $ (D+Q(16 \pi \ao \varphi^2)Q )^{-1}\gH_+ \subset H^1(\R^3)$ and $ (D+K )^{-1}\gH_+ \subset H^1(\R^3)$,  it suffices to show that 
$$
|\langle g, (2K- 16 \pi \ao \varphi^2)f \rangle_{L^2(\R^3)}| \leq C \ell^2 \|f\|_{H^1}\|g\|_{H^1}, \quad \forall f,g\in \gH_+\cap H^1(\R^3). 
$$
By the same argument leading to \eqref{eq:Phi-delta-Tc}, we have, for every  $f,g\in   \gH_+\cap H^1(\R^3)$, 
\begin{align*}
|\langle g, (2K- 16 \pi \ao \varphi^2)f \rangle_{L^2(\R^3)}| 
	&= 2 \left|\int \overline{g}\varphi \Big( N^3\eps_{\ell,N} \ast (\varphi f) - 8 \pi \ao  \varphi f \Big)  \right| \\
	&\leq 2 \left| \int \overline{ \widehat{g \varphi}} \Big(N^3 \widehat{\eps}_{\ell,N}(p) -  N^3 \widehat{\eps}_{\ell,N}(0)\Big) \widehat{f\varphi }\right| \\
	&\leq C   \|(1+|x|^2) N^3 \eps_{\ell,N}\|_{L^1}  \int |\widehat{\overline g \varphi}| p^2 |\widehat{\varphi f}| \\
	&\leq C \ell^2 \|\varphi\|_{H^2}^2 \|f\|_{H^1} \|g\|_{H^1}.
\end{align*}
The claim of the Lemma follows from  
\begin{align*}
| \lambda_L(E) - \lambda_L( E_{\infty})| 
	&= | \lambda_L(E_\infty)^{-2} - \lambda_L(E)^{-2}| \big|\frac{ \lambda_L(E) \lambda_L( E_{\infty})}{\lambda_L(E) +\lambda_L( E_{\infty})}\big| \\
	&\leq  C \lambda_L(E) \| E_\infty^{-2}- E^{-2} \|_{op}.
\end{align*}
\end{proof}

\subsection{Conclusion of Theorem \ref{thm:main}} 
Finally, using Lemma \ref{lem:1-body-ev}  in (\ref{eq:spectrum_E}) and using that $\ell \simeq N^{-1/8}$, we have for $\lambda(\dd\Gamma(E)) \leq \Lambda$, 
\begin{align*}
\left| \lambda_L(\dd\Gamma(E)) - \lambda_L(\dd\Gamma(E_\infty))\right| \leq \sum_{i \geq 1} n_i |\widetilde{e_i} - e_j| \leq C \ell^2 \sum_{i \geq 1} n_i \widetilde{e_i} \leq C \Lambda N^{-1/16}.
\end{align*}
This completes the proof of Theorem \ref{thm:main}.


\begin{thebibliography}{10}

\bibitem{AdhBreSch-21}
{\sc A.~Adhikari, C.~Brennecke, and B.~Schlein}, {\em {B}ose-{E}instein
  condensation beyond the {G}ross-{P}itaevskii regime}, Ann. Henri Poincar\'e,
  22 (2021), pp.~1163--1233.

\bibitem{Wieman-Cornell-95}
{\sc M.~H. Anderson, J.~R. Ensher, M.~R. Matthews, C.~E. Wieman, and E.~A.
  Cornell}, {\em Observation of {B}ose-{E}instein condensation in a dilute
  atomic vapor}, Science, 269 (5221) (1995), pp.~198--201.

\bibitem{BasCenSch-21}
{\sc G.~Basti, S.~Cenatiempo, and B.~Schlein}, {\em A new second order upper
  bound for the ground state energy of dilute Bose gases}, Forum Math. Sigma, Sigma, 9 (2021), E74 Cambridge University Press. 

\bibitem{BenOliSch-15b}
{\sc N.~{Benedikter}, G.~{de Oliveira}, and B.~{Schlein}}, {\em Quantitative
  derivation of the {G}ross-{P}itaevskii equation}, Comm. Pure App. Math., 68
  (2015), pp.~1399--1482.

\bibitem{BenPorSch-15}
{\sc N.~{Benedikter}, M.~{Porta}, and B.~{Schlein}}, {\em {Effective Evolution
  Equations from Quantum Dynamics}}, SpringerBriefs in Mathematical Physics,
  2016.

\bibitem{BocBreCenSch-17b}
{\sc C.~Boccato, C.~Brennecke, S.~Cenatiempo, and B.~Schlein}, {\em Complete
  {B}ose-{E}instein condensation in the {G}ross-{P}itaevskii regime}, Commun.
  Math. Phys., 359 (2018), pp.~975--1026.

\bibitem{BocBreCenSch-19}
\leavevmode\vrule height 2pt depth -1.6pt width 23pt, {\em {B}ogoliubov theory
  in the {G}ross-{P}itaevskii limit}, Acta Math., 222 (2019), pp.~219--335.

\bibitem{BocBreCenSch-17}
\leavevmode\vrule height 2pt depth -1.6pt width 23pt, {\em The excitation
  spectrum of Bose gases interacting through singular potentials}, J. Eur.
  Math. Soc., 22 (2020), pp.~2331--2403.

\bibitem{BocBreCenSch-20}
\leavevmode\vrule height 2pt depth -1.6pt width 23pt, {\em Optimal rate for
  {B}ose-{E}instein condensation in the {G}ross-{P}itaevskii regime}, Commun.
  Math. Phys., 376 (2020), pp.~1311--1395.

\bibitem{Bogoliubov-47}
{\sc N.~N. Bogoliubov}, {\em On the theory of superfluidity}, J. Phys. (USSR), 11 (1947), pp.~23-32.

\bibitem{Bose-24}
{\sc S.~Bose}, {\em {P}lancks {G}esetz und {L}ichtquantenhypothese}, Z. Phys.,
  26 (1924), pp.~178--181.

\bibitem{BosPetSei-20}
{\sc L.~Bossmann, S.~Petrat, and R.~Seiringer}, {\em Asymptotic expansion of
  low-energy excitations for weakly interacting bosons}, Forum Math. Sigma, 9 (2021), E28, pp. 1-61.
  
\bibitem{BreCapSch-21}
{\sc C.~Brennecke, M.~Caporaletti, and B.~Schlein}, {\em Excitation spectrum
  for {B}ose gases beyond the {G}ross-{P}itaevskii regime}, Rev. Math. Phys., 34 (2022) p.~2250027.
  
\bibitem{BreSch-19}
{\sc C.~Brennecke and B.~Schlein}, {\em Gross-Pitaevskii dynamics for
  Bose-Einstein condensates}, Analysis \& PDE, 12 (2019), pp.~1513--1596.

\bibitem{BreSchSch-21}
{\sc C.~Brennecke, B.~Schlein, and S.~Schraven}, {\em {B}ose-{E}instein
  condensation with optimal rate for trapped bosons in the {G}ross-{P}itaevskii
  regime}, Math. Phys. Anal. Geom., 25 (2022)
  
  \bibitem{BreSchSch-21b}
\leavevmode\vrule height 2pt depth -1.6pt width 23pt, {\em Bogoliubov Theory for Trapped Bosons
in the Gross-Pitaevskii Regime}, Ann. Henri Poincar\'{e}, 23 (2022), pp.~1583--1658

  \bibitem{CarCenSch-21}
{\sc 
C.~ Caraci, S.~Cenatiempo, and B.~Schlein}, {\em {B}ose--{E}instein condensation for two dimensional bosons in the {G}ross--{P}itaevskii regime}, Journal of Statistical Physics, 183 (3), (2021) pp.~1--72.

  \bibitem{CarCenSch-22}
  {\sc 
C.~ Caraci, S.~Cenatiempo, and B.~Schlein} {\em The excitation spectrum of two dimensional Bose gases in the Gross-Pitaevskii regime}, Ann. Henri Poincaré (2023) \url{https://doi.org/10.1007/s00023-023-01278-1}

\bibitem{Ketterle-95}
{\sc K.~B. Davis, M.~O. Mewes, M.~R. Andrews, N.~J. van Druten, D.~S. Durfee,
  D.~M. Kurn, and W.~Ketterle}, {\em {B}ose-{E}instein {C}ondensation in a
  {G}as of {S}odium {A}toms}, Phys. Rev. Lett., 75 (1995), pp.~3969--3973.

\bibitem{DerNap-14}
{\sc J.~{Derezi{\'n}ski} and M.~{Napi{\'o}rkowski}}, {\em {Excitation spectrum
  of interacting bosons in the mean-field infinite-volume limit}}, Ann. Henri
  Poincar\'e, 15 (2014), pp.~2409--2439.
  
  \bibitem{Dyson-57}
{\sc F.~J. Dyson}, {\em Ground-state energy of a hard-sphere gas}, Phys. Rev.,
  106 (1957), pp.~20--26.

\bibitem{Einstein-24}
{\sc A.~Einstein}, {\em Quantentheorie des einatomigen idealen {G}ases},
  Sitzber. Kgl. Preuss. Akad. Wiss., 1924, pp.~261--267.
  
  \bibitem{ErdSchYau-08} {\sc L.  Erdos, B. Schlein, and H.T. Yau}, {\em The ground state energy of a low density Bose gas: a second order upper bound}, Phys. Rev. A 78 (2008), 053627.  
  
  \bibitem{ErdYau-01} {\sc L. Erdos, and H.T. Yau}, {\em Derivation of the nonlinear Schr\" odinger equation from a many body Coulomb system}, Adv. Theor. Math. Phys. 5.6 (2001), pp.~1169--1205.

\bibitem{Fournais-20}
{\sc S.~Fournais}, {\em Length scales for {BEC} in the dilute {B}ose gas}, pp. 115–133, EMS Ser. Congr. Rep., EMS Press, Berlin, 2021.

\bibitem{FouSol-20}
{\sc S.~Fournais and J.~P. Solovej}, {\em The energy of dilute {B}ose gases},
  Ann. of Math., 192 (2020), pp.~893--976.
  
\bibitem{FouSol-21} S. Fournais and J. P. Solovej. The energy of dilute Bose gases II: The general case. {\em Invent. Math.} (2022). \url{https://doi.org/10.1007/s00222-022-01175-0}

\bibitem{GreSei-13}
{\sc P.~Grech and R.~Seiringer}, {\em The excitation spectrum for weakly
  interacting bosons in a trap}, Commun. Math. Phys., 322 (2013), pp.~559--591.

\bibitem{Hainzl-20}
{\sc C.~Hainzl}, {\em Another proof of {BEC} in the {GP}-limit}, J. Math. Phys, 62 (2021), pp.~459–485 
  
\bibitem{HaiSchTri-22}  
{\sc C.~Hainzl, B.~Schlein, and A.~Triay}, {\em Bogoliubov theory in the Gross-Pitaevskii limit: A simplified approach} Forum Math. Sigma, 10 (2022) E90.

\bibitem{Landau-41}
{\sc L.~Landau}, {\em {Theory of the Superfluidity of Helium II}}, Phys. Rev.,
  60 (1941), pp.~356--358.

\bibitem{LewNamSerSol-15}
{\sc M.~Lewin, P.~T. Nam, S.~Serfaty, and J.~P. Solovej}, {\em Bogoliubov
  spectrum of interacting {B}ose gases}, Comm. Pure Appl. Math., 68 (2015),
  pp.~413--471.

\bibitem{LieSei-02}
{\sc E.~H. Lieb and R.~Seiringer}, {\em {Proof of {B}ose-{E}instein
  Condensation for Dilute Trapped Gases}}, Phys. Rev. Lett., 88 (2002),
  p.~170409.

\bibitem{LieSei-06}
\leavevmode\vrule height 2pt depth -1.6pt width 23pt, {\em Derivation of the
  {G}ross-{P}itaevskii equation for rotating {B}ose gases}, Commun. Math.
  Phys., 264 (2006), pp.~505--537.

\bibitem{LieSeiSolYng-05}
{\sc E.~H. Lieb, R.~Seiringer, J.~P. Solovej, and J.~Yngvason}, {\em The
  mathematics of the {B}ose gas and its condensation}, Oberwolfach {S}eminars,
  Birkh{\"a}user, 2005.

\bibitem{LieSeiYng-00}
{\sc E.~H. Lieb, R.~Seiringer, and J.~Yngvason}, {\em Bosons in a trap: A
  rigorous derivation of the {G}ross-{P}itaevskii energy functional}, Phys.
  Rev. A, 61 (2000), p.~043602.

\bibitem{LieSol-01}
{\sc E.~H. Lieb and J.~P. Solovej}, {\em Ground state energy of the
  one-component charged {B}ose gas}, Commun. Math. Phys., 217 (2001),
  pp.~127--163.
  
  \bibitem{LieYng-98}
{\sc E.~Lieb and J.~Yngvason}, {\em Ground state energy of the low density
  {B}ose gas}, Phys. Rev. Lett., 80 (1998), pp.~2504--2507.

\bibitem{NamNap-21}
{\sc P.~T. Nam and M.~Napi\'orkowski}, {\em Two-term expansion of the ground
  state one-body density matrix of a mean-field {B}ose gas}, Calc. Var. PDE, 60:99 (2021),  pp. 1-30. 
  
\bibitem{NamNapRicTri-21}
{\sc P.~T. Nam, M.~Napi\'orkowski, J.~Ricaud, and A.~Triay}, {\em {Optimal rate
  of condensation for trapped bosons in the Gross-Pitaevskii regime}}, Anal.
  \& PDE,  pp.~1585–1616 (2022). 

\bibitem{NamNapSol-16}
{\sc P.~T. Nam, M.~Napi\'orkowski, and J.~P. Solovej}, {\em {Diagonalization of
  bosonic quadratic Hamiltonians by Bogoliubov transformations}}, J. Funct.
  Anal., 270 (2016), pp.~4340--4368.

\bibitem{NamRouSei-16}
{\sc P.~T. Nam, N.~Rougerie, and R.~Seiringer}, {\em Ground states of large
  {B}ose systems: {T}he {G}ross-{P}itaevskii limit revisited}, Analysis \& PDE,
  9 (2016), pp.~459--485.

\bibitem{NamSei-15}
{\sc P.~T. Nam and R.~Seiringer}, {\em Collective excitations of {B}ose gases
  in the mean-field regime}, Arch. Rational Mech. Anal., 215 (2015),
  pp.~381--417.
  
  \bibitem{NapReuSol-18} {\sc M. Napi\'orkowski, R. Reuvers, and J. P. Solovej}, {\em The Bogoliubov free energy functional II. The dilute limit}, Commun. Math. Phys. 360 (2018), 347-403. 
  
  
\bibitem{Pizzo-15}
{\sc A.~Pizzo}, {\em Bose particles in a box {I}. {A} convergent expansion of
  the ground state of a three-modes {B}ogoliubov {H}amiltonian}, Preprint
  arXiv:1511.07022,  (2015).

\bibitem{RouSpe-18}
{\sc N.~{R}ougerie and D.~{S}pehner}, {\em Interacting bosons in a double-well potential: Localization regime}, Commun.  Math. 
  Phys., 361 (2018), pp.~737--786.

\bibitem{Seiringer-11}
{\sc R.~Seiringer}, {\em The excitation spectrum for weakly interacting
  bosons}, Commun. Math. Phys., 306 (2011), pp.~565--578.

\bibitem{Solovej-ESI-2014}
{\sc J.~P. Solovej}, {\em Many body quantum mechanics}.
\newblock Lecture notes at the Erwin Schroedinger Institute 2014, available
  online at
  \url{http://www.math.ku.dk/~solovej/MANYBODY/mbnotes-ptn-5-3-14.pdf}.

\bibitem{Davidson-02}
{\sc J.~Steinhauer, R.~Ozeri, N.~Katz, and N.~Davidson}, {\em Excitation
  spectrum of a {B}ose-{E}instein condensate}, Phys. Rev. Lett., 88 (2002),
  p.~120407.

\bibitem{Ketterle-02}
{\sc J.~M. Vogels, K.~Xu, C.~Raman, J.~R. {Abo-Shaeer}, and W.~Ketterle}, {\em
  Experimental observation of the {B}ogoliubov transformation for a
  {B}ose-{E}instein condensed gas}, Phys. Rev. Lett., 88 (2002), p.~060402.

\bibitem{YauYin-09}
{\sc H.-T. Yau and J.~Yin}, {\em The second order upper bound for the ground
  energy of a {B}ose gas}, J. Stat. Phys., 136 (2009), pp.~453--503.

\end{thebibliography}
\end{document}